\documentclass[12pt]{article}
\usepackage{amssymb}

\usepackage{amsmath}

\newtheorem{theorem}{Theorem}
\newtheorem{acknowledgement}{Acknowledgement}

\newtheorem{lemma}[theorem]{Lemma}

\input{tcilatex}

\begin{document}

\title{Self-adjoint differential operators assosiated with self-adjoint
differential expressions}
\author{B.L. Voronov\thanks{%
Lebedev Physical Institute, Moscow, Russia; e-mail: voronov@lpi.ru}, D.M.
Gitman\thanks{%
Institute of Physics, University of Sao Paulo, Brazil; e-mail:
gitman@dfn.if.usp.br}, and I.V. Tyutin\thanks{%
Lebedev Physical Institute, Moscow, Russia; e-mail: tyutin@lpi.ru}}
\date{ }
\maketitle

\begin{abstract}
Considerable attention has been recently focused on quantum-mechanical
systems with boundaries and/or singular potentials for which the
construction of physical observables as self-adjoint (s.a.) operators is a
nontrivial problem. We present a comparative review of various methods of
specifying ordinary s.a. differential operators generated by formally s.a.
differential expressions based on the general theory of s.a. extensions of
symmetric operators. The exposition is untraditional and is based on the
concept of asymmetry forms generated by adjoint operators. The main
attention is given to a specification of s.a. extensions by s.a. boundary
conditions. All the methods are illustrated by examples of
quantum-mechanical observables like momentum and Hamiltonian. In addition to
the conventional methods, we propose a possible alternative way of
specifying s.a. differential operators by explicit s.a. boundary conditions
that generally have an asymptotic form for singular boundaries. A
comparative advantage of the method is that it allows avoiding an evaluation
of deficient subspaces and deficiency indices. The effectiveness of the
method is illustrated by a number of examples of quantum-mechanical
observables.
\end{abstract}

\section{Introduction}

Among the problems of quantum description of physical systems and its proper
interpretation, there is the problem of a correct definition of observables
as self-adjoint (s.a. in what follows) operators in an appropriate Hilbert
space. This problem is highly nontrivial for physical systems with
boundaries and/or with singular interactions (including QFT models); for
brevity, we call such systems nontrivial physical systems. The interest in
this problem is periodically revived in connection with one or another
particular physical system. The reason is that the solution of this problem
and consequently a consistent quantum-mechanical treatment of nontrivial
systems requires appealing to some nontrivial chapters of functional
analysis concerning the theory of unbounded linear operators, but the
content of such chapters is usually beyond the scope of the mathematical
apparatus given in standard text-books on quantum mechanics for physicists%
\footnote{%
The exceptions like \cite{Neuma32,FadYa80,BerSh91} are mainly intended for
mathematicians.}. A crucial subtlety is that an unbounded operator, in
particular, a quantum-mechanical observable, cannot be defined in the whole
Hilbert space, i.e., for any quantum-mechanical state. But ``there is no
operator without its domain of definition'', an operator is not only a
``rule of acting'', but also a domain in a Hilbert space, to which this rule
is applicable. In the case of unbounded operators, the same rule for
different domains defines different operators with sometimes crucially
different properties. It is a proper choice of the domain for a
quantum-mechanical observable that makes it a s.a. operator. The main
problems are related exactly with this task.

The formal rules of canonical quantization are of preliminary nature and
generally provide only, so to speak, ``candidates'' for unbounded
quantum-mechanical observables, for example, formally s.a. differential
expressions\footnote{%
S.a. according to Lagrange in mathematical terminology, see below sec.2.},
because their domains are not prescribed by the quantization rules and are
not even clear at the first stage of quantization, especially for nontrivial
physical systems, even though it is prescribed that observables must be s.a.
operators.

We would like to elucidate our understanding of this point. The choice of
domains providing the self-adjointness of all observables involved,
especially of primarily important observables like position, momentum,
Hamiltonian, symmetry generators, is a necessary part of quantization
resulting in a specification of quantum-mechanical description of a physical
system under consideration; this is actually a physical problem. Mathematics
can only help a physicist in this choice indicating various possibilities.

It appears that for physical systems whose classical description
incorporates infinite plane phase spaces like $\mathbb{R}^{2n}$ and
``regular'' interactions, quantization is practically unique: the most
important physical observables are defined as s.a. operators on some
``natural'' domains, in particular, classical symmetries can be conserved in
a quantum description. The majority of textbooks begin their exposition of
quantum mechanics exactly with the treatment of such physical systems. Of
course, nontrivial physical systems are also considered afterwards. But the
common belief is that no actual singularities exist in Nature. They are the
products of our idealization of reality, i.e., are of a model nature, for
example, related to our ignorance of the details of an interaction at small
distances. We formally extend an interaction law known for finite distances
between finite objects to infinitely small distances between point-like
objects. We treat boundaries as a result of infinite potential walls that
are actually always finite\footnote{%
To be true, a plane infinite space is also an idealization, as any infinity.}%
. The consequence is that singular problems in quantum mechanics are
commonly solved via some regularization that is considered natural and then
by a following limiting process of removing the regularization. However, in
some cases the so-called infinite renormalization (of ``charges'', for
example) is required. Moreover, in some cases there exists no reasonable
limit. (We should emphasize that we speak here about conventional quantum
mechanics, rather than about quantum field theory.) It can also happen that
physical results are unstable under regularization: different
regularizations yield different physical results. It is exactly in these
cases that mathematics can help a physicist with the theory of s.a.
extensions of symmetric operators. This was first recognized by Berezin and
Faddeev \cite{BerFa61} in connection with the three-dimensional $\delta $%
-potential problem.

The practice of the quantization of nontrivial systems shows that
preliminary candidates for observables can be quite easily assigned
symmetric operators defined on the domains that ``avoid'' the problems: they
do not ``touch'' boundaries and ``escape'' singularities of interaction;
this is a peculiar kind of ``mathematical regularization''. However such
symmetric operators are commonly non-self-adjoint. The main question then is
whether these preliminary observables can be assigned s.a. operators by
extensions which make the candidates real observables. The answer is simple
if a symmetric operator under consideration is bounded. But if it is
unbounded, the problem is generally nontrivial.

The theory of s.a. extensions of unbounded symmetric operators is the main
tool in solving this problem. It appears that in general these extensions
are highly nonunique if at all possible. For physics, this implies that
there are many quantum mechanical descriptions of the same nontrivial
physical system. The general theory shows all the possibilities that
mathematics can present to a physicist for his choice. Of course, the
physical interpretation of available s.a. extensions is a purely physical
problem. Any extension is a certain prescription for the behavior of a
physical system under consideration near boundaries and singularities. We
also believe that each extension can be understood through an appropriate
regularization and limiting process, although this in itself is generally a
complicated problem. But, in any case, the right of a final choice belongs
to a physicist.

The general theory of extensions of unbounded symmetric operators is mainly
due to von Neumann \cite{Neuma29} (An English exposition of von Neumann's
paper can be found in \cite{Stone32}). We expound only a necessary part of
this theory that concerns the case of s.a. extensions.

The following three theorems exhaust the content of the necessary part of
the theory. They bear no name in the conventional mathematical literature %
\cite{AkhGl81,Naima69}; instead, their crucial formulas are called the\emph{%
\ }von Neumann formulas. We call these theorems the respective first and\
second von Neumann theorems and the main theorem\footnote{%
A reader interested in the final statement (without the details of a strict
proof) can go directly to the main theorem, Theorem \ref{t3a.3}, and the
subsequent comments placed at the end of Sec. 2.}.

We attempt to make our exposition maximally self-contained as far as
possible and first remind a reader the basic notions and facts, but only
those that are absolutely necessary for understanding the main statements;
there are many books on the subject. We mainly refer to \cite%
{AkhGl81,Naima69} although follow an alternative way of describing s.a.
extensions of symmetric operators. The final statements are our guides in
constructing quantum-mechanical observables.

The article is organized as follows: In Sec. 2, we remind of the general
theory of symmetric extensions of unbounded symmetric operators. The
exposition is untraditional and is based on the notion of asymmetry forms
generated by adjoint operators. The basic statements concerning the
possibility and specification of s.a. extensions both in terms of isometries
between the deficient subspaces and in terms of the sesquilinear asymmetry
form are collected in the main theorem. (There follows a comment on a direct
application of the main theorem to physical problems of quantization.) We
outline a possible general scheme of constructing quantum-mechanical
observables as s.a. operators starting from initial formal expressions
supplied by canonical quantization rules. The scheme is illustrated by the
example of the momentum operator for a particle moving on different
intervals of the real axis (the whole real axis, a semiaxis, a finite
interval). Sec. 3 is devoted to the exposition of specific features and
appropriate modifications of the general theory as applied to ordinary s.a.
differential operators in Hilbert spaces $L^{2}(a,b)$ associated with formal
differential expressions s.a. according to Lagrange. For differential
operators, the isometries between deficient subspaces specifying s.a.
extensions can be converted to s.a. boundary conditions, explicit or
implicit, based on the fact that asymmetry forms are expressed in terms of
the (asymptotic) boundary values of functions and their derivatives. We
describe various ways of specifying s.a. operators associated with s.a.
differential expressions by s.a. boundary conditions depending on the
regularity or singularity of the boundaries of the interval under
consideration. All the methods are illustrated by examples of
quantum-mechanical observables like momentum and Hamiltonian. In addition to
the known conventional methods, we discuss a possible alternative way of
specifying s.a. differential operators by explicit s.a. boundary conditions
that generally have an asymptotic form for singular boundaries. A
comparative advantage of the method is that it allows avoiding the
evaluation of deficient subspaces and deficiency indices. Its effectiveness
is illustrated by a number of examples of quantum-mechanical observables.

\section{Basics of theory of symmetric operators}

\subsection{Generalities}

We begin with a notation.

Let $\mathcal{H}$ be a Hilbert space, its vectors are denoted by Greek
letters: $\xi ,\eta ,...,\psi \in \mathcal{H}$. The symbol $\left( \eta ,\xi
\right) $ denotes a scalar product in $\mathcal{H};$ by the physical
tradition, the scalar product is linear in the second argument and
anti-linear in the first one.

Let $M$ be a subspace in $\mathcal{H},\;M\subset \mathcal{H}$, then its
closure and its orthogonal complement are respectively denoted by $\overline{%
M}$ and $M^{\bot }\,,$ $M$ is a closed subspace if $M=\overline{M}.$ For any 
$M$, the decomposition $\mathcal{H}=\overline{M}\oplus M^{\bot }$ holds,
where $\oplus $ is the symbol of a direct orthogonal sum, i.e., any vector $%
\xi \in \mathcal{H}$ is uniquely represented as%
\begin{equation*}
\xi =\underline{\xi }+\xi ^{\bot }\,,\;\underline{\xi }\in \overline{M}%
\,,\;\xi ^{\bot }\in M^{\bot }\,.
\end{equation*}%
A subspace $M$ is called dense in $\mathcal{H}$ if $\overline{M}=\mathcal{H}%
, $ then $M^{\bot }=\left\{ 0\right\} .$

Operators in $\mathcal{H}$, we consider only linear operators, are denoted
by the Latin letters $\hat{f},\,\hat{g}\,,...$ with a hat above. Their
domains and ranges are subspaces in $\mathcal{H}$ and are respectively
denoted by $D_{f}\,,\,D_{g}\,,...$ and $R_{f},\,R_{g},...\,.$ The unit, or
identity, operator is denoted by $\hat{I}.$ An operator $\hat{f}$ is called
densely defined\footnote{%
Actually, only such operators are interesting for quantum mechanics.} if $%
\overline{D_{f}}=\mathcal{H}.$

An operator $\hat{f}$ is defined by its graph%
\begin{equation*}
\mathbb{G}_{f}=\left\{ \left( 
\begin{array}{c}
\xi \\ 
\eta =\hat{f}\,\xi%
\end{array}%
\right) \right\} \subset \mathbb{H}=\mathcal{H}\oplus \mathcal{H\,}%
,\;\forall \xi \in D_{f}\,,\;\eta \in R_{f}\,,
\end{equation*}%
a subspace in the direct orthogonal sum of two copies of $\mathcal{H},$ $\xi 
$ is an abscissa of the graph, $\eta $ is its ordinate. The scalar product
of two vectors $\mathbf{v}_{1}=\left( 
\begin{array}{c}
\xi _{1} \\ 
\eta _{1}%
\end{array}%
\right) \in \mathbb{H}$ and $\mathbf{v}_{2}=\left( 
\begin{array}{c}
\xi _{2} \\ 
\eta _{2}%
\end{array}%
\right) \in \mathbb{H}$ is defined by $\left( \mathbf{v}_{1},\mathbf{v}%
_{2}\right) =\left( \xi _{1},\xi _{2}\right) +\left( \eta _{1},\eta
_{2}\right) \,.$ Two operators $\hat{f}$ and $\hat{g}$ are equal if $\mathbb{%
G}_{f}=\mathbb{G}_{g}\,,$ in particular, $D_{f}=D_{g}\,.$

We assume that the notion of sum $\hat{f}+$ $\hat{g}$ of operators, $%
D_{f+g}=D_{f}\cap D_{g}\,,$ and the notion of the multiplication of an
operator by a complex number $z$, i.e., $\hat{f}\rightarrow z\hat{f}\,,$ $%
D_{zf}=D_{f}\,,$ are known; in particular, $D_{f-zI}=D_{f}\,.$

The kernel of an operator $\hat{f}$ is defined as the subspace of
null-vectors of the operator, $\ker \hat{f}=\left\{ \xi \in D_{f}:\hat{f}%
\,\xi =0\right\} .$ If $\ker \hat{f}=\left\{ 0\right\} ,$ the operator $\hat{%
f}$ is invertible, i.e., there exists the inverse operator, or simply
inverse, $\hat{f}^{-1}$ whose graph $\mathbb{G}_{f^{-1}}$ is%
\begin{equation*}
\mathbb{G}_{f^{-1}}=\left\{ \left( 
\begin{array}{c}
\eta =\hat{f}\,\xi \\ 
\hat{f}^{-1}\eta =\xi%
\end{array}%
\right) \right\} ,
\end{equation*}%
where abscissas and ordinates are interchanged with respect to $\mathbb{G}%
_{f}$ such that $D_{f^{-1}}=R_{f}$ and $R_{f^{-1}}=D_{f}\,.$ It is evident
that $\left( \hat{f}^{-1}\right) ^{-1}=\hat{f}\,.$

We assume that the notions of the operator norm, of bounded, or continuous,
and unbounded operators are known.

An operator $\hat{g}$ is called an extension of an operator $\hat{f}$ if $%
\mathbb{G}_{f}\subset \mathbb{G}_{g}\,,$ i.e., if $D_{f}\subset D_{g}$ and $%
\hat{g}\xi =\hat{f}\xi \,,$ $\forall \xi \in D_{f}\,,$ the operator $\hat{f}$
is respectively called the restriction of $\hat{g}$; this is written as $%
\hat{f}$ $\subset \hat{g}.$ A bounded continuous operator can be extended to
the whole $\mathcal{H}$ with the same norm.

For an unbounded operator, the notion of continuity is replaced by the
notion of closedness; for many purposes, it is sufficient that an operator
be closed. An operator $\hat{f}$ is called closed, which is written as $\hat{%
f}=\overline{\hat{f}},$ if its graph is closed, $\mathbb{G}_{f}=\overline{%
\mathbb{G}_{f}},$ as a subspace in $\mathbb{H},$ i.e., $\xi _{n}\rightarrow
\xi ,\;\hat{f}\xi _{n}\rightarrow \eta ,\;\left\{ \xi _{n}\right\}
_{1}^{\infty }\subset D_{f}\Longrightarrow \xi =D_{f},\;\eta =\hat{f}\xi \,.$
The difference between closedness and continuity is that not any convergent
sequence $\left\{ \xi _{n}\right\} _{1}^{\infty }\subset D_{f}$ yields a
convergent sequence $\left\{ \hat{f}\xi _{n}\right\} _{1}^{\infty },$ the
latter can diverge, but it is not allowed for two sequences $\left\{ \hat{f}%
\xi _{n}^{\left( 1\right) }\right\} _{1}^{\infty }$ and $\left\{ \hat{f}\xi
_{n}^{\left( 2\right) }\right\} _{1}^{\infty }$ to converge to different
limits if the sequences $\left\{ \xi _{n}^{\left( 1\right) }\right\}
_{1}^{\infty }$ and $\left\{ \xi _{n}^{\left( 2\right) }\right\}
_{1}^{\infty }$ have the same limit. If an operator $\hat{f}$ is bounded and
closed, its domain is a closed subspace, $D_{f}=\overline{D_{f}}\,$; if $%
\hat{f}$ is closed and invertible, its inverse is also closed, $\hat{f}^{-1}=%
\overline{\hat{f}^{-1}}\,$; for a closed operator, we also have $\overline{%
\hat{f}-z\hat{I}}=\hat{f}-z\hat{I}$. It is remarkable that a closed It is
remarkable that a closed operator defined everywhere is bounded (the theorem
on a closed graph), therefore, a closed unbounded operator defined
everywhere is impossible. An operator $\hat{f}$ \ by itself can be
nonclosed, but allow the closure, or be closable. A generally nonclosed
operator $\hat{f}$ is called closable if it allows a closed extension; the
minimum closed extension is called the closure\footnote{%
The fundamental notions of a closed operator and closability are usually
left aside in physical textbooks, probably because even though not any
operator allows a closure, such ``pathologic'' operators are not encountered
in physics.} of $\hat{f}$ and is denoted by $\overline{\hat{f}}\,,\;\hat{f}%
\subseteq \overline{\hat{f}}\,,$ its graph $\mathbb{G}_{\bar{f}}=\overline{%
\mathbb{G}_{\bar{f}}}\,,$ the closure of $\mathbb{G}_{f}$ in $\mathbb{H}.$
Of course, any graph can be made closed, $\mathbb{G}_{f}\rightarrow 
\overline{\mathbb{G}_{f}}$ but the closure $\overline{\mathbb{G}_{f}}$ must
remain a graph, i.e., a subspace in $\mathbb{H}$ where any abscissa uniquely
determines an ordinate, which is nontrivial.

Any densely defined (and only densely defined) operator$\;\hat{f}$ is
assigned the adjoint operator, or simply adjoint, $\hat{f}^{+}\,.$ Its graph%
\footnote{%
Here and elsewhere $\sigma ^{k},$ $k=1,2,3$ denote Pauli matrices.} $\mathbb{%
G}_{f^{+}}$ is $\mathbb{G}_{f^{+}}=\left( i\sigma ^{2}\mathbb{G}_{f}\right)
^{\bot }$ (the orthogonal complement is taken in $\mathbb{H}$);
equivalently, $\hat{f}^{+}$ is defined by the equation%
\begin{equation*}
\left( \xi _{\ast },\hat{f}\xi \right) -\left( \eta _{\ast }=\hat{f}^{+}\xi
_{\ast },\xi \right) \,,\;\forall \xi \in D_{f}\,,
\end{equation*}%
for the pairs of vectors $\xi _{\ast }\in D_{f^{+}}\,$and $\eta _{\ast }=%
\hat{f}^{+}\xi _{\ast }\in R_{f^{+}}$ constituting the graph of $\hat{f}^{+}$%
. We call this equation the defining equation for $\hat{f}^{+}$ and only
note that $\hat{f}^{+}$ must be evaluated. It is evident that\footnote{%
The bar over numerical quantities denotes complex conjugation.} $\left( z%
\hat{f}\right) ^{+}=\overline{z}\hat{f}^{+}.$ The adjoint $\hat{f}^{+}$ is
always closed because any orthogonal complement is a closed subspace. It is
important that an extension of a densely defined operator is accompanied by
a restriction of its adjoint: $\;\hat{f}\subset \hat{g}\Longrightarrow 
\hat{g}^{+}\subseteq \hat{f}^{+}\,.$ The closure of a densely defined
operator, if it exists, has the same adjoint, $\left( \overline{\hat{f}}%
\right) ^{+}=\hat{f}^{+}\,.$ A densely defined operator $\hat{f}$ is
closable iff\footnote{%
Iff means ``if and only if''.} its adjoint is also densely defined, $%
\overline{D_{f^{+}}}=\mathcal{H},$ and if so, the equality $\overline{\hat{f}%
}=\left( \hat{f}^{+}\right) ^{+}$ holds. We note that the generally accepted
equality $\left( \hat{f}^{+}\right) ^{+}=\hat{f}$ \ holds only for closed
operators. We also note that generally $\left( \hat{f}+\hat{g}\right)
^{+}\neq \hat{f}^{+}+\hat{g}^{+}$ for densely defined unbounded operators: $%
\left( \hat{f}+\hat{g}\right) ^{+}$ may not exist if $\overline{D_{f}\cap
D_{g}}\neq \mathcal{H},$ and even if $\overline{D_{f}\cap D_{g}}=\mathcal{H}%
, $ we generally have $\hat{f}^{+}+\hat{g}^{+}\subseteq \left( \hat{f}+%
\hat{g}\right) ^{+}\,.$ But if one of the operators, let it be $\hat{g},$ is
bounded and defined everywhere, the generally accepted equality $\left( \hat{%
f}+\hat{g}\right) ^{+}=\hat{f}^{+}+\hat{g}^{+}$ holds, in particular, $%
\left( \hat{f}-z\hat{I}\right) ^{+}=\hat{f}^{+}-\overline{z}I\,.$ For a
densely defined operator $\hat{f},$ the equality $R_{f}^{\bot }=\ker \hat{f}%
^{+}$ holds, which implies the decomposition $\mathcal{H}=\overline{R_{f}}%
\oplus \ker \hat{f}^{+}\,,$ in particular, $\mathcal{H}=\overline{R_{f-zI}}%
\oplus \ker \left( \hat{f}^{+}-\overline{z}I\,\right) .$ If $\hat{f}$ and $%
\hat{f}^{+}$ are invertible, the equality $\left( \hat{f}^{-1}\right)
^{+}=\left( \hat{f}^{+}\right) ^{-1}$ holds.

\subsection{Self-adjoint and symmetric operators, deficiency indices}

A densely defined operator $\hat{f}$ is called s.a. if it coincides with its
adjoint $\hat{f}^{+}\,,\;\hat{f}=\hat{f}^{+}\,,$ i.e., $\mathbb{G}_{f}=%
\mathbb{G}_{f^{+}}\,,$ in particular, $D_{f}=D_{f^{+}}\,.$ All
quantum-mechanical observables are s.a. operators. A s.a. operator is
evidently closed. Therefore, any bounded s.a. operator is defined
everywhere, but an unbounded s.a. operator cannot be defined everywhere.
This concerns the majority of quantum-mechanical observables and generates
one of the main problems of quantization. One of the obstacles is that the
sum of two unbounded s.a. operators $\hat{f}=\hat{f}^{+},$ and $\hat{g}=%
\hat{g}^{+}$ is generally non-s.a.: even if $\overline{D_{f}\cap D_{g}}=%
\mathcal{H},$ we generally have $\hat{f}+\hat{g}\subseteq \left( \hat{f}+%
\hat{g}\right) ^{+}.$ But if one of the operators, let it be $\hat{g},$ is a
bounded s.a. operator, the sum $\hat{f}+\hat{g}$ is a s.a. operator with the
domain $D_{f+g}=D_{f}\,,$ in particular, $\hat{f}-\lambda \hat{I}=\left( 
\hat{f}-\lambda \hat{I}\right) ^{+}$ for $\lambda =\overline{\lambda }\,.$
It follows from the previous remarks that a s.a. operator $\hat{f}$ does not
allow s.a. extensions, and if it is invertible, its inverse $\hat{f}$ $^{-1}$
is also a s.a. operator.

The requirement of self-adjointness is a rather strong requirement.

A less restrictive notion is the notion of symmetric operator\footnote{%
Another name, probably obsolete, is Hermitian operator.}. An operator $\hat{f%
}$ is called symmetric if $\hat{f}$ is densely defined, $\overline{D_{f}}=%
\mathcal{H},$ and if the equality%
\begin{equation}
\left( \eta ,\hat{f}\xi \right) =\left( \hat{f}\eta ,\xi \right)
\,,\;\forall \xi ,\eta \in D_{f}\,  \label{sym}
\end{equation}%
holds. An equivalent definition of a symmetric operator $\hat{f}$ is that it
is densely defined and its adjoint $\hat{f}^{+}$ is an extension of $\hat{f}$
, $\hat{f}\subseteq \hat{f}^{+}$, i.e., $\mathbb{G}_{f}\subseteq \mathbb{G}%
_{f^{+}}\,,$ in particular, $D_{f}\subseteq D_{f^{+}}$. A s.a. operator is a
symmetric operator with an additional property $D_{f}=D_{f^{+}}\,.$ The
problem, we are interested in all what follows is whether a given symmetric
operator allows s.a. extensions.

We list the basic properties of symmetric operators that are used below.
They directly follow from the aforesaid or can be found in \cite%
{AkhGl81,Naima69}.

Any symmetric operator $\hat{f}$ has a symmetric closure $\overline{\hat{f}}$
such that the chain of inclusions $\hat{f}\subseteq \overline{\hat{f}}%
=\left( \hat{f}^{+}\right) ^{+}\subseteq \left( \overline{\hat{f}}\right)
^{+}=\hat{f}^{+}$ holds, in particular, $\hat{f}^{+}\underline{\xi }=%
\overline{\hat{f}}\underline{\xi }$ for any vector $\underline{\xi }\in D_{%
\overline{f}}$ .

Therefore, when setting the problem of symmetric extensions, especially,
s.a. extensions, of a given symmetric operator $\hat{f}$ , we can assume
without loss of generality that the initial symmetric operator is closed,
which is usually adopted in the mathematical literature. But in physics, a
preliminary symmetric operator $\hat{f},$ a ''candidate to an observable'',
usually appears to be nonclosed, while constructing and describing the
closure $\overline{\hat{f}}$ of $\hat{f}$ is generally nontrivial. In what
follows, we therefore consider an initial symmetric operator $\hat{f}$ in
general nonclosed. If $\hat{f}$ is, or appears, closed, the statements that
follow are easily modified or simplified in an obvious way.

In general, the adjoint of a symmetric operator$\ \hat{f}$ is nonsymmetric,
but if $\hat{f}^{+}$ is symmetric, then it is s.a. as well as the closure $%
\overline{\hat{f}}$ because $\hat{f}^{+}\subseteq \left( \hat{f}^{+}\right)
^{+}$ implies the inclusions $\hat{f}^{+}=\overline{\hat{f}}^{+}\subseteq 
\overline{\hat{f}}=\left( \hat{f}^{+}\right) ^{+}$ inverse to the previous
ones. Such a symmetric operator, i.e., a symmetric operator whose closure is
s.a., is called an essentially s.a. operator. A unique s.a. extension of an
essentially s.a. operator $\hat{f}$ is its closure $\overline{\hat{f}}$ that
coincides with its adjoint $\hat{f}^{+}\,.$ This is certainly the case if $%
\hat{f}$ is bounded, then we have $D_{\overline{f}}=\mathcal{H}.$

In what follows, by a symmetric operator we mean an unbounded symmetric
operator.

If $\hat{f}_{\mathrm{ext}}$ is a symmetric extension of a symmetric operator 
$\hat{f},$ then the chain of inclusions, $\hat{f}\subseteq \hat{f}_{\mathrm{%
ext}}\subseteq \left( \hat{f}_{\mathrm{ext}}\right) ^{+}\subseteq \hat{f}%
^{+} $ holds, i.e., any symmetric extension of $\hat{f}$ is a symmetric
restriction of $\hat{f}^{+}$. This is one of the basic starting points of
the theory to follow: when a symmetric operator is extended symmetrically,
its extensions and their adjoints go to meet each other; if the meeting
occurs, we get a s.a. operator, but the meeting may be impossible, and if
possible, there may be a nonunique way for it. The problem of the theory is
to describe all the possibilities.

The closure $\overline{\hat{f}}$ is a minimum closed symmetric extension of
a nonclosed symmetric operator $\hat{f}:$ $\overline{\hat{f}}$ is contained
in any closed symmetric extension of $\hat{f}.$ For brevity, we call $%
\overline{\hat{f}}$ the trivial symmetric extension of the a symmetric
operator $\hat{f};$ if $\hat{f}_{\mathrm{ext}}$ contains the closure $%
\overline{\hat{f}}$ and is different from it, $\overline{\hat{f}}\subset 
\hat{f}_{\mathrm{ext}}$ (a strict inclusion), we call such an extension
nontrivial.

A closed symmetric operator $\hat{f},\;$ $\hat{f}=\overline{\hat{f}},$ is
called maximal, if it does not allow nontrivial symmetric extensions. Any
s.a. operator $\hat{f},$ $\hat{f}=\hat{f}^{+},$ is a maximal symmetric
operator.

Because we consider in general nonclosed symmetric operators, it is natural
to introduce a notion of an essentially maximal operator, similarly to the
notion of an essentially s.a. operator, as a symmetric operator $\hat{f}$
whose closure $\overline{\hat{f}}$ is a maximal operator, or simply, maximal.

Any symmetric operator $\hat{f},$ in particular, its closure $\overline{\hat{%
f}},$ can have only real eigenvalues\footnote{%
A proof is a standard one; it is well known to physicists as applied to s.a.
operator. We only note that a symmetric operator may have no eigenvalues,
whereas its symmetric extensions can have eigenvalues.}, i.e., $\hat{f}\xi
=\lambda \xi \Longrightarrow \lambda =\overline{\lambda }\,,$ or 
\begin{align*}
& \ker \left( \hat{f}-z\hat{I}\right) =\ker \left( \overline{\hat{f}}-z%
\hat{I}\right) =\left\{ 0\right\} \,,\;\forall z\in \mathbb{C}_{+}\cup 
\mathbb{C}_{-}\,, \\
& \mathbb{C}_{+}=\left\{ z=x+iy:\;y>0\right\} \,,\;\mathbb{C}_{-}=\left\{
z=x+iy:\;y<0\right\} \,.
\end{align*}%
It follows that for any $z\in \mathbb{C}_{+}\cup \mathbb{C}_{-}$\thinspace
,\ the closed\ operator $\overline{\hat{f}}-z\hat{I}$ is invertible and the
inverse operator $\hat{R}_{z}=\left( \overline{\hat{f}}-z\hat{I}\right)
^{-1} $ is a bounded closed operator, therefore the range $\Re _{z}$ of the
operator $\overline{\hat{f}}-z\hat{I},$%
\begin{equation*}
\Re _{z}=R_{\bar{f}-zI}=\left\{ \underline{\eta }=\left( \overline{\hat{f}}-z%
\hat{I}\right) \underline{\xi }\,,\;\forall \underline{\xi }\in D_{\overline{%
f}}\right\} \,,
\end{equation*}%
is a closed subspace in $\mathcal{H}$ as the domain of the closed bounded
operator $\hat{R}_{z}$.

By definition, the orthogonal complement (in $\mathcal{H}$ ) to the range $%
\Re _{z}$ as well as to the range $R_{f-zI}$ of the operator $\hat{f}-z%
\hat{I}\,,$ is called the deficient subspace $\aleph _{z}$ of a symmetric
operator $\hat{f}$ corresponding to a point $z\in \mathbb{C}_{+}\cup \mathbb{%
C}_{-}\,$, 
\begin{align*}
& \aleph _{z}=\left( R_{f-zI}\right) ^{\bot }=\left( R_{\bar{f}-zI}\right)
^{\bot }=\left( \Re _{z}\right) ^{\bot } \\
& =\ker \left( \hat{f}^{+}-\overline{z}\hat{I}\right) =\left\{ \xi _{%
\overline{z}}\in D_{f^{+}}:\;\hat{f}^{+}\xi _{\overline{z}}=\overline{z}\xi
_{\overline{z}}\right\} \,.
\end{align*}%
A deficient subspace $\aleph _{z}$ is a closed subspace.

It is important that the dimension of $\aleph _{z},$%
\begin{equation*}
\dim \aleph _{z}=\left\{ 
\begin{array}{c}
m_{+},\;z\in \mathbb{C}_{\_}\;\left( \bar{z}\in \mathbb{C}_{+}\right) \,, \\ 
m_{-},\;z\in \mathbb{C}_{+}\;\left( \bar{z}\in \mathbb{C}_{-}\right) \,,%
\end{array}%
\right.
\end{equation*}%
is independent of $z$ in the respective domains $\mathbb{C}_{-}$ and $%
\mathbb{C}_{+};$ $m_{+}$ and $m_{-}$ are called the deficiency indices of
the operator $\hat{f}$ . For a given $z$, we therefore distinguish the two
deficient subspaces $\aleph _{z}$ and $\aleph _{\bar{z}}=\left\{ \xi _{z}\in
D_{f^{+}}:\;\hat{f}^{+}\xi _{z}=z\xi _{z}\right\} ,$ such that if $z\in 
\mathbb{C}_{\_}\left( \mathbb{C}_{+}\right) $ then $\dim \aleph
_{z}=m_{+}\left( m_{-}\right) $ while\footnote{%
We point out that there exists an anticorrespondence $z\rightleftarrows \bar{%
z}$ between the subscript $z$ of $\aleph _{z}$ and the respective eigenvalue 
$\bar{z}$ and the subscript of the eigenvector $\xi _{\bar{z}}$ of $\hat{f}%
^{+}.$ Perhaps it would be more convenient to change the notation $\aleph
_{z}\rightleftarrows \aleph _{\bar{z}};$ the conventional notation is due to
tradition. The same is true for the subscripts of $m_{\pm }$ and $C_{\mp }$ .%
} $\dim \aleph _{\bar{z}}=m_{-}\left( m_{+}\right) $; both $m_{+}$ and $%
m_{-} $ can be infinite, if $m_{+}$, $m_{-}=\infty ,$ they are considered
equal, $m_{+}=m_{-}=\infty .$

Accordingly, the decomposition%
\begin{equation}
\mathcal{H}=\Re _{z}\oplus \aleph _{z}  \label{3a.1}
\end{equation}%
holds, which means that any vector $\xi \in \mathcal{H}$ can be represented
as%
\begin{equation}
\xi =\left( \overline{\hat{f}}-z\hat{I}\right) \underline{\xi }+\xi _{%
\overline{z}}\,,  \label{3a.2}
\end{equation}%
with some $\underline{\xi }\in D_{\overline{f}}$ and $\xi _{\overline{z}}\in
\aleph _{z}$ that are uniquely defined by $\xi $. We note that for in
general nonclosed operator $\hat{f}$ , its closure $\overline{\hat{f}}$
enters decompositions (\ref{3a.1}) and (\ref{3a.2}).

\subsection{First von Neumann theorem}

This theorem provides a basic starting point in studying symmetric and s.a.
extensions of symmetric operators.

\begin{theorem}
\label{t3a.1}(The first von Neumann theorem) For any symmetric operator $%
\hat{f}$, the domain $D_{f^{+}}$ of its adjoint $\hat{f}^{+}$ is the direct
sum of the three linear manifolds $D_{\overline{f}}\,,\,\aleph _{\bar{z}}$
and $\aleph _{z}:$%
\begin{equation}
D_{f^{+}}=D_{\overline{f}}+\aleph _{\bar{z}}+\aleph _{z}\,,\;\forall z\in 
\mathbb{C}_{+}\cup \mathbb{C}_{-}\,,  \label{3a.3}
\end{equation}%
where $+$ is the symbol of a direct nonorthogonal sum, such that any vector $%
\xi _{\ast }\in D_{f^{+}}$ is uniquely represented as%
\begin{equation}
\xi _{\ast }=\underline{\xi }+\xi _{z}+\xi _{\overline{z}}\,,  \label{3a.4}
\end{equation}%
where $\underline{\xi }\in D_{\overline{f}}\,,$ $\xi _{z}\in \aleph _{\bar{z}%
}\,$, and $\xi _{\overline{z}}\in \aleph _{z}\,,$ and%
\begin{equation}
\hat{f}^{+}\xi _{\ast }=\overline{\hat{f}}\underline{\xi }+z\xi _{z}+%
\overline{z}\xi _{\overline{z}}\,.  \label{3a.5}
\end{equation}
\end{theorem}

Formula (\ref{3a.4}) is called the first von Neumann formula, we assign the
same name to formula (\ref{3a.3}).

It should be emphasized that for in general nonclosed symmetric operator $%
\hat{f}$ , the domain $D_{\overline{f}}$ of its closure $\overline{\hat{f}}$
enters decompositions (\ref{3a.3})-(\ref{3a.5}).

Proof. The domain $D_{\overline{f}}$ and the deficient subspaces $\aleph _{%
\bar{z}}$ and $\aleph _{z}$ are linear manifolds belonging to $D_{f^{+}}$,
therefore, a vector $\xi _{\ast }=\underline{\xi }+\xi _{z}+\xi _{\overline{z%
}}$ belongs to $D_{f^{+}}$ with any $\underline{\xi }\in D_{\overline{f}}$, $%
\xi _{z}\in \aleph _{\bar{z}}$, and $\xi _{\overline{z}}\in \aleph _{z}$. By
the definition of a direct sum of linear manifolds, it remains to show that
for any vector $\xi _{\ast }\in D_{f^{+}}\,,$ a unique representation (\ref%
{3a.4}) holds.

Let $\xi _{\ast }\in D_{f^{+}}$. According to (\ref{3a.1}) and (\ref{3a.2}),
the vector $\left( \hat{f}^{+}-z\hat{I}\right) \xi _{\ast }\,$, $\forall
z\in \mathbb{C}_{+}\cup \mathbb{C}_{-}$, is represented as%
\begin{equation}
\left( \hat{f}^{+}-z\hat{I}\right) \xi _{\ast }=\left( \overline{\hat{f}}-z%
\hat{I}\right) \underline{\xi }+\left( \overline{z}-z\right) \xi _{\overline{%
z}}\,,  \label{3a.6}
\end{equation}%
with some $\underline{\xi }\in D_{\overline{f}}$ and $\xi _{\overline{z}}\in
\aleph _{z}$ that are uniquely defined by $\xi _{\ast }$ (the nonzero factor 
$\overline{z}-z$ in front of $\xi _{\overline{z}}$ is introduced for
convenience). But$\overline{\hat{f}}\underline{\xi }=\hat{f}^{+}\underline{%
\xi }$ and $\overline{z}\xi _{\overline{z}}=\hat{f}^{+}\xi _{\overline{z}}\,$%
, and (\ref{3a.6}) becomes%
\begin{equation*}
\left( \hat{f}^{+}-z\hat{I}\right) \xi _{\ast }=\left( \hat{f}^{+}-z\hat{I}%
\right) \underline{\xi }+\left( \hat{f}^{+}-z\hat{I}\right) \xi _{\overline{z%
}},\;\mathrm{or}\;\left( \hat{f}^{+}-z\hat{I}\right) \left( \xi _{\ast }-%
\underline{\xi }-\xi _{\overline{z}}\right) =0\,,
\end{equation*}%
which yields $\xi _{\ast }-\underline{\xi }-\xi _{\overline{z}}=\xi _{z}\,$%
,\ or $\xi _{\ast }=\underline{\xi }+\xi _{z}+\xi _{\overline{z}},$ where $%
\xi _{z}\in \aleph _{\bar{z}}$ and is evidently uniquely defined by $\xi
_{\ast }$, $\underline{\xi }$ , and $\xi _{\overline{z}},$ therefore by $\xi
_{\ast }$ alone, as well as $\underline{\xi }$ and $\xi _{\overline{z}}$.
This proves representation (\ref{3a.4}) for any vector $\xi _{\ast }\in
D_{f^{+}}.$

After this, formula (\ref{3a.5}) is evident.

We note that

i) representations (\ref{3a.3})-(\ref{3a.5}) hold for any complex, but not
real, number $z=x+iy,\;y\neq0;$

ii) these representations are explicitly $z$-dependent because the deficient
subspaces $\aleph _{\bar{z}}$ and $\aleph _{z}$ and therefore the sum $%
\aleph _{\bar{z}}+\aleph _{z}$ depend on $z$, but $\dim \left( \aleph _{\bar{%
z}}+\aleph _{z}\right) =m_{+}+m_{-}$ , as well as $m_{+}$ and $m_{-},$ is
independent of $z\,$\footnote{%
Although $\aleph _{\bar{z}}$ and $\aleph _{z}$ are closed subspaces in $%
\mathcal{H},$ we cannot in general assert that their direct sum $\aleph _{%
\bar{z}}+\aleph _{z}$ is also a closed subspace. The latter is always true
if one of the subspaces is finite-dimensional.};

iii) the sum in (\ref{3a.3}) is direct, but not orthogonal, it cannot be
orthogonal, at least, because $\overline{D_{\bar{f}}}=\mathcal{H}$ and
therefore $D_{\overline{f}}^{\bot }=\{0\}.$

It immediately follows from the first von Neumann theorem that a nonclosed
symmetric operator $\hat{f}$ is essentially s.a. (and a closed symmetric
operator is s.a.) iff $\aleph _{\bar{z}}=\aleph _{z}=\left\{ 0\right\} $,
i.e., iff its deficiency indices are equal to zero, $m_{+}=m_{-}=0$, because
in this case, $D_{f^{+}}=D_{\overline{f}}$ , therefore $\overline{\hat{f}}=%
\hat{f}^{+}$. In other words, the adjoint $\hat{f}^{+}$ is symmetric iff $%
m_{+}=m_{-}=0.$

But this theorem, namely, formulas (\ref{3a.4}) and (\ref{3a.5}), also
allows estimating the {}``asymmetricity'' of the adjoint $\hat{f}^{+}$ in
the case where the deficiency indices $m_{+}$ and $m_{-}$ are not equal to
zero (one of them or both) and analyzing the possibilities of symmetric and
s.a. extensions of $\hat{f}$. We now turn to this case, the case where $\max
\left( m_{+},m_{-}\right) \neq 0.$

\subsection{Asymmetry forms $\protect\omega _{\ast }$ and $\Delta _{\ast }$}

The consideration to follow is proceeding with some arbitrary, but fixed,
complex number $z=x+iy,\;y\neq 0.$ A choice of a specific $z$ is a matter of
convenience, all $z$ are equivalent; in the mathematical literature, it is a
tradition to take $z=i$ ($x=0,\;y=1$).

We recall that by definition, a symmetric operator $\hat{f}$ is a densely
defined operator, $\overline{D_{f}}=\mathcal{H}$, with the property (\ref%
{sym}). The criterion of symmetricity is that all diagonal matrix elements
(all means) of a symmetric operator are real\footnote{%
It is well known to physicists as applied to s.a. operators.},%
\begin{equation*}
2i\func{Im}\left( \xi ,\hat{f}\xi \right) =\left( \xi ,\hat{f}\xi \right) -%
\overline{\left( \xi ,\hat{f}\xi \right) }=\left( \xi ,\hat{f}\xi \right)
-\left( \hat{f}\xi ,\xi \right) =0\,,\;\forall \xi \in D_{f}\,.
\end{equation*}

For this reason, it is natural to introduce two forms defined by the adjoint 
$\hat{f}^{+}$ in its domain $D_{f^{+}}:$ the sesquilinear form $\omega
_{\ast }$ given by 
\begin{equation}
\omega _{\ast }\left( \eta _{\ast },\xi _{\ast }\right) =\left( \eta _{\ast
},\hat{f}^{+}\xi _{\ast }\right) -\left( \hat{f}^{+}\eta _{\ast },\xi _{\ast
}\right) \,,\;\xi _{\ast },\eta _{\ast }\in D_{f^{+}}\,,  \label{3a.7}
\end{equation}%
and the quadratic form $\Delta _{\ast }$ given by 
\begin{equation}
\Delta _{\ast }\left( \xi _{\ast }\right) =\left( \xi _{\ast },\hat{f}%
^{+}\xi _{\ast }\right) -\left( \hat{f}^{+}\xi _{\ast },\xi _{\ast }\right)
=2i\func{Im}\left( \xi _{\ast },\hat{f}^{+}\xi _{\ast }\right) \,,\;\xi
_{\ast }\in D_{f^{+}}\,.  \label{3a.8}
\end{equation}%
The form $\omega _{\ast }$ is anti-Hermitian, $\omega _{\ast }\left( \eta
_{\ast },\xi _{\ast }\right) =-\overline{\omega _{\ast }\left( \xi _{\ast
},\eta _{\ast }\right) }\,,$ and the form $\Delta _{\ast }$ is pure
imaginary $\overline{\Delta _{\ast }\left( \xi _{\ast }\right) }=-\Delta
_{\ast }\left( \xi _{\ast }\right) \,.$ The forms $\omega _{\ast }$ and $%
\Delta _{\ast }$ determine each other. Really, $\Delta _{\ast }$ is an
evident restriction of $\omega _{\ast }$ to the diagonal $\xi _{\ast }=\eta
_{\ast },$%
\begin{equation*}
\Delta _{\ast }\left( \xi _{\ast }\right) =\omega _{\ast }\left( \xi _{\ast
},\xi _{\ast }\right) \,,
\end{equation*}%
while $\omega _{\ast }$ is completely determined by $\Delta _{\ast }$ in
view of the equality%
\begin{equation*}
\omega _{\ast }\left( \eta _{\ast },\xi _{\ast }\right) =\frac{1}{4}\left\{ %
\left[ \Delta _{\ast }\left( \xi _{\ast }+\eta _{\ast }\right) -\Delta
_{\ast }\left( \xi _{\ast }-\eta _{\ast }\right) \right] -i\left[ \Delta
_{\ast }\left( \xi _{\ast }+i\eta _{\ast }\right) -\Delta _{\ast }\left( \xi
_{\ast }-i\eta _{\ast }\right) \right] \right\}
\end{equation*}%
(the so-called polarization formula).

Each of these forms is a measure of asymmetricity of the adjoint $\hat{f}%
^{+} $, i.e., a measure of to what extent the adjoint $\hat{f}^{+}$ is
nonsymmetric. We therefore call $\omega _{\ast }$ and $\Delta _{\ast }$ the
respective sesquilinear asymmetry form and quadratic asymmetry form. If $%
\omega _{\ast }\equiv 0,$ or equivalently $\Delta _{\ast }\equiv 0,$ the
adjoint $\hat{f}^{+}$ is symmetric and $\hat{f}$ is essentially s.a. .

\subsection{Closure of symmetric operator in terms of asymmetry form $%
\protect\omega _{\ast }$}

One of the immediate advantages of introducing the sesquilinear form $\omega
_{\ast }$ is that it allows simply determining the closure $\overline{\hat{f}%
}$ of an initial generally nonclosed symmetric operator $\hat{f}$ if the
adjoint $\hat{f}^{+}$ is determined. Really, we know that $\overline{\hat{f}}
$ is symmetric, $\overline{\hat{f}}\subseteq \left( \overline{\hat{f}}%
\right) ^{+}$ with the same adjoint, $\left( \overline{\hat{f}}\right) ^{+}=$
$\hat{f}^{+},$ and coincides with the adjoint to the adjoint $\left( \hat{f}%
^{+}\right) ^{+},$ such that $\overline{\hat{f}}=\left( \hat{f}^{+}\right)
^{+}\subseteq \left( \overline{\hat{f}}\right) ^{+}=\hat{f}^{+}\,,$
therefore, $\overline{\hat{f}}$ can be determined as $\left( \hat{f}%
^{+}\right) ^{+}.$ The defining equation for $\left( \hat{f}^{+}\right) ^{+}=%
\overline{\hat{f}},$ i.e., for a pair $\underline{\psi }\in D_{\bar{f}}$ and 
$\underline{\chi }=\overline{\hat{f}}\underline{\psi },$ is\footnote{%
Here, we use the notation $\underline{\psi }$ and $\underline{\chi }$
instead of the conventional $\underline{\xi }$ and $\underline{\eta }$ in
oder to avoid a possible confusion: $\underline{\xi }$ is also a
conventional notation for the $D_{\bar{f}}$-component of $\xi _{\ast }$ in
representation (\ref{3a.4}) that is used below.}%
\begin{equation}
\left( \underline{\psi },\hat{f}^{+}\xi _{\ast }\right) -\left( \underline{%
\chi },\xi _{\ast }\right) =0\,,\;\forall \xi _{\ast }\in D_{f^{+}}\,.
\label{3a.9}
\end{equation}%
But $\overline{\hat{f}}\subseteq \hat{f}^{+}$ means that $D_{\bar{f}%
}\subseteq D_{f^{+}}\,,$ i.e., $\underline{\psi }\in D_{f^{+}},$ and $%
\underline{\chi }=\overline{\hat{f}}\underline{\psi }=\hat{f}^{+}\underline{%
\psi }$ (we know the ''rule'' for $\overline{\hat{f}}$), therefore, defining
equation (\ref{3a.9}) for the closure $\overline{\hat{f}}$ reduces to the
equation $\left( \underline{\psi },\hat{f}^{+}\xi _{\ast }\right) -\left( 
\hat{f}^{+}\underline{\psi },\xi _{\ast }\right) =0\,,\;\forall \xi _{\ast
}\in D_{f^{+}},$ i.e., to the equation%
\begin{equation}
\omega _{\ast }\left( \underline{\psi },\xi _{\ast }\right) =0\,,\;\forall
\xi _{\ast }\in D_{f^{+}}\,,  \label{3a.10}
\end{equation}%
for $\underline{\psi }\in D_{\bar{f}}$ only, or equivalently, taking the
complex conjugation of (\ref{3a.10}), to%
\begin{equation}
\omega _{\ast }\left( \xi _{\ast },\underline{\psi }\right) =0\,,\;\forall
\xi _{\ast }\in D_{f^{+}}\,,  \label{3a.11}
\end{equation}%
which is the linear equation for the domain $D_{\bar{f}}\subseteq D_{f^{+}}$
of the closure.

The closure $\overline{\hat{f}}$ of a symmetric operator $\hat{f},$ $\hat{f}%
\subseteq \hat{f}^{+},$ is thus given by\footnote{%
We adopt this form of representing operators; it actually represents the
graph of an operator.}%
\begin{equation}
\overline{\hat{f}}:\left\{ 
\begin{array}{l}
D_{\bar{f}}=\left\{ \underline{\psi }\in D_{f^{+}}:\omega _{\ast }\left( \xi
_{\ast },\underline{\psi }\right) =0\,,\;\forall \xi _{\ast }\in
D_{f^{+}}\right\} \,, \\ 
\overline{\hat{f}}\underline{\psi }=\hat{f}^{+}\underline{\psi }\,.%
\end{array}%
\right.  \label{3a.12}
\end{equation}%
Formula (\ref{3a.12}) specifies the closure $\overline{\hat{f}}$ as an
evidently symmetric restriction of the adjoint $\hat{f}^{+}:\,\omega _{\ast
}\left( \xi _{\ast },\underline{\psi }\right) =0$ implies%
\begin{equation*}
\omega _{\ast }\left( \underline{\eta },\underline{\xi }\right) =\left( 
\underline{\eta },\overline{\hat{f}}\underline{\xi }\right) -\left( 
\overline{\hat{f}}\underline{\eta },\underline{\xi }\right) =0\,,\;\forall 
\underline{\eta },\underline{\xi }\in D_{f^{+}}\,,
\end{equation*}%
which confirms the fact that the closure of a symmetric operator is
symmetric.

Because $\omega _{\ast }$ vanishes on $D_{\bar{f}}$ and because of
representation (\ref{3a.4}) for $\xi _{\ast }\in D_{f^{+}},$ the nontrivial
content of eq. (\ref{3a.11}) for the domain $D_{\bar{f}}$ in (\ref{3a.12})
is only due to the presence of the deficient subspaces. Really, substituting
representation (\ref{3a.4}) for $\xi _{\ast },$ $\xi _{\ast }=\underline{\xi 
}+\xi _{z}+\xi _{\overline{z}}$ in (\ref{3a.11}), and using the fact that $%
\omega _{\ast }$ vanishes on $D_{\bar{f}}$, we reduce it to the equation%
\begin{equation}
\omega _{\ast }\left( \xi _{z}+\xi _{\overline{z}},\underline{\psi }\right)
=0\,,\;\forall \xi _{z}\in \aleph _{\bar{z}}\,,\;\forall \xi _{\bar{z}}\in
\aleph _{z}\,,  \label{3a.13}
\end{equation}%
which is equivalent to the set of equations%
\begin{equation*}
\omega _{\ast }\left( \xi _{z},\underline{\psi }\right) =0\,,\;\omega _{\ast
}\left( \xi _{\overline{z}},\underline{\psi }\right) =0\,,\;\forall \xi
_{z}\in \aleph _{\bar{z}}\,,\;\forall \xi _{\bar{z}}\in \aleph _{z}\,.
\end{equation*}

Let the deficient subspaces be finite-dimensional, $\dim \aleph _{\bar{z}%
}=m_{\bar{z}}<\infty $ and $\dim \aleph _{z}=m_{z}<\infty $ ($m_{\bar{z}}$
is equal to $m_{+}$ or $m_{-}$ and $m_{z}=m_{-}$ or $m_{+}$ for the
respective $z\in \mathbb{C}_{-}$ or $z\in \mathbb{C}_{+}$), and let $\left\{
e_{z,k}\right\} _{1}^{m_{\bar{z}}}$ and $\left\{ e_{\bar{z},k}\right\}
_{1}^{m_{z}}$ be some basises in the respective $\aleph _{\bar{z}}$ and $%
\aleph _{z}$ . Then the last set of equations can be replaced by a finite set%
\begin{equation*}
\omega _{\ast }\left( e_{z,k},\underline{\psi }\right) =0\,,\;\omega _{\ast
}\left( e_{\bar{z},l},\underline{\psi }\right) =0\,,\;k=1,...,m_{\bar{z}%
}\,,\;l=1,...,m_{z}\,.
\end{equation*}%
Taking all this into account, we can effectively replace eq. (\ref{3a.12})
specifying the closure $\hat{f}$ by%
\begin{equation}
\overline{\hat{f}}:\left\{ 
\begin{array}{l}
D_{\bar{f}}=\left\{ \underline{\psi }\in D_{f^{+}}:\omega _{\ast }\left( \xi
_{z},\underline{\psi }\right) =\omega _{\ast }\left( \xi _{\overline{z}},%
\underline{\psi }\right) =0,\;\forall \xi _{z}\in \aleph _{\bar{z}%
}\,,\;\forall \xi _{\bar{z}}\in \aleph _{z}\right\} , \\ 
\overline{\hat{f}}\underline{\psi }=\hat{f}^{+}\underline{\psi }\,,%
\end{array}%
\right.  \label{3a.14}
\end{equation}%
which in the case of finite-dimensional deficient subspaces is equivalent to%
\begin{equation}
\overline{\hat{f}}:\left\{ 
\begin{array}{l}
D_{\bar{f}}=\left\{ \underline{\psi }\in D_{f^{+}}:\omega _{\ast }\left(
e_{z,k},\underline{\psi }\right) =\omega _{\ast }\left( e_{\bar{z},l},%
\underline{\psi }\right) =0,\;k=1,...,m_{\bar{z}}\,,\;l=1,...,m_{z}\right\} ,
\\ 
\overline{\hat{f}}\underline{\psi }=\hat{f}^{+}\underline{\psi }\,,%
\end{array}%
\right.  \label{3a.15}
\end{equation}%
where $\left\{ e_{z,k}\right\} _{1}^{m_{\bar{z}}}$ and $\left\{ e_{\bar{z}%
,k}\right\} _{1}^{m_{z}}$ \ are some basises in the respective deficient
subspaces $\aleph _{\bar{z}}$ and $\aleph _{z}$ .

\subsection{Von Neumann formula. Symmetric extensions. Second von Neumann
Theorem.}

But the main blessing of the two asymmetry forms $\omega _{\ast }$ and $%
\Delta _{\ast }$ is that they allow effectively studying the possibilities
of describing symmetric and s.a. extensions of symmetric operators. The key
ideas formulated, so to say, in advance are as follows. Any symmetric
extension of a symmetric operator $\hat{f}$ is a restriction of its adjoint $%
\hat{f}^{+}$ to a subdomain in $D_{f^{+}}$ such that the restriction of $%
\omega _{\ast }$ and $\Delta _{\ast }$ to this subdomain vanishes. On the
other hand, $\omega _{\ast }$ allows comparatively simply evaluating the
adjoint of the extension, while $\Delta _{\ast }$ allows estimating the
measure of the closedness of the extension and the possibility of a further
extension. S.a. extensions, if they are possible, correspond to maximum
subdomains where $\omega _{\ast }$ and $\Delta _{\ast }$ vanish, maximum in
the sense that a further extension to a wider domain where $\omega _{\ast }$
and $\Delta _{\ast }$ vanish is impossible.

According to the aforesaid, the both $\omega _{\ast }$ and $\Delta _{\ast }$
vanish on the domain $D_{\bar{f}}\subset D_{f^{+}}$ of the closure $%
\overline{\hat{f}}\subseteq \hat{f}^{+},$ 
\begin{equation}
\omega _{\ast }\left( \underline{\eta },\underline{\xi }\right)
=0\,,\;\forall \underline{\eta },\underline{\xi }\in D_{\bar{f}}\,;\;\Delta
_{\ast }\left( \underline{\xi }\right) =0\,,\;\forall \underline{\xi }\in D_{%
\bar{f}}\,,  \label{3a.16}
\end{equation}%
and are nonzero only because of the presence of the deficient subspaces $%
\aleph _{\bar{z}}$ and $\aleph _{z}$ .

We now evaluate $\omega _{\ast }\left( \eta _{\ast },\xi _{\ast }\right) .$
According to the first von Neumann theorem \ref{t3a.1}, representation (\ref%
{3a.4}) holds for any $\eta _{\ast },\xi _{\ast }\in D_{f^{+}}$.
Substituting this representation for both $\eta _{\ast }$ and $\xi _{\ast }$
in $\omega _{\ast }\left( \eta _{\ast },\xi _{\ast }\right) $ using the
sesquilinearity of the form $\omega _{\ast }$ and taking the facts that $%
\omega _{\ast }\left( \underline{\eta },\xi _{\ast }\right) =0,$ see (\ref%
{3a.10}), and $\omega _{\ast }\left( \eta _{z}+\eta _{\bar{z}},\underline{%
\xi }\right) =0,$ see (\ref{3a.13}), into account, we obtain that$\;\omega
_{\ast }\left( \eta _{\ast },\xi _{\ast }\right) =\omega _{\ast }\left( \eta
_{z}+\eta _{\bar{z}},\xi _{z}+\xi _{\bar{z}}\right) \,.$ Then using
definition (\ref{3a.7}) of $\omega _{\ast }$ and the definition of the
deficient subspaces according to which%
\begin{equation*}
\hat{f}^{+}\xi _{z}=z\xi _{z}\,,\;\hat{f}^{+}\eta _{z}=z\eta _{z}\,,\;\;\hat{%
f}^{+}\xi _{\bar{z}}=\bar{z}\xi _{\bar{z}}\,,\;\hat{f}^{+}\eta _{\bar{z}}=%
\bar{z}\eta _{\bar{z}}\,,
\end{equation*}%
we finally find%
\begin{equation}
\omega _{\ast }\left( \eta _{\ast },\xi _{\ast }\right) =2iy\left[ \left(
\eta _{z},\xi _{z}\right) -\left( \eta _{\bar{z}},\xi _{\bar{z}}\right) %
\right] \,,\;2iy=\left( z-\bar{z}\right) \,.  \label{3a.17}
\end{equation}%
It follows a similar representation for $\Delta _{\ast }\left( \xi _{\ast
}\right) =\omega _{\ast }\left( \xi _{\ast },\xi _{\ast }\right) :$%
\begin{equation}
\Delta _{\ast }\left( \xi _{\ast }\right) =2iy\left( \left\| \xi
_{z}\right\| ^{2}-\left\| \xi _{\bar{z}}\right\| ^{2}\right) \,.
\label{3a.18}
\end{equation}%
Formula (\ref{3a.18}) is sometimes called the von Neumann formula (without a
number).

We really see that the asymmetricity of the adjoint $\hat{f}^{+}$ is due to
the deficient subspaces. What is more, $\omega _{\ast }$ and $\Delta _{\ast
} $ are of a specific structure: up to a nonzero factor $\left( z-\bar{z}%
\right) =2iy$, the contributions of the different deficient subspaces $%
\aleph _{\bar{z}}$ and $\aleph _{z}$ are of the opposite signs and, in
principle, can compensate each other under an appropriate correspondence
between $\xi _{z}$ and $\xi _{\bar{z}}$, the respective $\aleph _{\bar{z}}$-
and $\aleph _{z}$-components of vectors $\xi _{\ast }\in D_{f^{+}}.$

In our exposition, these formulas (\ref{3a.17}) and (\ref{3a.18}) together
with the first von Neumann theorem form a basis for estimating the
possibility and constructing, if possible, s.a. extensions of a symmetric
operator $\hat{f}$. Although the forms $\omega _{\ast }$ and $\Delta _{\ast
} $ and the respective formulas (\ref{3a.17}) and (\ref{3a.18}) are
equivalent, it is convenient to use the both of them, one or another in
dependence of the context.

An alternative method for studying and constructing symmetric and s.a.
extensions of symmetric operators is based on the so-called Cayley
transformation of a closed symmetric operator $\hat{f},$ $\hat{f}=\overline{%
\hat{f}},$ to an isometric operator $\hat{V}=\left( \hat{f}-z\hat{I}\right)
\left( \hat{f}-\bar{z}\hat{I}\right) ^{-1},$ with the domain $D_{V}=\Re _{%
\bar{z}}=R_{f-\bar{z}I}$ and the range $R_{V}=\Re _{z}=R_{f-zI}$, and vice
versa, $\hat{f}=\left( z\hat{I}-\bar{z}\hat{V}\right) \left( \hat{I}-\hat{V}%
\right) ^{-1};$ all that can be found in \cite{AkhGl81,Naima69}.

A nontrivial symmetric extension $\hat{f}_{\mathrm{ext}}$ of a symmetric
operator $\hat{f}$ , $\overline{\hat{f}}\subset \hat{f}_{\mathrm{ext}%
}\subseteq \hat{f}_{\text{\textrm{ext}}}^{+}\subset \hat{f}^{+}$ with the
domain $D_{f_{\mathrm{ext}}}$, $D_{\bar{f}}\subset D_{f_{\mathrm{ext}%
}}\subset D_{f^{+}}$ is possible only at the expense of deficient subspaces $%
\aleph _{\bar{z}}$ and $\aleph _{z}$:%
\begin{equation*}
D_{f_{\mathrm{ext}}}=\left\{ \xi _{\mathrm{ext}}=\underline{\xi }+\xi _{z,%
\mathrm{ext}}\,+\xi _{\bar{z},\mathrm{ext}}\,,\;\forall \underline{\xi }\in
D_{\bar{f}}\,,\;\xi _{z,\mathrm{ext}}\in \aleph _{\bar{z}}\,,\;\xi _{\bar{z},%
\mathrm{ext}}\in \aleph _{z}\,\right\} \,,
\end{equation*}%
(any$\;\underline{\xi }\in D_{\bar{f}}\,$\ and\ some$\;\xi _{z,\mathrm{ext}%
}\in \aleph _{\bar{z}}\;$and\ $\xi _{\bar{z},\mathrm{ext}}\in \aleph _{z}$%
)\thinspace , or $D_{f_{\mathrm{ext}}}=D_{\bar{f}}+\Delta D_{f_{\mathrm{ext}%
}}\,,$ where $\Delta D_{f_{\mathrm{ext}}}=\left\{ \Delta \xi _{\mathrm{ext}%
}=\xi _{z,\text{\textrm{ext}}}+\xi _{\bar{z},\text{\textrm{ext}}}\right\}
\subseteq \aleph _{\bar{z}}+\aleph _{z}\,,$ is nontrivial, $\Delta D_{f_{%
\mathrm{ext}}}\neq \{0\}.$

$\Delta D_{f_{\mathrm{ext}}}$ is a subspace as well as $D_{f_{\mathrm{ext}%
}}, $ therefore, the sets $\Delta D_{\bar{z},\mathrm{ext}}=\{\xi _{z,\text{%
\textrm{ext}}}\}\subset \aleph _{\bar{z}}$ and $\Delta D_{z,\mathrm{ext}%
}=\{\xi _{\bar{z},\text{\textrm{ext}}}\}\subset \aleph _{z}$ of the
respective $\xi $$_{z,\text{\textrm{ext}}}$ and $\xi _{\bar{z},\text{\textrm{%
ext}}}$ involved must also be subspaces. We caution against that $\Delta
D_{f_{\mathrm{ext}}}$ \ belonging to $\aleph _{\bar{z}}+\aleph _{z}$ be
considered a direct sum of $\Delta D_{\bar{z},\mathrm{ext}}$ and $\Delta
D_{z,\mathrm{ext}}\,,$ $\Delta D_{f_{\mathrm{ext}}}\neq \Delta D_{\bar{z},%
\mathrm{ext}}+\Delta D_{z,\mathrm{ext}}\,,$ see below.

The crucial remark is then that a symmetric extension $\hat{f}_{\mathrm{ext}%
} $ of $\hat{f}$ to $D_{f_{\mathrm{ext}}}=D_{\bar{f}}+\Delta D_{f_{\mathrm{%
ext}}}$ is simultaneously a symmetric restriction of the adjoint $\hat{f}%
^{+} $ to $D_{f_{\mathrm{ext}}}\subset D_{f^{+}}$. In particular, this
implies that we know the ''rule'' for $\hat{f}_{\mathrm{ext}}$ : according
to (\ref{3a.5}), it acts as $\overline{\hat{f}}$ on $D_{\bar{f}}$ and as a
multiplication by $z$ on $\Delta D$$_{\bar{z},\text{\textrm{ext}}}$ and by $%
\bar{z}$ on $\Delta D_{z,\text{\textrm{ext}}}$ .

The requirement that the restriction $\hat{f}_{\mathrm{ext}}$ of the adjoint 
$\hat{f}^{+}$ to a subspace $D_{f_{\mathrm{ext}}}\subset D_{f^{+}}$ be
symmetric is equivalent to the requirement that the restriction of the
asymmetry forms $\omega _{\ast }$ and $\Delta _{\ast }$ to $D_{f_{\mathrm{ext%
}}}$ vanish,%
\begin{equation}
\omega _{\ast }\left( \eta _{\mathrm{ext}},\xi _{\mathrm{ext}}\right)
=0,\;\forall \eta _{\mathrm{ext}},\xi _{\mathrm{ext}}\in D_{f_{\mathrm{ext}%
}};\;\Delta _{\ast }\left( \xi _{\mathrm{ext}}\right) =0\,,\;\forall \xi _{%
\mathrm{ext}}\in D_{f_{\mathrm{ext}}}\,.  \label{3a.19}
\end{equation}%
We now establish the necessary and sufficient conditions for the existence
of such nontrivial domain $D_{f_{\mathrm{ext}}}$ and describe their
structure. Each of conditions (\ref{3a.19}) is equivalent to another. In the
consideration to follow, we mainly deal with the quadratic asymmetry form $%
\Delta _{\ast }.$

According to von Neumann formula (\ref{3a.18}), the only nontrivial point in
the condition $\Delta _{\ast }\left( \xi _{\mathrm{ext}}\right) =0\,$ is
that the restriction of $\Delta _{\ast }$ to $\Delta D_{f_{\mathrm{ext}}}$
vanishes:%
\begin{equation}
\Delta _{\ast }\left( \Delta \xi _{\mathrm{ext}}=\xi _{z,\mathrm{ext}}+\xi _{%
\bar{z},\mathrm{ext}}\right) =2iy\left( \left\| \xi _{z,\mathrm{ext}%
}\right\| ^{2}-\left\| \xi _{\bar{z},\mathrm{ext}}\right\| ^{2}\right)
=0\,,\;\forall \Delta \xi _{\mathrm{ext}}\in \Delta D_{f_{\mathrm{ext}}}\,.
\label{3a.20}
\end{equation}

It immediately follows that if one of the deficient subspaces of the initial
symmetric operator $\hat{f}$ is trivial, i.e., if $\aleph _{\bar{z}}=\{0\}$
or $\aleph _{z}=\{0\},$ or, equivalently, if one of the deficiency indices
is equal to zero, i.e., if $m_{+}=0$ or $m_{-}=0,$ in short, $\min \left(
m_{+},m_{-}\right) =0,$ then there is no nontrivial symmetric extensions of
this operator. In other words, a symmetric operator $\hat{f}$ with one of
the deficiency indices equal to zero, $\min \left( m_{+},m_{-}\right) =0,$
is essentially maximal.

In what follows, we therefore consider the case where $\min \left(
m_{+},m_{-}\right) \neq 0$ and the both deficient subspaces $\aleph _{\bar{z}%
}$ and $\aleph _{z}$ of a symmetric operator $\hat{f}$ are nontrivial. We
show that in this case, nontrivial symmetric extensions of $\hat{f}$ do
exist. Without loss of generality, we assume that%
\begin{equation*}
0<\dim \aleph _{\bar{z}}=\min \left( m_{+},m_{-}\right) \leq \dim \aleph
_{z}=\max \left( m_{+},m_{-}\right) \,,
\end{equation*}%
we can always take an appropriate $z$. In the mathematical literature, it is
conventional to take $z\in \mathbb{C}_{+}\,,\;y>0\,,$ then if $0<m_{+}\leq
m_{-}\,,$ we fall into our condition; in the opposite case, the deficient
subspaces and deficiency indices are simply transposed in the consideration
to follow.

We first assume the existence of nontrivial symmetric extensions in the case
under consideration. Let $\hat{f}_{\mathrm{ext}}$ be a nontrivial symmetric
extension of a symmetric operator $\hat{f}$ with the both deficiency indices 
$m_{+}$ and $m_{-}$ different from zero. Formula (\ref{3a.20}) suggests that
the both deficient subspaces $\aleph _{\bar{z}}$ and $\aleph _{z}$ must be
involved in this extension, i.e., $\Delta D_{\bar{z},\mathrm{ext}}\neq \{0\}$
and $\Delta D_{z,\mathrm{ext}}\neq \{0\},$ and any involved $\xi _{z,\text{%
\textrm{ext}}}\in \Delta D_{\bar{z},\mathrm{ext}}\subseteq \aleph _{\bar{z}}$
must be assigned a certain $\xi _{\bar{z},\mathrm{ext}}\in \Delta D_{z,%
\mathrm{ext}}\subseteq \aleph _{z}$ of the same norm, $\left\| \xi _{z,%
\mathrm{ext}}\right\| =\left\| \xi _{\bar{z},\mathrm{ext}}\right\| ,$ for
their contributions to $\Delta _{\ast }$ compensate each other. We now note
that this assignment must be a one-to-one correspondence. Really, if, for
example, a vector $\Delta \xi _{\mathrm{ext}}=\xi _{z,\mathrm{ext}}+\xi _{%
\bar{z},\mathrm{ext}}$ and a vector $\Delta \xi _{\mathrm{ext}}^{\prime
}=\xi _{z,\mathrm{ext}}+\xi _{\bar{z},\mathrm{ext}}^{\prime }$ belong to $%
\Delta D_{f_{\mathrm{ext}}},$ then their difference $\Delta \xi _{\mathrm{ext%
}}^{\prime }-\Delta \xi _{\mathrm{ext}}=\xi _{\bar{z},\mathrm{ext}}^{\prime
}-\xi _{\bar{z},\mathrm{ext}}$ with the zero $\aleph _{\bar{z}}$-component
also belongs to $\Delta D_{f_{\mathrm{ext}}}$ because $\Delta D_{f_{\mathrm{%
ext}}}$ is a linear manifold. But then formula (\ref{3a.20}) implies that $%
\left| \left| \xi _{\bar{z},\mathrm{ext}}^{\prime }-\xi _{\bar{z},\mathrm{ext%
}}\right| \right| =0,$ i.e., $\xi _{\bar{z},\mathrm{ext}}^{\prime }=\xi _{%
\bar{z},\mathrm{ext}}\,.$ A similar consideration for a pair of vectors $%
\Delta \xi _{\mathrm{ext}}=\xi _{z,\mathrm{ext}}+\xi _{\bar{z},\mathrm{ext}%
}\in \Delta D_{f_{\mathrm{ext}}}$ and $\Delta \xi _{\mathrm{ext}}^{\prime
}=\xi _{z,\mathrm{ext}}^{\prime }+\xi _{\bar{z},\mathrm{ext}}\in \Delta
D_{f_{\mathrm{ext}}}$ results in the conclusion that there must be $\xi _{z,%
\mathrm{ext}}^{\prime }=\xi _{z,\mathrm{ext}}\,.$ In addition, this
correspondence must be a linear mapping of $\Delta D_{\bar{z},\mathrm{ext}}$
to $\Delta D_{z,\mathrm{ext}}\,$for $\Delta D_{f_{\mathrm{ext}}}$ to be a
linear manifold.

But this means that any nontrivial symmetric extension $\hat{f}_{\text{%
\textrm{ext}}}$ of $\hat{f}$ is defined by some linear isometric mapping, or
simply isometry,%
\begin{equation*}
\hat{U}:\;\aleph _{\bar{z}}\longrightarrow \aleph _{z}\,,
\end{equation*}%
with a domain $D_{U}=\Delta D_{\bar{z},\mathrm{ext}}\subseteq \aleph _{\bar{z%
}}$ and a range $R_{U}=\Delta D_{z,\mathrm{ext}}=\hat{U}\Delta D_{\bar{z},%
\mathrm{ext}}\subseteq \aleph _{z}\,.$ Because any isometry preserves
dimension, $\Delta D_{\bar{z},\mathrm{ext}}$ and $\Delta D_{z,\mathrm{ext}}$
must be of the same dimension,%
\begin{equation*}
\dim \Delta D_{\bar{z},\mathrm{ext}}=\dim \Delta D_{z,\mathrm{ext}%
}=m_{U}\leq \min \left( m_{+},m_{-}\right) ;
\end{equation*}%
$\Delta D_{f_{\mathrm{ext}}}$ is also of dimension $m_{U}$ because of the
one-to-one correspondence between the $\xi _{\bar{z},\mathrm{ext}}$ and $\xi
_{z,\mathrm{ext}}$ components in any vector $\Delta \xi _{\mathrm{ext}}\in
\Delta D_{f_{\mathrm{ext}}}\,.$

It is now reasonable to change the notation: we let $D_{U}$ denote $\Delta
D_{\bar{z},\text{\textrm{ext}}}$ and let $\hat{U}D_{U}$ denote $\Delta D_{z,%
\mathrm{ext}}$ and change the subscript ''$\mathrm{ext}$'' to the subscript
''$U$'' in other cases, such that $\hat{f}_{\text{\textrm{ext}}},\,D_{f_{%
\mathrm{ext}}},\,\Delta D_{f_{\mathrm{ext}}},$ and etc. are now denoted by $%
\hat{f}_{U}$\thinspace ,$\;D_{f_{U}}\,,$ $\Delta D_{f_{U}}$\thinspace , and
etc. In particular, $D_{f_{U}}$ is now written as%
\begin{align}
& D_{f_{U}}=D_{\bar{f}}+\Delta D_{f_{U}}=\left\{ \xi _{U}=\underline{\xi }%
+\Delta \xi _{U}\,,\;\forall \underline{\xi }\in D_{\bar{f}}\,,\;\forall
\Delta \xi _{U}\in \Delta D_{f_{U}}\right\} \,,  \notag \\
& \,\Delta D_{f_{U}}=\left( D_{U}+\hat{U}D_{U}\right) =\left( \hat{I}+\hat{U}%
\right) D_{U}=\left\{ \Delta \xi _{U}=\xi _{z,U}+\xi _{\bar{z},U}\,,\right. 
\notag \\
& \,\left. \xi _{z,U}\in D_{U}\subseteq \aleph _{\bar{z}}\,,\;\xi _{\bar{z}%
,U}=\hat{U}\xi _{z,U}\in \hat{U}D_{U}\subseteq \aleph _{z}\right\} \,,
\label{3a.21}
\end{align}%
where the parenthesis in the notation $\left( D_{U}+\hat{U}D_{U}\right) $
denotes that $\Delta D_{f_{U}}$ is not a direct sum of the linear manifolds $%
D_{U}$ and $\hat{U}D_{U}$ of equal dimension $m_{U}\leq \min \left(
m_{+},m_{-}\right) ,$ but a special linear manifold of dimension $m_{U}$
that can be considered a ''diagonal'' of the direct sum $D_{U}+\hat{U}%
D_{U}\,.$

We can now prove the existence of nontrivial symmetric extensions of a
symmetric operator $\hat{f}$ in the case where $\min\left(
m_{+},m_{-}\right) \neq0$ by reversing the above consideration. Namely, it
is now evident that if the deficient subspaces of $\hat{f},$ $\aleph_{\bar{z}%
}$ and $\aleph_{z}\,,$ are nontrivial, then any isometry $\hat{U}:\aleph_{%
\bar{z}}\rightarrow\aleph_{z}$ with the domain $D_{U}\subseteq\aleph_{\bar{z}%
}$ and the range $\hat{U}D_{U}\subseteq\aleph_{z}$ generates a nontrivial
symmetric extension $\hat{f}_{U}$ of $\hat{f}$ as the restriction of the
adjoint $\hat{f}^{+}$ to the domain $D_{U}$ given by (\ref{3a.21}) because
this restriction is evidently symmetric. It follows in particular that if $%
\hat{f}$ is an essentially maximal symmetric operator, then one of its
deficiency indices must be zero.

We collect all the aforesaid in a theorem.

\begin{theorem}
\label{t3a.2}(The second von Neumann theorem) A symmetric operator $\hat{f}$
is essentially s.a. iff its deficiency indices are equal to zero, $%
m_{+}=m_{-}=0$.

A symmetric operator $\hat{f}$ is essentially maximal iff one of its
deficiency indices is equal to zero, $\min \left( m_{+},m_{-}\right) =0$; if
its second deficiency index is also equal to zero, then $\hat{f}$ is
essentially s.a.; if the second deficiency index is nonzero, then $\hat{f}$
is only essentially maximal and does not allow s.a. extensions.

If the both deficiency indices of a symmetric operator $\hat{f}$ are
different from zero, $\min \left( m_{+},m_{-}\right) \neq 0$, i.e., the both
its deficient subspaces $\aleph _{\bar{z}}$ and $\aleph _{z}\,$are
nontrivial, then nontrivial symmetric extensions of $\hat{f}$ do exist. Any
symmetric extension $\hat{f}_{U}$ of $\hat{f}$ is defined by some isometric
operator $\hat{U}$ with a domain $D_{U}\subseteq \aleph _{\bar{z}}$ and a
range $\hat{U}D_{U}\subseteq \aleph _{z}$ and is given by\footnote{%
In this case, it seems more expressive to represent the graph of the
operator $\hat{f}_{U}$ by separate formulas.}%
\begin{align}
& D_{f_{U}}=D_{\underline{f}}+\left( \hat{I}+\hat{U}\right) D_{U}=\left\{
\xi _{U}=\underline{\xi }+\xi _{z,U}+\hat{U}\xi _{z,U}:\right.  \notag \\
& \left. \forall \underline{\xi }\in D_{\bar{f}}\,,\;\forall \xi _{z,U}\in
D_{U}\subseteq \aleph _{\bar{z}}\,,\;\hat{U}\xi _{z,U}\in \hat{U}%
D_{U}\subseteq \aleph _{z}\right\} \,,  \label{3a.22}
\end{align}%
and%
\begin{equation}
\hat{f}_{U}\xi _{U}=\overline{\hat{f}}\underline{\xi }+z\xi _{z,U}+\bar{z}%
\hat{U}\xi _{z,U}\,.  \label{3a.23}
\end{equation}%
Conversely, any isometric operator $\hat{U}$: $\aleph _{\bar{z}%
}\longrightarrow $ $\aleph _{z}$ with a domain $D_{U}\subseteq \aleph _{\bar{%
z}}$ and a range $\hat{U}D_{U}\subseteq \aleph _{z}$ defines a symmetric
extension $\hat{f}_{U}$ of $\hat{f}$ given by (\ref{3a.22}) and (\ref{3a.23}%
).
\end{theorem}

The formula $\xi _{U}=\underline{\xi }+\xi _{z,U}+\hat{U}\xi _{z,U}$ in (\ref%
{3a.22}) is called the second von Neumann formula.

We do not dwell on the theory of symmetric extensions of symmetric operators
in every detail because it hardly can find applications in constructing
quantum-mechanical observables and restrict ourselves to a few remarks on
the general properties of arbitrary symmetric extensions. All the details
can be found in \cite{AkhGl81,Naima69}.

i) It is evident that if $\hat{f}_{U}$ is a closed extension of a symmetric
operator $\hat{f},$ then $D_{U}$ and $\hat{U}D_{U}$ are closed subspaces in
the respective deficient subspaces $\aleph _{\bar{z}}$ and $\aleph _{z}$ and
vice versa.

ii) The deficient subspaces of an extension $\hat{f}_{U}$ are the respective
subspaces $\aleph_{\bar{z},U}=D_{U}^{\bot}=\aleph_{\bar{z}}\backslash 
\overline{D_{U}}$ and $\aleph_{z,U}=\left( \hat{U}D_{U}\right) ^{\bot
}=\aleph_{z}\backslash\overline{\hat{U}D_{U}}\,,$ the orthogonal complements
of $D_{U}$ and $\hat{U}D_{U}$\thinspace\ in the respective deficient
subspaces $\aleph_{\bar{z}}$ and $\aleph_{z}$ of the initial symmetric
operator $\hat {f},$ therefore, the deficiency indices of the extension $%
\hat{f}_{U}$ are the respective $m_{+,U}=m_{+}-m_{U}$ and $%
m_{-,U}=m_{-}-m_{U}\,,$ where $m_{U}=\dim D_{U}\,.$ The evaluation of the
deficient subspaces and deficiency indices in the particular case of a
maximal symmetric extension $\hat{f}_{U}$ is given below. Its modification
for the general case is evident.

iii) Any symmetric operator $\hat{f}$ with the both deficiency indices
different from zero can be extended to a maximal symmetric operator, see
below.

iv) The description of symmetric extensions of a symmetric operator $\hat{f}$
in terms of isometries $\hat{U}:$ $\aleph _{\bar{z}}\rightarrow \aleph _{z}$
is evidently $z$-dependent: for a given and fixed symmetric extension of $%
\hat{f}$, the corresponding isometry $\hat{U}$ changes with changing $z$
together with the deficient subspaces $\aleph _{\bar{z}}$ and $\aleph _{z}.$

\subsection{Self-adjoint extensions. Main Theorem.}

Our main interest here is with a possibility and a construction of s.a.
extensions of symmetric operators with nonzero deficiency indices.

We first note that any s.a. extension, if at all possible, is a maximal
symmetric operator. This implies (in our case where $\dim \aleph _{\bar{z}%
}\leq \dim \aleph _{z}$) that the deficient subspace $\aleph _{\bar{z}}$
must be involved in the extension as a whole, i.e., $D_{U}=\aleph _{\bar{z}}$%
, otherwise, a further symmetric extension is possible by extending the
isometry $\hat{U}$ to the whole $\aleph _{\bar{z}}\,.$ The domain of a
maximal symmetric extension $\hat{f}_{U}$ of $\hat{f}$ is thus given by%
\begin{align}
& D_{f_{U}}=D_{\overline{f}}+\left( \hat{I}+\hat{U}\right) \,\aleph _{\bar{z}%
}  \notag \\
& =\left\{ \xi _{U}=\underline{\xi }+\xi _{z}+\hat{U}\xi _{z}:\,\forall 
\underline{\xi }\in D_{\bar{f}}\,,\;\forall \xi _{z}\in \aleph _{\bar{z}%
}\,,\;\hat{U}\xi _{z}\in \aleph _{z}\right\} \,,  \label{3a.24}
\end{align}%
while $\aleph _{z}$ can be represented as $\aleph _{z}=\hat{U}\aleph _{\bar{z%
}}\oplus \left( \hat{U}\aleph _{\bar{z}}\right) ^{\bot }\,,$ where%
\begin{equation*}
\left( \hat{U}\aleph _{\bar{z}}\right) ^{\perp }=\left\{ \xi _{\bar{z}%
,U}^{\perp }\in \aleph _{z}:\,\left( \xi _{\bar{z},U}^{\perp },\hat{U}\xi
_{z}\right) =0\,,\;\forall \xi _{z}\in \aleph _{\bar{z}}\right\}
\end{equation*}%
is the orthogonal complement of a subspace $\hat{U}\aleph _{\bar{z}%
}\subseteq \aleph _{z}$ in the deficient subspace $\aleph _{z}.$

We now evaluate the adjoint $\hat{f}_{U}^{+}$. Because both $\hat{f}_{U}$
and $\hat{f}_{U}^{+}$ are the restrictions of the adjoint $\hat{f}^{+}$, $%
\hat{f}_{U}\subseteq \hat{f}_{U}^{+}\subset \hat{f}^{+},$ we can use
arguments similar to those in evaluating the closure $\overline{\hat{f}}$ of 
$\hat{f},$ see formulas (\ref{3a.9})-(\ref{3a.12}): the defining equation
for $\hat{f}_{U}^{+}$ is reduced to a linear equation for a domain $%
D_{f_{U}^{+}}\subset D_{f^{+}},$ i.e., for vectors $\eta _{\ast U}\in
D_{f_{U}^{+}}\,,$ namely,%
\begin{equation}
\omega _{\ast }\left( \xi _{U},\eta _{\ast U}\right) =0\,,\;\forall \xi
_{U}\in D_{f_{U}}\,.  \label{3a.25}
\end{equation}%
Let $\eta _{\ast U}=\underline{\eta }+\eta _{z}+\eta _{\bar{z}}$ be
representation (\ref{3a.4}) for $\eta _{\ast U},$ which we rewrite as%
\begin{equation*}
\eta _{\ast U}=\underline{\eta }+\eta _{z}+\hat{U}\eta _{z}+\left( \eta _{%
\bar{z}}-\hat{U}\eta _{z}\right) =\eta _{U}+\left( \eta _{\bar{z}}-\hat{U}%
\eta _{z}\right) \,,
\end{equation*}%
where $\eta _{U}\in D_{f_{U}},$ see (\ref{3a.24}), and $\eta _{\bar{z}}-%
\hat{U}\eta _{z}\in \aleph _{z}\,.$ Because $\omega _{\ast }$ vanishes on $%
D_{f_{U}},$ see (\ref{3a.19}), equation (\ref{3a.25}) reduces to the
equation for the component $\eta _{\bar{z}}-\hat{U}\eta _{z}\in \aleph _{z},$%
\begin{equation*}
\omega _{\ast }\left( \xi _{U},\eta _{\bar{z}}-\hat{U}\eta _{z}\right)
=0\,,\;\forall \xi _{U}\in D_{f_{U}}\,.
\end{equation*}%
Substituting now representation (\ref{3a.24}) for $\xi _{U},$ $\xi _{U}=%
\underline{\xi }+\xi _{z}+\hat{U}\xi _{z}\,,$ and using representation (\ref%
{3a.17}) for $\omega _{\ast },$ we finally obtain that $(\hat{U}\xi
_{z},\eta _{\bar{z}}-\hat{U}\eta _{z})=0\,,\;\forall \xi _{z}\in \aleph _{%
\bar{z}}\,,$ which implies that $\eta _{\bar{z}}-\hat{U}\eta _{z}=\eta _{%
\bar{z},U}^{\bot }\in \left( \hat{U}\aleph _{\bar{z}}\right) ^{\bot }\,.$
Any $\eta _{\ast U}\in D_{f_{U}^{+}}$ is thus represented as%
\begin{equation}
\eta _{\ast U}=\eta _{U}+\eta _{\bar{z},U}^{\bot }\,,  \label{3a.26}
\end{equation}%
with some $\eta _{U}\in D_{f_{U}}$ and some $\eta _{\bar{z},U}^{\bot }\in
\left( \hat{U}\aleph _{\bar{z}}\right) ^{\bot }\subset \aleph _{z}\,.$

Conversely, it is evident from the above consideration that a vector $\eta
_{\ast U}$ of form (\ref{3a.26}) with any $\eta _{U}\in D_{f_{U}}$ and any $%
\eta _{\bar{z},U}^{\bot }\in \left( \hat{U}\aleph _{\bar{z}}\right) ^{\bot }$
satisfies defining equation (\ref{3a.25}) and therefore belongs to $%
D_{f_{U}^{+}}\,.$

Naturally changing the notation $\eta _{\ast U}\rightarrow \xi _{\ast U},$
we thus obtain that%
\begin{equation*}
D_{f_{U}^{+}}=D_{f_{U}}+\left( \hat{U}\aleph _{\bar{z}}\right) ^{\perp
}=\left\{ \xi _{\ast U}=\xi _{U}+\xi _{\bar{z},U}^{\bot }\,,\;\forall \xi
_{U}\in D_{f_{U}}\,,\;\forall \xi _{\bar{z},U}^{\bot }\in \left( \hat{U}%
\aleph _{\bar{z}}\right) ^{\perp }\right\}
\end{equation*}%
and $\hat{f}_{U}^{+}\xi _{\ast U}=\hat{f}_{U}\xi _{U}+\bar{z}\xi _{\bar{z}%
,U}^{\bot }\,.$

This result allows answering the main question about possible s.a.
extensions of symmetric operators. If the subspace $\left( \hat{U}\aleph _{%
\bar{z}}\right) ^{\perp }$ is nontrivial, $\left( \hat{U}\aleph _{\bar{z}%
}\right) ^{\bot }=\aleph _{z}\backslash \hat{U}\aleph _{\bar{z}}\neq \left\{
0\right\} ,$ we have the strict inclusion $D_{f_{U}}\subset D_{f_{U}^{+}}\,,$
i.e., the extension $\hat{f}_{U}$ is only maximal, but not s.a., symmetric
operator; if this subspace is trivial, $\left( \hat{U}\aleph _{\bar{z}%
}\right) ^{\perp }=\left\{ 0\right\} ,$ we have $D_{f_{U}}=D_{f_{U}^{+}}\,,$
which implies the equality $\hat{f}_{U}=\hat{f}_{U}^{+},$ i.e., the maximal
extension $\hat{f}_{U}$ is s.a.. We now evaluate the dimension $\dim \left( 
\hat{U}\aleph _{\bar{z}}\right) ^{\bot }$ of the subspace $\left( \hat{U}%
\aleph _{\bar{z}}\right) ^{\bot }$ that is the evident criteria for $\left( 
\hat{U}\aleph _{\bar{z}}\right) ^{\bot }$ be nontrivial, $\dim \left( \hat{U}%
\aleph _{\bar{z}}\right) ^{\bot }\neq 0,$ or trivial, $\dim \left( \hat{U}%
\aleph _{\bar{z}}\right) ^{\bot }=0,$ and respectively for a maximal
symmetric extension $\hat{f}_{U}$ be non-s.a. or s.a.. It appears that $%
\left( \hat{U}\aleph _{\bar{z}}\right) ^{\bot }$ is essentially determined
by the deficiency indices of the initial symmetric operator.

If one of the (nontrivial) deficiency indices of the initial symmetric
operator $\hat{f}$ is finite, i.e., in our case, $\dim\aleph_{\bar{z}%
}=\min\left( m_{+},m_{-}\right) <\infty$ (we remind that we consider the
case where $\min\left( m_{+},m_{-}\right) \neq0$), while the other, $\dim
\aleph_{z}=\max\left( m_{+},m_{-}\right) $, can be infinite, then we have 
\begin{align*}
& \dim\left( \hat{U}\aleph_{\bar{z}}\right)
^{\bot}=\dim\aleph_{z}-\dim\left( \hat{U}\aleph_{\bar{z}}\right)
=\dim\aleph_{z}-\dim\aleph _{\bar{z}} \\
& \,=\max\left( m_{+},m_{-}\right) -\min\left( m_{+},m_{-}\right) =\left|
m_{+}-m_{-}\right| \,,
\end{align*}
where we use the equality $\dim\left( \hat{U}\aleph_{\bar{z}}\right)
=\dim\aleph_{\bar{z}}\,.$ If the both deficient subspaces $\aleph_{\bar{z}}$
and $\aleph_{z}$ are infinite dimensional, $m_{+}=m_{-}=\infty\,,$ we
encounter the uncertainty $\dim\left( \hat{U}\aleph_{\bar{z}}\right) ^{\bot
}=\infty-\infty,$ and a specific consideration is required. The point is
that in this case, the isometry $\hat{U}:\aleph_{\bar{z}}\rightarrow%
\aleph_{z}$ defining a maximal symmetric extension $\hat{f}_{U}$ can be
isometric mapping of the infinite-dimensional subspace $\aleph_{\bar{z}}$
both into and onto the infinite-dimensional subspace $\aleph_{z}$. In the
case ''into'', the subspace $\left( \hat{U}\aleph_{\bar{z}}\right) ^{\bot}$
is nontrivial, $\dim\left( \hat{U}\aleph_{\bar{z}}\right) ^{\bot}\neq0,$
while in the case ''onto'', the subspace $\left( \hat{U}\aleph_{\bar{z}%
}\right) ^{\bot}$ is trivial, $\dim\left( \hat{U}\aleph_{\bar{z}}\right)
^{\bot}=0.$

It follows that

i) a symmetric operator $\hat{f}$ with different deficiency indices, $%
m_{+}\neq m_{-}\,$, (which implies $\min \left( m_{+},m_{-}\right) <\infty $%
) has no s.a. extensions, but only maximal symmetric extensions;

ii) a symmetric operator $\hat{f}$ with equal and finite deficiency indices, 
$m_{+}=m_{-}=m<\infty ,$ has s.a. extensions, and what is more, any maximal
symmetric extension of such an operator is s.a.;

iii) a symmetric operator $\hat{f}$ with infinite deficiency indices, $%
m_{+}=m_{-}=\infty ,$ allows both a s.a. and non-s.a. maximal extensions.

Any s.a. extension is defined by an isometric mapping $\hat{U}$ of one of
the deficient subspaces, for example, $\aleph _{\bar{z}}$, to another
deficient subspace, $\aleph _{z},\;\hat{U}:\aleph _{\bar{z}}\rightarrow
\aleph _{z}.$ This mapping establishes an isomorphism between the deficient
subspaces. Conversely, any such an isometric mapping $\hat{U}:\aleph _{\bar{z%
}}\rightarrow \aleph _{z}$ defines a s.a. extension $\hat{f}_{U}$ of $\hat{f}
$ given by (\ref{3a.22}) and (\ref{3a.23}) with $D_{U}=\aleph _{\bar{z}}$
and $\hat{U}D_{U}=\aleph _{z}.$

We note that there is another way (maybe, more informative) of establishing
these results. It seems evident from (\ref{3a.24}) and can be proved using
arguments similar to those in proving the first von Neumann theorem that in
our case, the deficient subspaces of a maximal symmetric extension $\hat{f}%
_{U}$ are $\aleph _{\bar{z},U}=\left\{ 0\right\} $ and $\aleph _{z,U}=\left( 
\hat{U}\aleph _{\bar{z}}\right) ^{\bot }\subseteq \aleph _{z}$ and its
respective deficiency indices are $\dim \aleph _{\bar{z},U}=\min \left(
m_{+U},m_{-U}\right) =0$ and $\dim \aleph _{z,U}=\max \left(
m_{+U},m_{-U}\right) =\dim \left( \hat{U}\aleph _{\bar{z}}\right) ^{\bot }$
(which confirms that $\hat{f}_{U}$ is really a maximal symmetric operator).
It then remains to evaluate $\dim \left( \hat{U}\aleph _{\bar{z}}\right)
^{\bot }$ and to refer to the above-established relation between the
deficiency indices of a maximal symmetric operator and its self-adjointness:
a maximal symmetric operator is s.a. iff the both its deficient indices are
equal to zero.

The presented consideration seems more direct.

A s.a. extension of a symmetric operator $\hat{f}$ with equal deficiency
indices, i.e., with isomorphic deficient subspaces $\aleph _{\bar{z}}$ and $%
\aleph _{z},$ the extension specified by an isometry $\hat{U}:\aleph _{\bar{z%
}}\rightarrow \aleph _{z}$ and given by formulas (\ref{3a.22}) and (\ref%
{3a.23}) with $D_{U}=\aleph _{\bar{z}}$ and $\hat{U}D_{U}=\aleph _{z}\,,$
can be equivalently defined in terms of the sesquilinear asymmetry form $%
\omega _{\ast }$ similarly to the closure $\overline{\hat{f}}\,,$ see
formulas (\ref{3a.12}) and (\ref{3a.13}). Namely, $\hat{f}_{U}$ is such an
extension iff it is a restriction of the adjoint $\hat{f}^{+}$ to the domain 
$D_{f_{U}}$ that is defined by the linear equation%
\begin{equation}
\omega _{\ast }\left( \eta _{z}+\hat{U}\eta _{z},\xi _{U}\right) =0\,,\;\xi
_{U}\in D_{f_{U}}\subset D_{f^{+}}\,,\;\forall \eta _{z}\in \aleph _{\bar{z}%
}\,.  \label{3a.29}
\end{equation}

Necessity. Let $\hat{f}_{U}$ be a s.a. extension of $\hat{f}$ . Then the
restriction of the form $\omega _{\ast }$ to its domain $D_{f_{U}}$
vanishes, see (\ref{3a.19}), $\omega _{\ast }\left( \eta _{U},\xi
_{U}\right) =0$,\ $\forall \xi _{U},\eta _{U}\in D_{f}\,.$ Using now the
representation $\eta _{U}=\underline{\eta }+\eta _{z}+\hat{U}\eta _{z}$ and
the equality $\omega _{\ast }\left( \underline{\eta },\xi _{U}\right) =0\,,$
see (\ref{3a.10}) with $\underline{\psi }=\underline{\eta }$ and $\xi _{\ast
}=\xi _{U}\,,$ we reduce this equation to (\ref{3a.29}).

Sufficiency. Let $\hat{U}:\aleph _{\bar{z}}\rightarrow \aleph _{z}$ be an
isometry of one of the deficient subspaces onto another. We consider linear
equation (\ref{3a.29}) for a subspace $D_{f_{U}}=\{\xi _{U}\}\subset
D_{f^{+}}$ and show that its general solution is%
\begin{equation}
\xi _{U}=\underline{\xi }+\xi _{z}+\hat{U}\xi _{z}\,,\;\forall \underline{%
\xi }\in D_{\bar{f}}\,,\;\forall \xi _{z}\in \aleph _{\bar{z}}\,,\;\hat{U}%
\xi _{z}\in \aleph _{\bar{z}}\,.  \label{3a.30}
\end{equation}%
Really, a vector $\xi _{U}$ of form (\ref{3a.30}) evidently satisfies eq. (%
\ref{3a.29}):%
\begin{equation*}
\omega _{\ast }\left( \eta _{z}+\hat{U}\eta _{z},\underline{\xi }+\xi _{z}+%
\hat{U}\xi _{z}\right) =2iy\left[ \left( \eta _{z},\xi _{z}\right) -\left( 
\hat{U}\eta _{z},\hat{U}\xi _{z}\right) \right] =0\,,
\end{equation*}%
where we use eq. (\ref{3a.17}) and the fact that $\hat{U}$ is an isometry.
Conversely, let a vector $\xi _{U}\in D_{f^{+}}$ satisfies eq. (\ref{3a.29}%
), then representing it as%
\begin{equation*}
\xi _{U}=\underline{\xi }+\xi _{z}+\xi _{\bar{z}}=\underline{\xi }+\xi _{z}+%
\hat{U}\xi _{z}+\left( \xi _{\bar{z}}-\hat{U}\xi _{z}\right) \,,\;\underline{%
\xi }\in D_{\bar{f}}\,,\;\xi _{z}\in \aleph _{\bar{z}}\,,\;\xi _{\bar{z}},%
\hat{U}\xi _{z}\in \aleph _{z}\,,
\end{equation*}%
and using again formulas (\ref{3a.17}) and the isometricity of $\hat{U}$, we
reduce eq. (\ref{3a.29}) to $\left( \hat{U}\eta _{z},\xi _{\bar{z}}-\hat{U}%
\xi _{z}\right) =0\,,\;\forall \eta _{z}\in \aleph _{\bar{z}}\,,$ whence it
follows that $\xi _{\bar{z}}-\hat{U}\xi _{z}=0,$ or $\xi _{\bar{z}}=\hat{U}%
\xi _{z}\,,$ because the subspace $\left\{ \hat{U}\eta _{z},\forall \eta
_{z}\in \aleph _{\bar{z}}\right\} =\hat{U}\aleph _{\bar{z}}=\aleph _{z}\,.$

Actually, eq. (\ref{3a.29}) is the defining equation for the adjoint $\hat{f}%
_{U}^{+}$ of the operator $\hat{f}_{U}$ that is the restriction of the
adjoint $\hat{f}^{+}$ to the domain $D_{f_{U}}=D_{\bar{f}}+\left( \hat{I}+%
\hat{U}\right) \aleph _{\bar{z}}\,,$ the equation that we already encounter
above, see eq. (\ref{3a.25}), where the substitutions $\xi _{U}\rightarrow
\eta _{U}$ and $\eta _{\ast U}\rightarrow \xi _{U}$ must be made. Its
solution in the case where $\hat{U}\aleph _{\bar{z}}=\aleph _{z}$ shows that 
$\hat{f}_{U}^{+}=\hat{f}_{U}\,.$

In the case of a symmetric operator $\hat{f}$ with equal and finite
deficiency indices, $m_{+}=m_{-}=m<\infty ,$ an isometry $\hat{U}:\aleph _{%
\bar{z}}\rightarrow \aleph _{z},$ and thereby a s.a. extension $\hat{f}_{U}$%
, can be specified by a unitary $m\times m$ matrix. For this purpose, we
choose some orthobasis $\left\{ e_{z,k}\right\} _{1}^{m}$ in $\aleph _{\bar{z%
}},$ such that any vector $\xi _{z}\in \aleph _{\bar{z}}$ is represented as $%
\xi _{z}=\sum_{k=1}^{m}c_{k}e_{z,k}\,,\;c_{k}\in \mathbb{C\,},$ and some
orthobasis $\left\{ e_{\bar{z},l}\right\} _{1}^{m}$ in $\aleph _{z}\,.$ Then
any isometric operator $\hat{U}$ with the domain $\aleph _{\bar{z}}$ and the
range $\aleph _{z}$ is given by%
\begin{equation*}
\hat{U}e_{z,k}=\sum_{k=1}^{m}U_{lk}e_{\bar{z},l}\,,\;\mathrm{or\;}\hat{U}\xi
_{z}=\sum_{l=1}^{m}\left( \sum_{k=1}^{m}U_{lk}c_{k}\right) e_{\bar{z},l}\,,
\end{equation*}%
where $U=||U_{lk}||\,,\;l,k=1,...,m,$ is a unitary matrix. Conversely, any
unitary $m\times m$ matrix $U$ defines an isometry $\hat{U}$ given by the
above formulas. It is evident that for a given $\hat{U}$, the matrix $U$
changes appropriately with the change of the orthobasises$\left\{
e_{z,k}\right\} _{1}^{m}$ and $\left\{ e_{\bar{z},l}\right\} _{1}^{m}.$

It follows that in the case under consideration, the family $\left\{ \hat{f}%
_{U}\right\} $ of all s.a. extensions of a given symmetric operator $\hat{f}$
is a manifold of dimension $m^{2}$ that is a unitary group $U\left( m\right)
.$

This result can be extended to the case of infinite deficiency indices, $%
m=\infty ,$ but with a special assignment of a meaning for the indices $l$
and $k$ ranging from $1$ to $\infty $.

Because in the case where the both deficiency indices coincide, there is no
difference in the choice $z\in \mathbb{C}_{+}$ or $z\in \mathbb{C}_{-}\,,$
we take $z\in \mathbb{C}_{+}\,,$ i.e., $z=x+iy,\;y>0,$ in what follows, such
that from now on, $m_{+}=\dim \aleph _{\bar{z}}\,$and $m_{-}=\dim \aleph
_{z}\,.$

We now summarize all the relevant previous results in a theorem. This
theorem is of paramount importance: it is just what we need from mathematics
for our physical purposes. We therefore present the main theorem and the
subsequent comments in great detail, in fact, in an independent
self-contained way for ease of using without any further references.

\begin{theorem}
\label{t3a.3}(The main theorem) Let $\hat{f}$ be an (in general nonclosed)
symmetric operator with a domain $D_{f}$ in a Hilbert space $\mathcal{H},$ $%
\hat{f}\subseteq \hat{f}^{+},$ where $\hat{f}^{+}$ is the adjoint, let $%
\aleph _{\bar{z}}$ and $\aleph _{z}$ be the deficient subspaces of $\hat{f}$%
\thinspace ,%
\begin{equation*}
\aleph _{\bar{z}}=\ker \left( \hat{f}^{+}-z\hat{I}\right) =\left\{ \xi _{z}:%
\hat{f}^{+}\xi _{z}=z\xi _{z}\right\}
\end{equation*}%
and%
\begin{equation*}
\aleph _{z}=\ker \left( \hat{f}^{+}-\bar{z}\hat{I}\right) =\left\{ \xi _{%
\bar{z}}:\hat{f}^{+}\xi _{\bar{z}}=\bar{z}\xi _{\bar{z}}\right\} \,,
\end{equation*}%
where $z$ is an arbitrary, but fixed, complex number in the upper
half-plane, $z=x+iy,\;y>0,$ and let $m_{+}$ and $m_{-}$ be the deficiency
indices of $\hat{f},$%
\begin{equation*}
m_{+}=\dim \aleph _{\bar{z}}\,,\;m_{-}=\dim \aleph _{z}\,,
\end{equation*}%
$m_{+}$ and $m_{-}$ are independent of $z.$

The operator $\hat{f}$ has s.a. extensions $\hat{f}_{U}=$ $\hat{f}%
_{U}^{+}\,, $ $\hat{f}\subseteq $ $\hat{f}_{U}\,,$ iff the both its
deficient subspaces $\aleph _{\bar{z}}$ and $\aleph _{z}$ are isomorphic and
are therefore of the same dimension, i.e., iff its deficiency indices $m_{+}$
and $m_{-}$ are equal, $m_{+}=m_{-}=m$ .

If the deficient subspaces are trivial, $\aleph _{\bar{z}}=\aleph
_{z}=\left\{ 0\right\} ,$ i.e., if the both deficiency indices $m_{+}$ and $%
m_{-}$ are equal to zero, $m_{+}=m_{-}=0,$ the operator $\hat{f}$ is
essentially s.a., and its unique s.a. extension is its closure $\overline{%
\hat{f}}=\left( \hat{f}^{+}\right) ^{+}$ which coincides with its adjoint, $%
\overline{\hat{f}}=\left( \overline{\hat{f}}\right) ^{+}=\hat{f}^{+}$%
\thinspace .

If the deficient subspaces are nontrivial, i.e., if the deficiency indices
are different from zero, $m\neq 0,$ there exists an $m^{2}$-parameter family 
$\left\{ \hat{f}_{U}\right\} $ of s.a. extensions that is the manifold $%
U\left( m\right) $ , the unitary group.

Each s.a. extension $\hat{f}_{U}$ is defined by an isometric mapping $\hat{U}%
:\aleph _{\bar{z}}\rightarrow \aleph _{z}$ of one of the deficient subspaces
onto another, which establishes an isomorphism between the deficient
subspaces, and is given by 
\begin{equation}
D_{f_{U}}=D_{\bar{f}}+\left( \hat{I}+\hat{U}\right) \aleph _{\bar{z}%
}=\left\{ \xi _{U}=\underline{\xi }+\xi _{z}+\hat{U}\xi _{z},\forall 
\underline{\xi }\in D_{\bar{f}},\;\forall \xi _{z}\in \aleph _{\bar{z}},%
\text{\ }\hat{U}\xi _{z}\in \aleph _{z}\right\}  \label{3a.31}
\end{equation}%
where $D_{\bar{f}}$ is the domain of the closure $\overline{\hat{f}}\,$,\
and 
\begin{equation}
\hat{f}_{U}\xi _{U}=\overline{\hat{f}}\underline{\xi }+z\xi _{z}+\bar{z}%
\hat{U}\xi _{z}\,.  \label{3a.32}
\end{equation}

Conversely, any isometry $\hat{U}:\aleph _{\bar{z}}\rightarrow \aleph _{z}$
that establishes an isomorphism between the deficient subspaces defines a
s.a. extension $\hat{f}_{U}$ of $\hat{f}$ given by (\ref{3a.31}) and (\ref%
{3a.32}).

The s.a. extension $\hat{f}_{U}$ can be equivalently defined as a s.a.
restriction of the adjoint $\hat{f}^{+}$:%
\begin{equation}
\hat{f}_{U}:\left\{ 
\begin{array}{l}
D_{f_{U}}=\left\{ \xi _{U}\in D_{f^{+}}\,:\,\omega _{\ast }\left( \eta _{z}+%
\hat{U}\eta _{z},\xi _{U}\right) =0\,,\;\forall \eta _{z}\in \aleph _{\bar{z}%
}\right\} \,, \\ 
\hat{f}_{U}\xi _{U}=\hat{f}^{+}\xi _{U}\,.%
\end{array}%
\right.  \label{3a.33}
\end{equation}

If the deficient subspaces are finite-dimensional, i.e., if the deficiency
indices of $\hat{f}$ are finite, $0<m<\infty ,$ the s.a. extensions $\hat{f}%
_{U}$ are specified in terms of unitary matrices $U\in U\left( m\right) .$
Namely, let $\left\{ e_{z,k}\right\} _{1}^{m}$ and $\left\{ e_{\bar{z}%
,l}\right\} _{1}^{m}$ be some orthobasises in the respective deficient
subspaces $\aleph _{\bar{z}}$ and $\aleph _{z}$, then a s.a. extension $\hat{%
f}_{U}$ is defined by 
\begin{equation}
D_{f_{U}}=\left\{ \xi _{U}=\underline{\xi }+\sum_{k=1}^{m}c_{k}\left(
e_{z,k}+\sum_{l=1}^{m}U_{lk}e_{\bar{z},l}\right) \,,\;\forall \underline{\xi 
}\in D_{\bar{f}}\,,\;\forall c_{k}\in \mathbb{C}\right\} \,,  \label{3a.34}
\end{equation}%
and 
\begin{equation}
\hat{f}_{U}\xi _{U}=\overline{\hat{f}}\underline{\xi }+\sum_{k=1}^{m}c_{k}%
\left( ze_{z,k}+\bar{z}\sum_{l=1}^{m}U_{lk}e_{\bar{z},l}\right) \,,
\label{3a.35}
\end{equation}%
where\ $U=\left\| U_{lk}\right\| ,$ $l,k=1,...,m,$ is a unitary matrix.

The equivalent definition of $\hat{f}_{U}$ in terms of the adjoint $\hat{f}%
^{+}$ becomes%
\begin{equation}
\hat{f}_{U}\,:\left\{ 
\begin{array}{l}
D_{f_{U}}=\left\{ \xi _{U}\subset D_{f^{+}}:\omega _{\ast }\left(
e_{z,k}+\sum_{l=1}^{m}U_{lk}e_{\bar{z},l},\xi _{U}\right)
=0,\;k=1,...,m\right\} , \\ 
\hat{f}_{U}\xi _{U}=\hat{f}^{+}\xi _{U}\,.%
\end{array}%
\right. ,  \label{3a.36}
\end{equation}
\end{theorem}

Theorem \ref{t3a.3} finishes our exposition of the general theory of s.a.
extensions of symmetric operators. However, we would like to give some
comments and remarks of practical importance, without being afraid of
repeating ourselves, and to end this section with some practical
``instructions'', following from the general theory, for a quantizing
physicist.

\subsection{Comments and remarks}

\textbf{Comment 1:} In the case of finite-dimensional deficient subspaces of
equal dimensions, $0<m<\infty $, any maximal symmetric extension of a
symmetric operator $\hat{f}$ is s.a., while in the case of
infinite-dimensional deficient subspaces, there exists a possibility of both
s.a. and maximal non-s.a. extensions.

If the deficient indices of a symmetric operator $\hat{f}$ are nonequal,
then there exist no s.a. extensions of $\hat{f}$ .

\textbf{Comment 2:} Of course, s.a. extensions can be equivalently defined
in terms of isometric mappings of the deficient subspace $\aleph _{z}$ onto $%
\aleph _{\bar{z}}$. In the previous terms, they are described by isometric
operators $\hat{U}^{-1}$ and matrices $U^{-1}=\left\| \overline{U}%
_{kl}\right\| $.

\textbf{Comment 3:} The isometries $\hat{U}:\aleph _{\bar{z}}\rightarrow
\aleph _{z}$ in (\ref{3a.31}), (\ref{3a.32}), and (\ref{3a.33}) that define
s.a. extensions $\hat{f}_{U}$ of a symmetric operator $\hat{f}$ depend on $z$%
, as well as the deficient subspaces; for a given s.a. extension, they
change with changing $z$. The same is true for the matrix $U=\left\|
U_{lk}\right\| $ in (\ref{3a.34}), (\ref{3a.35}), and (\ref{3a.36}) in the
case of finite deficient indices, $0<m<\infty $. In addition, for a given
s.a. extension, this matrix changes in an obvious manner with a change of
the respective basises in the deficient subspaces\footnote{%
We emphasize once again that any s.a. extension is contained in the family
of s.a. extensions constructed with a chosen $z$ and certain orthobasises in 
$\aleph _{\bar{z}}$ and $\aleph _{z}.$}.

\textbf{Comment 4:} The last comment is a more extensive comment concerning
a possible application of the general theory of s.a. extensions of symmetric
operators to physical problems of quantization, namely, to a definition of
quantum-mechanical observables as s.a. operators. We give it a form of some
``instructions''. They are generally applied to both quantum mechanics and
quantum field theory. But here, we mainly address to the case where
observables are represented by differential operators, as in nonrelativistic
and relativistic quantum mechanics of particles, especially having in mind
physical systems with boundaries and/or singularities of interaction
(potentials) (the position of singularities can coincide with boundaries),
we call such systems nontrivial systems. As to differential operators, the
''instructions'' to follow are of a preliminary nature; a more detailed
discussion of s.a. differential operators is given in the next sec.3.

A ``preliminary candidate'' to an observable, supplied, for example, by the
canonical quantization rules for a classical observable $f(q,p)$), is
usually a formal expression like $f(\hat{q},\hat{p}),$ or more specifically,
a formal ``differential expression''\footnote{%
All notions written in inverted commas are defined more precisely in the
next section.}, $f(x,-i\hbar d/dx)$, that is ``s.a.'' only from a purely
algebraic standpoint, within a formal algebra of symbols $\hat{q}=x$ and $%
\hat{p}=-i\hbar d/dx$ with involution. But as we incessantly repeat, such an
expression is only a ''rule'' and is not an operator unless its domain in an
appropriate Hilbert space is indicated. As to differential expressions, in a
physical literature, in particular, in many textbooks on quantum mechanics
for physicists, such a differential expressions are considered a s.a.
differential operator in a Hilbert space of wave functions like $L^{2}(a,b)$
actually with an implicit assumption that its domain is the so called
''natural domain'' that allows the corresponding differential operations
within a given Hilbert space. But in the case of nontrivial systems, such a
differential operator is not only non-s.a., but even nonsymmetric. This
hidden defect can manifest itself when we proceed to the eigenvalue problem.
''Thus, with sufficiently singular potentials, the customary methods of
finding energy eigenvalues and eigenfunctions fail'' \cite{Case50}: an
unexpected indefiniteness in the choice of eigenfunctions or even
nonphysical complex eigenvalues can occur. The early history of quantum
mechanics knows such examples \cite{MotMa33,Tamm40,CorSc40}, which first led
to the apprehension that singular potentials '' do not fall into the formal
structure of the Schr\"{o}dinger equation and its conventional
interpretation'' \cite{Case50}. It was later realized that some additional
requirements on the wave functions are needed, for example in the form of
specific boundary conditions.

The main mathematical and quantum-mechanical problem is to construct a
really s.a. operator in an appropriate Hilbert space starting from a
preliminary formally s.a. algebraic expression $f(\hat{q},\hat{p}),$ in
particular, differential expression $f(x,-i\hbar d/dx)$, or as we propose to
speak, a s.a. operator associated with a given formal differential
expression.

\textbf{1. The first step.}

The first step of a standard programme for solving this problem is to give
the meaning of a symmetric operator $\hat{f}$ in an appropriate Hilbert
space $\mathcal{H}$ to the formal expression by indicating its domain $%
D_{f}\subseteq \mathcal{H}$ which must be dense, $\overline{D_{f}}=\mathcal{H%
}.$ In the case of differential expressions and nontrivial systems, this is
usually achieved by choosing a domain $D_{f}$ in a Hilbert space of
functions (wave functions in the conventional physical terminology) like $%
L^{2}(a,b)$ such that it avoids the problems associated with boundaries and
singularities by the requirement that wave functions in $D_{f}$ vanish fast
enough near the boundaries and singularities. The symmetricity of $\hat{f}$
is then easily verified by integrating by parts.

\textbf{2. The second step.}

We then must evaluate the adjoint $\hat{f}^{+}$, i.e., to find its ``rule''
and its domain $D_{f^{+}}\supseteq D_{f}\,,$ solving the defining equation
for $\hat{f}^{+}$. Generally, this is a nontrivial task. Fortunately, as to
differential operators, the solution for a rather general symmetric
operators is known in the mathematical literature, see, for example, \cite%
{Stone32,Naima69,AkhGl81,ReeSi72,Richt78}. It usually appears that the
``rule'' for $\hat{f}^{+}$ does not change and is given by the same
differential expression\footnote{%
An exception is provided by $\delta $-like potentials.} $f(x,-i\hbar d/dx)$,
but its domain is larger and is a natural domain, such that $\hat{f}^{+}$ is
a real extension of the initial symmetric operator, $\hat{f}\subset \hat{f}%
^{+},$ the extension that is generally nonsymmetric.

\textbf{3. The third step.}

This step consists in evaluating the deficient subspaces $\aleph _{\bar{z}}$
and $\aleph _{z}$ with some fixed $z=x+iy$, $y>0,$ as the sets of solutions
of the respective (differential) equations $\hat{f}^{+}\xi _{z}=z\xi
_{z},\;\xi _{z}\in D_{f^{+}},$ and $\hat{f}^{+}\xi _{\bar{z}}=\bar{z}\xi _{%
\bar{z}},\;\xi _{\bar{z}}\in D_{f^{+}},$ and determining the deficiency
indices $m_{+}=\dim \aleph _{\bar{z}}$ and $m_{-}=\dim \aleph _{z}$. This
problem can also present a labour-intensive task, in the case of
differential operators, it usually requires an extensive experience in
special functions.

An important remark here is in order. As we already mentioned above, in the
mathematical literature, there is a tradition to take $z=i$ and $\bar{z}=-i$
(we remind a reader that all $z\in \mathbb{C}_{+}$ (or $z\in \mathbb{C}_{-}$%
) are equivalent). But in physics, a preliminary symmetric operator $\hat{f}$
and its adjoint $\hat{f}^{+}$ are usually assigned a certain dimension%
\footnote{%
In conventional units, a certain degree of length or momentum (or energy).}.
Therefore, it is natural to choose $z=\kappa i$ and $\bar{z}=-\kappa i$,
where $\kappa $ is an arbitrary, but fixed, constant parameter of the
corresponding dimension. In constructing a physical observable as s.a.
extension of a preliminary symmetric operators $\hat{f}$, this dimensional
parameter enters the theory. In particular, if preliminarily a theory has no
dimensional parameter that defines a scale, a naive scale invariance of the
theory can be broken after a specification of the observable.

Let the deficiency indices be found. If the deficiency indices appear
unequal, $m_{+}\neq m_{-}$, our work stops with the conclusion that there is
no quantum-mechanical analogue for the given classical observable $f(q,p)$.
Such a situation, nonequal deficiency indices, is encountered in physics
thus preventing some classical observables to be transferred to the quantum
level (an example is the momentum operator for a particle on a semi-axis,
see below). We note in advance that for differential operators with real
coefficients, the deficiency indices are always equal.

If the deficiency indices appear to be zero, $m_{+}=m_{-}=0$, our work also
stops: an operator $\hat{f}$ is essentially s.a. and a uniquely defined
quantum-mechanical observable is its closure $\overline{\hat{f}}$ that
coincides with the adjoint $\hat{f}^{+}$, $\overline{\hat{f}}=\hat{f}^{+}$.

If the deficiency indices appear to be equal and nonzero, $m_{+}=m_{-}=m>0$,
the fourth step follows.

\textbf{4. The fourth step.}

At this step, we correctly specify all the $m^{2}$-parameter family $\left\{ 
\hat{f}_{U}\right\} $ of s.a. extensions $\hat{f}_{U}$ of $\hat{f}$ in terms
of isometries $\hat{U}:\aleph _{-i\kappa }\rightarrow \aleph _{i\kappa }$ or
in terms of unitary matrices $U=\left\| U_{lk}\right\| $, $l,k=1,\ldots ,m$.
The general theory provides the two ways of specification given by the main
theorem. The specification based on formulas (\ref{3a.31}) and (\ref{3a.32}%
), or (\ref{3a.34}) and (\ref{3a.35}) (and usually presented in the
mathematical literature) seems more explicit in comparison with the
specification based on formulas (\ref{3a.33}) or (\ref{3a.36}), which
requires solving the corresponding linear equation for the domain $%
D_{f_{U}}\,.$ But the first specification assumes the knowledge of the
closure $\overline{\hat{f}}$ if the initial symmetric operator is nonclosed%
\footnote{%
We would like to stress that at this point the general theory requires
evaluating the closure $\overline{\hat{f}},$ it is exactly $\overline{\hat{f}%
}$ and $D_{\bar{f}}$ that enter formulas (\ref{3a.31}), (\ref{3a.32}), (\ref%
{3a.34}), and (\ref{3a.35}), while in the physical literature, we can
sometimes see that when citing and using these formulas, $\hat{f}$ and $%
D_{f} $ \ stand for $\overline{\hat{f}}$ and $D_{\bar{f}}$ \ even for
non-closed symmetric operator $\hat{f},$ which is incorrect.}, which
requires solving linear equations in (\ref{3a.12}), or (\ref{3a.14}), or in (%
\ref{3a.15}) for the domain $D_{\bar{f}}\,.$ The second specification can
sometimes become more economical because it avoids the evaluation of the
closure $\overline{\hat{f}}$ and directly deals with $D_{f_{U}}.$ This
specifically concerns the case of differential operators where $\hat{f}^{+}$
is usually given by the same differential expression as $\overline{\hat{f}}$
and where the second specification allows eventually specifying the s.a.
extensions $\hat{f}_{U}$ in the customary form of s.a. boundary conditions.
This possibility is discussed below in sec.3. We say in advance that in
sec.3, we also propose the third possible way of s.a. extensions of
symmetric differential operators directly in terms of, in general
asymptotic, boundary conditions.

In the physical literature, there is a convention to let $D_{+}$ denote the
deficient subspace $\aleph _{-i\kappa }=\ker \left( \hat{f}^{+}-i\kappa 
\hat{I}\right) $ and let $D_{-}$ denote the deficient subspace $\aleph
_{i\kappa }=\ker \left( \hat{f}^{+}+i\kappa \hat{I}\right) $, such that the
isometry $\hat{U}$ is now written as $\hat{U}:D_{+}\rightarrow D_{-}$. The
elements of the deficient subspaces $D_{+}$ and $D_{-}$ are respectively
denoted by\footnote{%
It is the sign in front of ``i'' in the latter formulas that defines the
subscript $+\,$\ or $-$ in $D$, see footnote 10.
\par
{}} $\xi _{+}$, $\hat{f}^{+}\xi _{+}=i\kappa \xi _{+}$, and $\xi _{-}$, $%
\hat{f}^{+}\xi _{-}=-i\kappa \xi _{-}$, and the orthobasises in $D_{+}$ and $%
D_{-}$ are respectively denoted by $\left\{ e_{+,\kappa }\right\} ^{m}$ and $%
\left\{ e_{-,\kappa }\right\} ^{m}$. In these terms, formulas (\ref{3a.31})
and (\ref{3a.34}), and formulas (\ref{3a.32}) and (\ref{3a.35}) that define
a s.a. extension $\hat{f}_{U}$ of an initial symmetric operator $\hat{f}$ in
the case of $m>0$ become 
\begin{align}
& D_{f_{U}}=D_{\bar{f}}+\left( \hat{I}+\hat{U}\right) D_{+}=\left\{ \xi _{U}=%
\underline{\xi }+\xi _{+}+\hat{U}\xi _{+},\;\forall \underline{\xi }\in D_{%
\bar{f}},\;\forall \xi _{+}\in D_{+},\;\hat{U}\xi _{+}\in D_{-}\right\} 
\notag \\
& =\left\{ \xi _{U}=\underline{\xi }+\sum_{k=1}^{m}c_{k}\left(
e_{+,k}+\sum_{l=1}^{m}U_{lk}e_{-,l}\right) ,\;\forall \underline{\xi }\in D_{%
\bar{f}}\,,\;\forall c_{k}\in \mathbb{C}\right\}  \label{3a.37}
\end{align}%
and 
\begin{equation*}
\hat{f}_{U}\xi _{U}=\overline{\hat{f}}\underline{\xi }+i\kappa \xi
_{+}-i\kappa \hat{U}\xi _{+}=\hat{f}\underline{\xi }+i\kappa
\sum_{k=1}^{m}c_{k}\left( e_{+,k}-\sum_{l=1}^{m}U_{lk}e_{-,l}\right) \,,
\end{equation*}%
while formulas (\ref{3a.33}) and (\ref{3a.36}) become 
\begin{equation}
\hat{f}_{U}\,:\left\{ 
\begin{array}{l}
D_{f_{U}}=\left\{ \xi _{U}\in D_{f^{+}}\,:\omega _{\ast }\left( \xi _{+}+%
\hat{U}\xi _{+},\xi _{U}\right) =0\,,\;\forall \xi _{+}\in D_{+}\right\} \\ 
=\left\{ \xi _{U}\in D_{f^{+}}\,:\omega _{\ast }\left(
e_{+,k}+\sum_{l=1}^{m}U_{lk}e_{-,l}\,,\xi _{U}\right)
=0\,,\;k=1,...,m\,\right\} \,, \\ 
\hat{f}_{U}\xi _{U}=\hat{f}^{+}\xi _{U}\,.%
\end{array}%
\right.  \label{3a.39}
\end{equation}

At last, we should not forget that an isometry $\hat{U}$ and matrices $%
U_{lk} $, as well as $D_{+}$ and $D_{-}\,,$ depend on the real parameter $%
\kappa ,$ and for the same s.a. extension, they change with changing $\kappa
.$

There is a slightly modified method of finding s.a. operators associated
with formally s.a. differential expressions $f(x,-i\hbar d/dx)$, see \cite%
{AkhGl81,Naima69}. This method differs from the above-described one by some
transpositions of steps 1 and 2 and partly of steps 3 and 4 . We actually
can start with the end of step 2, namely with an operator $\hat{f}^{\ast }$
given by the initial differential expression $\check{f}$ and defined in $%
L^{2}\left( a,b\right) $ on a subspace of all functions $\psi _{\ast }\left(
x\right) $ such that $\left( f(x,-i\hbar d/dx)\psi _{\ast }\right) \left(
x\right) $ also belongs to $L^{2}\left( a,b\right) .$ This is the most wide
``natural'' domain for such an operator. The operator $\hat{f}^{\ast }$ is
generally non-s.a. and even nonsymmetric. Then we evaluate its adjoint $%
\left( \hat{f}^{\ast }\right) ^{+}=\hat{f}$ and find that $\hat{f}$ is
symmetric and that $\hat{f}^{\ast }$ is really the adjoint of $\hat{f},$ $%
\hat{f}^{+}=\hat{f}^{\ast }.$ It follows that $\hat{f}$ is a closed
symmetric operator. After this, we can proceed to the steps 3-5.

The method was far developed for a wide class of differential operators,
especially for ordinary even-order differential operators with real
coefficients. Unfortunately, arbitrary odd-order or mixed deferential
operators practically remained apart (see, however \cite{Shin43,Everi63}).
In addition, this method is inapplicable to the physically interesting case
where the coefficient functions of a differential expression $f$ are
singular at the inner points of the interval $\left( a,b\right) $, an
example is a $\delta $-like potential, whereas the first method does work in
this case.

This method is rather a method of s.a. restrictions of an initial most
widely defined differential operator that is generally nonsymmetric, and all
the more s.a., than the method of s.a. extensions of an initial symmetric
operator. We note that, in fact, the conventional practice in physics
implicitly follow this method, but, so to say, in an ``extreme'' form.
Namely, a s.a. differential expression is considered a s.a. operator in
appropriate Hilbert space of functions with implicitly assuming that its
domain is the most wide natural domain. Therefore, the standard physical
practice is to directly proceed to finding its spectrum and eigenfunctions
as the solutions of the eigenvalue problem for the corresponding
differential equation. Sometimes, this approach works: the only requirements
of the square-integrability of eigenfunctions or their ``normalization to $%
\delta $-function'' appears sufficient. From the mathematical standpoint,
this means that the operator under consideration is really s.a., or from the
standpoint of the first method, that an initial symmetric operator is
essentially s.a.. To be true, it sometimes appears that some additional
specific boundary conditions or conditions near the singularities of the
potential on the wave functions are necessary for fixing the eigenfunctions.
In some cases, these boundary conditions are so natural that are considered
unique although this is not true. But in some cases, it appears that there
is no evident way of choosing between different possibilities, and this
becomes a problem for the quantum-mechanical treatment of the corresponding
physical system. From the mathematical standpoint, such a situation means
that the initial operator is nonsymmetric, or, from the standpoint of the
first method, that an initial symmetric operator is not essentially s.a. and
allows different s.a. extensions, if these are at all possible, and there is
no physical arguments in favor of a certain choice.

We return to this subject once more in the next section devoted to
differential operators.

\textbf{5. The final step.}

The final step is the standard spectral analysis, i.e., finding the spectrum
and eigenvectors of the obtained s.a. extensions $\hat{f}_{U}$ and their
proper physical interpretation, in particular, the explanation of the
possible origin and the physical meaning of the new $m^{2}$ parameters
associated with the isometries $\hat{U},$ or unitary matrices $U=||U_{lk}||,$
in the case where the deficiency indices are different from zero. The
problem of the physical interpretation of these additional parameters that
are absent in the initial formal (differential) expression $f$ and in the
initial symmetric operator $\hat{f}$ is sometimes a most difficult one. The
usual attempts to solve this problem are related to the search for an
appropriate regularization of singularities in $f$ and a change of
boundaries by finite walls.

The most ambitious programme is to change the initial singular
(differential) expression $f$ by a regular expression $f_{\mathrm{reg}}$
with $m^{2}$ parameters of regularization, such that the initial symmetric
operator $\hat{f}_{\mathrm{reg}}$ is essentially s.a., and then reproduce
all the s.a. extensions $\hat{f}_{U}$ of a singular problem as a certain
limit of the regularized s.a. operator under properly removing the
regularization. This procedure is like a well-known renormalization
procedure in QFT, and the new $m^{2}$ parameters may be associated with
``conterterms''. Of course, the regularization can be partial if some
singularities and arbitrariness associated with them are well-interpreted.
In many cases, this problem remains unsolved.

The above-described general procedure for constructing quantum-mechanical
observables starting from preliminary formal expressions is not universally
obligatory because in particular cases more direct procedures are possible,
especially if there exist additional physical arguments.

For example, in some cases we can guess a proper domain $D_{f}$ for initial
symmetric operator $\hat{f}$ such that $\hat{f}$ appears to be essentially
s.a. from the very beginning.

In other cases, it can happen that an initial symmetric operator $\hat{f}$
may be represented as $\hat{f}=``\hat{a}^{+}"\hat{a}+\hat{b}$, where an
operator $\hat{a}$ is densely defined and an operator $``\hat{a}^{+}"$ is
its formal ``adjoint'' and is also densely defined, actually, $``\hat{a}%
^{+}" $ is a restriction of the really adjoint $\hat{a}^{+},$ and $\hat{b}=%
\hat{b}^{+}$ is a bounded operator, in particular, a constant.

There is one remarkable criterion for self-adjointness that is directly
applicable to this case, we call it the Akhiezer-Glazman theorem (see \cite%
{AkhGl81}).

\begin{theorem}
\label{AGT}Let $\hat{a}$ be a densely defined closed operator, $\overline{%
D_{a}}=\mathcal{H}\,,\;\hat{a}=\overline{\hat{a}}\,,$ therefore the adjoint $%
\hat{a}^{+}$ exists and is also densely defined. Then, the operator $\hat{f}=%
\hat{a}^{+}\hat{a}$ is s.a. , the same is true for the operator $\hat{g}=%
\hat{a}\hat{a}^{+}\,.$
\end{theorem}

This theorem must seem evident for physicists by the example of the harmonic
oscillator Hamiltonian. A subtlety is that $\hat{a}$ must be closed.

Based on the Akhiezer--Glazman theorem, we have at least one s.a. extension $%
\widetilde{\hat{f}}=\left( \widetilde{\hat{f}}\right) ^{+}$ of the initial
symmetric operator $\hat{f}$, given by $\widetilde{\hat{f}}=\hat{a}^{+}%
\overline{\hat{a}}+\hat{b}=\left( \overline{\hat{a}}\right) ^{+}\overline{%
\hat{a}}+\hat{b}=\left( \widetilde{\hat{f}}\right) ^{+}\,,$ where $\overline{%
\hat{a}}$ is the closure of $\hat{a}$. This extension may be nonunique, but
its existence guarantees that the deficiency indices of $\hat{f}$ are equal,
and we can search for other s.a. extensions of $\hat{f}$ without fail.

\subsection{Illustration by example of momentum operator}

To illustrate the above-given general scheme, we consider a simplest
one-dimensional quantum-mechanical system, a spinless particle moving on an
interval $\left( a,b\right) $ of a real axis $\mathbb{R}^{1}$, and a
well-known observable in this system, the momentum operator. The interval
can be (semi)open or closed, the ends $a$ and $b$ can be infinities $\left(
-\infty \;\mathrm{or}\;+\infty \right) .$ For the space of states of the
system, we conventionally take the Hilbert space $L^{2}\left( a,b\right) $
whose vectors are wave functions $\psi \left( x\right) ,\;x\in \left(
a,b\right) $ (the $x$-representation). If we set the Planck constant $\hbar $
to be unity, $\hbar =1,$ then the standard well-known expression for the
momentum operator is $\hat{p}=-id/dx\,.$ But as we now realize, for the
present, this formally s.a. ''operator'' is only a preliminary differential
expression\footnote{%
In what follows, we distinguish formal differential expressions from
operators by an inverted hat $\vee ,$ see sec.3.}%
\begin{equation}
\check{p}=-i\frac{d}{dx}  \label{3b.9}
\end{equation}%
because its domain is not prescribed in advance (by the known) quantization
rules. The problem of quantization in this particular case is to construct a
s.a. operator, an observable, associated with this differential expression.
It turns out that the solution of this problem crucially depends on the type
of the interval: whether it is a whole real axis, $\left( a,b\right) =\left(
-\infty ,+\infty \right) =\mathbb{R}^{1}\,,$ or a semiaxis $\left(
a,b\right) =[0,\infty )$ ($a$ is taken to be zero for convenience, it can be
any finite number) or $\left( a,b\right) =(-\infty ,0],$ or a finite segment%
\footnote{%
Because the finite ends of an interval have a zero measure, we can include
(or exclude) the finite ends in the interval. $L^{2}\left( \left( a,b\right)
\right) $ and $L^{2}\left( \left[ a,b\right] \right) $ are the same.} $\left[
a,b\right] ,\;-\infty <a<b<\infty .$

A most wide natural domain for a linear operator defined in $L^{2}\left(
a,b\right) $ and given by the differential operation $-id/dx$ is the
subspace $D_{\ast }$ of wave functions $\psi _{\ast }\left( x\right) \in
L^{2}\left( a,b\right) $ that are absolutely continuous on $\left(
a,b\right) ,$ the term ``on'' implies continuity up to the finite end or
ends of the interval $\left( a,b\right) ,$ and such that their derivative $%
\psi _{\ast }^{\prime }\left( x\right) $ also belongs\footnote{%
Of course, we could extend $D_{\ast }$ by step-functions that are also
differentiable almost everywhere, but then there would be no possibility for
integrating by parts and no chance for the symmetricity of the corresponding
operator.} to $L^{2}\left( a,b\right) .$ We let $\hat{p}^{\ast }$ denote
this operator, the above notation is justified below. The operator $\hat{p}%
^{\ast }$ is thus defined by\footnote{%
In what follows, we use the abbreviation (\textrm{a.c}.=is absolutely
continuous).}%
\begin{equation}
\hat{p}^{\ast }:\left\{ 
\begin{array}{l}
D_{p_{\ast }}=D_{\ast }=\left\{ \psi _{\ast }:\psi _{\ast }\;\mathrm{%
a.c.\;on\;}\left( a,b\right) ;\,\psi _{\ast },\psi _{\ast }^{\prime }\in
L^{2}\left( a,b\right) \right\} \,, \\ 
\hat{p}^{\ast }\psi _{\ast }=\check{p}\psi _{\ast }=-i\psi _{\ast }^{\prime
}\,.%
\end{array}%
\right.  \label{3a.40}
\end{equation}

We first check the symmetricity of this operator (i.e., whether the equality 
$\left( \chi _{\ast },\hat{p}^{\ast }\psi _{\ast }\right) -\left( \hat{p}%
^{\ast }\chi _{\ast },\psi _{\ast }\right) =0$ holds for any $\psi _{\ast
},\,\chi _{\ast }\in D_{_{\ast }}$) and consider the difference 
\begin{equation}
\omega _{\ast }\left( \chi _{\ast },\psi _{\ast }\right) =\left( \chi _{\ast
},\hat{p}^{\ast }\psi _{\ast }\right) -\left( \hat{p}^{\ast }\chi _{\ast
},\psi _{\ast }\right) =-i\int_{a}^{b}dx\overline{\chi _{\ast }}\psi _{\ast
}^{\prime }-i\int_{a}^{b}dx\overline{\chi _{\ast }^{\prime }}\psi _{\ast
},\;\forall \psi _{\ast },\chi _{\ast }\in D_{_{\ast }}\,.  \label{3a.41}
\end{equation}%
A reader easily recognizes the sesquilinear asymmetry form of the operator $%
\hat{p}^{\ast }$ in $\omega _{\ast }$. Integrating by parts in the second
term, we find 
\begin{equation}
\omega _{\ast }\left( \chi _{\ast },\psi _{\ast }\right) =\left. \left[ \chi
_{\ast },\psi _{\ast }\right] \right| _{a}^{b}=\left[ \chi _{\ast },\psi
_{\ast }\right] \left( b\right) -\left[ \chi _{\ast },\psi _{\ast }\right]
\left( a\right) \,,  \label{3a.42}
\end{equation}%
where we introduce a local sesquilinear form $\left[ \chi _{\ast },\psi
_{\ast }\right] $ defined by 
\begin{equation}
\left[ \chi _{\ast },\psi _{\ast }\right] =-i\overline{\chi _{\ast }\left(
x\right) }\psi _{\ast }\left( x\right) \,,  \label{3a.43}
\end{equation}%
and where $\left[ \chi _{\ast },\psi _{\ast }\right] \left( a\right) $ and $%
\left[ \chi _{\ast },\psi _{\ast }\right] \left( b\right) $ are the
respective limits of this form as $x\rightarrow b,a$, 
\begin{equation}
\left[ \chi _{\ast },\psi _{\ast }\right] \left( a\right)
=\lim_{x\rightarrow a}\left[ \chi _{\ast },\psi _{\ast }\right] \left(
x\right) ,\;\left[ \chi _{\ast },\psi _{\ast }\right] \left( b\right)
=\lim_{x\rightarrow b}\left[ \chi _{\ast },\psi _{\ast }\right] \left(
x\right) \,.  \label{3a.44}
\end{equation}%
We call these limits the boundary values of the local form, or simply
boundary term. These limits certainly exist because the integrals in r.h.s.
of (\ref{3a.41}) do exist, they are the sesquilinear forms in the
(asymptotic) boundary values of the wave functions in $D_{\ast }$. Eqs. (\ref%
{3a.42},\ref{3a.43}) manifest that the sesquilinear asymmetry form of the
differential operator $\hat{p}^{\ast }$ is reduced to the boundary values of
the local form and the asymmetricity of $\hat{p}^{\ast }$ is defined by the
asymptotic boundary values of the wave functions in $D_{\ast }$ . At the
moment, we have no ideas on the values of $\left[ \chi _{\ast },\psi _{\ast }%
\right] \left( \pm \infty \right) $ in the case of infinite intervals. We
must note that in the physical literature we can meet the assertion that the
square-integrability of $\psi \left( x\right) $ at infinity, for example, $%
\psi \in L^{2}\left( -\infty ,+\infty \right) \,,\;\int_{-\infty }^{+\infty
}dx\left| \psi \right| ^{2}<\infty \,,$ implies that $\psi $ vanishes at
infinity, $\psi \left( x\right) \rightarrow 0$ as $x\rightarrow \pm \infty $.

This is incorrect:\ it is a simple exercise to find a continuous function
that is square-integrable at infinity but can take arbitrarily large values
at arbitrarily large $x$. (To be true, in the following section we show that 
$\left[ \chi _{\ast },\psi _{\ast }\right] \left( \pm \infty \right) =0$
because $\psi _{\ast }^{\prime }$ is also square-integrable.)

On the other hand, what we certainly know is that in the case where one or
both ends of an interval $\left( a,b\right) $ are finite, for example, $%
\left| a\right| <\infty $ and/or $\left| b\right| <\infty $, we generally
have $\left[ \chi _{\ast },\psi _{\ast }\right] \left( a\right) =-i\overline{%
\chi _{\ast }}\left( a\right) \psi _{\ast }\left( a\right) \neq 0$ and/or $%
\overline{\chi _{\ast }}\left( b\right) \psi _{\ast }\left( b\right) \neq 0$%
, which implies that the operator $\hat{p}^{\ast }$ in this case is
nonsymmetric.

There are two conclusions from this preliminary (perhaps excessively
detailed and seemingly boring) consideration of this simple example, the
conclusions that prove to be valid for more general differential
expressions. First, a natural domain for a differential expression does not
provide a symmetric operator in the case of finite boundaries. Second, the
asymmetry form of a formally s.a. differential operator is defined by the
boundary terms. It is a sesquilinear form in the (asymptotic) boundary
values of functions involved (and their derivatives in the case of
differential operators of higher order). Therefore, in order to guarantee
the existence an initial symmetric operator associated with a given
differential expression, it is necessary to take a more restricted domain of
functions vanishing fast enough at the boundaries (and singularities) and
yielding no contributions to the boundary terms.

In our case, we therefore restart with a domain $D\left( a,b\right) $ of
finite smooth functions\footnote{%
This choice may seem too cautious in our case; however, $D\left( a,b\right) $
allows a universal consideration of symmetric operators with smooth
coefficients of arbitrary order \ (see the following section).}, $\overline{%
D\left( a,b\right) }=L^{2}\left( a,b\right) $, and respectively with a
symmetric operator $\hat{p}^{\left( 0\right) }$ defined by%
\begin{equation}
\hat{p}^{\left( 0\right) }:\left\{ 
\begin{array}{l}
D_{p^{\left( 0\right) }}=D\left( a,b\right) =\left\{ \varphi \left( x\right)
:\,\varphi \in C^{\infty }\,,\;\mathrm{supp\,}\varphi \subset \left(
a,b\right) \right\} \,, \\ 
\hat{p}^{\left( 0\right) }\varphi =\check{p}\varphi =-i\varphi ^{\prime }\,.%
\end{array}%
\right.  \label{3a.45}
\end{equation}%
The operator $\hat{p}^{\left( 0\right) }$ is a restriction of the operator $%
\hat{p}^{\ast }$ to $D\left( a,b\right) $ and is evidently symmetric:\ the
boundary terms $\left. \left[ \chi ,\psi \right] \right| _{a}^{b}$ vanishes
for any $\chi ,\varphi \in D\left( a,b\right) $ because of the requirements
on the support of functions in $D\left( a,b\right) $: they must vanish in a
vicinity of the boundaries.

The first step of the general programme is thus completed.

We now must evaluate the adjoint $\left( \hat{p}^{\left( 0\right) }\right)
^{+}$. The defining equation for a pair $\psi _{\ast }\in D_{\left(
p^{\left( 0\right) }\right) ^{+}}$ and $\chi _{\ast }=\left( \hat{p}^{\left(
0\right) }\right) ^{+}\psi _{\ast }$,%
\begin{equation*}
\left( \psi _{\ast },\hat{p}^{\left( 0\right) }\varphi \right) -\left( \chi
_{\ast },\varphi \right) =0\,,\;\forall \varphi \in D_{p^{\left( 0\right)
}}=D\left( a,b\right) \,,
\end{equation*}%
is%
\begin{equation}
i\int_{a}^{b}dx\overline{\psi _{\ast }}\varphi ^{\prime }+\int_{a}^{b}dx%
\overline{\chi _{\ast }}\varphi =0\,,\;\forall \varphi \in D\left(
a,b\right) \,.  \label{3a.46}
\end{equation}%
We solve it using the following observation. We introduce an absolutely
continuous function%
\begin{equation}
\widetilde{\psi _{\ast }}\left( x\right) =i\int_{c}^{x}d\xi \chi _{\ast
}\left( \xi \right) \,,\;a\leq c\leq b  \label{3a.47}
\end{equation}%
such that $\chi _{\ast }=-i\widetilde{\psi _{\ast }}^{\prime }\,.$
Substituting (\ref{3a.47}) in (\ref{3a.46}) and integrating by parts in the
second term, we reduce eq. (\ref{3a.46}) to%
\begin{equation*}
\int_{a}^{b}dx\left( \psi _{\ast }-\widetilde{\psi _{\ast }}\right) \varphi
^{\prime }=0\,,\;\forall \varphi \in D\left( a,b\right) \,.
\end{equation*}%
(the boundary terms vanish because of $\varphi \left( x\right) $). By the
known du Boi--Reymond lemma, it follows that $\psi _{\ast }-\widetilde{\psi
_{\ast }}=c=\mathrm{const\,},$ or%
\begin{equation}
\psi _{\ast }\left( x\right) =i\int_{c}^{x}d\xi \chi _{\ast }\left( \xi
\right) +c\,,  \label{3a.49}
\end{equation}%
which implies that $\psi _{\ast }$ is absolutely continuous on $\left(
a,b\right) $ and $\chi _{\ast }=\check{p}\psi _{\ast }=-i\psi _{\ast
}^{\prime }\,.$ Conversely, any such function given by (\ref{3a.49})
evidently satisfies the defining equation (\ref{3a.46}).

This means that the adjoint $\left( \hat{p}^{\left( 0\right) }\right) ^{+}$
coincides with the above-introduced operator $\hat{p}^{\ast }$ given by (\ref%
{3a.40}), i.e., it is given by the same differential expression (\ref{3b.9})
and its domain is a natural one.

The second step of the general programme is also completed.

We now must evaluate the deficient subspaces and deficiency indices. It is
this step where the difference in the type of the interval $\left(
a,b\right) $ manifests itself. The deficient subspaces $D_{\pm }$ are
defined by the differential equations%
\begin{equation*}
-i\psi _{\pm }^{\prime }\left( x\right) =\pm i\kappa \psi _{\pm }\left(
x\right) \,,\;\psi _{\pm }\in D_{\ast }\subset L^{2}\left( a,b\right) \,,
\end{equation*}%
\ $\kappa $ is an arbitrary, but fixed, parameter with the dimensionality of
inverse length. The respective general solutions of differential equations (%
\ref{3a.46}) by itself are%
\begin{equation}
\psi _{\pm }\left( x\right) =c_{\pm }e^{\mp \kappa x}\,,  \label{3a.51}
\end{equation}%
where $c_{\pm }\in \mathbb{C}$ are constants.

Let $\left( a,b\right) =\left( -\infty ,+\infty \right) =\mathbb{R}^{1}$,
then both $\psi _{\pm }$ in (\ref{3a.51}) are non-square-integrable, $\psi
_{+}$ is on $-\infty $ and $\psi _{-}\,$is on $+\infty $ unless $c_{\pm
}\neq 0$. Therefore, in this case, the deficient subspaces are trivial, $%
D_{\pm }=\left\{ 0\right\} $, and the deficiency indices are zero, $%
m_{+}=m_{-}=0$, and the operator $\left( \hat{p}^{\left( 0\right) }\right)
^{+}=\hat{p}^{\ast }$ (\ref{3a.40}) turns out to be symmetric (as we already
mentioned above,\ the corresponding boundary terms are equal to zero). The
operator $\hat{p}^{\left( 0\right) }$ (\ref{3a.45}) is thus essentially
s.a., and its unique s.a. extension is its closure, $\overline{\hat{p}%
^{\left( 0\right) }}=\hat{p}=\left( \hat{p}^{\left( 0\right) }\right) ^{+}=%
\hat{p}^{\ast }$; we let $\hat{p}$ denote the closure $\overline{\hat{p}%
^{\left( 0\right) }}$.

The conclusion is that in the case $\left( a,b\right) =\left( -\infty
,+\infty \right) ,$ there is only one s.a. operator associated with the
differential expression $\check{p}$ (\ref{3b.9}). Passing to the physical
language, we assert that for a spinless particle moving along the real axis $%
\mathbb{R}^{1},$ there is a unique momentum operator $\hat{p}$, an
observable given by (we actually rewrite (\ref{3a.40}))%
\begin{equation}
\hat{p}:\left\{ 
\begin{array}{l}
D_{p}=\left\{ \psi :\psi \;\mathrm{a.c.\;in\;}\left( -\infty ,+\infty
\right) ;\,\psi ,\,\psi ^{\prime }\in L^{2}\left( \mathbb{R}^{1}\right)
\right\} \,, \\ 
\hat{p}\psi =\check{p}\psi =-i\psi ^{\prime }\,.%
\end{array}%
\right.  \label{3a.52}
\end{equation}

A forth step is unnecessary. The spectrum, eigenfunctions and the physical
interpretation of this operator are well-known.

Let $\left( a,b\right) =[0,\infty )=\mathbb{R}_{+}^{1}$, a semiaxis, then $%
\psi _{+}$ in (\ref{3a.51}) is square-integrable, while $\psi _{-}$ is not,
unless $c_{-}=0$. We obtain that the deficiency indices of $\hat{p}^{\left(
0\right) }$ in this case are $m_{+}=1$ and $m_{-}=0$ (in the case of $\left(
a,b\right) =\left( -\infty ,0\right] $, they interchange). This implies that
in the case of a semiaxis, there is no s.a. operator associated with the
differential expression $\check{p}$ (\ref{3b.9}). In the physical language,
this means that for a particle moving on a semiaxis, the notion of momentum
as a quantum-mechanical observable is absent. In particular, this implies
the absence of the notion of radial momentum.

The general programme in the case of a semiaxis terminates at the third step.

Let now $\left( a,b\right) =\left[ a,b\right] $, $0<a<b<\infty $, a finite
segment, without loss of generality we take $\left[ a,b\right] =\left[ 0,l%
\right] $, $l<\infty $. Then both $\psi _{+}$ and $\psi _{-}$ in (\ref{3a.51}%
) are square-integrable. This implies that in the case of a finite interval,
the both deficient subspaces $D_{\pm }=\left\{ c_{\pm }e_{\pm }\left(
x\right) \right\} $, with $e_{+}=e^{-\kappa x}$ and $e_{-}=e^{-\kappa \left(
l-x\right) }$ being the respective basis vectors of the same norm, are
one-dimensional, such that the equal nonzero deficiency indices are $%
m_{+}=m_{-}=1$. According to the main theorem, this means that in the case
of a finite interval, we have a one-parameter $U\left( 1\right) $-family of
s.a. operators associated with the differential expression $\check{p}$ (\ref%
{3b.9}) (the group $U\left( 1\right) $ is a circle $\left\{ e^{i\theta
}\right\} ,$ $0\leq \theta \leq 2\pi ,\;0\sim 2\pi ,$ the symbol $\backsim $
is the symbol of equivalence, or identification), and the fourth step is
necessary.

We consider the both ways of specification given by the main theorem.

The first way requires evaluating the closure $\hat{p}=\overline{\hat{p}%
^{\left( 0\right) }}$ of $\hat{p}^{\left( 0\right) }$ (\ref{3a.45}), which
reduces to finding its domain $D_{p}$. Equivalent defining equations for $%
D_{p}$ are given in (\ref{3a.12}) and (\ref{3a.14}), or (\ref{3a.15}). We
use the defining equation in (\ref{3a.12}), which in our case is $\omega
_{\ast }\left( \psi _{\ast },\underline{\psi }\right) =0\,$,\ $\underline{%
\psi }\in D_{p}\,$,\ $\forall \psi _{\ast }\in D_{\ast }\,$. According to (%
\ref{3a.42}), (\ref{3a.43}), this equation reduces to%
\begin{equation*}
i\left. \left[ \psi _{\ast },\underline{\psi }\right] \right| _{0}^{l}=%
\overline{\psi _{\ast }\left( l\right) }\underline{\psi }\left( l\right) -%
\overline{\psi _{\ast }\left( 0\right) }\underline{\psi }\left( 0\right)
=0\,,\;\forall \psi _{\ast }\in D_{\ast }\,,
\end{equation*}%
a linear equation for the boundary values of functions in $D_{p}$. Because $%
\psi _{\ast }\left( 0\right) $ and $\psi _{\ast }\left( l\right) $ can take
arbitrary values independently, which, in particular, follows from
representation (\ref{3a.49}), this yields $\underline{\psi }\left( 0\right) =%
\underline{\psi }\left( l\right) =0\,.$

We obtain the same result considering the defining equation for $D_{p}$ in (%
\ref{3a.15}), because the determinant of the boundary values of the basis
vectors $e_{\pm }$ is nonzero,%
\begin{equation*}
\det \left( 
\begin{array}{cc}
e_{+}\left( l\right) & e_{+}\left( 0\right) \\ 
e_{-}\left( l\right) & e_{-}\left( 0\right)%
\end{array}%
\right) =e^{-2\kappa l}-1\neq 0\,.
\end{equation*}

The closure $\hat{p}$ is thus specified by additional zero boundary
conditions on the functions $\underline{\psi }$ in $D_{p}$ in comparison
with the functions $\psi _{\ast }$ in $D_{\ast }$ that can take arbitrary
boundary values:%
\begin{equation}
\hat{p}=\overline{\hat{p}^{\left( 0\right) }}:\left\{ 
\begin{array}{l}
D_{p}=\left\{ \underline{\psi }:\underline{\psi }\;\mathrm{a.c.\;on\;}%
[0,l];\,\underline{\psi },\underline{\psi }^{\prime }\in L^{2}\left(
0,l\right) ,\,\underline{\psi }\left( 0\right) =\underline{\psi }\left(
l\right) =0\right\} \,, \\ 
\hat{p}\underline{\psi }=\check{p}\underline{\psi }=-i\underline{\psi }%
^{\prime }\,.%
\end{array}%
\right.  \label{3a.53}
\end{equation}%
The isometries $\hat{U}:D_{+}\rightarrow D_{-}$ are given by a complex
number of unit module, $\hat{U}e_{+}=e^{i\theta }e_{-}\,$, and are labelled
by an angle $\theta $, $0\leq \theta \leq 2\pi $, $0$ $\backsim 2\pi $, $%
\hat{U}=\hat{U}\left( \theta \right) $. Respectively, the $U\left( 1\right) $%
-family $\left\{ \hat{p}_{\theta }\right\} $ of s.a. extensions of $\hat{p}%
^{\left( 0\right) }$ (\ref{3a.45}), and $\hat{p}$ (\ref{3a.53}), is given by%
\begin{equation}
\hat{p}_{\theta }:\left\{ 
\begin{array}{l}
D_{\theta }=D_{p_{\theta }}=\left\{ \psi _{\theta }=\underline{\psi }%
+c\left( e^{-\kappa x}+e^{i\theta }e^{-\kappa \left( l-x\right) }\right)
\right\} \,, \\ 
\hat{p}_{\theta }\psi _{\theta }=\check{p}\psi _{\theta }=-i\psi _{\theta
}^{\prime }\,,%
\end{array}%
\right.  \label{3a.54}
\end{equation}%
where\ $\underline{\psi }$\ is\ given\ by\ (\ref{3a.53}).

The second way of specification of s.a. extensions of $\hat{p}^{\left(
0\right) }$, and $\hat{p}$, requires solving the defining equation for $%
D_{p_{\theta }}$ in (\ref{3a.33}), or (\ref{3a.36}). In our case, this
equation, $\omega _{\ast }\left( e_{+}+e^{i\theta }e_{-},\psi _{\theta
}\right) =0,$ reduces to%
\begin{equation*}
\left. \left[ e_{+}+e^{i\theta }e_{-},\psi _{\theta }\right] \right|
_{0}^{l}=-i\left( e^{-\kappa l}+e^{-i\theta }\right) \psi _{\theta }\left(
l\right) +i\left( 1+e^{-i\theta }e^{-\kappa l}\right) \psi _{\theta }\left(
0\right) =0\,,
\end{equation*}%
the equation relating the boundary values of functions in $D_{\theta }$, and
yields%
\begin{equation}
\psi _{\theta }\left( l\right) =e^{i\vartheta }\psi _{\theta }\left(
0\right) \,,  \label{3a.55}
\end{equation}%
where the angle $\vartheta $ is%
\begin{equation}
\vartheta =\theta -2\arctan \left( \frac{\sin \theta }{e^{\kappa l}+\cos
\theta }\right) \,.  \label{3a.56}
\end{equation}%
The angle $\vartheta $ ranges from $0$ to $2\pi $ when $\theta $ goes from $%
0 $ to $2\pi $, $0\leq \vartheta \leq 2\pi $, $0\backsim 2\pi $, and is in
one-to-one correspondence with the angle $\theta $ (it is sufficient to show
that $\vartheta \left( \theta \right) $ is a monotonic function, $d\vartheta
/d\theta >0$); therefore, the angle $\vartheta $ equivalently labels the $%
U\left( 1\right) $-family of s.a. extensions, which we write as $\hat{p}%
_{\theta }=\hat{p}_{\vartheta }$.

Eq. (\ref{3a.55}) is an additional boundary condition for the functions $%
\psi _{\theta }=\psi _{\vartheta }$ in $D_{\theta }=D_{\vartheta }$ in
comparison with the functions $\psi _{\ast }\in D_{\ast }$. It is easy to
verify that this boundary condition is equivalent to the representation $%
\psi _{\theta }$ in (\ref{3a.54}); therefore, this boundary condition is a
s.a. boundary condition specifying the s.a. extensions as%
\begin{equation}
\hat{p}_{\vartheta }:\left\{ 
\begin{array}{l}
D_{\vartheta }=D_{p_{\vartheta }}=\left\{ \psi _{\vartheta }:\psi
_{\vartheta }\;\mathrm{a.c.\;on\;}\left[ 0,l\right] \,;\,\psi _{\vartheta
},\psi _{\vartheta }^{\prime }\in L^{2}(0,l);\,\psi _{\vartheta }\left(
l\right) =e^{i\vartheta }\psi _{\vartheta }\left( 0\right) \right\} \,, \\ 
\hat{p}_{\vartheta }\psi _{\vartheta }=\check{p}\psi _{\vartheta }=-i\psi
_{\vartheta }^{\prime }\,,%
\end{array}%
\right.  \label{3a.57}
\end{equation}%
where $0\leq \vartheta \leq 2\pi \,,\;0\backsim 2\pi \,.$ The second
specification seems more direct and explicit than the first one\footnote{%
Although in the general form of the main theorem this appears to be the
opposite.} because it specifies the s.a. extensions in the customary form of
s.a. boundary conditions that are more suitable for spectral analysis.

The conclusion is that for a particle moving on a finite segment $\left[ a,b%
\right] ,$ there is a one-parameter $U\left( 1\right) $-family (a circle) of
s.a. operators $\hat{p}_{\vartheta }=-id/dx$, that can be considered the
momentum of a particle. These operators are labelled by an angle $\vartheta $%
, and are specified by the s.a. boundary conditions $\psi _{\vartheta
}\left( l\right) =e^{i\vartheta }\psi _{\vartheta }\left( 0\right) $. In
short, the momentum operator for a particle on a finite segment is defined
nonuniquely.

The final step, the spectral analysis of these operators and the elucidation
of their physical meaning, is postponed to a special publication.

We now turn to the general s.a. differential operators (in terms of which
many observables in the quantum mechanics of particles are represented). We
only note in advance that many key points of the above consideration of the
momentum operator are characteristic for the general case.

\section{Differential operators}

\subsection{Introduction}

This section is devoted to differential operators, more specifically, to
constructing s.a. differential operators associated with formal s.a.
differential expressions\footnote{%
These notions are defined more precisely below.}. We try to make it as
self-contained as possible and therefore don't afraid to repeat some items
in the previous text. A reader who is acquainted with the end of the
pervious section will see that some of the key points and remarks of the
exposition to follow were already encountered in the above considerations.

We begin the section with remarks of the general character.

We restrict ourselves to ordinary differential operators in Hilbert spaces $%
L^{2}\left( a,b\right) $, $-\infty \leq a\leq b\leq \infty $ (scalar
operators) with a special attention to examples from the nonrelativistic
quantum mechanics of a one-dimensional motion (in particular, the radial
motion) of spinless particles. But an extension to matrix differential
operators in Hilbert spaces of vector-functions like $L^{2}\left( a,b\right)
\oplus \ldots L^{2}\left( a,b\right) $ is direct. Therefore, the main
results and conclusions of this section allow applying, with evident
modifications, to the quantum mechanics of the radial motion of particles
with spin, both nonrelativistic and relativistic, in particular, to the
quantum mechanics of Dirac particles of spin $1/2$.

As to partial differential operators, we refer to

\cite{Stone32,Titch58,KosKrS64,FadMa64,Berez65,Kato66,Richt78,BerSh91}; for
physicists, we strongly recommend references \cite{FadMa64} where
three-dimensional Hamiltonians are classified and \cite{BerSh91}.
Foundations of the general theory of ordinary differential operators were
laid by Weyl \cite{Weyl09,Weyl10,Weyl10a}. A somewhat different approach to
the theory was developed by Titchmarsh \cite{Titch46,Titch58}.

In view of many fundamental treatises on differential operators, our
exposition is of a qualitative character in some aspects, a number of items
is given under simplifying assumptions only to give basic ideas. But we try
formulate the main statements and results for the general case as far as
possible. By the mathematical tradition, we present them in the form of
theorems. A physicist may find this manner superfluously mathematical, while
a mathematician may find drawbacks in our formulations and proofs, but it
provides a suitable system of references and facilitate applications.

All theorems are illustrated by simple, but we hope, instructive, examples
of the well-known quantum mechanical operators like the momentum and
Hamiltonian.

We additionally restrict ourselves to the case where possible singularities
of the coefficient functions in a differential operator are on the
boundaries (which is natural for radial Hamiltonians). If a singularity is
located in the inner point $c$ of an interval $\left( a,b\right) $, like in
the case of $\delta $-potentials, the consideration must be appropriately
modified. We here refer to the extensive treatise \cite{AlbGeH88} on the
subject.

And finally, the remarks directly related to our subject.

The general method of s.a. extensions of symmetric operators presented in
the previous section and based on the main theorem is universal, i.e., it is
universally applicable to symmetric operators of any nature. But as any
universal method, it can turn out unsuitable as applied to some particular
problems with their own specific features and therefore requires appropriate
modifications. For example, in quantum mechanics for particles,
nonrelativistic and relativistic, quantum-mechanical observables are usually
defined in terms of s.a. differential operators, and the spectral problem is
formulated as an eigenvalue problem for the corresponding differential
equations\footnote{%
Of course, this does not concern the spin degrees of freedom and spin
systems where observables are represented by Hermitian matrices.}. In the
presence of boundaries and/or singularities of the potential, we are used to
accompany these equations with one or another boundary conditions on the
wave functions. This means that we additionally specify the domain of the
corresponding observables by the boundary conditions that provide the
self-adjointness of the differential operators under consideration. It is
natural to call such boundary conditions the s.a. boundary conditions, this
is a standard term in the mathematical literature.

A revealing of the specific features of s.a. extensions of differential
symmetric operators is just the subject if this section.

It appears that in the case of differential operators, the isometries $%
\hat{U}:D_{+}\rightarrow D_{-}$ of one deficient subspace to another
specifying s.a. extensions of symmetric operators can be converted into s.a.
boundary conditions, explicit or implicit. This possibility is based on the
fact that the asymmetry forms $\omega _{\ast }$ and $\Delta _{\ast }$ are
expressed in terms of asymptotic boundary values of functions and their
derivatives. In addition to conventional methods, we discuss a possible
alternative way of specifying s.a. differential operators in terms of
explicit boundary conditions. It is based on direct modification of the
arguments resulting in the main theorem. The method does not require
evaluating the deficient subspaces $D_{+}$ and $D_{-}$ and the deficiency
indices, the latter are determined in passing. Its effectiveness is
illustrated by a number of examples of quantum-mechanical operators.
Unfortunately, this method is not universal at present. Its applicability
depends on to what extent we can establish the boundary behavior of
functions involved. In general, it depends on specific features of
boundaries, in particular, whether they are regular or singular.\footnote{%
These notions are explained below.}

\subsection{Differential expressions}

Let $\left( a,b\right) $ be an interval of the real axis $\mathbb{R}^{1}$.
By $\left( a,b\right) $ we mean an interval in a generalized sense: the ends 
$a$ and $b$ of the interval can be infinite, $a=-\infty$ and/or $b=+\infty$;
if they are finite, $\left| a\right| <\infty$ and/or $\left| b\right|
<\infty $, they can be included in the interval such that we can have a pure
interval $\left( a,b\right) $, semi-interval $\left[ a,b\right) $ or $\left(
a,b\right] $, or a segment $\left[ a,b\right] $. This depends on the
regularity of the coefficients of a differential operator under
consideration.

Each interval $\left( a,b\right) $ is assigned the Hilbert space $%
L^{2}\left( a,b\right) $ of functions, wave functions in the physical
terminology. We recall that from the standpoint of Hilbert spaces, the
inclusion of the finite end points $a$ and/or $b$ in $\left( a,b\right) $ is
irrelevant: the Hilbert spaces $L^{2}\left( (a,b)\right) $ and $L^{2}\left( %
\left[ a,b\right] \right) $ for the respective pure interval $\left(
a,b\right) $ and segment $\left[ a,b\right] $ are the same because the
Lebesgue measure of a point is zero.

A differential expression, or a differential operation, $\check{f}$
associated with an interval $\left( a,b\right) $ is an expression of the form%
\begin{equation}
\check{f}=f_{n}\left( x\right) \left( \frac{d}{dx}\right) ^{n}+f_{n-1}\left(
x\right) \left( \frac{d}{dx}\right) ^{n-1}+\cdots +f_{1}\left( x\right) 
\frac{d}{dx}+f_{0}\left( x\right) \,,\;x\in \left( a,b\right) \,,
\label{3b.1}
\end{equation}%
where $f_{k}\left( x\right) $, $k=0,1,\ldots ,n$, are some functions on $%
\left( a,b\right) $ that are called the coefficient functions, or simply
coefficients, of the differential expression, $f_{n}\left( x\right) \neq 0;$
an integer $n\geq 1$ is called the order of $\check{f}$.

The differential expression $\check{f}$ naturally defines a linear
differential operator over functions on $\left( a,b\right) $, whence an
alternative name ``differential operation'' for $\check{f},$%
\begin{equation}
\left( \check{f}\psi \right) \left( x\right) =f_{n}\left( x\right) \psi
^{\left( n\right) }\left( x\right) +f_{n-1}\left( x\right) \psi ^{\left(
n-1\right) }\left( x\right) +\cdots +f_{1}\left( x\right) \psi ^{\prime
}\left( x\right) +f_{0}\left( x\right) \psi \left( x\right) \,,  \label{3b.2}
\end{equation}%
under the natural assumption that $\psi $ is absolutely continuous together
with its $n-1$ derivatives\footnote{$\psi ^{\left( k\right) }$ is a
conventional symbol of the derivative of order $k$.} $\psi ^{\left( 1\right)
}=\psi ^{\prime },\ldots ,\psi ^{\left( n-1\right) }$. Formula (\ref{3b.2})
defines the ''rule of acting'' for future operators in $L^{2}\left(
a,b\right) .$

An intermediate remark is in order here.

The consideration to follow are directly extended to matrix deferential
expressions, i.e., to deferential expressions with matrix coefficients, that
generate systems of differential equations and differential operators in
Hilbert space of vector-functions like $L^{2}\left( a,b\right) \oplus \cdots
\oplus L^{2}\left( a,b\right) $ where vector-functions are columns of
square-integrable functions. Such matrix differential expressions are
inherent in nonrelativistic and relativistic quantum mechanics of particles
with spin, in particular, Dirac particles (we mean the radial motion of
particles).

As is well known, an ordinary differential equation of order $n$ can be
reduced to a system of $n$ first-order differential equations, and vice
versa. What is more, this reduction is useful in analyzing homogeneous an
inhomogenous differential equations, in particular, in establishing the
structure of their general solution. Respectively, any differential
expression $\check{f}$ (\ref{3b.1}) is assigned a first-order matrix
differential expression with $n\times n$ matrix coefficients.

The regularity conditions for the coefficients $f_{k}$ (integrability,
continuity, differentiability, etc.) depend on the context. The standard
conditions are that $f_{k}$, $k=1,\ldots ,n-1$, has $k$ derivatives in $%
\left( a,b\right) $, $f_{n}\neq 0$, and $f_{0}$ is locally integrable%
\footnote{%
It is integrable on any finite interval inside $(a,b).$} in $\left(
a,b\right) $; the coefficients, for example, $f_{0},$ can be infinite as $%
x\rightarrow a$ and/or $x\rightarrow b$. These conditions are sufficient for
the function $\check{f}\psi $ to allow integrating by parts and a given
differential expression $\check{f}$ to have an adjoint differential
expression $\check{f}^{\ast },$ see below, and for the functions $%
f_{0},f_{1},\ldots ,f_{n-1}$ and $1/f_{n}$ to be locally integrable in $%
\left( a,b\right) $, which is necessary for the theory of usual differential
equations generated by a given differential expression:\ the homogenous
equation $\check{f}\psi =0$ and the inhomogenous equation $\check{f}\psi
=\chi $, see below. The conditions on the coefficients sometimes can be
considerably weakened for another representation of differential
expressions, see below. For the first reading, one can consider the
coefficients $f_{k}$ smooth functions. If the coefficients have
singularities in $\left( a,b\right) $, a separate special consideration is
required.

In the physical language, $\check{f}$ (\ref{3b.1}) can be considered an
element of a formal algebra generated by the ``operators'' $\hat{q}=x$ (the
position operator) and $\hat{p}=-id/dx$ (the momentum operator\footnote{%
The Planck constant $\hbar $ is set to unity, $\hbar =1$. In the
mathematical language, $\hat{q}$ and $\hat{p}$ are called the generators of
the algebra, or ``symbols''. We no longer use the physical symbol $f\left( 
\hat{q},\hat{p}\right) =f\left( x,-i\frac{d}{dx}\right) ,$ or more briefly $%
f\left( x,\frac{d}{dx}\right) $, for $\check{f}$, because its origin is
irrelevant here.
\par
\bigskip}), satisfying the canonical commutation relation $\left[ \hat{q},%
\hat{p}\right] =i$,%
\begin{equation*}
\check{f}=f_{n}\left( \hat{q}\right) \left( i\hat{p}\right)
^{n}+f_{n-1}\left( \hat{q}\right) \left( i\hat{p}\right) ^{n-1}+\cdots
+f_{1}\left( \hat{q}\right) i\hat{p}+f_{0}\left( \hat{q}\right) \,,
\end{equation*}%
with the so-called $qp$-ordering \cite{BerSh91}.

The differential expression (\ref{3b.1}) is called the regular differential
expression if the interval $\left( a,b\right) $ is finite and if the
coefficients $f_{0},\ldots ,f_{n-1},$ and the function $f_{n}^{-1}$ are
integrable\footnote{%
This condition does not exclude that the coefficients, for example, $%
f_{0}\left( x\right) $, can be infinite as $x\rightarrow a$ and/or $%
x\rightarrow b$.} on $\left( a,b\right) $, the term ``on $\left( a,b\right) $%
'' means on the whole $\left( a,b\right) $, including the ends $a$ and $b$;
in this case, we consider $\left( a,b\right) $ as a segment $\left[ a,b%
\right] $. In the opposite case, $\check{f}$ is called the singular
differential expression. The left end $a$ is called the regular end if $%
a>-\infty $, and the indicated functions are integrable on any segment $%
\left[ a,\beta \right] $, $\beta <b$. In the opposite case, i.e., if $%
a=-\infty $ and/or the integrability condition on $\left[ a,\beta \right] $
for the coefficients does not hold, the end $a$ is called the singular end.
Similar notions are introduced for the right end.

Let $\varphi \left( x\right) $ and $\phi \left( x\right) $ be smooth finite
functions, $\varphi ,\phi \in D\left( a,b\right) $, then the function $%
\check{f}\varphi $ is square-integrable\footnote{%
Because of our conditions for the coefficients of $\check{f}$ and because of
a finite support of $\varphi $ and therefore of $\check{f}\varphi $.} on $%
\left( a,b\right) ,$ as well as $\varphi ,$ and the scalar product $\left(
\phi ,f\varphi \right) =\int_{a}^{b}dx\overline{\phi }\check{f}\varphi $ has
a sense. We consider this integral. Integrating by parts and taking into
account that the standard boundary terms vanish because of finite supports
of $\varphi $ and $\phi $, we have%
\begin{equation}
\left( \phi ,\check{f}\varphi \right) =\int_{a}^{b}dx\overline{\phi }\check{f%
}\varphi =\int_{a}^{b}dx\overline{\check{f}^{\ast }\phi }\varphi =\left( 
\check{f}^{\ast }\phi ,\varphi \right) \,,  \label{3b.3}
\end{equation}%
where the function $\check{f}^{\ast }\varphi $ is given by%
\begin{equation}
\check{f}^{\ast }\phi =\left( -\frac{d}{dx}\right) ^{n}\left( \overline{f_{n}%
}\phi \right) +\left( -\frac{d}{dx}\right) ^{n-1}\left( \overline{f_{n-1}}%
\phi \right) +\cdots +\left( -\frac{d}{dx}\right) \left( \overline{f_{1}}%
\phi \right) +\overline{f_{0}}\phi \,,  \label{3b.4}
\end{equation}%
and defines the differential expression%
\begin{equation}
\check{f}^{\ast }=\left( -\frac{d}{dx}\right) ^{n}\overline{f_{n}}+\left( -%
\frac{d}{dx}\right) ^{n-1}\overline{f_{n-1}}+\cdots +\left( -\frac{d}{dx}%
\right) \overline{f_{1}}+\overline{f_{0}}\,,  \label{3b.5}
\end{equation}%
a differential operation each term $\left( -\frac{d}{dx}\right) ^{k}%
\overline{f_{k}}$ of which implies first multiplying a function by the
function $\overline{f_{k}}$ and then differentiating the result $k$ times,
which has a sense on the above-given set of functions because $f_{k}\left(
x\right) $ is $k$-time-differentiable. The differential expression $\check{f}%
^{\ast }$ (\ref{3b.5})\ is called the the adjoint differential expression
(to $\check{f}$), or simply the adjoint, or the adjoint by Lagrange. In the
physical language, the adjoint is defined by%
\begin{equation}
\check{f}^{\ast }=\left( -i\hat{p}\right) ^{n}\overline{f_{n}}\left( \hat{q}%
\right) +\left( -i\hat{p}\right) ^{n-1}\overline{f_{n-1}}\left( \hat{q}%
\right) +\cdots -i\hat{p}\overline{f_{1}}\left( \hat{q}\right) +\overline{%
f_{0}}\left( \hat{q}\right) \,,  \label{3b.6}
\end{equation}%
it is the adjoint in the above-mentioned formal algebra with involution (the
standard rule for taking the adjoint:\ reversing the order of ``operators''
and the complex conjugation of the numerical coefficients, which is denoted
by a bar over the function symbol. It naturally arises as a $pq$-ordered
expression. The adjoint $\check{f}^{\ast }$ (\ref{3b.5}) can be reduced to
form (\ref{3b.1}),%
\begin{align}
& \check{f}^{\ast }=\overline{f_{n}}\left( -\frac{d}{dx}\right) ^{n}+\left[ 
\overline{f_{n-1}}-n\overline{f_{n}^{\prime }}\right] \left( -\frac{d}{dx}%
\right) ^{n-1}+\cdots +\left[ \overline{f_{1}}+\cdots +\left( -1\right)
^{n-2}\left( n-1\right) \overline{f_{n-1}^{\left( n-2\right) }}\right. 
\notag \\
& +\left. \left( -1\right) ^{n-1}n\overline{f_{n}^{\left( n-1\right) }}%
\right] \left( -\frac{d}{dx}\right) +\left[ \overline{f_{0}}+\cdots +\left(
-1\right) ^{n-1}\overline{f_{n-1}^{\left( n-1\right) }}+\left( -1\right) ^{n}%
\overline{f_{n}^{\left( n\right) }}\right] \,,  \label{3b.7}
\end{align}%
by subsequently differentiating in r.h.s. of (\ref{3b.4}) and using the
Leibnitz rule, or by rearranging the $pq$-ordering in (\ref{3b.6}) to the $%
qp $-ordering by subsequently commuting all $\hat{p}$'s in $\hat{p}^{k}=\hat{%
p}\cdots \hat{p}$ with $f_{k}\left( \hat{q}\right) $ with the rule%
\begin{equation*}
\hat{p}\overline{f_{k}^{\left( l\right) }}\left( \hat{q}\right) =\overline{%
f_{k}^{\left( l\right) }}\left( \hat{q}\right) \hat{p}+\left[ \hat{p},%
\overline{f_{k}^{\left( l\right) }}\left( \hat{q}\right) \right] =\overline{%
f_{k}^{\left( l\right) }}\left( \hat{q}\right) \hat{p}+\left( -i\right) 
\overline{f_{k}^{\left( l+1\right) }}\left( \hat{q}\right) \,,\;l=0,1,\ldots
,k-1\,.
\end{equation*}%
A reader can easily write a detailed formula.

A differential expression $\check{f}$ is called a s.a. differential
expression, or s.a. by Lagrange , if it coincides with its adjoint, $\check{f%
}=\check{f}^{\ast }\,.$

Any differential expression $\check{f}$ can be assigned a differential
operator in $L^{2}\left( a,b\right) $ if an appropriate domain in $%
L^{2}\left( a,b\right) $ for this operator with the ''rule of acting'' given
by $\check{f}$ is indicated. But only a s.a. differential expression can
generate a s.a. differential operator in $L^{2}\left( a,b\right) $, which is
of interest from the standpoint of quantum mechanics. We refer to such an
operator as a s.a. operator associated with a given s.a. differential
expression. The self-adjointness of a differential expression is only
necessary for the existence of the respective s.a. operator and in general
is not sufficient: the main problem is to indicate the proper domain in $%
L^{2}\left( a,b\right) $ such that $\check{f}$ becomes a s.a. operator,
sometimes, it appears impossible; in addition, different s.a. operators can
be associated with the same differential expression as we already know from
the example at the end of the previous section.

We now describe the general structure of s.a. differential expressions of
any finite order that makes its self-adjointness obvious.

The coefficients of a s.a. differential expression $\check{f}$ (\ref{3b.1}), 
$\check{f}=\check{f}^{\ast }$, satisfy the following conditions with respect
to complex conjugation:%
\begin{align*}
& \overline{f_{n}}=\left( -\right) ^{n}f_{n}\,, \\
& \overline{f_{n-1}}=\left( -\right) ^{n-1}f_{n}+\left( -\right)
^{n}nf_{n}^{\left( 1\right) }\,, \\
& \vdots \\
& \overline{f_{1}}=-f_{1}+\cdots +\left( -\right) ^{n-1}\left( n-1\right)
f_{n-1}^{\left( n-2\right) }+\left( -\right) ^{n}nf_{n}^{\left( n-1\right)
}\,, \\
& \overline{f_{0}}=f_{0}+\cdots +\left( -\right) ^{n-1}f_{n-1}^{\left(
n-1\right) }+\left( -\right) ^{n}f_{n}^{\left( n\right) }\,,
\end{align*}%
that follow from the comparison of r.h.s. in (\ref{3b.1}) with r.h.s. in (%
\ref{3b.7}) and the subsequent complex conjugation. These conditions can be
resolved, which leads to the so-called canonical form of a s.a. differential
expression. The canonical form of a s.a. differential expression is a sum of
s.a. odd binomials,

\begin{eqnarray}
\hspace{-0.7cm}\check{f}_{\left( 2k-1\right) } &=&\frac{i}{2}\left[ \left( 
\frac{d}{dx}\right) ^{k-1}f_{2k-1}\left( -\frac{d}{dx}\right) ^{k}+\left( -%
\frac{d}{dx}\right) ^{k}f_{2k-1}\left( \frac{d}{dx}\right) ^{k-1}\right] \,,
\label{3b.7a} \\
\;\hspace{-0.7cm}f_{2k-1} &=&\overline{f_{2k-1}},\;k=1,2,\ldots ,  \notag
\end{eqnarray}
and s.a. even monomials, 
\begin{equation}
\check{f}_{\left( 2k\right) }=\left( -\frac{d}{dx}\right) ^{k}f_{2k}\left( 
\frac{d}{dx}\right) ^{k}\,,\;f_{2k}=\overline{f_{2k}}\,,\;k=0,1,\ldots \;,
\label{3b.7b}
\end{equation}
with the real coefficient functions $f_{l}$ ($f_{\left( 0\right)
}=f_{0}\left( x\right) $ is here considered a differential expression of
order zero); for brevity, we use the same notation for the coefficient
functions as for those in (\ref{3b.1}).

In terms of the formal algebra, these are the respective ``operators'' 
\begin{equation*}
\check{f}_{\left( 2k-1\right) }=\frac{1}{2}\left[ \hat{p}^{k-1}f_{2k-1}%
\left( \hat{q}\right) \hat{p}^{k}+\hat{p}^{k}f_{2k-1}\left( \hat{q}\right) 
\hat{p}^{k-1}\right] \,,\;k=1,2\,,...,
\end{equation*}
and 
\begin{equation*}
\check{f}_{\left( 2k\right) }=\hat{p}^{k}f_{2k}\left( \hat{q}\right) \hat{p}%
^{k}\,,\;k=0,1,\ldots \;,
\end{equation*}
with the properly symmetrized $pq$-ordering; they are well-known to
physicists \cite{BerSh91}.

The canonical form of a s.a. differential expression $\check{f}=\check{f}%
^{\ast }$ of order $n\geq 1$ is thus given by

\begin{equation}
\,\check{f}=\sum_{k}\check{f}_{\left( 2k\right) }+\sum_{k}\check{f}_{\left(
2k+1\right) }  \label{3b.8}
\end{equation}

In this form (\ref{3b.8}) for a s.a. differential expression, the regularity
conditions for the coefficient functions $f_{l}\left( x\right) $ can be
weakened:\ there is no need in the $l$-time-differentiability of $%
f_{l}\left( x\right) $, a natural sufficient requirement is that $%
f_{2k}\left( x\right) $ and $f_{2k-1}\left( x\right) $ be only $k$%
-time-differentiable.

The simplest first-order s.a. differential expression is $\check{f}=\check{p}
$ given by (\ref{3b.9}) that is identified in physics with the quantum
mechanical momentum of a particle moving on an interval $\left( a,b\right) $
of a real axis; it was considered at the end of the previous section.

The even second-order differential expression with the conventional notation 
$f_{2}\left( x\right) =p\left( x\right) ,$ $f_{0}\left( x\right) =q\left(
x\right) $ is the Sturm-Lioville differential expression 
\begin{equation*}
\check{f}=-\frac{d}{dx}p\left( x\right) \frac{d}{dx}+q\left( x\right)
\,,\;p\left( x\right) =\overline{p\left( x\right) }\,,\;q\left( x\right) =%
\overline{q\left( x\right) }\,.
\end{equation*}%
With $p\left( x\right) =1$ and $q\left( x\right) =V\left( x\right) $, we let 
$\check{H}$ denote $\check{f}$ and obtain the second-order s.a. differential
expression%
\begin{equation}
\check{H}=-\frac{d^{2}}{dx^{2}}+V\left( x\right)  \label{3b.10}
\end{equation}%
that is identified in physics with the quantum-mechanical Hamiltonian%
\footnote{%
To be true, this identification assumes appropriate units, where, for
example, the Planck constant $\hbar =1$ and the mass of a particle $m=1/2$;
with the usual units, differential expression (\ref{3b.10}) corresponds to
the Hamiltonian multiplied by a numerical factor $\frac{2m}{\hbar ^{2}}.$}
for a nonrelativistic particle moving on an interval $\left( a,b\right) $ of
the real axis in the potential field $V\left( x\right) .$ In what follows,
we mainly deal with this simplest, but physically interesting, differential
expression (\ref{3b.10}) when illustrating the general assertions.

The general even s.a. differential expression of order $n$,%
\begin{equation}
\check{f}=\sum_{k=0}^{n/2}\left( -\frac{d}{dx}\right) ^{k}f_{2k}\left( \frac{%
d}{dx}\right) ^{k}\,,\;f_{2k}=\overline{f_{2k}}\,,  \label{3b.11}
\end{equation}%
can be rewritten in terms of differential operations $D^{\left[ k\right] }$, 
$k=1,\ldots ,n$, that are defined recursively and separately for each $%
\check{f}$ by%
\begin{align*}
& D^{\left[ k\right] }=\left( \frac{d}{dx}\right) ^{k}\,,\;k=1,\ldots
,n/2-1\,,\;D^{\left[ n/2\right] }=f_{n}\left( \frac{d}{dx}\right) ^{n/2}\,,
\\
& D^{\left[ n/2+k\right] }=f_{n-2k}\left( \frac{d}{dx}\right) ^{n-2k}-\frac{d%
}{dx}D^{\left[ n/2+k-1\right] }\,,\;k=1,\ldots ,n/2\,,
\end{align*}%
and define the respective so-called quasi-derivatives \cite{AkhGl81,Naima69}
by\footnote{%
In \cite{AkhGl81,Naima69}, an even $n$ is denoted by $2n,$ and the
coefficient functions $f_{2k}\left( x\right) $ are denoted by $p_{n-k}\left(
x\right) $: $p_{n-k}\left( x\right) =f_{2k}\left( x\right) $.}%
\begin{equation*}
\psi ^{\left[ k\right] }=D^{\left[ k\right] }\psi \Longrightarrow \left\{ 
\begin{array}{l}
\psi ^{\left[ k\right] }=\psi ^{\left( k\right) }\,,\;k=1,\ldots
,n/2-1\,,\;\psi ^{\left[ n/2\right] }=f_{n}\psi ^{\left( n/2\right) }\,, \\ 
\psi ^{\left[ n/2+k\right] }=f_{n-2k}\psi ^{(n-2k)}-\frac{d}{dx}\psi ^{\left[
n/2+k-1\right] }\,,\;k=1,\ldots ,n/2\,.%
\end{array}%
\right.
\end{equation*}%
Then the differential expression (\ref{3b.11}) is simply written as%
\begin{equation}
\check{f}=D^{\left[ n\right] }\,,  \label{3b.15}
\end{equation}%
and%
\begin{equation}
\check{f}\psi =\psi ^{\left[ n\right] }\,.  \label{3b.16}
\end{equation}%
With this form (\ref{3b.15}) for $\check{f}$ (\ref{3b.11}) and (\ref{3b.16})
for $\check{f}\psi $, the regularity conditions for the coefficient
functions $f_{2k}$ can be essentially weakened:\ it is not necessary that $%
f_{2k}$ be $k$-time differentiable; it is sufficient that $\psi ^{\left[ k%
\right] }$, $k=1,\ldots ,n-1,$ be absolutely continuous in $\left(
a,b\right) $ for $\psi ^{\left[ n\right] }$ to have a sense and that the
functions $f_{0},\ldots ,f_{n-2},1/f_{n}\neq 0$ be locally integrable for
the homogenous and inhomogenous differential equations $\check{f}\psi =0$
and $\check{f}\psi =\chi $ to be solvable with usual properties of their
general solution. The notions of regular and singular ends are modified
respectively.

Any even s.a. expression $\check{f}$ (\ref{3b.11}), (\ref{3b.15}) is
assigned at least one associated s.a. operator (see below). The notion of
quasi-derivatives allows highly elaborating the theory of even s.a.
differential operators with real coefficients \cite{AkhGl81,Naima69}. To our
knowledge, there is no similar notion for odd s.a. differential expressions
and for the respective s.a. differential operators with imaginary
coefficients. For any s.a. differential expression (\ref{3b.8}) of any order 
$n$, the so called Lagrange identity%
\begin{equation}
\overline{\chi }\check{f}\psi -\left( \overline{\check{f}\chi }\right) \psi =%
\frac{d}{dx}\left[ \chi ,\psi \right]  \label{3b.17}
\end{equation}%
holds, where $\left[ \chi ,\psi \right] $ is a local sesquilinear form in
functions and their derivatives of order up to $n-1$:%
\begin{align}
& \,\left[ \chi ,\psi \right] =-i\sum_{k=1}^{\left[ \frac{n+1}{2}\right] }%
\overline{\chi ^{\left( k-1\right) }}f_{2k-1}\psi ^{\left( k-1\right) }+%
\frac{i}{2}\sum_{k=2}^{\left[ \frac{n+1}{2}\right] }\sum_{l=0}^{k-2}\left\{ 
\overline{\chi ^{\left( l\right) }}\left( -\frac{d}{dx}\right)
^{k-2-l}\right.  \notag \\
& \,\left. \times \left[ \left( f_{2k-1}\psi ^{\left( k\right) }\right)
+\left( f_{2k-1}\psi ^{\left( k-1\right) }\right) ^{\prime }\right] +\left( -%
\frac{d}{dx}\right) ^{k-2-l}\left[ \left( f_{2k-1}\overline{\chi ^{\left(
k\right) }}\right) +\left( f_{2k-1}\overline{\chi ^{\left( k-1\right) }}%
\right) ^{\prime }\right] \psi ^{\left( l\right) }\right\}  \notag \\
& \,+\sum_{k=1}^{\left[ \frac{n}{2}\right] }\sum_{l=0}^{k-1}\left[ \psi
^{\left( k\right) }\left( -\frac{d}{dx}\right) ^{k-l-1}\left( f_{2k}%
\overline{\chi ^{\left( k\right) }}\right) -\overline{\chi ^{\left( l\right)
}}\left( -\frac{d}{dx}\right) ^{k-l-1}\left( f_{2k}\psi ^{\left( k\right)
}\right) \right] \,.  \label{3b.18}
\end{align}

Equalities (\ref{3b.17}), (\ref{3b.18}) can be derived by the standard
procedure of subsequently extracting a total derivative in the l.h.s. of (%
\ref{3b.17}) used in integrating by parts or can be verified directly by
differentiating $\left[ \chi,\psi\right] $ (\ref{3b.18}) in the r.h.s. of (%
\ref{3b.17}).

It follows the integral Lagrange identity%
\begin{equation}
\int_{\alpha }^{\beta }dx\overline{\chi }\check{f}\psi -\int_{\alpha
}^{\beta }dx\psi \overline{\check{f}\chi }=\left. \left[ \chi ,\psi \right]
\right| _{\alpha }^{\beta }\,,  \label{3b.19}
\end{equation}%
where $\left[ \alpha ,\beta \right] $ is any finite segment of $\left(
a,b\right) $, $\left[ \alpha ,\beta \right] \subset \left( a,b\right) $,
and, by definition, $\left. \left[ \chi ,\psi \right] \right| _{\alpha
}^{\beta }$ is the difference of the form$\ \left[ \chi ,\psi \right] $ at
the respective points $\beta $ and $\alpha $: 
\begin{equation*}
\left. \left[ \chi ,\psi \right] \right| _{\alpha }^{\beta }=\left[ \chi
,\psi \right] (\beta )-\left[ \chi ,\psi \right] (\alpha ).
\end{equation*}

As simple examples, for first-order differential expressions (\ref{3b.9}),
we have $\left[ \chi ,\psi \right] (x)=-i\overline{\chi \left( x\right) }%
\psi \left( x\right) \,,$ while for the second-order deferential expression (%
\ref{3b.10}), we have%
\begin{equation}
\left[ \chi ,\psi \right] (x)=-\left[ \overline{\chi \left( x\right) }\psi
^{\prime }\left( x\right) -\overline{\chi ^{\prime }\left( x\right) }\psi
\left( x\right) \right] \,.  \label{3b.22}
\end{equation}

We point out some properties of the form $\left[ \chi ,\psi \right] $. We
first note that the conventional symbol $\left[ \cdot ,\cdot \right] $ for
this form is identical to the symbol of a commutator (perhaps, this is
because the l.h.s. of (\ref{3b.17}) is similar to a commutator and because
in the framework of the commutative algebra of functions there is no need in
the symbol of a true commutator, so that a confusion is avoided). But $\left[
\chi ,\psi \right] $ is not a commutator, and, in particular, $\left[ \psi
,\psi \right] \neq 0$ in general.

For even s.a. expressions $\check{f}$ (\ref{3b.11}), (\ref{3b.15}) of order $%
n$, the form $\left[ \chi ,\psi \right] $ is a simple sesquilinear form in
quasi-derivatives:%
\begin{equation}
\left[ \chi ,\psi \right] =-\sum_{k=0}^{n/2-1}\left( \overline{\chi ^{\left[
k\right] }}\psi ^{\left[ n-k-1\right] }-\overline{\chi ^{\left[ n-k-1\right]
}}\psi ^{\left[ k\right] }\right) \,.  \label{3b.23}
\end{equation}%
We note that because the coefficient functions of even s.a. expressions are
real, we have $\left[ \overline{\psi },\psi \right] \equiv 0$, $\forall \psi 
$, while $\left[ \psi ,\psi \right] \neq 0$ in general unless $\psi $ is
real up to a constant factor of module unity, $\overline{\psi }=e^{i\theta
}\psi $, $\theta =\mathrm{const.}$

The form $\left[ \chi ,\psi \right] $ (\ref{3b.18}) is evidently
antisymmetric, $\overline{\left[ \chi ,\psi \right] }=-\left[ \psi ,\chi %
\right] \,,$ and its reduction to the diagonal $\chi =\psi $, the quadratic
form $\left[ \psi ,\psi \right] $, is purely imaginary,$\;\overline{\left[
\psi ,\psi \right] }=-\left[ \psi ,\psi \right] \,.$ Let the functions $\psi 
$ and $\chi $ in $\left[ \chi ,\psi \right] $ satisfy the same homogenous
linear differential equation generated by a s.a. expression $\check{f}$, $%
\check{f}\psi =0$ and $\check{f}\chi =0$; we note that if $\check{f}$ is odd
with pure imaginary coefficients or even with real coefficients, the complex
conjugate functions $\overline{\psi }$ and $\overline{\chi }$ satisfy the
same equation. It simply follows from (\ref{3b.17}) that the form $\left[
\chi ,\psi \right] $ (\ref{3b.18}) for solutions of the homogenous equation
does not depend on $x$, i.e., $\left[ \chi ,\psi \right] =\mathrm{const}$.
For second-order differential expressions this is a well-known fact:\ a
reader can easily recognize the Wronskian for $\overline{\chi }$ and $\psi $
in (\ref{3b.22}). We only note that this is the Wronskian for $\overline{%
\chi }$, which is also a solution of the homogenous equation, and $\psi $,
but not for $\chi $ and $\psi $ in particular, $\left[ \psi ,\psi \right] $
which is the Wronskian for $\overline{\psi }$ and $\psi $, is generally not
equal to zero; this is a specific feature of the complex linear space under
consideration.

The same is true if $\psi $ and $\chi $ are the solutions of the respective
spectral equations $\check{f}\psi =\lambda \psi $ and $\check{f}\chi =%
\overline{\lambda }\chi $ with complex conjugated parameters. We again note
that if $\check{f}\psi =\lambda \psi $ then for an odd s.a. differential
expressions, we have $\check{f}\overline{\psi }=-\overline{\lambda }%
\overline{\psi }$, while for an even s.a. differential expressions, we have $%
\check{f}\overline{\psi }=\overline{\lambda }\overline{\psi }$.

\subsection{Differential equations}

Before we turn to differential operators generated by s.a. differential
expression, we must recall some facts of the theory of ordinary differential
equations, homogenous and inhomogenous.

The theory of s.a. differential operators in $L^{2}\left( a,b\right) $ is
based on the theory of ordinary linear differential equations, especially on
the theory of its general solutions including the so-called generalized
ones. We recall the basic points of this theory as applied to homogenous and
inhomogenous differential equations generated by the above-introduced s.a.
differential expressions. We present them by the simple examples of
differential expressions (\ref{3b.9}) of the first order and (\ref{3b.10})
of the second order. On the one hand, these expressions are of physical
interest and are widely used in physical applications, on the other hand,
they allow demonstrating the common key points of the general consideration.

As to the simplest first-order differential expression (\ref{3b.9}), this
programme has been accomplished above, in the end of the previous section.
The general solutions of the corresponding equation $-i\frac{d}{dx}y\left(
x\right) =0$ and $-i\frac{d}{dx}y\left( x\right) =h\left( x\right) $ are so
obvious that this allows completely solving all the problems related to this
differential expression, including s.a. operators. The consideration was so
easy that some general points could prove to be somewhat hidden.

We therefore proceed to differential expression (\ref{3b.10}). This
differential expression is correctly defined as a differential operator on
the complex linear space of functions on $\left( a,b\right) $ that are
absolutely continuous in $\left( a,b\right) $ together with their first
derivatives. We change the notation of functions from $\psi \left( x\right)
,\chi \left( x\right) ,\ldots ,$ which is usually adopted in physics for
functions in $L^{2}\left( a,b\right) ,$ to $u\left( x\right) ,y\left(
x\right) ,\ldots ,$ which is usually adopted in the theory of differential
equations, in particular, because these functions are generally
non-square-integrable on an arbitrary interval $\left( a,b\right) \subseteq 
\mathbb{R}^{1}$.

On this space, we consider the homogenous differential equation%
\begin{equation}
\check{H}u=-u^{\prime \prime }+Vu=0  \label{3b.26}
\end{equation}%
and the inhomogenous differential equation%
\begin{equation}
\check{H}y=-y^{\prime \prime }+Vy=h\,,  \label{3b.27}
\end{equation}%
where $h\left( x\right) $ is assumed to be locally integrable.

It is known from the theory of ordinary differential equations that if $V$
is locally integrable, eq. (\ref{3b.26}) has two linearly independent
solutions $u_{1}$ and $u_{2}$, $\check{H}u_{1,2}=0,$ that form a fundamental
system of eq. (\ref{3b.26}) in the sense that the general solution of eq. (%
\ref{3b.26}) is%
\begin{equation}
u=c_{1}u_{1}+c_{2}u_{2}\,,  \label{3b.28}
\end{equation}%
where $c_{1}$ and $c_{2}$ are arbitrary complex constants, these constants
are fixed by initial conditions on $u$ and $u^{\prime }$ at some inner point
in $\left( a,b\right) $ or at a regular end. The linear independence of $%
u_{1}$ and $u_{2}$ is equivalent to the requirement that their Wronskian $%
W\left( u_{1},u_{2}\right) =u_{1}\left( x\right) u_{2}^{\prime }\left(
x\right) -u_{2}\left( x\right) u_{1}^{\prime }\left( x\right) \,,$ which is
a constant for any two solutions of eq. (\ref{3b.26}), be nonzero, $W\left(
u_{1},u_{2}\right) =\mathrm{const}\neq 0\,.$ Of course, the fundamental
system $u_{1},u_{2}$ is defined up to a nonsingular linear transformation.
For real potentials, $V=\overline{V}$, the functions $u_{1}$ and $u_{2}$ can
also be taken to be real . If the end $a$ of the interval $\left( a,b\right) 
$ is regular, i.e., if $-\infty <a$ and $V$ is integrable up to $a$, i.e., $%
\int_{a}^{\beta }dx\left| V\right| <\infty $, $\beta <b$, then any solution (%
\ref{3b.28}) has a finite limit at this end together with its first
derivative. The same is true for a regular right end $b$. In the case of
singular ends, one or both of fundamental solutions, i.e., $%
u_{1},u_{1}^{\prime }$ and/or $u_{2},u_{2}^{\prime }$, can be infinite at
such ends. If the potential $V$ is smooth in $\left( a,b\right) $, $V\in
C^{\infty }\left( a,b\right) $, which does not exclude that $V$ is infinite
at the ends, then any solution $u$ (\ref{3b.28}) is also smooth in $\left(
a,b\right) $.

The general solution of inhomogenous equation (\ref{3b.27}) is given by%
\begin{equation*}
y\left( x\right) =\frac{1}{W\left( u_{1},u_{2}\right) }\left[ u_{1}\left(
x\right) \int_{\alpha }^{x}d\xi u_{2}h+u_{2}\left( x\right) \int_{x}^{\beta
}d\xi u_{1}h\right] +c_{1}u_{1}\left( x\right) +c_{2}u_{2}\left( x\right) \,,
\end{equation*}%
where $\alpha $ and $\beta $ are arbitrary, but fixed, inner points in $%
\left( a,b\right) $, in particular, we can choose $\alpha =\beta ,$ and $%
c_{1}$ and $c_{2}$ are arbitrary constants that are fixed by initial
conditions on $y$ and $y^{\prime }$ at some inner point in $\left(
a,b\right) $ or at a regular end. If the left end $a$ of the interval $%
\left( a,b\right) $ is regular, we can always take $\alpha =a$, we can also
do this in the case where the end $a$ is singular if the respective integral
is certainly convergent, for example, if the functions $u_{2}$ and $h$ are
square-integrable on the segment $\left[ a,x\right] $; the same is true for
the right end $b$.

We now consider the question about the so-called generalized solutions of
homogenous equation (\ref{3b.26}), i.e., the question about functions $u$
that satisfy the linear functional equation%
\begin{equation}
\int_{a}^{b}dx\overline{u}\check{H}\varphi =0\,,\;\forall \varphi \in
D\left( a,b\right) \,.  \label{3b.31}
\end{equation}%
Generally speaking, $u$ in (\ref{3b.31}) can be considered a generalized
function (a distribution), then the integral in (\ref{3b.31}) is symbolical,
but for our purposes, it appears sufficient that $u$ be a function\footnote{%
In the theory of distributions, $u\left( x\right) $ usually stands for $%
\overline{u\left( x\right) }$ in (\ref{3b.31}). For s.a. differential
expressions $\check{H}$ with real coefficients, $\overline{u\left( x\right) }
$ in (\ref{3b.31}) can be equivalently replaced by $u\left( x\right) $, as
for any even s.a. differential expression $\check{f}$, or for any odd s.a.
expression $\check{f}$ with pure imaginary coefficients. Form (\ref{3b.31})
\ with $\overline{u\left( x\right) }$ is more convenient here because the
following consideration is applicable to any mixed s.a. differential
expression $\check{f}$ and because for (locally) square integrable $u\left(
x\right) \,$,\ the integral in (\ref{3b.31}) becomes a scalar product $\
\left( u,\check{H}\varphi \right) $ in $L^{2}\left( a,b\right) $.}. It is
evident that any usual solution (\ref{3b.28}) of homogenous equation (\ref%
{3b.26}) is a generalized solution, i.e., satisfies eq. (\ref{3b.31})
because of the equality%
\begin{equation}
\int_{a}^{b}dx\overline{y}\check{H}\varphi =\int_{a}^{b}dx\varphi \overline{%
\check{H}y}\,,\;\forall \varphi \in D\left( a,b\right) \,,  \label{3b.32}
\end{equation}%
for any function $y$ absolutely continuous in $\left( a,b\right) $ together
with its derivative $y^{\prime }$, which follows from integrating by parts
in l.h.s. in (\ref{3b.32}) with vanishing boundary terms because of a finite
support of $\varphi $, $\mathrm{supp}\,\varphi \subseteq \left[ \alpha
,\beta \right] \subset \left( a,b\right) $, i.e., because $\varphi $
vanishes in a neighborhood of the limits of integration. Actually, eq. (\ref%
{3b.32}) is a particular case of the extension of eq. (\ref{3b.3}) for s.a.
differential expressions $\check{f}=\check{f}^{\ast }$ of any order $n$ from
functions $\phi \in D\left( a,b\right) $ to functions $y$ absolutely
continuous in $\left( a,b\right) $ together with its $n-1$ derivatives,%
\begin{equation}
\int_{a}^{b}dx\overline{y}\check{f}\varphi =\int_{a}^{b}dx\varphi \overline{%
\check{f}y}\,,  \label{3b.33}
\end{equation}%
for the validity of equality (\ref{3b.33}), it is sufficient that only $%
\varphi \in D\left( a,b\right) $. We would like to show that conversely, any
generalized solution of homogenous equation is a usual solution, i.e., any
solution $u\left( x\right) $ of eq. (\ref{3b.31}) is given by eq. (\ref%
{3b.28}).

Here, we make a simplifying technical assumption that the potential $V$ is a
smooth function, $V\in C^{\infty }\left( a,b\right) $, and $\check{H}\varphi
\in D\left( a,b\right) $ as well as $\varphi $, which allows making use of
the well-developed theory of distributions (strictly speaking, $u\left(
x\right) $ in (\ref{3b.31}) can be considered a distribution only in this
case).

This assumption is in fact technical; the main result can be extended to the
general case, see below. We also note that many potentials encountered in
physics satisfy this condition. But if $V$ is nonsmooth, no practical loss
of generality from the standpoint of constructing s.a. operators associated
with $\check{H}$ occurs. Let the potential $V$ be a locally bounded
function, i.e., it is bounded in any finite segment $\left[ \alpha ,\beta %
\right] \subset \left( a,b\right) $, with possible finite jumps, such that
step-like potentials or barriers are admissible. Any such potential can be
smoothed out, i.e., approximated by a smooth potential $V_{\mathrm{reg}%
}\left( x\right) $, such that the difference $\delta V=V\left( x\right) -V_{%
\mathrm{reg}}\left( x\right) $ is uniformly bounded. Then the operators $%
\hat{H}$ and $\hat{H}_{\mathrm{reg}}$ in $L^{2}\left( a,b\right) $
associated with the respective differential expressions (\ref{3b.10}) and $%
\check{H}_{\mathrm{reg}}=-\frac{d^{2}}{dx^{2}}+V_{\mathrm{reg}}\left(
x\right) $ differ by a bounded s.a. multiplication operator $\widehat{\delta
V}=\delta V\left( x\right) $ defined everywhere, $\hat{H}-\hat{H}_{\mathrm{%
reg}}=\widehat{\delta V}$, and, therefore, are s.a. or non-s.a.
simultaneously, more precisely, any s.a. operator $\hat{H}_{\mathrm{reg}}$
is assigned a s.a. operator $\hat{H}=\hat{H}_{\mathrm{reg}}+\widehat{\delta V%
}$ with the same domain, and vice-versa.

Let thus the potential $V$ be smooth, and we return to the problem of the
generalized solutions of homogenous equation (\ref{3b.26}), i.e., the
solutions of eq. (\ref{3b.31}). We actually need a generalization of the du
Boi--Reymond lemma used at the end of the previous section when constructing
s.a. operators associated with the first-order differential expression $%
\check{p}$ (\ref{3b.9}). We obtain this generalization based on two
auxiliary lemmas. In the process, it becomes clear how the result on the
generalized solutions can be extended to differential expressions of any
order.

\begin{lemma}
\label{l3b.1}\textbf{\ }A function $\chi \left( x\right) \in D\left(
a,b\right) $ is represented as%
\begin{equation*}
\chi =\check{H}\phi \,,\;\phi \in D\left( a,b\right) \,,
\end{equation*}%
iff $\chi $ is orthogonal to solutions $u$ of homogenous equation (\ref%
{3b.26}),%
\begin{equation}
\left( u,\chi \right) =\int_{a}^{b}dx\overline{u\left( x\right) }\chi \left(
x\right) =0\,,\;\forall u:\;\check{H}u=0\,,  \label{3b.35}
\end{equation}%
which is evidently equivalent to the requirement that $\chi $ be orthogonal%
\footnote{%
Although $u\left( x\right) $ is generally non-square-integrable, the symbol $%
\left( \;,\right) $ of a scalar product in (\ref{3b.35}) is proper because
of a finite support of $\chi \left( x\right) $.} to fundamental solutions $%
u_{1}$ and $u_{2}$ of eq. (\ref{3b.26}), $\left( u_{1},\chi \right) =\left(
u_{2},\chi \right) =0$.
\end{lemma}

The necessity immediately follows from equality (\ref{3b.32}) with $y=u$.

Sufficiency. Let $\chi \in D\left( a,b\right) $ and satisfy condition (\ref%
{3b.35}). For this $\chi $, we take a specific solution $\phi $ of
inhomogenous eq. (\ref{3b.27}) with $h=\chi ,\;\check{H}\phi =\chi \,,$
given by (\ref{3b.28}) with $c_{1}=c_{2}=0$ and $\alpha =a$, $\beta =b$%
\begin{equation*}
\phi \left( x\right) =\frac{1}{W\left( u_{1},u_{2}\right) }\left[
u_{1}\left( x\right) \int_{a}^{x}d\xi u_{2}\chi +u_{2}\left( x\right)
\int_{x}^{b}d\xi u_{1}\chi \right] \,.
\end{equation*}%
We can set $\alpha =a$ and $\beta =b$ even if the interval $\left(
a,b\right) $ is infinite because of a finite support of $\chi $. Because $%
u_{1},u_{2}$, and $\chi $ are smooth, the function $\phi $ is also smooth, $%
\phi \in C^{\infty }\left( a,b\right) $, and because of condition (\ref%
{3b.35}) and $\mathrm{supp}\,\chi \left( x\right) \subseteq \left[ \gamma
,\delta \right] \subset \left( a,b\right) ,$ we have $\phi =0$ for $x<\gamma 
$ and $x>\delta $, i.e., $\phi \in D\left( a,b\right) $, which proves the
lemma.

\begin{lemma}
\label{l3b.2} Any finite function $\varphi \left( x\right) \in D\left(
a,b\right) $ can be represented as%
\begin{equation*}
\varphi =c_{1}\left( \varphi \right) \varphi _{1}+c_{2}\left( \varphi
\right) \varphi _{2}+\check{H}\phi \,,\;c_{i}\left( \varphi \right) =\left(
u_{i},\varphi \right) \,,\;i=1,2\,,
\end{equation*}%
where $u_{1}$ and $u_{2}$ are fundamental solutions of homogenous equation (%
\ref{3b.26}), and $\varphi _{1}$, $\varphi _{2}$, and $\phi $ are some
finite functions, $\varphi _{1}$, $\varphi _{2}$, $\phi \in D\left(
a,b\right) ,$ such that%
\begin{equation}
\left( u_{i},\varphi _{j}\right) =\delta _{ij}\,,\;i,j=1,2\,,  \label{3b.37}
\end{equation}%
the functions, $\varphi _{1}$, $\varphi _{2}$ can be considered some fixed
functions independent of $\varphi .$
\end{lemma}

We first prove the existence of a pair $\varphi _{1}$, $\varphi _{2}$ of
finite functions with property (\ref{3b.37}) (although somebody may consider
this evident). It is sufficient to show that there exists a pair $\phi _{1}$%
, $\phi _{2}$ of finite functions such that the matrix $A_{ij}=\left(
u_{i},\phi _{j}\right) $ is nonsingular, $\det A\neq 0$, and, therefore, has
the inverse $A^{-1}$. Then the functions $\varphi _{i}=\left( A^{-1}\right)
_{ji}\phi _{j}$ form the required pair. We now show qualitatively that the
pair $\phi _{1}$, $\phi _{2}$ does exist. Let $\left( \alpha ,\beta \right) $
be any finite interval in the initial interval $\left( a,b\right) $. The
restrictions of fundamental solutions $u_{1}$ and $u_{2}$ to this interval,
i.e., $u_{1}$ and $u_{2}$ considered only for $x\in \left( \alpha ,\beta
\right) $, belong to $L^{2}\left( \alpha ,\beta \right) $. The linear
independence of fundamental solutions $u_{1}$ and $u_{2}$ implies that the
matrix $U_{ij}=\int_{\alpha }^{\beta }dx\overline{u_{i}}u_{j}$, is
nonsingular. Because $D\left( \alpha ,\beta \right) $ is dense in $%
L^{2}\left( \alpha ,\beta \right) $, we can find finite functions $\phi _{1}$
and $\phi _{2}$ that are arbitrarily close to the respective $u_{1}$ and $%
u_{2}$ on the interval $\left( \alpha ,\beta \right) $. This implies that
the matrix $A_{ij}=\int_{\alpha }^{\beta }dx\overline{u_{i}}\phi _{j}\,,$ is
also arbitrarily close to the matrix $U$, therefore, $\det A\neq 0$, and $A$
is nonsingular. A reader can easily give a rigorous form to these
qualitative arguments.

It then remains to note that the function $\varphi -c_{1}\left( \varphi
\right) \varphi _{1}-c_{2}\left( \varphi \right) \varphi _{2}$ satisfies the
conditions of Lemma \ref{l3b.1}.

We can now prove a lemma generalizing the du-Boi-Reymond lemma.

\begin{lemma}
\label{l3b.3}A locally integrable function $u\left( x\right) $ satisfies the
condition (\ref{3b.31})%
\begin{equation*}
\left( u,H\varphi \right) =\int_{a}^{b}dx\overline{u}\check{H}\varphi
=0\,,\;\forall \varphi =D\left( a,b\right) \,,
\end{equation*}%
iff $u$ is absolutely continuous in $\left( a,b\right) $ together with its
first derivative $u^{\prime }$ and satisfies homogenous equation (\ref{3b.26}%
) $\check{H}u=0.$ This means that any generalized solution of the homogenous
equation is a usual solution.
\end{lemma}

As to sufficiency, it was already proved above based on eq. (\ref{3b.32})
(and actually repeats the proof of necessity in Lemma \ref{l3b.1}.

The necessity is proved using Lemma \ref{l3b.2}. For convenience of
references, we let $\phi $ denote $\varphi $ in (\ref{3b.31}), after which
it becomes%
\begin{equation}
\left( u,\check{H}\phi \right) =0\,,\;\forall \phi =D\left( a,b\right) \,.
\label{3b.38}
\end{equation}%
Let $\varphi $ be an arbitrary finite function, $\varphi \in D\left(
a,b\right) $. By Lemma \ref{l3b.2}, we have the representation%
\begin{equation*}
\varphi -\left( u_{1},\varphi \right) \varphi _{1}-\left( u_{2},\varphi
\right) \varphi _{2}=\check{H}\phi
\end{equation*}%
with some finite functions $\varphi _{1},\varphi _{2},\phi \in D\left(
a,b\right) $, $u_{1}$ and $u_{2}$ are fundamental solutions of eq. (\ref%
{3b.26}). Substituting this representation of $\check{H}\phi $ in l.h.s. of (%
\ref{3b.38}) and appropriately rearranging it, we have%
\begin{align*}
& \,\left( u,\check{H}\phi \right) =\left( u,\varphi -\left( u_{1},\varphi
\right) \varphi _{1}-\left( u_{2},\varphi \right) \varphi _{2}\right)
=\left( u,\varphi \right) -\left( u,\varphi _{1}\right) \left( u_{1},\varphi
\right) -\left( u,\varphi _{2}\right) \left( u_{2},\varphi \right) \\
& \,=\left( u-\overline{\left( u,\varphi _{1}\right) }u_{1}-\overline{\left(
u,\varphi _{2}\right) }u_{2},\varphi \right) =\int_{\alpha }^{\beta }dx%
\overline{\left( u-c_{1}u_{1}-c_{2}u_{2}\right) }\varphi =0\,,\;\forall
\varphi =D\left( a,b\right) \,,
\end{align*}%
where $c_{i}=\left( \varphi _{i},u\right) $, $i=1,2$, are constants, which
yields $u=c_{1}u_{1}+c_{2}u_{2}\,,$ representation (\ref{3b.28}) for a
solution of eq. (\ref{3b.26}), and thus proves the lemma.

This lemma as well as the du-Boi-Rymond lemma are particular cases of the
universal general theorem in the theory of distributions: the generalized
solution of a homogenous differential equation of any order generated by a
s.a. differential expression with smooth coefficients is a smooth function
that is a usual solution of the same equation \cite{Shilo65}. We only note
that it is evident how the method for proving the above lemma, the method
based on using the fundamental system of a homogenous equation, is extended
to the general case.

As to the case of nonsmooth coefficients, a similar assertion on the
generalized solutions of a homogenous equation also holds under the
above-mentioned standard condition on the coefficients of the corresponding
differential expression $\check{f}$ (\ref{3b.8}), (\ref{3b.7a}), (\ref{3b.7b}%
) and with an appropriate change of the space of finite functions in terms
of which the generalized solution is defined. We recall that the standard
conditions are that the coefficients $f_{2k-1}$ and $f_{2k}$ in (\ref{3b.8}%
), (\ref{3b.7a}), (\ref{3b.7b}) are $k$-time-differentiable and $f_{0}$ is
locally integrable. Under these conditions, the homogenous equation $\check{f%
}u=0$ is solvable and has a system $\left\{ u_{i}\right\} ^{n}$ of linearly
independent fundamental solutions whose linear combination $%
u=\sum_{i=1}^{n}c_{i}u_{i}$ with arbitrary complex constants $c_{i}$, $%
i=1,\ldots ,n$, yields the general solution of the homogenous equation and
in terms of which the general solution of the inhomogenous equation $\check{f%
}y=h$ is canonically expressed as a sum of a particular solution and the
general solution of the homogenous equation; the constants $c_{i}$, $%
i=1,\ldots ,n$, are fixed by initial conditions on the respective $u$ and $y$
together with its $n-1$ derivatives at some inner point in $\left(
a,b\right) $ or at a regular end.

The only difference is that the space $D\left( a,b\right) $ of smooth finite
functions that is universally suitable for differential expressions with
smooth coefficients of any order is inappropriate in this case because $%
\check{f}\varphi $ is no longer smooth and has to be replaced for each
differential expression of any order $n$ by its own space $D_{n}\left(
a,b\right) $ of functions $\varphi $ with a compact support in $\left(
a,b\right) $ and absolutely continuous together with its $n-1\,$derivatives%
\begin{equation}
D_{n}\left( a,b\right) =\left\{ \varphi :\varphi \in C^{n}\left( a,b\right)
\,,\;\mathrm{supp\,}\varphi \subseteq \left[ \alpha ,\beta \right] \subset
\left( a,b\right) \right\} \,;  \label{3b.39}
\end{equation}%
of course, $D\left( a,b\right) \subset D_{n}\left( a,b\right) $. It is
natural to keep the name ``finite functions'' for such functions. In the
case of a regular end $a$ where a solution of a homogenous equation has a
finite limit together with its $n-1$ derivatives, the space $D_{n}\left(
a,b\right) $ can be extended to functions vanishing at this regular end
together with its $n-1$ derivatives. The same is true for a regular end $b$.
It is easy to see that above Lemmas \ref{l3b.1} and \ref{l3b.2} are directly
extended to such finite functions, and therefore, the extension Lemma \ref%
{l3b.3} to s.a. differential expressions of any order also holds.

For even s.a. expressions, the corresponding assertion holds under the
weakened above-mentioned conditions on the coefficients in terms of
quasiderivatives, see \cite{AkhGl81,Naima69}.

This result is the main ingredient in evaluating the adjoint of a
preliminary symmetric operator associated with a given s.a. expression, see
below.

\subsection{Natural domain. Operator $\hat{f}^{\ast }.$}

We are now ready to proceed to constructing s.a. operators in $L^{2}\left(
a,b\right) $ associated with a given s.a. differential expression $\check{f}$
(\ref{3b.8}) based on the general theory of s.a. extensions of symmetric
operators presented in the previous section. For simplicity, we consider the
case of smooth coefficients which allows universally considering
differential expressions and associated operators of any order. The results
are naturally extended to the general case of nonsmooth coefficients under
the above-mentioned conditions on the coefficients.

We begin with the so-called natural domain for a s.a. differential
expression $\check{f}$ (\ref{3b.8}).

Let $D_{\ast }$ be a subspace of square-integrable functions\footnote{%
The expediency of this notation is justified below.} $\psi _{\ast }$ that
are absolutely continuous\footnote{%
When we say that some property of functions under consideration holds in $%
\left( a,b\right) $, we mean that this property holds for any finite segment 
$\left[ \alpha ,\beta \right] \subset \left( a,b\right) $.} in $\left(
a,b\right) $ together with its $n-1$ derivatives and such that $\check{f}%
\psi _{\ast }$ is square-integrable as well as $\psi _{\ast }$:%
\begin{equation}
D_{\ast }=\left\{ \psi _{\ast }:\psi _{\ast },\psi _{\ast }^{\prime },\ldots
,\psi _{\ast }^{\left( n-1\right) }\;\mathrm{a.c.\;in\;}\left( a,b\right)
;\,\psi _{\ast },\,\check{f}\psi _{\ast }\in L^{2}\left( a,b\right) \right\}
.  \label{3b.40}
\end{equation}%
It is evident that $D_{\ast }$ is the largest linear subspace in $%
L^{2}\left( a,b\right) $ on which a differential operator in $L^{2}\left(
a,b\right) $ can be defined with the ``rule of acting'' given by $\check{f}$%
: the requirement that $\psi _{\ast },\psi _{\ast }^{\prime },\ldots ,\psi
_{\ast }^{\left( n-1\right) }$ be absolutely continuous in $\left(
a,b\right) $ is necessary for $\check{f}\psi _{\ast }$ to have a sense of
function, the requirement that $\psi _{\ast }$ and $\check{f}\psi _{\ast }$
belong to $L^{2}\left( a,b\right) \,\ $is necessary for $\psi _{\ast }$ and $%
\check{f}\psi _{\ast }$ be the respective pre-image and image of an operator
in $L^{2}\left( a,b\right) $ defined by $\check{f}$. We call the domain $%
D_{\ast }$ (\ref{3b.40}) the natural domain for a s.a. differential
expression $\check{f}$ (\ref{3b.8}) and let $\hat{f}^{\ast }$ denote the
respective operator in $L^{2}\left( a,b\right) $\thinspace associated with
the differential expression $\check{f}$ and defined on the natural domain $%
D_{\ast }$, such that%
\begin{equation}
\hat{f}^{\ast }=\left\{ 
\begin{array}{l}
D_{f^{\ast }}=D_{\ast }\,, \\ 
\hat{f}^{\ast }\psi _{\ast }=\check{f}\psi _{\ast }\,.%
\end{array}%
\right.  \label{3b.41}
\end{equation}

It is evident that the space $D\left( a,b\right) $ of finite smooth
functions belongs to $D_{\ast }$, $D\left( a,b\right) \subset D_{\ast }$%
\thinspace , and because $D\left( a,b\right) $ is dense in $L^{2}\left(
a,b\right) \,,\;$ $\overline{D\left( a,b\right) }=L^{2}\left( a,b\right) ,$
the domain $D_{\ast }$ is all the more dense in $L^{2}\left( a,b\right) \,,$%
\ $\overline{D_{\ast }}=L^{2}\left( a,b\right) ,$ such that the operator $%
\hat{f}^{\ast }$ is densely defined.

As we already mentioned above in Comment 4 in the previous section, in the
physical literature and even in some textbooks on quantum mechanics for
physicists, s.a. differential expression (\ref{3b.8}) is identified with a
s.a. operator in $L^{2}\left( a,b\right) $ without any reservation on its
domain, and the spectrum and eigenfunctions of this operator are immediately
looked for. Although its domain is not indicated, but actually, the natural
domain for $\check{f}$ is implicitly meant by this domain:\ it is believed
that the only requirements are the requirement of square integrability for
the respective eigenfunctions of bound eigenstates and the requirement of
local square integrability and the ``normalization to $\delta $-function''
for (generalized) eigenfunctions of the continuous spectrum. In some cases,
this appears sufficient, but sometimes, is not:\ possible situations are
shortly described in Comment 4 in the previous section.

As we show later, to verify that $\hat{f}^{\ast }$ is s.a., it is sufficient
to verify that it is symmetric, the necessary and sufficient conditions for
which are that its sesquilinear asymmetry form $\omega _{\ast }$ or its
quadratic asymmetry form $\Delta _{\ast }$ defined on its domain $D_{\ast }$
respectively by, see (\ref{3a.7}), (\ref{3a.8}),%
\begin{equation}
\omega _{\ast }\left( \chi _{\ast },\psi _{\ast }\right) =\int_{a}^{b}dx%
\overline{\chi _{\ast }}\check{f}\psi _{\ast }-\int_{a}^{b}dx\psi _{\ast }%
\overline{\check{f}\chi _{\ast }},\;\forall \chi _{\ast },\psi _{\ast }\in
D_{\ast },  \label{3b.42}
\end{equation}%
and%
\begin{equation}
\,\Delta _{\ast }\left( \psi _{\ast }\right) =\int_{a}^{b}dx\overline{\psi
_{\ast }}\check{f}\psi _{\ast }-\int_{a}^{b}dx\psi _{\ast }\overline{\check{f%
}\psi _{\ast }}\,,\;\forall \psi _{\ast }\in D_{\ast }\,,  \label{3b.43}
\end{equation}%
vanish; because $\omega _{\ast }$ and $\Delta _{\ast }$ define each other,
see the previous section, it is sufficient to do this for only one of this
form.

We now show that the values of asymmetry forms $\omega _{\ast }$ (\ref{3b.42}%
) and $\Delta _{\ast }$ (\ref{3b.43}) are defined by the behavior of
functions belonging to $D_{\ast }$ near the ends $a$ and $b$ of the interval 
$\left( a,b\right) $ because the both $\omega _{\ast }$ and $\Delta _{\ast }$
are determined by the boundary values of the respective sesquilinear form $%
\left[ \chi _{\ast },\psi _{\ast }\right] $ (\ref{3b.18}) and quadratic form 
$\left[ \psi _{\ast },\psi _{\ast }\right] $, its reduction to the diagonal,
that are local forms in functions and its derivatives of order up to $n-1$.
Really, by the integral Lagrange identity (\ref{3b.19}), we have%
\begin{equation}
\omega _{\ast }\left( \chi _{\ast },\psi _{\ast }\right) =\left. \left[ \chi
_{\ast },\psi _{\ast }\right] \right| _{a}^{b}=\left[ \chi _{\ast },\psi
_{\ast }\right] \left( b\right) -\left[ \chi _{\ast },\psi _{\ast }\right]
\left( a\right) \,,\;\forall \chi _{\ast },\psi _{\ast }\in D_{\ast }\,,
\label{3b.44}
\end{equation}%
where, by definition,%
\begin{equation}
\left[ \chi _{\ast },\psi _{\ast }\right] \left( a\right) =\underset{%
x\rightarrow a}{\lim }\left[ \chi _{\ast },\psi _{\ast }\right] \,,\;\left[
\chi _{\ast },\psi _{\ast }\right] \left( b\right) =\underset{x\rightarrow b}%
{\lim }\left[ \chi _{\ast },\psi _{\ast }\right] \,;  \label{3b.45}
\end{equation}%
the boundary values $\left[ \chi _{\ast },\psi _{\ast }\right] \left(
b\right) $ and $\left[ \chi _{\ast },\psi _{\ast }\right] \left( a\right) $
of the form $\left[ \chi _{\ast },\psi _{\ast }\right] $ do exist for any $%
\chi _{\ast },\psi _{\ast }\in D_{\ast }$ because the integrals in r.h.s. in
(\ref{3b.42}) defining $\omega _{\ast }\left( \chi _{\ast },\psi _{\ast
}\right) $ exist. We note that the existence of limits (\ref{3b.45}) does
not imply that the functions in $D_{\ast }$ have the respective boundary
values at $a$ and $b$ together with its $n-1$ derivatives; in general, these
may not exist.

Similarly, for the quadratic asymmetry form $\Delta _{\ast }$ (\ref{3b.43}),
we have%
\begin{equation}
\Delta _{\ast }\left( \psi _{\ast }\right) =\left. \left[ \psi _{\ast },\psi
_{\ast }\right] \right| _{a}^{b}=\left[ \psi _{\ast },\psi _{\ast }\right]
\left( b\right) -\left[ \psi _{\ast },\psi _{\ast }\right] \left( a\right)
\,,  \label{3b.46}
\end{equation}%
where%
\begin{equation}
\left[ \psi _{\ast },\psi _{\ast }\right] \left( a\right) =\underset{%
x\rightarrow a}{\lim }\left[ \psi _{\ast },\psi _{\ast }\right] \,,\;\left[
\psi _{\ast },\psi _{\ast }\right] \left( b\right) =\underset{x\rightarrow b}%
{\lim }\left[ \psi _{\ast },\psi _{\ast }\right] \,.  \label{3b.47}
\end{equation}

We note that the boundary values (\ref{3b.45}) and (\ref{3b.47}) of local
forms are independent in the following sense. Let we evaluate $\left[ \chi
_{\ast },\psi _{\ast }\right] \left( a\right) $ for some functions $\chi
_{\ast },\psi _{\ast }\in D_{\ast }$. For any function $\chi _{\ast },$
there exists another function $\widetilde{\chi }_{\ast }\in D_{\ast }$ that
coincides with $\chi _{\ast }$ near the end $a$ and vanishes near the end $b$%
, more strictly $\widetilde{\chi }_{\ast }=\chi _{\ast }$, $a\leq x<\alpha
<b $ and $\widetilde{\chi }_{\ast }=0$, $\alpha <\beta <x\leq b$. In the
case of a differential expression with differentiable coefficients\footnote{%
This is a short name for a differential expression with coefficients
satisfying the standard differentiability conditions, in particular, with
smooth coefficients.}, such a function can be obtained by multiplying $\chi
_{\ast }$ by a smooth step-like function $\tilde{\theta}\left( x\right) $
equal to unity near $x=a$ and zero near $x=b$. In the case of an even
differential expression with nondifferentiable coefficients, the
multiplication by $\widetilde{\theta }$ in general makes $\widetilde{\chi
_{\ast }}$ to leave the domain $D_{\ast }$, but the existence of functions $%
\widetilde{\chi }$ with the required properties can be proved \cite%
{AkhGl81,Naima69}. We then have $\left[ \widetilde{\chi }_{\ast },\psi
_{\ast }\right] \left( a\right) =\left[ \chi _{\ast },\psi _{\ast }\right]
\left( a\right) $ while $\left[ \widetilde{\chi }_{\ast },\psi _{\ast }%
\right] \left( b\right) =0$. The same is true for the end $b$. It follows
that the conditions of vanishing the asymmetry form $\omega _{\ast }$ with
an arbitrary first argument, i.e., the condition $\omega _{\ast }\left( \chi
_{\ast },\psi _{\ast }\right) =0$, $\forall \chi _{\ast }\in D_{\ast }$, is
equivalent to the condition of separately vanishing boundary values (\ref%
{3b.45}), i.e., to the boundary conditions%
\begin{equation}
\left[ \psi _{\ast },\psi _{\ast }\right] \left( a\right) =\left[ \psi
_{\ast },\psi _{\ast }\right] \left( b\right) =0\,,\;\forall \psi _{\ast
}\in D_{\ast }\,.  \label{e.9}
\end{equation}%
It is evident that we can interchange the first and second arguments $\chi
_{\ast }$ and $\psi _{\ast }$ in the above consideration.

All the above-said is true for boundary values (\ref{3b.47}). In particular,
the condition $\Delta _{\ast }\left( \psi _{\ast }\right) =0$, $\forall \psi
_{\ast }\in D_{\ast }$, for an operator $\hat{f}^{\ast }$ associated with a
differential expression $\check{f}$ is equivalent to the boundary conditions
(\ref{e.9}).

It follows that an answer to the question of whether the operator $\hat{f}%
^{\ast }$ (\ref{3b.41}) is symmetric, and therefore s.a., or not, is defined
by possible boundary values (\ref{3b.45}) and (\ref{3b.47}) for the
respective asymmetry forms $\omega _{\ast }$ and $\Delta _{\ast }$ for all
functions in $D_{\ast }$, namely, whether they vanish identically or not. We
shortly discuss the possibility to answer the question. For definiteness, we
speak about boundary values (\ref{3b.47}). The natural domain $D_{\ast }$ (%
\ref{3b.40}) can be defined as the space of square-intergrable solutions $%
\psi _{\ast }$ of the differential equation%
\begin{equation}
\check{f}\psi _{\ast }=\chi _{\ast }\,,\;\forall \chi _{\ast }\in
L^{2}\left( a,b\right) \,.  \label{3b.48}
\end{equation}%
Therefore, boundary values (\ref{3b.47}) can be evaluated by analyzing the
behavior of the general solution $\psi _{\ast }$ of eq. (\ref{3b.48}) near
the ends $a$ and $b$ of the interval $\left( a,b\right) $ with the
additional condition that $\psi _{\ast }$ must be square integrable up to
the ends.

If we succeeded in proving that boundary values (\ref{3b.47}) vanish for all
functions in $D_{\ast }$, we thus prove that the operator $\hat{f}^{\ast }$ (%
\ref{3b.41}) associated with a given differential expression $\check{f}$ and
defined on the natural domain $D_{\ast }$ (\ref{3b.40}) is s.a.. We show
later that it is a unique s.a. operator associated with $\check{f}$.
Therefore, it seems evident that the first thing we should do is to attempt
to prove that boundary values (\ref{3b.47}) of the quadratic local form $%
[\psi _{\ast },\psi _{\ast }]$ vanish for all functions $\psi _{\ast }$ in
the natural domain $D_{\ast }$ (\ref{3b.40}). But if we can indicate a
function $\psi _{\ast }\in D_{\ast }$ such that, for example, $\left[ \psi
_{\ast },\psi _{\ast }\right] \left( a\right) \neq 0$, we thus prove that
the operator $\hat{f}^{\ast }$ (\ref{3b.41}) is nonsymmetric and, all the
more, non-s.a..

In general, the set of possible boundary values (\ref{3b.47}) depends on the
type of the interval $(a,b)$, namely, whether it is a whole axis $\mathbb{R}%
^{1}$, or a semiaxis, or a finite interval, and on the behavior of the
coefficients of $\check{f}$ as $x\rightarrow a$ and $x\rightarrow b$. We
illustrate possible situations by simple examples related to differential
expression $\check{H}$ (\ref{3b.10}).

Let $(a,b)=(-\infty ,\infty )=\mathbb{R}^{1}$, and let $\check{f}=\check{H}$
given by (\ref{3b.10}) with the zero potential, $V=0$. We conventionally let 
$\check{H}_{0}$ denote this differential expression,%
\begin{equation}
\check{H}_{0}=-\frac{d^{2}}{dx^{2}}\,,  \label{3b.48a}
\end{equation}%
it is identified with the Hamiltonian of a free nonrelativistic particle
moving along the real axis $\mathbb{R}^{1}$. The natural domain $D_{0\ast }$
for $\check{H}_{0}$ is%
\begin{equation}
D_{0\ast }=\left\{ \psi _{\ast }:\,\psi _{\ast },\,\psi _{\ast }^{\prime }\;%
\text{\textrm{a.c. in\ }}\mathbb{R}^{1};\,\psi _{\ast },\,\psi _{\ast
}^{\prime \prime }\in L^{2}(\mathbb{R}^{1})\right\} \,.  \label{3b.49}
\end{equation}%
As we already mentioned above, $\psi _{\ast }\in L^{2}(\mathbb{R}^{1})$ does
not imply that $\psi _{\ast }\rightarrow 0$ as $x\rightarrow \pm \infty $.
Therefore, the proof of the self-adjointness (actually, the symmetricity) of
the free Hamiltonian based on the opposite assertion in some textbooks for
physicists is incorrect. But it can be shown, and we show this later, that $%
\psi _{\ast }\in D_{0\ast }$ (\ref{3b.49}) implies that $\psi _{\ast },\psi
_{\ast }^{\prime }\rightarrow 0$ as $x\rightarrow \pm \infty $, and
therefore, the quadratic local form $\left[ \psi _{\ast },\psi _{\ast }%
\right] =-\left[ \overline{\psi _{\ast }}\psi _{\ast }^{\prime }-\overline{%
\psi _{\ast }^{\prime }}\psi _{\ast }\right] $ for $\check{H}_{0}$, see (\ref%
{3b.22}) with $\chi _{\ast }=\psi _{\ast }$, vanishes at infinities, $\left[
\psi _{\ast },\psi _{\ast }\right] \rightarrow 0\,,\;x\rightarrow \pm \infty
\,,$ i.e., boundary values (\ref{3b.47}) vanish for all $\psi _{\ast }$ in
this case. It follows that the operator $\hat{H}_{0}$ (we conventionally
omit the upperscript $\ast $) associated with the differential expression $%
\check{H}_{0}$ and defined on the natural domain, the free Hamiltonian, is
really s.a., which we know from textbooks. As we show later, the same is
true for the potential $V(x)=x^{2},$ we then deal with the differential
expression $\check{H}=-d^{2}/dx^{2}+x^{2}$, which is identified with the
Hamiltonian for a quantum oscillator: the same local form $\left[ \psi
_{\ast },\psi _{\ast }\right] $ vanishes at infinities also in this case.
This implies that the operator $\hat{H}$ associated with this differential
expression $\check{H}=-d^{2}/dx^{2}+x^{2}$ and defined on the natural domain%
\begin{equation*}
D_{\ast }=\left\{ \psi _{\ast }:\,\psi _{\ast },\psi _{\ast }^{\prime }\;%
\text{\textrm{a.c. in\ }}\mathbb{R}^{1}\,;\,\psi _{\ast },-\psi _{\ast
}^{\prime \prime }+x^{2}\psi _{\ast }\in L^{2}(\mathbb{R}^{1})\right\} \,.
\end{equation*}%
is s.a., which we also know from textbooks.

But let now $V(x)$ be a rather exotic potential rapidly going to $-\infty $
as $x\rightarrow \pm \infty $, for example, let $V=-x^{4}$, such that the
''Hamiltonian'' is $\check{H}=-d^{2}/dx^{2}-x^{4}$. Let $\psi _{\ast }$ be a
square-integrable smooth function that exponentially vanishes as $%
x\rightarrow -\infty $ and such that%
\begin{equation*}
\psi _{\ast }=\frac{1}{x}\exp \left( \frac{i}{3}x^{3}\right) \,,\;x>N>0\,.
\end{equation*}%
It is easy to verify that $\psi _{\ast }$ belongs to the natural domain $%
D_{\ast }$ for this $\check{H}$:%
\begin{equation*}
-\psi _{\ast }^{\prime \prime }-x^{4}\psi _{\ast }=-\frac{2}{x^{3}}\exp
\left( \frac{i}{3}x^{3}\right) \,,\;x>N\,,
\end{equation*}%
and is square-integrable at $+\infty $, as well as $\psi _{\ast }$, while
the left end, $-\infty $, is evidently safe. It is also easy to evaluate the
form $\left[ \psi _{\ast },\psi _{\ast }\right] $ for $x>N$, it is $\left[
\psi _{\ast },\psi _{\ast }\right] =-2i.$ It follows that for this function,
the boundary value $\left[ \psi _{\ast },\psi _{\ast }\right] (+\infty
)=-2i\neq 0$, which implies that the operator $\hat{H}^{\ast }$ associated
with the differential expression $\check{H}=-d^{2}/dx^{2}-x^{4}$ and defined
on the natural domain%
\begin{equation*}
D_{\ast }=\left\{ \psi _{\ast }:\,\psi _{\ast },\psi _{\ast }^{\prime }%
\mathrm{\;}\text{\textrm{a.c. in\ }}\mathbb{R}^{1};\psi _{\ast },-\psi
_{\ast }^{\prime \prime }-x^{4}\psi _{\ast }\in L^{2}(\mathbb{R}^{1})\right\}
\end{equation*}%
is non-s.a., and even nonsymmetric,and,therefore, it cannot be considered a
quantum-mechanical Hamiltonian for a particle in the potential field $%
V=-x^{4}$. The correct Hamiltonian in this case requires an additional
specification. We only note in advance that this is possible, but
nonuniquely. It is also interesting that the spectrum of such a Hamiltonian
is discrete, although it may seem unexpected at the first glance.

If the interval $\left( a,b\right) $ is a semiaxis, for example, the
positive semiaxis $\left( 0,\infty \right) $, and the left end $a=0$ is
regular, then $\psi _{\ast },\psi _{\ast }^{\prime }$ are continuous up to
this end and can take arbitrary complex values, which implies that $\left[
\psi _{\ast },\psi _{\ast }\right] \left( 0\right) =-\left[ \overline{\psi
_{\ast }\left( 0\right) }\psi _{\ast }^{\prime }\left( 0\right) -\overline{%
\psi _{\ast }^{\prime }\left( 0\right) }\psi _{\ast }\left( 0\right) \right] 
$ can also take arbitrary nonzero imaginary values and, therefore, the
operator $\hat{H}^{\ast }$ is not s.a..

An important remark concerning real quantum mechanics is in order here.

In physics, the differential expressions like (\ref{3b.10}) on the positive
semiaxis usually have a three-dimensional origin. Their standard source is a
problem of a space motion of a quantum particle in spherically symmetric or
axially symmetric fields.

Let we consider a space motion, for example, the scattering or bound states,
of a nonrelativistic spinless particle in a spherically symmetric field. The
quantum states of the particle are described by wave function $\psi (\mathbf{%
r}),\;\mathbf{r}$ is the radius-vector, $\psi (\mathbf{r})\in L^{2}(R^{3})$,
and the motion is governed by the ``Hamiltonian'' $\mathbf{H}=-\Delta +V(r)$%
, where $\Delta $ is the Laplacian, $V(r)$ is a potential, and $r=\left| 
\mathbf{r}\right| $ (the appropriate units are assumed, in particular, $%
\hbar =1$). The problem is usually solved by separating the variables $%
\mathbf{r}\rightarrow r,\theta ,\varphi $, where $\theta ,\varphi $ are
spherical angles. When passing from the three-dimensional wave function $%
\psi (\mathbf{r})$ to its partial waves $u_{l}(r,\theta ,\varphi
)=u_{l}(r)Y_{lm}(\theta ,\varphi ),$ where $Y_{lm}$ are spherical harmonics, 
$\psi (\mathbf{r})\overset{\infty }{\underset{l=0}{=\sum }}%
(2l+1)u_{l}(r)Y_{lm}(\theta ,\varphi )$, the differential expressions like (%
\ref{3b.10}) naturally arise as the so called radial ``Hamiltonians'' $%
\check{H}_{l}$ for the radial motion with the angular momentum $l=0,1,...$:%
\begin{equation}
\check{H}_{l}=-\frac{d^{2}}{dr^{2}}+V_{l}(r)\,,  \label{3b.50}
\end{equation}%
where the partial potential $V_{l}(r)$ is%
\begin{equation}
V_{l}(r)=V(r)+\frac{l(l+1)}{r^{2}}  \label{3b.51}
\end{equation}%
and includes the so-called centrifugal term $l(l+1)/r^{2}$. The radial
motion is described in terms of the radial wave functions $\psi _{l}(r)\in
L^{2}(0,\infty )$ that differ from the partial amplitudes $u_{l}(r)$ by the
factor $r$, $\psi _{l}(r)=ru_{l}(r)$, which is essential. If the initial
three-dimensional potential $V(r)$ is nonsingular at the origin or have
rather weak singularity (we do not define an admissible singularity for $%
V(r) $ at $r=0$ more precisely here, see \cite{FadMa64}, the natural domain
for the three-dimensional Hamiltonian $\mathbf{H}$ consists of functions $%
\psi _{\ast }(\mathbf{r})$ that are sufficiently regular in the neighborhood
of the origin, such that the partial amplitudes $u_{l}\left( r\right) $ are
finite at $r=0$ and, therefore, the radial wave functions $\psi _{l}\left(
r\right) $ must vanish at $r=0$. In this setting, the natural domain $%
D_{\ast l}$ for $\check{H}_{l}$ is reduced to a domain $D_{l}$ that differs
form $D_{\ast l}$ by the additional boundary condition $\psi _{l}\left(
0\right) =0$. This boundary condition is the well-known conventional
condition in physics for the radial wave-functions. If, in addition, $\left[
\psi _{l},\psi _{l}\right] \left( \infty \right) =0$, which holds if $%
V\left( r\right) \rightarrow 0$ as $r\rightarrow 0,$ as we show later, then
the operator $\hat{H}_{l}$ associated with the differential expression $%
\check{H}_{l}$ (\ref{3b.50}), (\ref{3b.51}) and defined on the domain $D_{l}$
is a s.a. operator and can be considered a quantum-mechanical observable
that we know from textbooks. To be true, the zero boundary condition at $r=0$
for the radial wave functions is critical only for the $s$-wave, $l=0$,
because for $l=1,2,\ldots $ the natural domain $D_{\ast l}$ coincides with $%
D_{l}$.

These arguments fail if the potential is strongly singular, for example, in
the cases where $V=-\alpha /r^{2}$, $\alpha >-\frac{1}{4}$, or $V=-\alpha
/r^{\beta }$, $\alpha >0$, $\beta >2$, and where the so-called phenomenon of
``fall to the center'' occurs.

A similar consideration can be carried out for a motion of a particle in an
axially symmetric potential field $V\left( \rho \right) $, $\rho $ is the
distance to the axis, with the same conclusion for the partial radial
Hamiltonians%
\begin{equation*}
\check{H}_{m}=-\frac{d^{2}}{d\rho ^{2}}+V_{m}\left( \rho \right) \,,
\end{equation*}%
where the partial potential is%
\begin{equation*}
V_{m}\left( \rho \right) =V\left( \rho \right) +\frac{m^{2}}{\rho ^{2}}\,,
\end{equation*}%
and $m=0,1,\ldots $ is the projection of the angular momentum to the axis.
The reason is that the radial wave functions $\psi _{m}\left( \rho \right)
\in L^{2}\left( 0,\infty \right) $ differ from the original partial
amplitudes $u_{m}\left( \rho \right) $ square integrable with the measure $%
\rho d\rho $ by the factor $\rho ^{1/2}$, $\psi _{m}\left( \rho \right)
=\rho ^{1/2}u_{m}\left( \rho \right) $, and if the initial potential $%
V\left( \rho \right) $ is not too singular, the natural domain for $\check{H}%
_{m}$ is supplied by the additional boundary condition $\psi _{m}\left(
0\right) =0$.

If an interval $\left( a,b\right) $ is finite and one of its ends $a$ and $b$
or the both are regular, then one of the boundary values (\ref{3b.47}) or
the both can be nonzero, and, therefore, the operator $\hat{H}^{\ast }$
associated with the differential expression $\check{H}$ and defined on the
natural domain is non-s.a. in this case. For example, this assertion holds
for the differential expression (\ref{3b.48a}), the ``Hamiltonian'' for a
free particle on a finite interval of the real axis. The physical reason for
this is evident:\ a particle can ``escape'' from or enter the interval
through the ends, which results in the nonunitarity of evolution. Only
additional physical arguments preventing these possibilities by additional
boundary conditions that make the asymmetry form $\Delta _{\ast }$ (\ref%
{3b.43}) to be zero result in the self-adjointness of the real Hamiltonian $%
\hat{H}_{0}$ associated with the differential expression $\check{H}_{0}$.
The most known s.a. boundary conditions are $\psi \left( a\right) =\psi
\left( b\right) =0$, which corresponds to a particle in an ``infinite
potential well'', and the periodic boundary conditions $\psi \left( a\right)
=\psi \left( b\right) $, $\psi ^{\prime }\left( a\right) =\psi ^{\prime
}\left( b\right) $ (the latter condition is usually hidden in textbooks),
which corresponds to ``quantization in a box'' conventionally used in
statistical physics.

\subsection{Initial symmetric operator and its adjoint. Deficiency indices.}

We now return to the general consideration. Because the operator $\hat{f}%
^{\ast }$ (\ref{3b.41}) associated with the s.a. differential expression $%
\check{f}$ (\ref{3b.8}) and defined on the natural domain $D_{\ast }$ is
generally non-s.a., we proceed to the general programme of constructing s.a.
operators presented in the previous section. In the case of differential
operators in $L^{2}\left( a,b\right) $, it is mainly based on a possibility
to represent their asymmetry forms $\omega _{\ast }$ and $\Delta _{\ast }$
in terms of (asymptotic) boundary values of the local form $\left[ \cdot
,\cdot \right] $ (\ref{3b.18}) similar to the respective (\ref{3b.44}), (\ref%
{3b.45}), and (\ref{3b.46}), (\ref{3b.47}).

As the first step, we must define a symmetric operator in $L^{2}\left(
a,b\right) $ associated with a given s.a. differential expression $\check{f}$
of order $n$. In the case of smooth coefficients, it is natural to take the
subspace $D\left( a,b\right) $ of smooth functions, $D\left( a,b\right)
\subset L^{2}\left( a,b\right) $, for a domain of such an operator and thus
to start with a symmetric operator $\hat{f}^{\left( 0\right) }$ defined in $%
L^{2}\left( a,b\right) $ by%
\begin{equation}
\hat{f}^{\left( 0\right) }:\ \left\{ 
\begin{array}{l}
D_{f^{\left( 0\right) }}=D\left( a,b\right) \,, \\ 
\hat{f}^{\left( 0\right) }\varphi =\check{f}\varphi \,,\;\forall \varphi \in
D\left( a,b\right) \,.%
\end{array}%
\right.  \label{3b.52}
\end{equation}%
It is evident that $\check{f}\varphi \in L^{2}\left( a,b\right) $ as well as 
$\varphi $ because of a finite support of $\varphi $. It is also evident
that $\hat{f}^{\left( 0\right) }$ is symmetric because it is densely
defined, $\overline{D\left( a,b\right) }=L^{2}\left( a,b\right) $, and the
equality%
\begin{equation*}
\left( \phi ,\hat{f}^{\left( 0\right) }\varphi \right) =\left( \phi ,\hat{f}%
^{\left( 0\right) }\varphi \right) \,,\;\forall \varphi ,\phi \in
D_{f^{\left( 0\right) }}=D\left( a,b\right)
\end{equation*}%
holds because it coincides with eq. (\ref{3b.3}) $\left( \phi ,\check{f}%
\varphi \right) =\left( \check{f}\phi ,\varphi \right) $ for the s.a. $%
\check{f}=\check{f}^{\ast }$, the latter equality is simply the
manifestation of the self-adjointness of $\check{f}$ as a differential
expression, see also eq. (\ref{3b.33}) with $y=\phi $. We emphasize once
more that because of the self-adjointness of $\check{f}$ as a differential
expression, $\hat{f}^{\left( 0\right) }$ is generally only a symmetric, but
not s.a., operator in $L^{2}\left( a,b\right) $.

The second step is evaluating the adjoint $\left( \hat{f}^{\left( 0\right)
}\right) ^{+}$ by solving the defining equation%
\begin{equation*}
\left( \psi _{\ast },\hat{f}^{\left( 0\right) }\varphi \right) -\left( \chi
_{\ast },\varphi \right) =0\,,\;\forall \varphi \in D_{f^{\left( 0\right) }}
\end{equation*}%
for a pair of vectors $\psi _{\ast }\in D_{\left( f^{\left( 0\right)
}\right) ^{+}}\subset L^{2}\left( a,b\right) $ and $\chi _{\ast }=\left( 
\hat{f}^{\left( 0\right) }\right) ^{+}\psi _{\ast }\in L^{2}\left(
a,b\right) $, see subsec.2.1. In our case, this is the equation%
\begin{equation}
\int_{a}^{b}dx\overline{\psi _{\ast }}\check{f}\varphi -\int_{a}^{b}dx%
\overline{\chi _{\ast }}\varphi =0\,,\;\forall \varphi \in D\left(
a,b\right) \,,  \label{3b.53}
\end{equation}%
for a pair of square-integrable functions $\psi _{\ast }$ and $\chi _{\ast }$%
. We assert that the adjoint $\left( \hat{f}^{\left( 0\right) }\right) ^{+}$
coincides with the above-introduced operator $\hat{f}^{\ast }$ (\ref{3b.41}%
), $\left( \hat{f}^{\left( 0\right) }\right) ^{+}=\hat{f}^{\ast }$, in
particular, its domain $D_{\left( f^{\left( 0\right) }\right) ^{+}}$ is the
natural domain $D_{\ast }$ (\ref{3b.40}). In other words, we assert that
functions $\psi _{\ast }\in L^{2}\left( a,b\right) $ and $\chi _{\ast }\in
L^{2}\left( a,b\right) $ solve eq. (\ref{3b.53}) iff $\psi _{\ast }$ is
absolutely continuous in $\left( a,b\right) $ together with its derivatives
of order up to $n-1$ and $\chi _{\ast }=\check{f}\psi _{\ast }$.

Sufficiency is evident because of eq. (\ref{3b.33}) with $y=\psi _{\ast }$.

Necessity is proved as follows. Let $\psi _{\ast }\in L^{2}\left( a,b\right) 
$ and $\chi _{\ast }\in L^{2}\left( a,b\right) $ solve eq. (\ref{3b.53}),
and let $\widetilde{\psi }_{\ast }$ be some solution of the inhomogenous
differential equation $\check{f}\,\widetilde{\psi }_{\ast }=\chi _{\ast }\,.$
Such a function certainly exists because the square integrability of $\chi
_{\ast }(x)$ implies its local integrability in $(a,b)$; in addition, $%
\tilde{\psi}_{\ast }$ is absolutely continuous in $(a,b)$ together with its
derivatives of order up to $n-1$. We then have%
\begin{equation*}
\int_{a}^{b}dx\,\overline{\chi _{\ast }}\varphi =\int_{a}^{b}dx\varphi \,%
\overline{\check{f}\tilde{\psi}_{\ast }}=\int_{a}^{b}dx\overline{\tilde{\psi}%
_{\ast }}\,\check{f}\varphi
\end{equation*}%
because of the same eq. (\ref{3b.33}) with $y=\tilde{\psi}_{\ast }$, and the
defining equation (\ref{3b.53}) becomes%
\begin{equation*}
\int_{a}^{b}dx\,\overline{u}\check{f}\varphi =0\,,\;\forall \varphi \in
D(a,b),
\end{equation*}%
where $u=\psi _{\ast }-\tilde{\psi}_{\ast }$.

By the above-cited distribution theory theorem on the generalized solution
of the homogeneous equation $\check{f}u=0$, it follows that $%
u=\sum_{i=1}^{n}c_{i}u_{i}\,,$ where $\left\{ u_{i}\right\} _{1}^{n}$, is a
fundamental system of this homogeneous equation, and we finally obtain that $%
\psi _{\ast }=\tilde{\psi}_{\ast }+\sum_{i=1}^{n}c_{i}u_{i}\,,$ which
implies, that $\psi _{\ast }$ is absolutely continuous in $\left( a,b\right) 
$ together with its derivatives of order up to $n-1$ and $\chi _{\ast }=%
\check{f}\psi _{\ast }$. This completes the proof of the above assertion.

This assertion is evidently extended to the general case of nonsmooth
coefficients under the standard conditions on the coefficients of a
differential expression with a change of the domain $D_{f^{(0)}}$ of the
initial \ symmetric operator $\hat{f}^{(0)}$ from the space $D(a,b)$ of
finite smooth functions to the space $D_{n}(a,b)$ (\ref{3b.39}) of finite
functions. For even s.a. expressions, the requirements on the coefficients
can be weakened up to the similar conditions on the quasiderivatives, see %
\cite{AkhGl81,Naima69}.

The main conclusion is that in any case, the adjoint $\left( \hat{f}%
^{(0)}\right) ^{+}$ of the initial symmetric operator $\hat{f}^{(0)}$
associated with a s.a. differential expression $\check{f}$ (\ref{3b.8}) is
given by the same differential expression $\check{f}$ and defined on the
natural domain. Under the the standard conditions on the coefficients of $%
\check{f}$, the natural domain $D_{\ast }$ is given by (\ref{3b.40}) and the
adjoint $\left( \hat{f}^{(0)}\right) ^{+}$ coincides with the operator $\hat{%
f}^{\ast }$ (\ref{3b.41}). For even s.a. differential expression, the
condition of absolute continuity for derivatives can be weakened to the same
condition on quasiderivatives.

Therefore, the asymmetry forms $\omega _{\ast }$ and $\Delta _{\ast }$ of
the adjoint $\left( f^{\left( 0\right) }\right) ^{+}$ coincide with the
respective forms $\omega _{\ast }$ (\ref{3b.42}) and $\Delta _{\ast }$ (\ref%
{3b.43}) and are represented respectively by (\ref{3b.44}), (\ref{3b.45})
and (\ref{3b.46}), (\ref{3b.47}) in terms of boundary values of the local
form $\left[ \cdot ,\cdot \right] $ (\ref{3b.18}).

According to the general theory, if the adjoint $\left( f^{\left( 0\right)
}\right) ^{+}$ appears to be symmetric, which is equivalent to identically
vanishing boundary values (\ref{3b.45}) and (\ref{3b.47}), then $\left(
f^{\left( 0\right) }\right) ^{+}$ is s.a., the initial symmetric operator $%
\hat{f}^{\left( 0\right) }$ is essentially s.a. and its unique s.a.
extension is its closure $\overline{\hat{f}^{\left( 0\right) }}=\hat{f}$
coinciding with its adjoint $\left( \hat{f}^{(0)}\right) ^{+}$. This
justifies our preliminary statement that if the operator $\hat{f}^{\ast }$ (%
\ref{3b.41}) associated with a s.a. differential expression $\check{f}$ (\ref%
{3b.8}) and defined on the natural domain $D_{\ast }$ (\ref{3b.40}) is
symmetric, then it is s.a. and is a unique s.a. operator associated with a
given differential expression.

But according to the previous discussion, the adjoint $\left( \hat{f}%
^{(0)}\right) ^{+}=\hat{f}^{\ast }$ is generally nonsymmetric and we must
continue our programme of constructing s.a. operators associated with a
given differential expression $\check{f}$ by extending the initial symmetric
operator $\hat{f}^{\left( 0\right) }$ and restricting the adjoint\footnote{%
We note once again that an additional specification of a ``Hamiltonian'' $%
\check{H}$ by some boundary conditions for wave functions (which is a
standard practice in physics) is actually a self-adjoint restriction of $%
\hat{H}^{\ast }$ when it becomes clear that the Hamiltonian under
consideration is non-self-adjoint on the natural domain.}\ $\left( \hat{f}%
^{(0)}\right) ^{+}=\hat{f}^{\ast }$. According to this programme, the next
step is evaluating the deficient subspaces $D_{+}$ and $D_{-}$ of the
initial symmetric operator $\hat{f}^{\left( 0\right) },$%
\begin{equation*}
D_{\pm }=\left\{ \psi _{\pm }\in D_{\ast }:\;\hat{f}^{\ast }\psi _{\pm }=\pm
i\kappa \psi _{\pm }\right\} \,,
\end{equation*}%
and its deficiency indices $m_{+}$ and $m_{-}$, $m_{\pm }=\dim D_{\pm }$. In
our case, $D_{+}$ and $D_{-}$ are the spaces of square-integrable solutions $%
\psi _{+}$ and $\psi _{-}$ of the respective homogeneous differential
equations%
\begin{equation}
\check{f}\psi _{\pm }=i\kappa \psi _{\pm }\,,  \label{3b.54}
\end{equation}%
where $\kappa $ is an arbitrary, but fixed, dimensional parameter whose
dimension is the dimension of $\check{f}$.

We must find the complete systems $\left\{ e_{+,k}\right\} _{1}^{m_{+}}$ and 
$\left\{ e_{-,k}\right\} _{1}^{m_{-}}$ of linearly independent
square-integrable solutions of respective eqs. (\ref{3b.54}):%
\begin{equation}
\check{f}e_{+,k}=i\kappa e_{+,k}\,,\;k=1,\ldots ,m_{+}\,,\;\check{f}%
e_{-,k}=-i\kappa e_{-,k}\,,\;k=1,\ldots ,m_{-}\,;  \label{3b.56}
\end{equation}%
for the future, it is convenient to orthonormalize them,%
\begin{equation}
\left( e_{+,k},e_{+,l}\right) =\delta _{kl}\,,\;k,l=1,\ldots
,m_{+}\,,\;\left( e_{-,k},e_{-,l}\right) =\delta _{kl}\,,\;k,l=1,\ldots
,m_{-}\,,  \label{3b.58}
\end{equation}%
then $\left\{ e_{+,k}\right\} _{1}^{m_{+}}$ and $\left\{ e_{-,k}\right\}
_{1}^{m_{-}}\,$form the orthobasises in the respective $D_{+}$ and $D_{-}$,%
\begin{equation*}
\psi _{+}=\sum_{k=1}^{m_{+}}c_{+,k}e_{+,k}\,,\;c_{+,k}=\left( e_{+,k},\psi
_{+}\right) \,;\;\psi
_{-}=\sum_{k=1}^{m_{+}}c_{-,k}e_{-,k}\,,\;c_{-,k}=\left( e_{-,k},\psi
_{-}\right) \,.
\end{equation*}

As to possible values of deficiency indices, the following remarks of the
general nature can be useful.

We first note that the deficiency indices $m_{+}$ and $m_{-}$ of a symmetric
ordinary differential operator of order $n$ are always finite and do not
exceed $n$:\ for a differential expression $\check{f}$ of order $n$, the
whole number of linearly independent solutions, fundamental solutions $%
u_{\pm i}\left( x\right) $, of each of homogenous equations (\ref{3b.54}),
is equal to $n$, the additional requirement of square integrability of
solutions can only reduce this number, such that we generally have the
restriction $0\leq m_{+},m_{-}\leq n$.

As is clear from the above discussion of the operator $\hat{f}^{\ast }$, the
deficiency indices depend both on the type of the interval $\left(
a,b\right) $ and on the type of its ends $a$ and $b$, whether they are
regular or singular. If some end, $a$ or $b$, is regular, the general
solution of each of eqs. (\ref{3b.54}) is square integrable at this end, and
the square integrability of $\psi _{+}$ and $\psi _{-}$ is thus defined by
their square integrability at singular ends.

It follows that in the case where the interval $\left( a,b\right) $ is
finite and the both its ends are regular, we have $m_{+}=m_{-}=n$ for any
symmetric operator $\hat{f}^{\left( 0\right) }$ of order $n$. According to
the main theorem in the previous section, this implies that there is a $%
n^{2} $-parameter $U\left( n\right) $-family of s.a. operators associated
with a given differential expression $\check{f}$ of order $n$. For example,
the differential expression (\ref{3b.9}) generates a one-parameter $U\left(
1\right) $-family of s.a. operators each of which can be considered the
quantum-mechanical momentum of a particle on a finite interval of the real
axis (we already know this fact from the previous section), while the
differential expression (\ref{3b.48a}) generates a four-parameter $U\left(
2\right) $-family of s.a. operators each of which can be considered the
quantum-mechanical energy of a free particle on a finite interval. This
means that for a particle on a finite interval, an explicitly s.a.
differential expression does not yet define uniquely a quantum-mechanical
observable, and a further specification of the observable is required. We
show later that this specification is achieved by s.a. boundary conditions
on the wave functions in the domain of the observable. The optimistic point
of the conclusion is that s.a. operators associated with any s.a.
differential expression do exist in this case.

As to the case where one or both ends are singular, the situation is not so
optimistic in general. In particular, it is different for even s.a.
differential expressions with real coefficients and for odd s.a.
differential expressions with pure imaginary coefficients, all the more for
mixed differential expressions.

For even s.a. differential expressions $\check{f},$ the deficiency indices
of the associated symmetric operator $\hat{f}^{\left( 0\right) }$ are always
equal, $m_{+}=m_{-}=m$, independently of the type of an interval and its
ends. Really, because of the real coefficients of $\check{f}$, any
square-integrable solution $\psi _{+}$ of eq. (\ref{3b.54}) is assigned a
square-integrable solution $\psi _{-}=\overline{\psi _{+}}$, while the
linear independence of solutions preserves under complex conjugation. In
particular, for basis vectors $e_{+,k}$ in $D_{+}$ and $e_{-,k}$ in $D_{-}$
defined by (\ref{3b.56}), we can take complex conjugated functions such that 
$e_{-,k}=\overline{e_{+,k}},\;k=1,...,m$. Therefore, any even s.a.
expression always generates at least one s.a. operator in $L^{2}\left(
a,b\right) $ in contrast to odd s.a. differential expressions, as we already
know from the previous section by the example of the first-order
differential expression $\check{p}$ (\ref{3b.9}). In particular, for any
interval $\left( a,b\right) $, the energy of a nonrelativistic particle
associated with a differential expression $\check{H}$ (\ref{3b.10}) can
always be defined as a quantum-mechanical observable, although in general
nonuniquely.

The last two assertions on deficiency indices concern symmetric operators
associated with even s.a. expressions\footnote{%
Although it is quite probable that similar assertions hold for any s.a.
differential expressions, perhaps under some additional conditions for the
coefficients.}, see \cite{AkhGl81,Naima69}; for brevity, we call them even
symmetric operators. These assertions are based on the notion of the
dimension of a linear space modulo its subspace, on the boundary properties
of the functions in the domain of the closure of an even symmetric operator
at a regular end, on first von Neumann formula (\ref{3a.3}), and on second
von Neumann formula (\ref{3a.22}) and the remark to the second von Neumann
theorem on the relation between the deficiency indices of a symmetric
operator and its symmetric extension.

Let $L$ be some linear space, and let $M$ be its subspace, $M\subset L$. We
consider the factor space $L/M$, or the space $L$ modulo the subspace $M$,
that is a linear space whose vectors are equivalent classes of vectors in $L$
with respect to the equivalence relation where two vectors $\xi \in L$ and $%
\eta \in L$ are considered equivalent if their difference belongs to $M$, $%
\xi -\eta \in M$. The dimension of the factor space $L/M$ is denoted by $%
\dim _{M}L$ and is called the dimension of $L$ modulo $M$. Linearly
independent vectors $\xi _{1},\xi _{2},\ldots ,\xi _{k}\in L$ are called
linearly independent modulo $M$ if none of their nontrivial linear
combinations $\sum_{i=1}^{k}c_{i}\xi _{i}$ belongs to $M$: $%
\sum_{i=1}^{k}c_{i}\xi _{i}\in M\Longrightarrow \forall c_{i}=0$. If $\dim
_{M}L=n$, then the maximum number of vectors in $L$ linearly independent
modulo $M$ is equal to $n$, such that $k\leq n$. Let a space $L$ be a direct
sum of two its subspaces $L_{1}$ and $L_{2}$, $L=L_{1}+L_{2}$, then its
dimension is a sum of the dimensions of the subspaces, $\dim L=\dim
L_{1}+\dim L_{2}$, and $\dim _{L_{1}}L=\dim L_{2}$ and $\dim _{L_{2}}L=\dim
L_{1}$.

We discuss the closures of symmetric operators $\hat{f}^{\left( 0\right) }$
a bit later, and here, we only need one preliminary remark on this subject.
Let $\hat{f}^{\left( 0\right) }$ be an even symmetric operator of order $n$
with a regular end, let it be $a$, let $\hat{f}$ be its closure, $\hat {f}=%
\overline{\hat{f}^{\left( 0\right) }}$, with a domain $D_{f}$. It appears
that at a regular end, the functions in $D_{f}$ vanish together with their $%
n-1$ quasiderivatives: $\psi_{f}\in D_{f}\Longrightarrow\psi _{f}^{\left[ k%
\right] }\left( a\right) =0$, $k=0,\ldots,n-1$.

After this retreat, we return to the deficiency indices of even symmetric
operators.

If one of the ends of an interval $(a,b)$ is regular, let it be $a$, while
the second, $b$, is singular, the deficiency indices of even symmetric
operator $\hat{f}^{\left( 0\right) }$ of order $n$, being equal, $%
m_{+}=m_{-}=m$, and bounded from above, $m\leq n$, are also bounded from
below by $n/2$, such that the double-sided restriction%
\begin{equation}
\frac{n}{2}\leq m\leq n  \label{3b.60}
\end{equation}%
holds. In particular, the symmetric operator $\hat{H}^{\left( 0\right) }$
associated with the differential expression $\check{H}$ (\ref{3b.10}) for
the energy of a nonrelativistic particle on a semiaxis $(0,\infty )$ in a
potential field $V,$ where $V$ is regular at $x=0$, can have the deficiency
indices $m=1$ and $m=2$ in dependence on the behavior of $V$ at infinity,
but not zero. This implies that the quantum-mechanical Hamiltonian for such
a particle cannot be defined uniquely as a s.a. operator in $L^{2}(0,\infty
) $ without additional arguments. This fact is known since Weyl \cite{Weyl10}%
, where the cases $m=1$ and $m=2$ were respectively called the case of a \
''limit point'' and the case of ''limit circle'' due to a method of embedded
circles used by Weyl.

To prove the lower bound, we turn to the representation of the domain $%
D_{\ast }$ of the adjoint $\left( \hat{f}^{\left( 0\right) }\right) ^{+}=%
\hat{f}^{\ast }$ as a direct sum of $D_{f},\;D_{+}$ and $D_{-}\,,$ $D_{\ast
}=D_{f}+D_{+}+D_{-}\,,$ according to first von Neumann formula (\ref{3a.3}).
This formula implies that the \ maximum number of functions in $D_{\ast }$
linearly independent modulo $D_{f}$ is equal to $2m$ because%
\begin{equation*}
\dim _{D_{f}}D_{\ast }=\dim (D_{+}+D_{-})=\dim D_{+}+\dim D_{-}=2m\,.
\end{equation*}

If we prove that there exists a set $\left\{ \psi _{\ast l}\right\} _{1}^{n}$
of functions in $D_{\ast }$ linearly independent modulo $D_{f}\,$, we would
have $n\leq 2m$, which is required. But we know that the functions $\psi
_{\ast }$ in $D_{\ast }$ together with their quasiderivatives $\psi _{\ast
}^{\left[ k\right] }$ of order up to $\ n-1$ are finite at a regular end and
can take arbitrary values. Therefore, in our case of the regular end $a$,
there exists a set $\left\{ \psi _{\ast l}\right\} _{1}^{n}$ of linearly
independent functions such that the matrix $A,\;A_{l}^{k}=\psi _{\ast l}^{%
\left[ k\right] }(a)$, is nonsingular, $\ \det A\neq 0$. We assert that
these functions are also linearly independent modulo $D_{f}$ $.$ Really, let 
$\sum_{l}c_{l}\psi _{\ast l}=\psi \in D_{f}$. Then by the above remark, $%
\psi ^{\left[ k\right] }(a)=0,$ or $\sum_{l}c_{l}\psi _{\ast l}^{\left[ k%
\right] }(a)=\sum_{l}A_{l}^{k}c_{l}=0$, whence it follows that all $c_{l}=0$%
, $l=1,..,n,\,$\ because of the nonsingularity of the matrix $A$, which
completes the proof.

In the case where the both ends $a$ and $b$ of an interval $\left(
a,b\right) $ are singular, the evaluation of deficiency indices is reduced
to the case of one regular and one singular end by a specific symmetric
restriction of an initial symmetric operator $\hat{f}^{\left( 0\right) }$
and a comparison of the respective closures of the restriction and $\hat {f}%
^{\left( 0\right) }$ itself.

Let $\check{f}$ be an even s.a. differential expression of order $n$ on an
interval $\left( a,b\right) $ with the both singular ends, let $\hat{f}%
^{\left( 0\right) }$ be a symmetric operator in $L^{2}\left( a,b\right) $
associated with $\check{f}$, let $m_{+}=m_{-}=m$ be its deficiency indices,
and let $\hat{f}$ be its closure, $\hat{f}=\overline{\hat{f}^{\left(
0\right) }}$. Let $c$ be an intermediate point in the interval $\left(
a,b\right) $, $a<c<b$, such that $\left( a,b\right) =\left( ac\right) \cup
\left( cb\right) $. We note that the Hilbert space $L^{2}\left( a,b\right) $
is a direct sum of the Hilbert spaces $L^{2}\left( a,c\right) $ and $%
L^{2}\left( c,b\right) $, $L^{2}\left( a,b\right) =L^{2}\left( a,c\right)
\oplus L^{2}\left( c,b\right) $.

We consider the s.a. restrictions $\check{f}_{-}$ and $\check{f}_{+}$ of the
initial s.a. expression $\check{f}$ to the respective intervals $\left(
a,c\right) $ and $\left( c,b\right) $; the end $c$ for both differential
expressions $\check{f}_{-}$ and $\check{f}_{+}$ is evidently regular. Let $%
\hat{f}_{-}^{\left( 0\right) }$ and $\hat{f}_{+}^{\left( 0\right) }$ be the
symmetric operators in the respective $L^{2}\left( a,c\right) $ and $%
L^{2}\left( c,b\right) $ associated with the respective s.a. expressions $%
\check{f}_{-}$ and $\check{f}_{+}$ of order $n$ and defined on the
respective domains $D\left( a,c\right) \subset L^{2}\left( a,c\right) $ and $%
D\left( c,b\right) \subset L^{2}\left( c,b\right) $; let their deficiency
indices be respectively $m_{+}^{\left( -\right) }=m_{-}^{\left( -\right)
}=m^{\left( -\right) }$ and $m_{+}^{\left( +\right) }=m_{-}^{\left( +\right)
}=m^{\left( +\right) },$ and let $\hat{f}_{-}$ and $\hat{f}_{+}$ be their
closures in the respective $L^{2}\left( a,c\right) $ and $L^{2}\left(
c,b\right) $, $\hat{f}_{-}=\overline{\hat{f}_{-}^{\left( 0\right) }}$ and $%
\hat{f}_{+}=\overline{\hat{f}_{+}^{\left( 0\right) }}$, with the respective
domains $D_{f_{-}}\subset L^{2}\left( a,c\right) $ and $D_{f_{+}}\subset
L^{2}\left( c,b\right) $. Because the end $c$ is regular for the both $f_{-}$
and $f_{+}$, the functions in the both domains $D_{f_{-}}$ and $D_{f_{+}}$
vanish at the end $c$ together with their derivatives of order up to $n-1$.

We now consider a new symmetric operator $\hat{f}_{c}^{\left( 0\right) }$ in 
$L^{2}\left( a,b\right) $ associated with the initial differential
expression $\check{f}$ and defined on the domain $D_{f_{c}^{\left( 0\right)
}}$ that is a direct sum of \ $D\left( a,c\right) $ and $D\left( c,b\right) $%
, $D_{f_{c}^{\left( 0\right) }}=D\left( a,c\right) \oplus D\left( c,b\right) 
$. It is evident that $\overline{D_{f_{c}^{\left( 0\right) }}}=L^{2}\left(
a,b\right) $ and $D_{f_{c}^{\left( 0\right) }}\subset D\left( a,b\right)
=D_{f^{\left( 0\right) }}$, such that $\hat{f}_{c}^{\left( 0\right) }$ is a
symmetric operator in $L^{2}\left( a,b\right) $ that is a specific symmetric
restriction of the symmetric operator $\hat{f}^{\left( 0\right) }$, $\hat{f}%
_{c}^{\left( 0\right) }\subset \hat{f}^{\left( 0\right) }.$ Let its
deficiency indices be $m_{c_{+}}=m_{c_{-}}=m_{c}$, and let $\hat{f}_{c}$ be
its closure in $L^{2}\left( a,b\right) $, $\hat{f}_{c}=\overline{\hat{f}%
_{c}^{\left( 0\right) }}$, it is evident that $\hat{f}_{c}\subset \hat{f}$.

The crucial remark is that $\hat{f}_{c}^{\left( 0\right) }$ is a direct sum
of the operators $\hat{f}_{-}^{\left( 0\right) }$ and $\hat{f}_{+}^{\left(
0\right) }$, $\hat{f}_{c}^{\left( 0\right) }=\hat{f}_{-}^{\left( 0\right) }+%
\hat{f}_{+}^{\left( 0\right) }\,.$ It follows, first, that its deficiency
indices are the sums of the deficiency indices of the summands, i.e.,%
\begin{equation}
m_{c}=m^{\left( -\right) }+m^{\left( +\right) }\,,  \label{3b.61}
\end{equation}%
and, second, that its closure $\hat{f}_{c}$ is a direct sum of the closures $%
\hat{f}_{-}$ and $\hat{f}_{+}$, $\hat{f}_{c}=\hat{f}_{-}+\hat{f}_{+}\,,$
which implies that $\hat{f}_{c}$ is the restriction of $\hat{f}$ to the
domain $D_{f_{c}}\subset D_{f}$ that differs from $D_{f}$ by the only
additional condition on the functions $\psi \in D_{f}$ that $\psi ^{\left[ k%
\right] }\left( c\right) =0$, $k=0,1,\ldots ,n-1$, which in turn implies
that there exist exactly $n$, and not more, linearly independent functions
in $D_{f}$ that do not satisfy this condition and are linearly independent
modulo $D_{f_{c}}$, i.e.,%
\begin{equation}
\dim _{D_{f_{c}}}D_{f\text{ }}=n\,.  \label{3b.62}
\end{equation}

On the other hand, the second von Neumann theorem is applicable to $\hat{f}$
as a nontrivial symmetric extension of $\hat{f}_{c}^{\left( 0\right) }$.
According to this theorem, namely to second von Neumann formula (\ref{3a.22}%
) and to the remark ii) to the theorem, the dimension of $D_{f\text{ }}$
modulo $D_{f_{c}}$ is equal to the difference of the deficiency indices of $%
\hat{f}_{c}^{\left( 0\right) }$ and $\hat{f}^{\left( 0\right) }$\footnote{%
We recall that the deficiency indices of a symmetric operator and its
closure coincide.},%
\begin{equation}
\dim _{D_{f_{c}}}D_{f\text{ }}=m_{c}-m\,.  \label{3b.63}
\end{equation}%
The comparison of (\ref{3b.61}), (\ref{3b.62}), and (\ref{3b.63}) yields the
relation 
\begin{equation}
m=m^{\left( +\right) }+m^{\left( -\right) }-n  \label{3b.64}
\end{equation}%
between the deficiency indices of $\hat{f}^{\left( 0\right) }$ and $\hat{f}%
_{-}^{\left( 0\right) },\;$ $\hat{f}_{+}^{\left( 0\right) }$. We note that
because $n/2\leq m^{\left( -\right) },\;m^{\left( +\right) }\leq n,$ this
relation is compatible with \ the general restriction on the deficiency
indices of $\hat{f}^{\left( 0\right) }$,$\;0\leq m\leq n$. It is known that
in the case where the both ends are singular, the deficiency indices can
take any value from $0$ to $n$ \cite{AkhGl81,Naima69}.

Let we evaluate the deficient subspace $D_{+}$ and $D_{-}$ and the
respective deficiency indices $m_{+}$ and $m_{-}$ of an initial symmetric
operator $\hat{f}^{\left( 0\right) }$ associated with a given s.a.
differential expression $\check{f}$.

By the main theorem in the previous section, we know that three
possibilities for the\ s.a. extensions of $\hat{f}^{\left( 0\right) }$ can
occur.

Let the deficiency indices be different, $m_{+}\neq m_{-}$ which can happen
only for odd or mixed s.a. expressions with at least one singular end. In
this case, \ there exist no s.a. extensions of $\hat{f}^{\left( 0\right) }$,
i.e., there is no s.a. differential operators associated with a given s.a.
differential expression $\check{f}$.

Let the both deficiency indices be equal to zero, $m_{+}=m_{-}=0$, for even
s.a. differential expressions, this can happen only if the both ends are
singular\footnote{%
A natural hypothesis is that the same is true for any s.a. differential
expression.}. In this case, the initial symmetric operator $\hat{f}^{\left(
0\right) }$ is essentially s.a., and its unique s.a. extension is it closure 
$\hat{f}$ that coincides with its adjoint $\left( \hat{f}^{\left( 0\right)
}\right) ^{+}=\hat{f}^{\ast }.$ In other words, there is only one s.a.
differential operator in $L^{2}\left( a,b\right) $ associated with a given
differential expression $\check{f}$. As we already mentioned above, this
fact can become clear without evaluating the deficient subspaces and
deficient indices if the asymmetry form $\Delta _{\ast },$ or $\omega _{\ast
},$ is easily evaluated and appears to be zero.

Let the both deficiency indices be different from zero and equal, $%
m_{+}=m_{-}=m>0,$ which is always the case if the both ends are regular. In
this case, there exists an $m^{2}$-parameter family of s.a. extensions of $%
\hat{f}^{\left( 0\right) }.$ In other words, there is an $U\left( m\right) $%
-family $\left\{ \hat{f}_{U}\right\} $ of s.a. operators $\hat{f}_{U},\;U\in
U\left( m\right) ,$ the group, associated with a given differential
expression $\check{f}$, and the problem of their proper and convenient, if
possible, specification arises.

\subsection{Specification of self-adjoint extensions in terms of deficient
subspaces.}

Two simple preliminary remarks are useful. First, any s.a. extension $\hat{f}%
_{U}$ of an initial symmetric operator $\hat{f}^{\left( 0\right) }$ is
simultaneously a s.a. extension of its closure $\hat{f}$ with a domain $%
D_{f} $ and a symmetric restriction of the adjoint $\left( \hat{f}^{\left(
0\right) }\right) ^{+}=\hat{f}^{\ast }$ with a domain $D_{\ast }\,.$ All
these operators are given by the same initial differential expression $%
\check{f}$, but defined on different domains such that $D_{f}\subset
D_{f_{U}}\subset D_{\ast }\,,$ where $D_{f_{U}}$ is the domain of $\hat{f}%
_{U}$. Therefore, a specification of a s.a. operator $\hat{f}_{U}$ is
completely defined by a specification of its domain $D_{f_{U}}$\thinspace ,
second, because the deficiency indices of the symmetric operator $\hat{f}%
^{\left( 0\right) }$ of any finite order $n$ are finite, $m<\infty $, the
isometries $\hat{U}:D_{+}\rightarrow D_{-}$ defining the s.a. extensions $%
\hat{f}_{U}$ in the main theorem are defined by $m\times m$ unitary matrices 
$U=\left\| U_{lk}\right\| $, $l,k=1,2,\ldots ,m$, $U^{+}=U^{-1}$.

The main theorem furnishes the two ways of specification.

The first way is based on formulas (\ref{3a.34}), (\ref{3a.35}) for $%
D_{f_{U}}$ and requires the knowledge of the domain $D_{f}$ of the closure $%
\hat{f}$ apart from the deficient subspaces $D_{+}$ and $D_{-}$. The domain $%
D_{f}$ is defined by formula (\ref{3a.12}) with the appropriate change of
notation $D_{\bar{f}}\rightarrow D_{f}$, $D_{f^{+}}\rightarrow D_{\ast }$, $%
\underline{\psi }\rightarrow \psi ,$ and $\xi _{\ast }\rightarrow \psi
_{\ast }$, or equivalently by formulas (\ref{3a.14}) or (\ref{3a.15}) with
the additional change of notation $\xi _{z},\xi _{\bar{z}}\rightarrow \psi
_{+},\psi _{-}$ (\ref{3b.54}) and $e_{z,k},\,e_{\bar{z},k}\rightarrow
e_{+,k},\,e_{-,k}$ (\ref{3b.56}) with $m_{+}=m_{-}=m$.

We use the definition of $D_{f}$ by (\ref{3a.12}): $D_{f}=\left\{ \psi \in
D_{\ast }:\omega _{\ast }\left( \psi _{\ast },\psi \right) =0\,,\;\forall
\psi _{\ast }\in D_{\ast }\right\} \,,$ where $\omega _{\ast }\left( \psi
_{\ast },\psi \right) $ is given by (\ref{3b.44}), $\omega _{\ast }\left(
\psi _{\ast },\psi \right) =\left. \left[ \psi _{\ast },\psi \right] \right|
_{a}^{b}$ in terms of boundary values (\ref{3b.45}) of a local bilinear form 
$\left[ \psi _{\ast },\psi \right] $ which certainly exist. Taking the above
remarks (after formula (\ref{3b.47})) on the independence of these boundary
values, we can reduce the condition $\omega _{\ast }\left( \psi _{\ast
},\psi \right) =0$, $\forall \psi _{\ast }\in D_{\ast }$, to the independent
boundary conditions%
\begin{equation}
\left[ \psi _{\ast },\psi \right] \left( a\right) =\left[ \psi _{\ast },\psi %
\right] \left( b\right) =0\,,\;\forall \psi _{\ast }\in D_{\ast }.
\label{3b.65}
\end{equation}%
We formulate the result as a lemma.

\begin{lemma}
\label{l3b.4}The domain $D_{f}$ of the closure $\hat{f}$ of a symmetric
operator $\hat{f}$ $^{\left( 0\right) }$ associated with a s.a. differential
expression $\check{f}$ is specified by two boundary conditions (\ref{3b.65})
and is given by%
\begin{equation}
D_{f}=\left\{ \psi \in D_{\ast }:\,\left[ \psi _{\ast },\psi \right] \left(
a\right) =0\,,\;\left[ \psi _{\ast },\psi \right] \left( b\right)
=0\,,\;\forall \psi _{\ast }\in D_{\ast }\right\} \,.  \label{3b.66}
\end{equation}
\end{lemma}

In some cases, boundary conditions (\ref{3b.65}) in (\ref{3b.66}) can be
explicitly represented in terms of boundary conditions on the functions $%
\psi $ and their (quasi)derivatives of order up to $n-1$, where $n$ is the
order of $\check{f},$ at the end $a\,$and/or $b$. For example, let $\check{f}
$ be an even differential expression of order $n$ on an interval $\left(
a,b\right) $ and let the left end $a$ be regular. Then $\psi $ and its
quasiderivatives of order up to $n-1$ have finite values $\psi ^{\left[ k%
\right] }\left( a\right) $, $k=0,1,\ldots ,n-1$, at the end $a$, as well as
any $\psi _{\ast }\in D_{\ast }$, and the condition $\left[ \psi _{\ast
},\psi \right] \left( a\right) =0$ becomes%
\begin{equation*}
\sum_{k=0}^{\frac{n}{2}-1}\left( \overline{\psi _{\ast }^{\left[ k\right]
}\left( a\right) }\psi ^{\left[ n-k-1\right] }\left( a\right) -\overline{%
\psi _{\ast }^{\left[ n-k-1\right] }\left( a\right) }\psi ^{\left[ k\right]
}\left( a\right) \right) =0\,,
\end{equation*}%
see (\ref{3b.23}). Because $\psi _{\ast }^{\left[ k\right] }\left( a\right) $%
, $k=0,1,\ldots ,n-1$, can take arbitrary values, the boundary condition $%
\left[ \psi _{\ast },\psi \right] \left( a\right) =0$, $\forall \psi _{\ast
}\in D_{\ast }\,,$ reduces to zero boundary conditions $\psi ^{\left[ k%
\right] }\left( a\right) =0\,,\;k=0,1,\ldots ,n-1$ for functions $\psi \in
D_{f}$ and their quasiderivatives at the left regular end $a$. The same is
true for the regular end $b$.

We thus obtain that, in the presence of regular ends, a more explicit form
can be given to Lemma \ref{l3b.4}.

\begin{lemma}
\label{l3b.5}If $\check{f}$ is an even s.a. differential expression of order 
$n$ with both regular ends, then the domain $D_{f}$ is given by%
\begin{equation}
D_{f}=\left\{ \psi \left( x\right) \in D_{\ast }:\,\psi ^{\left[ k\right]
}\left( a\right) =\psi ^{\left[ k\right] }\left( b\right)
=0\,,\;k=0,1,\ldots ,n-1\right\} \,,  \label{3b.68}
\end{equation}%
if only one end, let it be $a$, regular, then the domain $D_{f}$ is given by%
\begin{equation}
D_{f}=\left\{ \psi \left( x\right) \in D_{\ast }:\psi ^{\left[ k\right]
}\left( a\right) =0,\;k=0,1,\ldots ,n-1;\left[ \psi _{\ast },\psi \right]
\left( b\right) =0,\;\forall \psi _{\ast }\in D_{\ast }\right\} .
\label{3b.69}
\end{equation}
\end{lemma}

It is evident that this result can be extended to any s.a. differential
expression $\check{f}$ with differentiable coefficients and regular ends
with the change of quasiderivatives to usual derivatives if a local form $%
\left[ \chi _{\ast },\psi _{\ast }\right] $ in functions and their
derivatives up to order $n-1$ is nondegenerate at regular ends.

As an illustration, we consider two simple s.a. differential expressions $%
\check{p}$ (\ref{3b.9}) and $\check{H}_{0}$ (\ref{3b.48a}) on a finite
interval $\left[ 0,l\right] $, the both ends of which are evidently regular.
The domain $D_{p}$ of the closure $\hat{p}$ of the initial symmetric
operator $\hat{p}^{\left( 0\right) }\,$with the domain $D_{p^{\left(
0\right) }}=D\left( 0,l\right) $ is given by\footnote{%
The condition $\psi \in L_{2}\left( a,b\right) $ is not independent; it is
automatically fulfilled in view of the first condition of the absolute
continuity of $\psi $ on the whole $[0,l]$; we give it for completeness.}%
\begin{equation}
D_{p}=\left\{ \psi :\psi \;\mathrm{a.c.\,on\,}\left[ 0,l\right] \,;\;\psi
,\,\psi ^{\prime }\in L^{2}\left( 0,l\right) \,;\;\psi \left( 0\right) =\psi
\left( l\right) =0\right\} \,,  \label{3b.70}
\end{equation}%
we already know this result from the previous section, see (\ref{3a.53}),
while the domain $D_{H_{0}}$ of the closure $\hat{H}_{0}$ of the initial
symmetric operator $\hat{H}_{0}^{\left( 0\right) }$, $D_{H_{0}^{\left(
0\right) }}=D\left( 0,l\right) $, is given by%
\begin{equation}
D_{H_{0}}=\left\{ \psi :\,\psi ,\,\psi ^{\prime }\mathrm{\,a.c.\,on\,}\left[
0,l\right] ;\;\psi ,\,\psi ^{\prime \prime }\in L^{2}\left( 0,l\right)
;\;\psi \left( 0\right) =\psi \left( l\right) =\psi ^{\prime }\left(
0\right) =\psi ^{\prime }\left( l\right) =0\right\} .  \label{3b.71}
\end{equation}

We note that the same domain evidently has the symmetric operator $\hat{H}$
associated with the s.a. differential expression $\check{H}$ (\ref{3b.10})
in the case where the potential $V$ is bounded $\left| V\left( x\right)
\right| <c<\infty $. If $V$ is nonbounded but locally integrable, the domain 
$D_{H}$ for the corresponding $\hat{H}$ is changed in comparison with $%
D_{H_{0}}$ (\ref{3b.70}) by the only replacement of the condition $\psi
^{\prime \prime }\in L^{2}\left( 0,l\right) $ by the condition $-\psi
^{\prime \prime }+V\psi \in L^{2}\left( 0,l\right) $.

We also note that both $\hat{H}_{0}$ and $\hat{H}$ are evidently symmetric,
but not s.a., because of the additional zero boundary conditions on the
derivatives.

After the specification of the domain $D_{f}$ of the closure $\hat{f}$, we
can formulate a theorem describing all s.a. operators associated with a
given s.a. differential expression $\check{f}$.

This theorem is a paraphrase of the main theorem in the part related to
formulas (\ref{3a.34}), (\ref{3a.37}).

\begin{theorem}
\label{t3b.1}The set of all s.a. differential operators associated with a
given s.a. differential expression $\check{f}$ in the case where the initial
symmetric operator $\hat{f}^{\left( 0\right) }$ has nonzero equal deficiency
indices $m_{+}=m_{-}=m>0$ is the $m^{2}$-parameter $U\left( m\right) $%
-family $\left\{ \hat{f}_{U}\right\} $ parametrized by elements of the
unitary group $U\left( m\right) $, $U\in U\left( m\right) $. Namely, each
s.a. operator $\hat{f}_{U}$ is in one-to-one correspondence with a unitary
matrix $U=\left\| U_{lk}\right\| $, $l,k=1,2,\ldots ,m$, $U^{+}=U^{-1}$, and
is given by%
\begin{equation}
\hat{f}_{U}:\,\left\{ 
\begin{array}{l}
D_{f_{U}}=\left\{ \psi _{U}=\psi +\sum_{k=1}^{m}c_{k}\left[
e_{+,k}+\sum_{l=1}^{m}U_{lk}e_{-,k}\right] ,\,\forall \psi \in
D_{f},\;\forall c_{k}\in \mathbb{C}\right\} , \\ 
\hat{f}\psi _{U}=\check{f}\psi _{U}\,,%
\end{array}%
\right.  \label{3b.72}
\end{equation}%
where $D_{f}$ is the domain of the closure $\hat{f}$ of $\hat{f}$ $^{\left(
0\right) }$ specified by (\ref{3b.66}), or (\ref{3b.68}), or (\ref{3b.69}), $%
\left\{ e_{+,k}\right\} ^{m}$ and $\left\{ e_{-,k}\right\} ^{m}$ are
orthobasises in the respective deficient subspaces $D_{+}$ and $D_{-}$
defined by (\ref{3b.54}),(\ref{3b.56}), and (\ref{3b.58}). In the case of an
even differential expression with real coefficients, we can take $e_{-,k}=%
\overline{e_{+,k}}\,$.
\end{theorem}

As an illustration, we consider the simple examples of differential
expressions $\check{p}$ (\ref{3b.9}) and $\check{H}_{0}$ (\ref{3b.48a}) on a
finite interval $\left[ 0,l\right] $. Both ends are regular, which implies
that the deficiency indices $\left( m_{+},m_{-}\right) $ are the respective $%
\left( 1,1\right) $, i.e., $m=1$, and $\left( 2,2\right) $, i.e., $m=2$.
Therefore, for the differential expressions $\check{p}$, we have a
one-parameter $U\left( 1\right) $-family $\left\{ \hat{p}_{\theta }\right\} $
of associated s.a. operators $\hat{p}_{U}=\hat{p}_{\theta }$ because in this
case, $U=e^{i\theta }$, $0\leq \theta \leq 2\pi $, $0\backsim 2\pi $; this
family is completely described in the previous section. For the differential
expression $\check{H}_{0}$, we have a four-parameter $U\left( 2\right) $%
-family $\left\{ \hat{H}_{0,U}\right\} $ of associated s.a. operators $%
\hat{H}_{0,U}$, $U\in U\left( 2\right) $, which we describe below.

To simplify the description, it is convenient to choose the dimensional
parameter $\kappa $ in (\ref{3b.54}) to be $\kappa =2\left( \pi /l\right)
^{2}$. For the orthobasis vectors in the deficient two-dimensional subspaces 
$D_{+}$ and $D_{-}$, we can take the respective functions%
\begin{eqnarray}
e_{+,1} &=&\sigma \exp \rho \,,\;e_{+,2}=\sigma \exp \left( \pi -\rho
\right) \,,\;\rho =\left( 1-i\right) \pi \frac{x}{l}\,,  \notag \\
e_{-,1} &=&\overline{e_{+,1}\,},\;\;e_{-,2}=\overline{e_{+,2}}\,,\;\sigma
=\left( e^{2\pi }-1\right) ^{-1/2}\left( 2\pi /l\right) ^{1/2}\,,
\label{3b.55}
\end{eqnarray}%
where $\sigma $ is a normalization factor. In view of (\ref{3b.70}), the
s.a. operator $\hat{H}_{0U}$ associated with the differential expression $%
\check{H}_{0}$ is then given by\footnote{%
We change the notation of indices in (\ref{3b.73}) in comparison with (\ref%
{3b.71}) to avoid a confusion with the index $l$ and the symbol $l$ for the
right end of the interval.}%
\begin{equation}
\hat{H}_{0U}:\left\{ 
\begin{array}{l}
D_{H_{0U}}=\left\{ \psi _{U}=\psi +\sum_{j=1}^{2}c_{j}e_{U,j}\right. :\,\psi
,\psi ^{\prime }\,\mathrm{a.c.\,on\,}\left[ 0,l\right] ;\,\psi ,\psi
^{\prime \prime }\in L^{2}\left( 0,l\right) \,,\,\psi \left( 0\right) \\ 
\left. =\psi \left( l\right) =\psi ^{\prime }\left( 0\right) =\psi ^{\prime
}\left( l\right) =0\,;\;e_{U,j}=e_{+,j}+\sum_{k=1}^{2}U_{kj}\overline{e_{+,k}%
}\,,\;j=1,2;\;\forall c_{j}\in \mathbb{C}\right\} , \\ 
\hat{H}_{0U}\psi _{U}=-\psi _{U}^{\prime \prime }\,,%
\end{array}%
\right.  \label{3b.73}
\end{equation}%
where $U=\left\| U_{kj}\right\| \,,\;k,j=1,2,\;$is\ a\ unitary\ matrix.

The normalization factor $\sigma $ in $e_{+,1}$, $e_{+,2}$ (\ref{3b.55}) can
be absorbed in $c_{1}$, $c_{2}$ and is irrelevant.

As we already mentioned above, this specification of the domain $D_{H_{0U}}$
by specifying the functions $\psi _{U}$ in $D_{H_{0U}}$ as a sum of
functions $\psi \in D_{H_{0}}$ and an arbitrary linear combination of
vectors $e_{U,j}$, $j=1,2$, that are the basis vectors in the
two-dimensional subspace $\left( D_{+}+\hat{U}D_{+}\right) $ seems
inconvenient for spectral analysis of $\hat{H}_{0U}$ and is unaccustomed in
physics where we used to appeal to (s.a.) boundary conditions for functions $%
\psi _{U}$ in $D_{H_{0U}}$, these conditions are relations between the
boundary values of the functions and their first derivatives, without
mentioning the domain $D_{H_{0}}$.

The main observation is that according to formula%
\begin{equation}
\psi _{U}\left( x\right) =\psi \left( x\right)
+\sum_{k=1}^{2}c_{j}e_{U,j}\left( x\right) \,,  \label{3b.74}
\end{equation}%
the four boundary values of the absolutely continuous functions $\psi _{U}$
and $\psi _{U}^{\prime }$ are defined by the only second term in r.h.s. in (%
\ref{3b.74}) because of the zero boundary values of $\psi $ and $\psi
^{\prime }$, namely, by the certain boundary values of $e_{U,j}$ and $%
e_{U,j}^{\prime }$ and only two arbitrary constants $c_{1}$ and $c_{2}$,
which result in two relations between the boundary values of $\psi _{U}$ and 
$\psi _{U}^{\prime }$, the relations defined by the unitary matrix $U$. To
demonstrate this fact, it is convenient to proceed in terms of two-columns
and $2\times 2$ matrices. Formula (\ref{3b.74}) yields%
\begin{equation}
\left( 
\begin{array}{c}
\psi _{U}\left( 0\right) \\ 
\psi _{U}^{\prime }\left( 0\right)%
\end{array}%
\right) =E_{U}\left( 0\right) \left( 
\begin{array}{c}
c_{1} \\ 
c_{2}%
\end{array}%
\right) \,,\;\left( 
\begin{array}{c}
\psi _{U}\left( l\right) \\ 
\psi _{U}^{\prime }\left( l\right)%
\end{array}%
\right) =E_{U}\left( l\right) \left( 
\begin{array}{c}
c_{1} \\ 
c_{2}%
\end{array}%
\right) \,.  \label{3b.75}
\end{equation}%
where the $2\times 2$ matrices $E_{U}\left( 0\right) =\left\| E_{U,kj}\left(
0\right) \right\| $ and $E_{U}\left( l\right) =\left\| E_{U,kj}\left(
l\right) \right\| $ are given by%
\begin{equation*}
E_{U,kj}\left( 0\right) =e_{U,j}^{\left( k-1\right) }\left( 0\right)
\,,\;E_{U,kj}\left( l\right) =e_{U,j}^{\left( k-1\right) }\left( l\right) \,.
\end{equation*}

It turns out that the rank of the rectangular $4\times 2$ matrix $\left( 
\begin{array}{c}
E_{U}\left( 0\right) \\ 
E_{U}\left( l\right)%
\end{array}%
\right) $ is maximal and equal to $2$. Therefore, we could express constants 
$c_{1}$ and $c_{2}$ in terms of $\psi _{U}\left( 0\right) ,\ldots ,\psi
_{U}^{\prime }\left( l\right) $ from some two relations in (\ref{3b.75}),
then substitute the obtained expressions in the remaining two relations and
thus obtain two linear relations between the boundary values of functions in 
$D_{H_{0U}}$ and their first derivatives that are defined by the matrix $U.$
But it is more convenient to proceed as follows. We multiply the first and
the second relation in (\ref{3b.75}) by the respective matrices $%
E_{U}^{+}\left( 0\right) \mathcal{E}$ and $E_{U}^{+}\left( l\right) \mathcal{%
E}$, where the matrix $\mathcal{E}$=$\sigma ^{2}/i$ and obtain that%
\begin{align*}
& E_{U}^{+}\left( 0\right) \mathcal{E}\left( 
\begin{array}{c}
\psi _{U}\left( 0\right) \\ 
\psi _{U}^{\prime }\left( 0\right)%
\end{array}%
\right) =E_{U}^{+}\left( 0\right) \mathcal{E}E_{U}\left( 0\right) \left( 
\begin{array}{c}
c_{1} \\ 
c_{2}%
\end{array}%
\right) \,, \\
& E_{U}^{+}\left( l\right) \mathcal{E}\left( 
\begin{array}{c}
\psi _{U}\left( l\right) \\ 
\psi _{U}^{\prime }\left( l\right)%
\end{array}%
\right) =E_{U}^{+}\left( l\right) \mathcal{E}E_{U}\left( l\right) \left( 
\begin{array}{c}
c_{1} \\ 
c_{2}%
\end{array}%
\right) \,.
\end{align*}%
The crucial remark is that the matrix%
\begin{equation*}
R=E_{U}^{+}\left( l\right) \mathcal{E}E_{U}\left( l\right) -E_{U}^{+}\left(
0\right) \mathcal{E}E_{U}\left( 0\right)
\end{equation*}%
is the null matrix:\ taking (\ref{3b.44}), (\ref{3b.45}), and (\ref{3b.22})
for $\check{f}=\check{H}_{0}$,%
\begin{equation*}
\left[ \chi _{\ast },\psi _{\ast }\right] =\sum_{k,j=1}^{2}\overline{\chi
_{\ast }^{\left( k-1\right) }\left( x\right) }\mathcal{E}_{kj}\psi _{\ast
}^{\left( j-1\right) }\left( x\right) \,,
\end{equation*}%
into account, it is easy to see that matrix elements of $R$ are%
\begin{equation*}
R_{kj}=\left. \left[ e_{U,k},e_{U,j}\right] \right| _{0}^{l}=\omega _{\ast
}\left( e_{U,k},e_{U,j}\right) =0
\end{equation*}%
because the reduction of the sesquilinear antisymmetric form $\omega _{\ast
} $ to $D_{H_{0U}}$ is equal to zero. It follows that%
\begin{equation}
E_{U}^{+}\left( l\right) \mathcal{E}\left( 
\begin{array}{c}
\psi _{U}\left( l\right) \\ 
\psi _{U}^{\prime }\left( l\right)%
\end{array}%
\right) -E_{U}^{+}\left( 0\right) \mathcal{E}\left( 
\begin{array}{c}
\psi _{U}\left( 0\right) \\ 
\psi _{U}^{\prime }\left( 0\right)%
\end{array}%
\right) =0\,,  \label{3b.76}
\end{equation}%
which is equivalent to%
\begin{equation}
\left. \left[ e_{U,j},\psi _{U}\right] \right| _{0}^{l}=0\,,\;j=1,2\,.
\label{3b.77}
\end{equation}%
Relations (\ref{3b.76}), (\ref{3b.77}) are the boundary conditions
specifying the s.a. extension $\hat{H}_{0U}$, i.e., the s.a. boundary
conditions. It is clear how the representation (\ref{3b.74}) for $\psi
_{U}\in D_{H_{0U}}$ is restored from boundary conditions (\ref{3b.76}), (\ref%
{3b.77}) by reversing the above procedure. It is also clear how this
consideration is generalized to s.a. operators associated with even
differential expressions of any order in the case where both ends are
regular.

\subsection{Specification of self-adjoint extensions in terms of
self-adjoint boundary conditions}

The second alternative way of the specification of s.a. differential
operators $\hat{f}_{U}$ in $L^{2}\left( a,b\right) $ associated with a given
s.a. differential expression $\check{f},$ the operators that are s.a.
extensions of the initial symmetric operator $\hat{f}^{\left( 0\right) }$,
is based on formulas (\ref{3a.36}), (\ref{3a.39}) in the main theorem and
formulas (\ref{3b.44}), (\ref{3b.45}) for the asymmetry form $\omega _{\ast
} $. It avoids the evaluation of the domain $\hat{f}=\overline{\hat{f}%
^{\left( 0\right) }}$ of the closure $D_{f}$ and directly leads to the
specification of the s.a. operators $\hat{f}_{U}$ in terms of s.a. boundary
conditions. A corresponding theorem is alternative to Theorem 4; it is a
paraphrase of the main theorem in the part related to formulas (\ref{3a.36}%
), (\ref{3a.39}) with due regard to formulas (\ref{3b.44}), (\ref{3b.45})
and the appropriate change of notation in (\ref{3a.39}): $\xi
_{U}\rightarrow \psi _{U}$. For brevity, we do not repeat the first general
assertion and the explanation of symbols that are common to the both
theorems. We also introduce the abbreviated notation $e_{U,k}$ for the basis
functions $e_{+,k}+\sum_{l=1}^{m}U_{lk}e_{-,k}$ in the subspace $\left(
D_{+}+\hat{U}D_{+}\right) \subset D_{f_{U}}\subset D_{\ast }$.

\begin{theorem}
\label{t3b.2}Each s.a. operator $\hat{f}_{U}$ in $L^{2}\left( a,b\right) $
associated with a given s.a. differential expression $\check{f}$ is given by%
\begin{equation}
\hat{f}_{U}:\,\left\{ 
\begin{array}{l}
D_{f_{U}}=\left\{ \psi _{U}\in D_{\ast }:\left. \,\left[ e_{U,k},\psi _{U}%
\right] \right| _{a}^{b}=0\right\} \,, \\ 
\hat{f}_{U}\psi _{U}=\check{f}\psi _{U}\,,%
\end{array}%
\right.  \label{3b.78}
\end{equation}%
where$\;e_{U,k}=e_{+,k}+\sum_{l=1}^{m}U_{lk}e_{-,l}\,.$
\end{theorem}

We make two remarks on Theorem \ref{t3b.2}. First, this theorem explicitly
specifies $\hat{f}_{U}$ as a restriction of the adjoint $\hat{f}^{\ast }$ to
the domain $D_{f_{U}}$ defined by s.a. boundary conditions. These boundary
conditions considered as additional linear equations for functions $\psi
_{\ast }\in D_{\ast }$ are linearly independent. Really, let the relation%
\begin{equation*}
\left. \sum_{k=1}^{m}c_{k}\left[ e_{U,k},\psi _{\ast }\right] \right|
_{a}^{b}=0\,,\;\forall \psi _{\ast }\in D_{\ast }\,,
\end{equation*}%
holds, with some constants $c_{k}$. This relation is equivalent to $\left. %
\left[ \psi _{\ast },\sum_{k=1}^{m}\overline{c_{k}}e_{U,k}\right] \right|
_{a}^{b}=0$ and by Lemma \ref{l3b.4}, see (\ref{3b.66}), implies that $%
\sum_{k=1}^{m}\overline{c_{k}}e_{U,k}\in D_{f}$, which is possible only if
all $c_{k}=0$, $k=1,\ldots ,m$, because $D_{f}\cap \left( D_{+}+\hat{U}%
D_{+}\right) =\left\{ 0\right\} $, or, in other words, because the functions 
$e_{U,k}$ are linearly independent modulo $D_{f}$. Second, the basis
functions $e_{U,k}$ in $\left( D_{+}+\hat{U}D_{+}\right) $ belong to $%
D_{f_{U}}$, therefore, the relation%
\begin{equation}
\left. \left[ e_{U,k},e_{U,l}\right] \right| _{a}^{b}=0\,,\;k,l=1,\ldots ,m,
\label{3b.79}
\end{equation}%
holds; its particular realization for $\check{f}=\check{H}_{0}$ with $m=2$
was already encountered above.

In some particular cases, boundary conditions (\ref{3b.78}) in Theorem \ref%
{t3b.2} become explicit boundary conditions in terms of boundary values of
functions and their (quasi)derivatives. We here present two such cases. The
first is the case of even s.a. differential expressions of order $n$ on a
finite interval $\left( a,b\right) $ with the both regular ends, the case
where the functions in $D_{\ast }$ and their quasiderivatives of order up to 
$n-1$ have finite boundary values and where the deficiency indices are
maximum, $m_{+}=m_{-}=n$. By formula (\ref{3b.23}), the s.a. boundary
conditions become%
\begin{equation*}
\left. \left[ e_{U,k},\psi _{U}\right] \right| _{a}^{b}=-\left. \sum_{l=0}^{%
\frac{n}{2}-1}\left[ \overline{e_{U,k}^{\left[ l\right] }}\psi _{U}^{\left[
n-l-1\right] }-\overline{e_{U,k}^{\left[ n-l-1\right] }}\psi _{U}^{\left[ l%
\right] }\right] \right| _{a}^{b}=0\,,\;k=1,\ldots ,n\,,
\end{equation*}%
or, shifting up the summation index by unity,%
\begin{equation}
\sum_{l,m=1}^{n}\left[ \overline{e_{U,k}^{\left[ l-1\right] }\left( b\right) 
}\mathcal{E}_{lm}\psi _{U}^{\left[ m-1\right] }\left( b\right) -\overline{%
e_{U,k}^{\left[ l-1\right] }\left( a\right) }\mathcal{E}_{lm}\psi _{U}^{%
\left[ m-1\right] }\left( a\right) \right] =0\,,\;k=1,\ldots ,n\,,
\label{3b.80}
\end{equation}%
where%
\begin{equation}
\mathcal{E}_{lm}=\delta _{l,n+1-m}\epsilon \left( l-\frac{n+1}{2}\right)
\,,\;l,m=1,\ldots ,n\,,  \label{3b.81}
\end{equation}%
and $\epsilon \left( x\right) $ is the well-known odd step function, $%
\epsilon \left( -x\right) =-\epsilon \left( x\right) $ and $\epsilon \left(
x\right) =1$ for $x>0$. Boundary conditions (\ref{3b.80}) can be
conveniently represented in condensed terms of the matrix $\mathcal{E=}%
\left\| \mathcal{E}_{lm}\right\| ,$ where $\mathcal{E}_{lm}$ are given by (%
\ref{3b.81}), the two $n\times n$ matrices of boundary values of the basis
functions $e_{U,k}$ and their quasiderivatives,%
\begin{align}
E_{U}\left( a\right) & =\left\| E_{U,lk}\left( a\right) \right\|
\,,\;E_{U,lk}\left( a\right) =e_{U,k}^{\left[ l-1\right] }\left( a\right) \,,
\notag \\
E_{U}\left( b\right) & =\left\| E_{U,lk}\left( b\right) \right\|
\,,\;E_{U,lk}\left( b\right) =e_{U,k}^{\left[ l-1\right] }\left( b\right)
\,,\;l,k=1,\ldots ,n\,,  \label{3b.83}
\end{align}%
and the two $n$-columns of boundary values of functions and their
quasiderivatives,%
\begin{equation}
\Psi _{U}\left( a\right) =\left( 
\begin{array}{c}
\psi _{U}\left( a\right) \\ 
\psi _{U}^{\left[ 1\right] }\left( a\right) \\ 
\vdots \\ 
\psi _{U}^{\left[ n\right] }\left( a\right)%
\end{array}%
\right) \,,\;\Psi _{U}\left( b\right) =\left( 
\begin{array}{c}
\psi _{U}\left( b\right) \\ 
\psi _{U}^{\left[ 1\right] }\left( b\right) \\ 
\vdots \\ 
\psi _{U}^{\left[ n\right] }\left( b\right)%
\end{array}%
\right) \,.  \label{3b.84}
\end{equation}%
Their realization for $\check{f}=\check{H}_{0}$ was already encountered
above. It seems useful to give a separate version of Theorem \ref{t3b.2} for
this case in the introduced condensed notation.

\begin{theorem}
\label{t3b.3}Any s.a. operator $\hat{f}_{U}$\ in $L^{2}\left( a,b\right) $\
associated with an even s.a. differential expression $\check{f}$\ of order $%
n $\ with the both regular ends is given by%
\begin{equation}
\hat{f}_{U}:\left\{ 
\begin{array}{l}
D_{f_{U}}=\left\{ \psi _{U}\in D_{\ast }:\,E_{U}^{+}\left( b\right) \mathcal{%
E}\Psi \left( b\right) -E_{U}^{+}\left( a\right) \mathcal{E}\Psi \left(
a\right) =0\right\} \,, \\ 
\hat{f}_{U}\psi _{U}=\check{f}\psi _{U}\,,%
\end{array}%
\right.  \label{3b.85}
\end{equation}%
where the matrices$\mathrm{\;}\mathcal{E},\,E_{U}\left( a\right)
,\,E_{U}\left( b\right) \;$and the$\;$columns$\;\Psi \left( a\right) ,\Psi
\left( b\right) \;$are\ given\ by\ the\ respective (\ref{3b.81}),\ (\ref%
{3b.83}), and (\ref{3b.84}).
\end{theorem}

The modified version of the two remarks to Theorem \ref{t3b.2} in this case
is

1) s.a. boundary conditions (\ref{3b.85}) are linearly independent, which is
equivalent to the property of the matrices $E_{U}\left( a\right) $ and $%
E_{U}\left( b\right) $ that the $2n\times n$ matrix $\mathbb{E}$ has the
maximum rank,%
\begin{equation}
\mathbb{E}=\left( 
\begin{array}{c}
E_{U}\left( a\right) \\ 
E_{U}\left( b\right)%
\end{array}%
\right) ,\;\mathrm{rank}\mathbb{E}=n\,;  \label{3b.85a}
\end{equation}%
really the above given proof of the linear independence of boundary
conditions was based on the property that $\sum_{k=1}^{m}c_{k}e_{U,k}\in
D_{f}\Longrightarrow c_{k}=0,$ $k=1,...,m,$ but in our case where $m=n$, in
view of Lemma \ref{l3b.5}, formula (\ref{3b.68}), this is equivalent to the
property that%
\begin{equation*}
\sum_{k=1}^{n}e_{U,k}^{[l-1]}\left( a\right)
c_{k}=0\,,\;\sum_{k=1}^{n}e_{U,k}^{[l-1]}\left( b\right)
c_{k}=0\,,\;l=1,...,n\Longrightarrow c_{k}=0,\;k=1,...,n\,;
\end{equation*}

2) relation (\ref{3b.79}) is written as%
\begin{equation}
E_{U}^{+}\left( b\right) \mathcal{E}E_{U}\left( b\right) -E_{U}^{+}\left(
a\right) \mathcal{E}E_{U}\left( a\right) =0\,.  \label{3b.86}
\end{equation}

Of course, in practical applications, the condensed notation requires
decoding, see below an example of the differential expression $\check{H}_{0}$%
.

We also note that matrices $E_{U}\left( a\right) $ and $E_{U}\left( b\right) 
$ with a given unitary matrix $U$ depend on the choice of the dimensional
parameter $\kappa $ in (\ref{3b.54}) and on the choice of the orthobasises $%
\left\{ e_{+,k}\right\} _{1}^{n}$ and $\left\{ e_{-,k}\right\} _{1}^{n}$ in
the respective deficient subspaces $D_{+}$ and $D_{-}$. For example, if we
change the orthobasises,%
\begin{equation*}
\left\{ e_{+,k}\right\} _{1}^{n}\rightarrow \left\{ \widetilde{e}%
_{+,k}=\sum_{l=1}^{n}V_{+lk}e_{+,l}\right\} _{1}^{n}\,,\;\left\{
e_{-,k}\right\} _{1}^{n}\rightarrow \left\{ \widetilde{e}_{-,k}=%
\sum_{l=1}^{n}V_{-lk}e_{-,l}\right\} _{1}^{n}\,,
\end{equation*}%
where matrices $V_{\pm }$ are unitary, and it is not obligatory that $V_{-}=%
\overline{V_{+}}\,,$ then the matrix $U$ for the same s.a. extension is
replaced according to the rule $U\rightarrow \tilde{U}=V_{-}^{-1}UV_{+}\,$.

Again , as after Lemma \ref{l3b.5}, we can add that a similar theorem holds
for any s.a. differential expression $\check{f}$ of any order with
differentiable coefficients and the both regular ends with the change of
quasiderivatives by usual derivatives if boundary values (\ref{3b.45}) are
finite forms in the boundary values of functions and their derivatives.

As an illustration, we consider our simple examples of differential
expressions $\check{p}$ (\ref{3b.9}) and $\check{H}_{0}$ (\ref{3b.48a}) on a
finite interval $\left[ 0,l\right] $ and compare the descriptions of the
respective one-parameter set of s.a. operators $\hat{p}_{U}$, $U\in U\left(
1\right) $, and four-parameter set of s.a. operators $\hat{H}_{0U}$, $U\in
U\left( 2\right) $, according to Theorem \ref{t3b.1} and to Theorem \ref%
{t3b.3} respectively. For the operators $\hat{p}_{U}$, this was already done
in the previous section, and it was demonstrated that the two descriptions
are equivalent. As to $\hat{H}_{0U}$, we must preliminarily evaluate the
domain $D_{0\ast }$ of $\hat{H}_{0}^{\ast }$. This is a natural domain for $%
\check{H}_{0}$ and is evidently given by (\ref{3b.49}) with the only change $%
\mathbb{R}^{1}\rightarrow \left[ 0,l\right] $. After this, any s.a. operator 
$\hat{H}_{0U}$ is given by%
\begin{equation*}
\hat{H}_{0U}:\,\left\{ 
\begin{array}{l}
D_{H_{0U}}=\left\{ \psi _{U}:\psi _{U},\,\psi _{U}^{\prime }\,\mathrm{%
\,a.c.\,on\,}\left[ 0,l\right] ;\;\psi _{U},\,\psi _{U}^{\prime \prime }\in
L^{2}\left( 0,l\right) \,;\right. \\ 
\left. E_{U}^{+}\left( l\right) \mathcal{E}\Psi _{U}\left( l\right)
=E_{U}^{+}\left( 0\right) \mathcal{E}\Psi _{U}\left( 0\right) \right\} \,,
\\ 
\hat{H}_{0U}\psi _{U}=-\psi _{U}^{\prime \prime }\,,%
\end{array}%
\right.
\end{equation*}%
where $\mathcal{E}$ and $E_{U}\left( l\right) ,$ $E_{U}\left( 0\right) $ are
the matrices given by the respective \ (\ref{3b.81}) and (\ref{3b.83}) with $%
n=2$ and the usual first derivatives of the basis vectors $e_{U,k}$ given by
(\ref{3b.55}), (\ref{3b.73}), while the two-columns $\Psi _{U}\left(
0\right) $ and $\Psi _{U}\left( l\right) $ are

\begin{equation*}
\Psi _{U}\left( 0\right) =\left( 
\begin{array}{c}
\psi _{U}\left( 0\right) \\ 
\psi _{U}^{\prime }\left( 0\right)%
\end{array}%
\right) ,\;\Psi _{U}\left( l\right) =\left( 
\begin{array}{c}
\psi _{U}\left( l\right) \\ 
\psi _{U}^{\prime }\left( l\right)%
\end{array}%
\right) \,,
\end{equation*}%
see (\ref{3b.84}) with $n=2,$ all these were already encountered above. If
we compare this description of $\hat{H}_{0U}$ according to Theorem \ref%
{t3b.3} with that obtained from Theorem \ref{t3b.1} and given by (\ref{3b.76}%
), we find that they are identical.

It is interesting to give examples of s.a. operators $\hat{H}_{0U}$
associated with the differential expression $\check{H}_{0}$ (\ref{3b.48a})
on $\left[ 0,l\right] $ and corresponding to particular choices of the
unitary matrix $U.$ Each of them is a candidate to the quantum mechanical
Hamiltonian for a free particle on the interval $\left[ 0,l\right] .$

Choosing $U=I,$ the unit matrix, we obtain the Hamiltonian $\hat{H}_{0I}$
specified by the s.a. boundary conditions that being decoded and presented
in conventional form\footnote{%
When writing boundary conditions with a specific $U$ separately, we
conventionally omit the subscript $U$ in the notation of the respective
functions.} looks rather exotic:%
\begin{align}
& \psi \left( l\right) =-\cosh \pi \,\psi \left( 0\right) -\frac{l}{\pi }%
\sinh \pi \,\psi ^{\prime }\left( 0\right) \,,  \notag \\
& \psi ^{\prime }\left( l\right) =-\frac{\pi }{l}\sinh \pi \,\psi \left(
0\right) -\cosh \pi \,\psi ^{\prime }\left( 0\right) \,,  \label{3b.87}
\end{align}

Choosing $U=-I,$ we obtain the Hamiltonian $\hat{H}_{0\,-I}$ specified by
the well-known s.a. boundary conditions 
\begin{equation}
\psi \left( 0\right) =\psi \left( l\right) =0  \label{3b.88}
\end{equation}%
corresponding to a particle in an infinite potential well.

Choosing $U=iI,$ we obtain the Hamiltonian $\hat{H}_{0\,iI}$ specified by
the s.a. boundary conditions 
\begin{equation}
\psi ^{\prime }\left( 0\right) =\psi ^{\prime }\left( l\right) =0\,.
\label{3b.89}
\end{equation}%
Choosing $U=-\frac{1}{2}\left[ \left( 1-i\right) I+\left( 1+i\right) \sigma
^{1}\right] ,$ we obtain the Hamiltonian $\hat{H}_{0U}$ specified by the
periodic boundary conditions\footnote{%
To be true, in this case we actually solve the inverse problem of finding a
proper $U$ for periodic boundary conditions.} 
\begin{equation}
\psi \left( 0\right) =\psi \left( l\right) \,,\;\psi ^{\prime }\left(
0\right) =\psi ^{\prime }\left( l\right) \,,  \label{3b.90}
\end{equation}%
conventionally adopted in statistical physics when quantizing an ideal gas
in a box.

The second case where the s.a. boundary conditions in Theorem \ref{t3b.2}
become explicit in terms of boundary values of functions and their
(quasi)derivatives of order up to $n-1$ is the case of even s.a.
differential expression $\check{f}$ of order $n$ with one regular and one
singular end for which the associated initial symmetric operator $\hat{f}%
^{\left( 0\right) }$ has minimum possible deficiency indices\footnote{%
We recall that the deficiency indices are always equal in the case of an
even s.a. differential expression.} $m_{+}=m_{-}=n/2,$ see (\ref{3b.60}).
This follows from some general assertion on differential symmetric operators.

\begin{lemma}
\label{l3b.6}Let $\hat{f}^{\left( 0\right) }$ be a symmetric operator
associated with an even s.a. differential expression $\check{f}$ of order $n$
on an interval $\left( a,b\right) $ with the regular end $a$ and the
singular end $b$, and let the deficiency indices of $\hat{f}^{\left(
0\right) }$ be $m_{+}=m_{-}=n/2\,.$ Then the equality%
\begin{equation}
\left[ \chi _{\ast },\psi _{\ast }\right] \left( b\right) =0\,,\;\forall
\chi _{\ast },\psi _{\ast }\in D_{\ast }\,,  \label{3b.91}
\end{equation}%
where $D_{\ast }$ is the domain of the adjoint $\hat{f}^{\ast }=\left( \hat{f%
}^{\left( 0\right) }\right) ^{+}$ holds. If the end $a$ is singular while
the end $b$ is regular, then $b$ in (\ref{3b.91}) is changed to $a.$
\end{lemma}

We show later that conversely, if the boundary values $\left[ \chi_{\ast
},\psi_{\ast}\right] \left( b\right) $ vanish for all $\chi_{\ast},\psi_{%
\ast}\in D_{\ast},$ the deficiency indices of $\hat{f}^{\left( 0\right) }$
are minimum, $m_{+}=m_{-}=n/2\,.$

The proof of the Lemma is based on the arguments already known and used
above in the proof of the independence of boundary values (\ref{3b.45}) and
in the proof of the lower bound in (\ref{3b.60}). Therefore, we don't repeat
them and only formulate two initial assertions following from the conditions
of the Lemma by these arguments. On the one hand, because the end $a$ is
regular, there exist $n$ functions $w_{k}\in D_{\ast },$ $k=1,...,n,$
vanishing near the singular end $b$ and linearly independent modulo $%
D_{f}\,, $ where $D_{f}$ is the domain of the closure $\hat{f}$ of $\hat{f}%
^{\left( 0\right) },$ $\hat{f}=\overline{\hat{f}^{\left( 0\right) }}\,.$ On
the other hand, because the deficiency indices of $\hat{f}^{\left( 0\right)
} $, and therefore of $\hat{f}$ are equal to $n/2,$ we have $\dim
_{D_{f}}D_{\ast }=n, $ whence it follows that any function $\psi _{\ast }\in
D_{\ast }$ can be represented as $\psi _{\ast }=\psi
+\sum_{k=1}^{n}c_{k}w_{k}\,,$ where $\psi \in D_{f}$ and $c_{k}$ are some
number coefficients. The boundary value $\left[ \chi _{\ast },\psi _{\ast }%
\right] \left( b\right) $ with any $\chi _{\ast },\psi _{\ast }\in D_{\ast }$
is then represented as%
\begin{equation*}
\left[ \chi _{\ast },\psi _{\ast }\right] \left( b\right) =\left[ \chi
_{\ast },\psi \right] \left( b\right) +\sum_{k=1}^{n}c_{k}\left[ \chi _{\ast
},w_{k}\right] \left( b\right) \,.
\end{equation*}%
But the first term in the last equality vanishes by Lemma \ref{l3b.5}, see
the second equality in (\ref{3b.69}) with the change $\psi _{\ast
}\rightarrow \chi _{\ast }\,,$ and the second term also vanishes because all 
$w_{k}$ vanish near the singular end $b,$ which proves the Lemma.

According to this Lemma, the term $\left[ e_{U,k},\psi _{U}\right] \left(
b\right) $ in boundary conditions (\ref{3b.78}) in Theorem \ref{t3b.2}
vanishes, and they reduces to $\left[ e_{U,k},\psi _{U}\right] \left(
a\right) .$ Because the end $a$ is regular, these s.a. boundary conditions
are explicit in terms of boundary values of functions and their
quasiderivatives at the end $a$,%
\begin{equation}
\sum_{l,m=1}^{n}e_{U,k}^{[l-1]}\left( a\right) \mathcal{E}_{lm}\psi
^{\lbrack m-1]}\left( a\right) =0\,,\;k=1,...,n/2\,,  \label{3b.92}
\end{equation}%
where $\mathcal{E}_{lm}$ are given by (\ref{3b.81}). If we introduce the
rectangular $n\times n/2$ matrix%
\begin{equation}
E_{1/2,U}\left( a\right) =\left| \left| E_{1/2,U,lk}\left( a\right) \right|
\right| \,,\;E_{1/2,U,lk}\left( a\right) =e_{U,k}^{[l-1]}\left( a\right)
\,,\;l=1,...,n\,,\;k=1,...,n/2\,,  \label{3b.93}
\end{equation}%
s.a. boundary conditions (\ref{3b.92}) are written in the condensed form as $%
E_{1/2,U}^{+}\left( a\right) \mathcal{E}\Psi \left( a\right) =0\,.$ It seems
useful to give a separate version of Theorem \ref{t3b.2} for this case in
the condensed notation.

\begin{theorem}
\label{t3b.4}Any s.a. operator $\hat{f}_{U}$ associated with an even s.a.
differential expression $\check{f}$ of order $n$ on an interval $\left(
a,b\right) $ with the regular and $a$ and the singular end $b$ in the case
where the initial symmetric operator $\hat{f}^{\left( 0\right) }$ has the
deficiency indices $m_{+}=m_{-}=n/2\,,$ $U\in U\left( n/2\right) ,$ is given
by%
\begin{equation}
\hat{f}_{U}:\left\{ 
\begin{array}{l}
D_{f_{U}}=\left\{ \psi _{f_{U}}\in D_{\ast }:E_{1/2,U}^{+}\left( a\right) 
\mathcal{E}\Psi _{U}\left( a\right) =0\right\} \,, \\ 
\hat{f}_{U}\psi _{U}=\check{f}\psi _{U}\,,%
\end{array}%
\right.  \label{3b.94}
\end{equation}%
where the matrix$\mathrm{\;}\mathcal{E}$ is given\ by (\ref{3b.81}), the
matrix $E_{1/2,U}\left( a\right) $ is given by$\;$(\ref{3b.93}), and $\Psi
_{U}\left( a\right) $ is given by (\ref{3b.84}).

If the end $a$ is singular while the end $b$ is regular, $a$ in (\ref{3b.94}%
) is changed to $b.$
\end{theorem}

The modified version of the two remarks to Theorem \ref{t3b.2} in this case
is

1)\ s.a. boundary conditions (\ref{3b.94}) are linearly independent, which
is equivalent to the property that the rectangular $n\times n/2$ matrix $%
E_{1/2,U}\left( a\right) $ is of maximum rank,%
\begin{equation}
\mathrm{rank}E_{1/2,U}\left( a\right) =\frac{n}{2}\,,  \label{3b.94a}
\end{equation}%
an analogue of (\ref{3b.85a});

2)\ relation (\ref{3b.79}) is written as%
\begin{equation}
E_{1/2,U}^{+}\left( a\right) \mathcal{E}E_{1/2,U}\left( a\right) =0
\label{3b.95}
\end{equation}%
which is an analogue of (\ref{3b.86}).

Of course, in applications, the condensed notation must be decoded.

As an illustration of Theorem \ref{t3b.4}, we consider the example of the
deferential expression $\check{H}_{0}$ (\ref{3b.48a}) on the semiaxis $\left[
0,\infty \right) $. As to the deferential expression $\check{p}$ (\ref{3b.9}%
), we know from the previous section that there are no s.a. operators
associated with $\check{p}$ on the semiaxis. The domain $D_{\ast }$ in this
case is the natural domain $D_{0\ast }$ for $\check{H}_{0}$, it is given by (%
\ref{3b.49}) with the only change $\mathbb{R}^{1}\rightarrow \mathbb{R}%
_{+}^{1}=\left[ 0,\infty \right) $. The deficient subspaces $D_{\pm }$ as
square-integrable solutions of eqs. (\ref{3b.54}),

$-\psi _{\pm }^{\prime \prime }=\pm i\kappa \psi _{\pm },$ are easily
evaluated. It is sufficient to find $D_{+}$, then $D_{-}$ is obtained by
complex conjugation. Among the two linearly independent solutions%
\begin{equation*}
\psi _{+1,2}=\exp \left[ \pm \left( 1-i\right) \sqrt{\frac{\kappa }{2}}x%
\right]
\end{equation*}%
of the equation for $D_{+}$, only one, $\exp \left[ \left( i-1\right) \sqrt{%
\frac{\kappa }{2}}x\right] $, is square integrable on $\left[ 0,\infty
\right) $. This means that the deficiency indices $\left( m_{+},m_{-}\right) 
$ in our case are $\left( 1,1\right) $, and we have a one-parameter $U\left(
1\right) $-family $\left\{ \hat{H}_{0U}\right\} $, $U\in U\left( 1\right) $,
of s.a. operators in $L^{2}\left( 0,\infty \right) $ associated with the
differential expression $\check{H}_{0}$. Their specification by s.a.
boundary conditions is performed in direct accordance with Theorem \ref%
{t3b.4}. The orthobasis vectors in $D_{\pm }$ are%
\begin{equation*}
e_{\pm }=\sqrt[4]{2\kappa }\exp \left[ \left( \pm i-1\right) \sqrt{\frac{%
\kappa }{2}}x\right] \,.
\end{equation*}%
The group $U\left( 1\right) $ is a circle and is naturally parametrized by
an angle $\theta $: $U=e^{i\theta }$, $-\pi \leq \theta \leq \pi $, $-\pi
\backsim \pi $, therefore, $\hat{H}_{0U}=\hat{H}_{0\theta }$, and the single
basis vector $e_{U}=e_{\theta }$ is%
\begin{equation*}
e_{\theta }=\sqrt[4]{2\kappa }\left\{ \exp \left[ \left( i-1\right) \sqrt{%
\frac{\kappa }{2}}x\right] +e^{i\theta }\exp \left[ -\left( 1+i\right) \sqrt{%
\frac{\kappa }{2}}x\right] \right\} \,.
\end{equation*}%
The matrix $E_{1/2,U}\left( a\right) $ in (\ref{3b.94}) in our case is a $%
2\times 1$ matrix, i.e., a column%
\begin{equation*}
E_{1/2,\theta }\left( 0\right) =\sqrt[4]{2\kappa }\left( 
\begin{array}{c}
1+e^{i\theta } \\ 
\sqrt{\frac{\kappa }{2}}\left[ \left( i-1\right) -\left( 1+i\right)
e^{i\theta }\right]%
\end{array}%
\right)
\end{equation*}%
therefore, s.a. boundary conditions (\ref{3b.94}) become%
\begin{equation*}
\left( 1+e^{-i\theta }\right) \psi _{\theta }^{\prime }\left( 0\right) -%
\sqrt{\frac{\kappa }{2}}\left[ -\left( 1+i\right) -\left( 1-i\right)
e^{i\theta }\right] \psi _{\theta }\left( 0\right) =0\,,
\end{equation*}%
or $\psi _{\theta }^{\prime }\left( 0\right) =\lambda \psi _{\theta }\left(
0\right) \,,$ where $\lambda =\lambda \left( \theta \right) $ is a $\psi $%
-independent dimensional parameter of dimension of inverse length%
\begin{equation*}
\lambda =\frac{\kappa }{2}\left( \tan \frac{\vartheta }{2}-1\right) \,.
\end{equation*}%
When $\theta $ ranges from $-\pi $ to $\pi $, $\lambda $ ranges from $%
-\infty $ to $+\infty $, and $\lambda =\pm \infty $ ($\theta =\pm \pi $)
equivalently describe the s.a. boundary condition $\psi \left( 0\right) =0$.
It is natural to introduce the notation $\hat{H}_{0\lambda }=\hat{H}%
_{0\theta }$ and $\psi _{\lambda }=\psi _{\theta }$, with this notation, we
finally obtain that any s.a. operator $\hat{H}_{0\lambda }$ associated with
s.a. expression $\check{H}_{0}$ on the semiaxis $\left[ 0,\infty \right) $
is given by%
\begin{equation}
\hat{H}_{0\lambda }:\left\{ 
\begin{array}{l}
D_{H_{0\lambda }}=\left\{ \psi _{\lambda }:\psi _{\lambda },\,\psi _{\lambda
}^{\prime }\;\mathrm{a.c.\,on}\;\left[ 0,\infty \right) ;\;\psi _{\lambda
},\,\psi _{\lambda }^{\prime \prime }\in L^{2}\left( 0,\infty \right)
\,,\right. \\ 
\left. \psi ^{\prime }\left( 0\right) =\lambda \psi \left( 0\right)
\,,\;-\infty \leq \lambda \leq \infty \right\} \,, \\ 
\hat{H}_{0\lambda }\psi _{\lambda }=-\psi _{\lambda }^{\prime \prime }\,.%
\end{array}%
\right.  \label{3b.96}
\end{equation}%
Both$\;\lambda =\pm \infty \;$yield\ the\ same\ boundary\ condition$\;\psi
\left( 0\right) =0.$

Each of these $\hat{H}_{0\lambda }$ is a candidate to the quantum-mechanical
Hamiltonian for a free particle on a semiaxis. The boundary condition $\psi
\left( 0\right) =0$ is conventional in physics, but the boundary conditions $%
\psi ^{\prime }\left( 0\right) =\lambda \psi \left( 0\right) $ with a finite 
$\lambda $ are also encountered. We note that a specific choice of the
dimensional parameter $\kappa $ appeared irrelevant as well as the
normalization factor $\sqrt{\frac{\kappa }{2}}$, but if $\lambda \neq \pm
\infty $, the dimensional parameter $\lambda $ absent in $\check{H}_{0}$
enters the quantum theory as an additional specifying parameter.

By the way, the correctness of calculation which is rather simple in this
case is confirmed by verifying that necessary conditions (\ref{3b.94a}) and (%
\ref{3b.95}), $E_{1/2,\theta }^{+}\left( 0\right) \mathcal{E}E_{1/2,\theta
}\left( 0\right) =0\,,$ hold.

After the example, we return to the general questions. The specification of
the s.a. differential operators $\hat{f}_{U}$ in terms of s.a. boundary
conditions according to Theorems \ref{t3b.2},\ref{t3b.3}, and \ref{t3b.4}
requires evaluating the orthobasis functions $\left\{ e_{+,k}\right\}
_{1}^{m}$ and $\left\{ e_{-,k}\right\} _{1}^{m}$ in the respective deficient
subspaces $D_{+}$ and $D_{\_}$ , but only their boundary behavior is
essential. In addition, there is an arbitrariness in the choice of the
orthobasis functions, and the last example demonstrates that their specific
boundary values do not actually enter the answer. All this allows suggesting
that many analytical details are irrelevant from the standpoint of the
general specification. And indeed, there is another way of specifying s.a.
boundary conditions where the analytic task is replaced by some algebraic
task avoiding the evaluation of the deficient subspaces provided that the
deficient indices are known and equal, $m_{+}=m_{-}=m>0$. This way can be
more convenient from the application standpoint. It is based on a modified
version of the main theorem in the part related to formulas (\ref{3a.36}), (%
\ref{3a.39}). We therefore return to the main theorem and to the notation in
the previous section, in particular, in (\ref{3a.39}), where an initial
symmetric operator, its adjoint, and its closure are respectively denoted by 
$\hat{f}$,\thinspace $\hat{f}^{+}$, and $\overline{\hat{f}}$and the vectors
in their domains and in the domain $D_{f_{U}}$ are denoted by $\xi $ with
appropriate subscripts.

We first note the evident fact that the vectors $e_{U,k}=e_{+,k}+%
\sum_{l=1}^{m}U_{lk}e_{-,l}$ in (\ref{3a.39}), forming a basis in the
subspace $\left( D_{+}+\hat{U}D_{+}\right) \subset D_{f^{+}}$ of dimension $%
m $ are linearly independent modulo $D_{\bar{f}}$. It is also evident that
because all $e_{U,k}$ belong to $D_{f_{U}}\,,$ the relation%
\begin{equation}
\omega _{\ast }\left( e_{U,k},e_{U,l}\right) =0\,,\;k,l=1,\ldots ,m\,,
\label{3b.97}
\end{equation}%
holds; in the case of s.a. differential operators, it becomes the already
known relation (\ref{3b.79}). It appears that the really essential points
are the linear independence of the $m$ vectors $\left\{ e_{U,k}\right\}
_{1}^{m}$, modulo $D_{f}$ and relation (\ref{3b.97}) for them.

We then note that the vectors $e_{U,k}$ in (\ref{3a.39}) can be equivalently
replaced by their nondegenerate linear combinations, $e_{U,k}\rightarrow
w_{U,k}=\sum_{a=1}^{m}X_{ak}e_{U,a}\,,\;$provided the matrix $X=\left\|
X_{ak}\right\| $, $a,k=1,\ldots ,m$, is nonsingular, $\det X\neq 0$. As $%
e_{U,k}$, the vectors $w_{U,k}$ form a basis in the subspace $\left( D_{+}+%
\hat{U}D_{+}\right) $ and are linearly independent modulo $D_{\bar{f}}$. Of
course, relation (\ref{3b.97}) is extended to $\left\{ w_{U,k}\right\}
_{1}^{m}$ as the relation $\omega _{\ast }\left( w_{U,k},w_{U,l}\right) =0.$
What is more, we can add arbitrary vectors $\underline{\xi }_{k}$ belonging
to the domain $D_{\bar{f}}$ of the closure $\bar{f}$ to any vector $w_{U,k}$%
, 
\begin{equation*}
w_{U,k}\rightarrow w_{k}=w_{U,k}+\underline{\xi }_{k}=%
\sum_{a=1}^{m}X_{ak}e_{U,k}+\underline{\xi }_{k}\,,\;\underline{\xi }_{k}\in
D_{\bar{f}}\,,\;k=1,\ldots ,m\,,
\end{equation*}%
and obtain the equivalent description of the domain $D_{f_{U}}$ of the s.a.
extension $\hat{f}_{U}$ in terms of the $m$ new vectors $w_{k}$,%
\begin{equation}
D_{f_{U}}=\left\{ \psi _{U}\in D_{f^{+}}:\omega _{\ast }\left( w_{k},\psi
_{U}\right) =0\,,\;k=1,\ldots ,m\right\} \,,  \label{3b.98}
\end{equation}%
because $\omega _{\ast }\left( \underline{\xi }_{k},\psi _{U}\right) =0$ by (%
\ref{3a.10}), see also (\ref{3a.12}). By the same reason, relation (\ref%
{3b.97}) is also extended to the set $\left\{ w_{k}\right\} _{1}^{m}$,%
\begin{equation}
\omega _{\ast }\left( w_{k},w_{l}\right) =0\,,\;k,l=1,\ldots ,m\,.
\label{3b.99}
\end{equation}%
It is also evident that the $m$ new vectors $w_{k}$ are linearly independent
modulo $D_{\bar{f}}$.

It appears that the converse is true. Let $\hat{f}$ be a symmetric operator
with the adjoint $\hat{f}^{+}$ and the closure $\overline{\hat{f}}$, and let
its deficiency indices be nonzero and equal, $m_{+}=m_{-}=m>0$, such that $%
D_{f}\subseteq D_{\bar{f}}\subset D_{f^{+}}$ and $\dim _{D_{\bar{f}%
}}D_{f^{+}}=2m$. Let $\left\{ w_{k}\right\} _{1}^{m}$ be a set of vectors
with the following properties:

1)\ $w_{k}\in D_{f^{+}}\,,\;k=1,\ldots,m\,$;

2)$\;$they are linearly independent modulo\textrm{\ }\ $D_{\bar{f}}\,,\;$%
i.e.,%
\begin{equation*}
\sum_{k=1}^{m}c_{k}w_{k}\in D_{\bar{f}\,}\,,\;\forall c_{k}\in \mathbb{C\,}%
\Longrightarrow c_{k}=0\,,\;k=1,\ldots ,m\,;
\end{equation*}

3) relation (\ref{3b.99}), $\omega_{\ast}\left( w_{k},w_{l}\right)
=0\,,\;k,l=1,\ldots,m\,,$ holds for vectors $w_{k}$.

We then assert that the set $\left\{ w_{k}\right\} _{1}^{m}$ defines some
s.a. extension $\hat{f}_{U}$ of $\hat{f}$ as a s.a. restriction of the
adjoint $\hat{f}^{+}$, $\hat{f}\subset\hat{f}_{U}=\hat{f}_{U}^{+}\subset\hat{%
f}^{+}$ to the domain $D_{f_{U}}\subset D_{f^{+}}$ given by (\ref{3b.98}).

To prove this assertion, it is sufficient to prove that all the vectors $%
w_{k}$ can be uniquely represented as%
\begin{equation*}
w_{k}=X_{ak}\left( e_{+,a}+\sum_{a=1}^{m}U_{ba}e_{-,b}\right) +\underline{%
\xi }_{k}\,,
\end{equation*}%
where $\left\{ e_{+,k}\right\} _{1}^{m}$ and $\left\{ e_{-,k}\right\}
_{1}^{m}$ are some orthobasises in the respective deficient subspaces $D_{+}$
and $D_{-}$ of the symmetric operator $\hat{f}$, $X_{ak}$ and $U_{ba}$ are
some coefficients such that the matrix $X$ is nonsingular, and the matrix $U$
is unitary, and the vectors $\underline{\xi }_{k}$ belong to $D_{\bar{f}}$, $%
\underline{\xi }_{k}\in D_{\bar{f}}$, $k=1,\ldots ,m$.

We first address to the condition 1). According to first von Neumann formula
(\ref{3a.4}), any vector $w_{k}\in D_{f^{+}}$ is uniquely represented as%
\begin{equation*}
\,w_{k}=\xi _{+,k}+\xi _{-,k}+\underline{\xi }_{k}=%
\sum_{a=1}^{m}X_{ak}e_{+,a}+\sum_{a=1}^{m}Y_{ak}e_{-,a}+\underline{\xi }%
_{k}\,,
\end{equation*}%
where $\xi _{+,k}\in D_{+}\,,\;\xi _{-,k}\in D_{-}\,,$ and $\underline{\xi }%
_{k}\in D_{\bar{f}}\,,$ while $X_{ak}$ and $Y_{ak}\,,\;a,k=1,...,m,$ are the
expansion coefficients of $\xi _{+,k}$ and $\xi _{-,k}$ with respect to the
respective orthobasises $\left\{ e_{+,k}\right\} _{1}^{m}$ and $\left\{
e_{-,k}\right\} _{1}^{m}.$ We now address to the conditions 2) and 3). The
crucial remark is that these conditions imply that the matrices $X$ and $Y$
are nonsingular. The proof of that is by contradiction. Let, for example,
the rank of $X$ is nonmaximal, \textrm{rank}$X<m,$ this means that there
exist a set $\left\{ c_{k}\right\} _{1}^{m}$ of nontrivial complex constants 
$c_{k}$ such that at least one of them is nonzero, but $%
\sum_{k=1}^{m}X_{ak}c_{k}=0,$ $a=1,...,m.$ We thus have%
\begin{equation*}
\sum_{k=1}^{m}c_{k}\xi _{+,k}=\sum_{a=1}^{m}\left(
\sum_{k=1}^{m}X_{ak}c_{k}\right) e_{+,a}=0
\end{equation*}%
and the vector $w=\sum_{k=1}^{m}c_{k}w_{k}$ is represented as%
\begin{equation*}
w=\xi _{-}+\underline{\xi }\,,\;\xi _{-}=\sum_{k=1}^{m}c_{k}\xi _{-,k}\,,\;%
\underline{\xi }=\sum_{k=1}^{m}c_{k}\underline{\xi }_{k}\,.
\end{equation*}%
On the other hand, it follows from the condition 3) that%
\begin{equation*}
\omega _{\ast }\left( w,w\right) =\Delta _{\ast }\left( w\right)
=\sum_{k,l=1}^{m}\bar{c}_{k}c_{l}\omega _{\ast }\left( w_{k},w_{l}\right)
=0\,.
\end{equation*}%
By von Neumann formula (\ref{3a.18}), we then have $\Delta _{\ast }\left(
w\right) =-2i\kappa \left| \left| \xi _{-}\right| \right| ^{2}=0\,,$ or $\xi
_{-}=0,$ whence it follows that $w=\underline{\xi }\in D_{\bar{f}}\,.$ But
by the condition 2), the latter implies that all coefficients $c_{k}$ are
zero, which is a contradiction.

The proof of the nonsingularity of the matrix $Y$ is similar.

The nonsingularity of the matrix $X$ allows representing the vectors $w_{k}$
as%
\begin{equation*}
w_{k}=\sum_{a=1}^{m}X_{ak}\left( e_{+,a}+\sum_{b=1}^{m}U_{ba}e_{-,b}\right) +%
\underline{\xi }_{k}\,,
\end{equation*}%
where the nonsingular matrix $U$ is given by $U=YX^{-1}\,,\;\mathrm{or}%
\;Y=UX\,.$ Again appealing to condition 3) and to formula (\ref{3a.17}), we
find%
\begin{equation*}
\omega _{\ast }\left( w_{k},w_{l}\right) =\omega _{\ast }\left( \xi
_{+,k}+\xi _{-,k}+\underline{\xi }_{k}\,,\xi _{+,l}+\xi _{-,l}+\underline{%
\xi }_{l}\right) =2i\kappa \left[ \left( \xi _{+,k},\xi _{+,l}\right)
-\left( \xi _{-,k},\xi _{-,l}\right) \right] =0\,,
\end{equation*}%
or%
\begin{equation*}
\sum_{a,b=1}^{m}\left[ \overline{X}_{ak}\left( e_{+,a},e_{+,b}\right) X_{bl}-%
\overline{Y}_{ak}\left( e_{-,a},e_{-,b}\right) Y_{bl}\right]
\,=\sum_{a,b=1}^{m}\left[ \overline{X}_{ak}X_{al}-\overline{Y}_{ak}Y_{al}%
\right] =0\,,
\end{equation*}%
$k,l=1,...,m\,,$ where we use the condition that the sets $\left\{
e_{+,k}\right\} _{1}^{m}$ and $\left\{ e_{-,k}\right\} _{1}^{m}$ are
orthonormalized, $\left( e_{+,a},e_{+,b}\right) =\left(
e_{-,a},e_{-,b}\right) =\delta _{ab\,,\;}a,b=1,...,m.$ The last equality can
be written in the matrix form as%
\begin{equation*}
X^{+}X-Y^{+}Y=X^{+}\left[ I-\left( \left( X^{+}\right) ^{-1}Y^{+}\right)
\left( YX^{-1}\right) \right] X=X^{+}\left( I-U^{+}U\right) X=0\,.
\end{equation*}%
Because $X$ is nonsingular, it follows that $U^{+}U=I\,$,\ i.e., the matrix $%
U$ is unitary.

It is also seen how the unitary matrix $U$ is uniquely restored from the
given set of vectors $\left\{ w_{k}\right\} _{1}^{m}$ under a certain choice
of the orthobasises $\left\{ e_{+,k}\right\} _{1}^{m}$ and $\left\{
e_{-,k}\right\} _{1}^{m}$ in the respective deficient subspaces $D_{+}$ and $%
D_{-}$ of the initial symmetric operator $\hat{f},$ which accomplish the
proof of the above assertion.

We formulate the results of the above consideration as an addition to the
main theorem which is a modification of the main theorem in the part related
to formulas (\ref{3a.36}), (\ref{3a.39}).

\begin{theorem}
\label{t3b.5}(Addition to the main theorem.)

Any s.a. extension $\hat{f}_{U}$ of a symmetric operator $\hat{f}$ with the
deficiency indices $m_{+}=m_{-}=m>0\,,\;\hat{f}\subseteq $ $\overline{\hat{f}%
}\subset \hat{f}_{U}=\hat{f}_{U}^{+}\subset \hat{f}^{+}\,,$ can be defined as%
\begin{equation}
\hat{f}_{U}:\left\{ 
\begin{array}{l}
D_{f_{U}}=\left\{ \psi _{U}\in D_{f^{+}}:\,\omega _{\ast }\left( w_{k},\psi
_{U}\right) =0\,,\;k=1,...,m\right\} \,, \\ 
\hat{f}_{U}\psi _{U}=\hat{f}^{+}\psi _{U}\,,%
\end{array}%
\right.  \label{3b.100}
\end{equation}%
where $\left\{ w_{k}\right\} _{1}^{m}$ is some set of vectors in $%
D_{f^{+}}\,,$ $w_{k}\in D_{f^{+}}\,,\;k=1,...,m\,,$ linearly independent
modulo $D_{\bar{f}}$ and satisfying relation (\ref{3b.99}), $\omega _{\ast
}\left( w_{k},w_{l}\right) =0\,,\;k,l=1,...,m.$

Conversely, any set $\left\{ w_{k}\right\} _{1}^{m}$ of vectors in $%
D_{f^{+}}\,,$ linearly independent modulo $D_{\bar{f}}$ and satisfying
relation (\ref{3b.99}) defines some s.a. extension of the symmetric operator 
$\hat{f}\,$\ by (\ref{3b.100}).
\end{theorem}

To be true, the $U\left( m\right) $ nature of the set $\left\{ \hat{f}%
_{U}\right\} $ of all s.a. extensions is disguised in this formulation. This
manifests itself in the fact that two sets $\left\{ w_{k}\right\} _{1}^{m}$
and $\left\{ \tilde{w}_{k}\right\} _{1}^{m}$ of vectors related by a
nondegenerate linear transformation $\tilde{w}_{k}=\sum_{l=1}^{m}Z_{lk}w_{l}%
\,,\;$where the matrix $Z=\left| \left| Z_{lk}\right| \right| $ is
nonsingular, defines the same s.a. extension. We can say that the
description of s.a. extensions according to the addition to the main theorem
is a description with some ''excess'', irrelevant, but controllable.

When applied to differential operators in $L^{2}\left( a,b\right) ,$ the
addition to the main theorem yields an evident modification of Theorem \ref%
{t3b.2}. Formulating this modification, we return to the notation adopted in
this section and omit the explanation of the conventional symbols.

\begin{theorem}
\label{t3b.6}Any s.a. operator $\hat{f}_{U}$ in $L^{2}\left( a,b\right) $
associated with a given s.a. differential expression $\check{f}$ in the case
where the initial symmetric operator $\hat{f}^{\left( 0\right) }$ with the
closure $\hat{f}$ has the nonzero equal deficiency indices $m_{+}=m_{\_}=m$
can be defined as 
\begin{equation}
\hat{f}_{U}:\left\{ 
\begin{array}{l}
D_{f_{U}}=\left\{ \psi _{U}\in D_{\ast }:\,\left. \left[ w_{k},\psi _{U}%
\right] \right| _{a}^{b}=0\,,\;k=1,...,m\right\} , \\ 
\hat{f}_{U}\psi _{U}=\check{f}\psi _{U}\,,%
\end{array}%
\right.  \label{3b.101}
\end{equation}%
where $\left\{ w_{k}\right\} _{1}^{m}$ is the set of functions belonging to $%
D_{\ast }\,,$ $w_{k}\in D_{\ast }\,,\;k=1,...,m\,,$ linearly independent
modulo $D_{f}$ and satisfying the relations%
\begin{equation}
\left. \left[ w_{k},w_{l}\right] \right| _{a}^{b}=0\,,\;k,l=1,...,m.
\label{3b.102}
\end{equation}

Conversely, any set $\left\{ w_{k}\right\} _{1}^{m}$ of functions belonging
to $D_{\ast },$ linearly independent modulo $D_{f}$ and satisfying relations
(\ref{3b.102}) defines some s.a. operator associated with differential
expression $\check{f}$\thinspace\ by (\ref{3b.101}).
\end{theorem}

The remark following the addition to the main theorem is completely
applicable to Theorem \ref{t3b.6}.

Theorem \ref{t3b.6} yields a modified version of Theorem \ref{t3b.3} for the
case of an even differential expression with the both regular ends where the
deficiency indices are maximum. The modification consists in the replacement
of the matrices $E_{U}\left( a\right) $ and $E_{U}\left( b\right) $ (\ref%
{3b.83}) of the boundary values of the basis functions $e_{U,k}$ and their
quasiderivatives of order up to $n-1$ by the similar matrices%
\begin{equation*}
W\left( a\right) =\left| \left| W_{lk}\left( a\right) =w_{k}^{[l-1]}\left(
a\right) \right| \right| \,,\;W\left( b\right) =\left| \left| W_{lk}\left(
b\right) =w_{k}^{[l-1]}\left( b\right) \right| \right| \,,
\end{equation*}%
generated by the functions $w_{k}\in D_{\ast }$ satisfying the conditions of
Theorem \ref{t3b.6}. We assert that these conditions, the linear
independence of the functions $w_{k}$ modulo $D_{f}$ and relation (\ref%
{3b.102}), are equivalent to the two respective conditions on the matrices $%
W\left( a\right) $ and $W\left( b\right) :$

1) the rank of a rectangular $2n\times n$ matrix $\mathbb{W}$ is maximum and
equal to $n,$%
\begin{equation}
\mathbb{W}=\left( 
\begin{array}{c}
W\left( a\right) \\ 
W\left( b\right)%
\end{array}%
\right) ,\;\mathrm{rank}\mathbb{W}=n,  \label{3b.103}
\end{equation}%
this property is a complete analogue of (\ref{3b.95});

2) the relation%
\begin{equation}
W^{+}\left( b\right) \mathcal{E}W\left( b\right) =W^{+}\left( a\right) 
\mathcal{E}W\left( a\right)  \label{3b.104}
\end{equation}
holds.

The necessity of condition (\ref{3b.103}) is proved by contradiction. Let $%
\mathrm{rank}\mathbb{W}<n.$ This means that there exists a set $\left\{
c_{k}\right\} _{1}^{n}$ of nontrivial numbers, i.e., at least one of $c_{k}$
is nonzero, such that%
\begin{equation*}
\sum_{k=1}^{n}W_{lk}\left( a\right) c_{k}=\sum_{k=1}^{n}w_{k}^{[l-1]}\left(
a\right) c_{k}=0\,,\;\sum_{k=1}^{n}W_{lk}\left( b\right)
c_{k}=\sum_{k=1}^{n}w_{k}^{[l-1]}\left( b\right) c_{k}=0\,.
\end{equation*}%
By Lemma \ref{l3b.5}, this implies that the function $w=%
\sum_{k=1}^{n}c_{k}w_{k}$ belongs to $D_{f}\,,$ the domain of the closure $%
\hat{f}$ of the initial symmetric operator $\hat{f}^{\left( 0\right) },$ but
because the functions $w_{k}$ are linearly independent modulo $D_{f},$ this
in turn implies that all $c_{k}$ are zero which is a contradiction. We
actually repeat the arguments leading to (\ref{3b.95}).

Conversely, let $W\left( a\right) =\left| \left| W_{lk}\left( a\right)
\right| \right| $ and $W\left( b\right) =\left| \left| W_{lk}\left( b\right)
\right| \right| $ be arbitrary matrices satisfying condition (\ref{3b.103}).
Because the functions in $D_{\ast }$ together with their quasiderivatives of
order up to $n-1$ can take arbitrary values at the regular ends $a$ and $b$,
there exist a set $\left\{ w_{k}\right\} _{1}^{n}$ of functions $w_{k}\in
D_{\ast }$ such that%
\begin{equation*}
W_{lk}\left( a\right) =w_{k}^{[l-1]}\left( a\right) \,,\;W_{lk}\left(
b\right) =w_{k}^{[l-1]}\left( b\right) \,,\;l,k=1,...,n\,;
\end{equation*}%
the functions $w_{k}$ are evidently linearly independent modulo $D_{f}\,.$

As to relation (\ref{3b.104}), this relation is equivalent to relation (\ref%
{3b.102}) in view of formulas (\ref{3b.44}), (\ref{3b.45}) where $\chi
_{\ast }$ and $\psi _{\ast }$ are replaced by the respective $w_{k}$ and $%
w_{l}$ , $l,k=1,...,n\,,$ and formula (\ref{3b.23}); it is the copy and
extension of relation (\ref{3b.86}). Because the functions $w_{k}$\ are
represented in this context only by the boundary values of their
quasiderivatives of order from $0$ up to $n-1,$ it is natural to introduce
the notation%
\begin{equation*}
A=\left| \left| a_{lk}\right| \right| =W\left( a\right) \,,\;B=\left| \left|
b_{lk}\right| \right| =W\left( b\right)
\end{equation*}%
and formulate a modification of Theorem \ref{t3b.6} as follows:

\begin{theorem}
\label{t3b.7}Any s.a. operator $\hat{f}_{U}$ in $L^{2}\left( a,b\right) $
associated with an even s.a. differential expression $\check{f}$ of order $n$
with both regular ends can be defined as 
\begin{equation}
\hat{f}_{U}:\left\{ 
\begin{array}{l}
D_{f_{U}}=\left\{ \psi _{U}\in D_{\ast }:\,B^{+}\mathcal{E}\Psi _{U}\left(
b\right) =A^{+}\mathcal{E}\Psi _{U}\left( a\right) \right\} \,, \\ 
\hat{f}_{U}\psi _{U}=\check{f}\psi _{U}\,,%
\end{array}%
\right.  \label{3b.105}
\end{equation}%
where $A$ and $B$ are some $n\times n$ matrices satisfying the conditions%
\begin{equation}
\mathrm{rank}\left( 
\begin{array}{c}
A \\ 
B%
\end{array}%
\right) =n\,,\;B^{+}\mathcal{E}B=A^{+}\mathcal{E}A\,,  \label{3b.107}
\end{equation}%
the matrix $\mathcal{E}$ and the columns $\Psi _{U}\left( b\right) $ and $%
\Psi _{U}\left( a\right) $ are respectively given by (\ref{3b.81}), and (\ref%
{3b.84}).

Conversely, any two matrices $A$ and $B$ satisfying conditions (\ref{3b.107}%
) define some s.a. operator associated with the s.a. differential expression 
$\check{f}$ by (\ref{3b.105}).
\end{theorem}

Again, as after Lemma \ref{l3b.4} and after Theorem \ref{t3b.3}, we can add
that a similar theorem holds for any s.a. differential expression $\check{f}$
of any order with differentiable coefficients and the both regular ends with
the change quasiderivatives to usual derivatives if boundary values (\ref%
{3b.45}) are finite forms in the boundary values of functions and their
derivatives.

The remarks after the addition to the main theorem \ref{t3b.5} and Theorem %
\ref{t3b.6} on the hidden $U\left( n\right) $-nature of $n$ s.a. boundary
conditions (\ref{3b.105}) become the remark that the matrices $\tilde{A}=AZ$
and $\tilde{B}=BZ\,,$ where the matrix $Z=\left| \left| z_{lk}\right|
\right| \,,\;l,k=1,...,n\,,$ is nonsingular, define the same s.a. operator.

Actually, this arbitrariness in the choice of the matrices $A$ and $B$ is
unremovable only if their ranks are not maximum\footnote{%
Of course, this condition is compatible with condition (\ref{3b.107}).}, $\ 
\mathrm{rank}A<n,\;\mathrm{rank}B<n,$ i.e., if they are singular, $\det
A=\det B=0$ (we note that condition (\ref{3b.107}) implies that the matrices 
$A$ and $B$ are singular or nonsingular simultaneously). If these matrices
are nonsingular, which is a general case, the arbitrariness can be
eliminated. Really, let $\det B\neq 0,$ therefore, $\det A\neq 0$ also.
Then, with taking the property $\mathcal{E}^{-1}=-\mathcal{E}$ of the
nonsingular matrix $\mathcal{E}$ into account, s.a. boundary conditions (\ref%
{3b.105}) can be represented as%
\begin{equation}
\Psi \left( b\right) =S\Psi \left( a\right) \,,\;\mathrm{or}\;\Psi \left(
a\right) =S^{-1}\Psi \left( b\right) \,,  \label{3b.108}
\end{equation}%
where the nonsingular matrix $S$ is $S=-\mathcal{E}\left( AB^{-1}\right) ^{+}%
\mathcal{E}\,.$ Because the matrix $\mathcal{E}$ is anti-Hermitian, $%
\mathcal{E}^{+}=-\mathcal{E}$, the adjoint $S^{+}$ is $S^{+}=-\mathcal{E}%
\left( AB\right) ^{-1}\mathcal{E}$ and the second condition in (\ref{3b.107}%
) is represented in terms of $S$ as%
\begin{equation}
S^{+}\mathcal{E}S=\mathcal{E}\,,  \label{3b.110}
\end{equation}%
otherwise, $S$ is arbitrary.

The algebraic sense of relation (\ref{3b.110}) is clear: it means that the
linear transformations%
\begin{equation}
\Psi \rightarrow S\Psi  \label{3b.111}
\end{equation}%
defined in the $n$-dimensional linear space of $n$-columns $\Psi $ with
elements $\psi _{i}\,,\;i=1,...,n,$ preserve the Hermitian sesquilinear form 
$\chi ^{+}\left( \frac{1}{i}\mathcal{E}\right) \Psi ,$ or equivalently, the
Hermitian quadratic form $\Psi ^{+}\left( \frac{1}{i}\mathcal{E}\right) \Psi
.$ The Hermitian matrix $\frac{1}{i}\mathcal{E}$ can be easily diagonalized
by a unitary transformation,%
\begin{equation}
\frac{1}{i}\mathcal{E}=T^{+}\Sigma T\,,  \label{3b.112}
\end{equation}%
where the diagonal $n\times n$ matrix $\Sigma $ is%
\begin{equation}
\Sigma =\mathrm{diag}\left( I,-I\right) \,,  \label{e.5}
\end{equation}%
$I$ is the $n/2\times n/2$ unit matrix, and the unitary matrix $T=\left|
\left| T_{lm}\right| \right| ,\;l,m=1,...,n\,,\;T^{+}T=\mathbf{I}\,,$ is%
\begin{eqnarray}
&&T_{lm}=\frac{1}{\sqrt{2}}\left\{ \delta _{l,m}\left[ \theta \left( \frac{%
n+1}{2}-m\right) -i\theta \left( m-\frac{n+1}{2}\right) \right] \right. 
\notag \\
&&\left. \delta _{l,n+1-m}\left[ \theta \left( \frac{n+1}{2}-m\right)
+i\theta \left( m-\frac{n+1}{2}\right) \right] \right\} \,,  \label{e.6}
\end{eqnarray}%
and $\theta \left( x\right) $ is the well-known step function,%
\begin{equation*}
\theta \left( x\right) =\left\{ 
\begin{array}{c}
1\,,\;x>0 \\ 
0\,,\;x<0%
\end{array}%
\right. ,
\end{equation*}%
and we see that the signature of the matrix $\frac{1}{i}\mathcal{E}$ is $%
\left( \frac{n}{2},\frac{n}{2}\right) ,$ which implies that the
transformations $S$ given by (\ref{3b.111}) and satisfying (\ref{3b.110})
form the group $U\left( \frac{n}{2},\frac{n}{2}\right) .$ We thus find that
some of s.a. boundary conditions, to be true, the most of them, are
parametrized by elements of the group $U\left( \frac{n}{2},\frac{n}{2}%
\right) ,$ which defines an embedding of the group $U\left( \frac{n}{2},%
\frac{n}{2}\right) $ into the group $U\left( n\right) $ parameterizing all
the s.a. boundary conditions. This embedding is an embedding ''into'', but
not ''onto'': although $U\left( \frac{n}{2},\frac{n}{2}\right) $ \ is an $%
n^{2}$-parameter manifold as $U\left( n\right) ,$ the group $U\left( \frac{n%
}{2},\frac{n}{2}\right) $ is noncompact, whereas $U\left( n\right) $ is
compact; it is also clear from the aforesaid that the s.a. boundary
conditions (\ref{3b.105}) with the singular matrices $A$ and $B$ cannot be
represented in form (\ref{3b.108}). Such boundary conditions can be obtained
by some limit procedure where some matrix element of $S$ tends to infinity
while others vanish (we note that $\left| \det S\right| =1$). This procedure
corresponds to a compactification of $U\left( \frac{n}{2},\frac{n}{2}\right) 
$ to $U\left( n\right) $ by adding some limit points.

We must note that looking at the representation of the sesquilinear
asymmetry form $\omega _{\ast }$ in terms of boundary values of functions
and their quasiderivatives in the case of an even s.a. differential
expression with the both regular ends\footnote{%
This representation based on formulas (\ref{3b.44}), (\ref{3b.45}), (\ref%
{3b.23}), and (\ref{3b.84}) was actually used above in the consideration
related to formulas (\ref{3b.80})-(\ref{3b.86}).}%
\begin{equation}
\omega _{\ast }\left( \chi _{\ast },\psi _{\ast }\right) =\chi _{\ast
}^{+}\left( b\right) \mathcal{E}\Psi _{\ast }\left( b\right) -\chi _{\ast
}^{+}\left( a\right) \mathcal{E}\Psi _{\ast }\left( a\right) \,,
\label{3b.114}
\end{equation}%
where the $n$-columns $\chi _{\ast }\left( a\right) ,\chi _{\ast }\left(
b\right) ,$ and $\Psi _{\ast }\left( a\right) ,\Psi _{\ast }\left( b\right) $
are given by (\ref{3b.84}) with the respective changes $\psi _{U}\rightarrow
\chi _{\ast }$ and $\psi _{U}\rightarrow \psi _{\ast }\,,$ it is easy to see
from the very beginning that boundary conditions (\ref{3b.108}) with any
fixed matrix $S$ satisfying (\ref{3b.110}) result in vanishing the asymmetry
form $\omega _{\ast }$ and thus define a symmetric restriction of the
adjoint $\hat{f}^{\ast }\,.$ Using the standard technique of evaluating the
adjoint in terms of $\omega _{\ast }$ (\ref{3b.114}), it is also easy to
prove that boundary condition (\ref{3b.108}), (\ref{3b.110}) are actually
s.a. boundary conditions defining a s.a. restriction of $\hat{f}^{\ast }$.
Unfortunately, these are not all possible s.a. boundary conditions.

It seems instructive to illustrate Theorem \ref{t3b.7} and also s.a.
boundary conditions (\ref{3b.108}), (\ref{3b.110}) based on representation (%
\ref{3b.114}) for $\omega _{\ast }$ and their extensions to s.a.
differential expressions of any order with the both regular ends by our
examples of the s.a. differential expressions $\check{p}$ (\ref{3b.9}) and $%
\check{H}_{0}$ (\ref{3b.48a}) on an interval $\left( 0,l\right) $.

As to $\check{p}$, an analogue of (\ref{3b.114}) for $\hat{p}^{\ast }$ (\ref%
{3a.40})\ is%
\begin{equation}
\omega _{\ast }\left( \chi _{\ast },\psi _{\ast }\right) =-i\left( \overline{%
\chi _{\ast }}\left( l\right) \psi \left( l\right) -\overline{\chi }\left(
0\right) \psi \left( 0\right) \right) \,,  \label{3b.115}
\end{equation}%
see (\ref{3a.41})--(\ref{3a.44}). It immediately follows the s.a. boundary
conditions%
\begin{equation}
\psi _{\vartheta }\left( l\right) =e^{i\vartheta }\psi _{\vartheta }\left(
0\right) \,,  \label{3b.116}
\end{equation}%
with arbitrary but fixed angle $\vartheta $, $0\leq \vartheta \leq 2\pi $, $%
0\backsim 2\pi $, which coincides with boundary conditions (\ref{3a.55})
defining s.a. operators $\hat{p}_{\vartheta }$ (\ref{3a.57}). In this case,
thus obtained boundary conditions (\ref{3b.116}) yield all the $U\left(
1\right) $-family of s.a. operators associated with the s.a. deferential
expression $\check{p}$ (\ref{3b.9}) on an interval $[0,l]$.

As to $\check{H}_{0}$, we show how the already known s.a. boundary
conditions (\ref{3b.87})--(\ref{3b.90}) are obtained without evaluating the
deficient subspaces.

Let%
\begin{equation*}
A=\left( 
\begin{array}{cc}
0 & 0 \\ 
0 & 1%
\end{array}%
\right) \,,\;B=\left( 
\begin{array}{cc}
0 & 0 \\ 
1 & 0%
\end{array}%
\right) \,,
\end{equation*}%
it is easy to verify that they satisfy conditions (\ref{3b.107}), then
formula (\ref{3b.105}) yields the s.a. boundary conditions $\psi \left(
0\right) =\psi \left( l\right) =0$ coinciding with (\ref{3b.88}). These
boundary conditions can be obtained from boundary conditions (\ref{3b.108})
with%
\begin{equation*}
S\left( \varepsilon \right) =\left( 
\begin{array}{cc}
0 & \varepsilon \\ 
-1/\varepsilon & 0%
\end{array}%
\right)
\end{equation*}%
in the limit $\varepsilon \rightarrow 0$, such $S\left( \varepsilon \right) $
arises if we slightly deform the initial $A$ and $B$,%
\begin{equation*}
A\mathbb{\rightarrow }A\left( \varepsilon \right) =\left( 
\begin{array}{cc}
\varepsilon & 0 \\ 
0 & 1%
\end{array}%
\right) \,,\;B\mathbb{\rightarrow }B\left( \varepsilon \right) =\left( 
\begin{array}{cc}
0 & -\varepsilon \\ 
1 & 0%
\end{array}%
\right) \,,
\end{equation*}%
removing their singularity without violating conditions (\ref{3b.107}).

Let now%
\begin{equation*}
A=\left( 
\begin{array}{cc}
0 & 1 \\ 
0 & 0%
\end{array}%
\right) \,,\;B=\left( 
\begin{array}{cc}
1 & 0 \\ 
0 & 0%
\end{array}%
\right) \,,
\end{equation*}%
these matrices also satisfy (\ref{3b.107}), then formula (\ref{3b.105})
yields the s.a. boundary conditions $\psi ^{\prime }\left( 0\right) =\psi
^{\prime }\left( l\right) =0$ coinciding with (\ref{3b.89}). Again, these
boundary conditions can be obtained from boundary conditions (\ref{3b.108})
with%
\begin{equation*}
S\left( \varepsilon \right) =\left( 
\begin{array}{cc}
0 & -1/\varepsilon \\ 
\varepsilon & 0%
\end{array}%
\right)
\end{equation*}%
in the limit $\varepsilon \rightarrow 0$, this $S\left( \varepsilon \right) $
arises as a result of a deformation%
\begin{equation*}
A\rightarrow A\left( \varepsilon \right) =\left( 
\begin{array}{cc}
0 & 1 \\ 
-\varepsilon & 0%
\end{array}%
\right) \,,\;B\rightarrow B\left( \varepsilon \right) =\left( 
\begin{array}{cc}
1 & 0 \\ 
0 & \varepsilon%
\end{array}%
\right) \,.
\end{equation*}

If we take%
\begin{equation*}
A=\left( 
\begin{array}{cc}
0 & a_{2} \\ 
0 & a_{4}%
\end{array}%
\right) \,,\;B=\left( 
\begin{array}{cc}
b_{1} & 0 \\ 
b_{3} & 0%
\end{array}%
\right) \,,
\end{equation*}%
where at least one number in the pairs $a_{2},a_{4}$ and $b_{1},b_{3}$ is
nonzero, which is required by the first condition in (\ref{3b.107}) and also 
$a_{2}\overline{a_{4}}=\overline{a_{2}}a_{4}$ and $b_{1}\overline{b_{3}}=%
\overline{b_{1}}b_{3}$, which is required by second condition in (\ref%
{3b.107}), we obtain the so called splitted boundary conditions%
\begin{equation}
\psi ^{\prime }\left( 0\right) =\lambda \psi \left( 0\right) \,,\;\psi
^{\prime }\left( l\right) =\mu \psi \left( l\right) \,,  \label{3b.117}
\end{equation}%
where $\lambda $ and $\mu $ are arbitrary numbers, $-\infty \leq \lambda $, $%
\mu \leq +\infty $, $-\infty \backsim +\infty $, and the both $\lambda =\pm
\infty $ yield $\psi \left( 0\right) =0$ while $\mu =\pm \infty $ yield $%
\psi \left( l\right) =0$.

Taking $S=I$ in (\ref{3b.108}), we obtain the periodic boundary conditions%
\begin{equation*}
\psi \left( 0\right) =\psi \left( l\right) \,,\;\psi ^{\prime }\left(
0\right) =\psi ^{\prime }\left( l\right)
\end{equation*}%
coinciding with (\ref{3b.90}). If we take $S=e^{i\vartheta }I$, $0\leq
\vartheta \leq 2\pi $, $0\backsim 2\pi $, we obtain the modified periodic
s.a. boundary conditions%
\begin{equation}
\psi \left( l\right) =e^{i\vartheta }\psi \left( 0\right) \,,\;\psi ^{\prime
}\left( l\right) =e^{i\vartheta }\psi ^{\prime }\left( 0\right)
\label{3b.118}
\end{equation}%
including periodic, $\vartheta =0$, and antiperiodic, $\vartheta =\pi $,
boundary conditions. It is interesting to note that these s.a. boundary
conditions define the s.a. operator $\hat{H}_{0\vartheta }$ which can also
be represented as%
\begin{equation}
\hat{H}_{0\vartheta }=\hat{p}_{\vartheta }^{2}\,.  \label{3b.119}
\end{equation}%
The \ one-parameter family $\left\{ H_{0\vartheta }\right\} \,$of s.a.
operators (\ref{3b.119}) among the whole four-parameter family of s.a.
operators associated with the s.a. differential expression $\check{H}_{0}$
immediately follows from the Akhiezer-Glazman theorem (Theorem \ref{AGT})
with $\hat{a}=\hat{a}^{+}=\hat{p}_{\vartheta }$, after constructing s.a.
operator $\hat{p}_{\vartheta }$ (\ref{3a.57}).

As to the ``entangled''\ s.a. boundary conditions (\ref{3b.87}), it is easy
to verify that they can be represented in form (\ref{3b.108}), $\Psi \left(
l\right) =S\Psi \left( 0\right) $,, where the matrix%
\begin{equation*}
S=-\left( 
\begin{array}{cc}
\cosh \mathrm{\,}\pi & \frac{l}{\pi }\mathrm{\sinh }\pi \\ 
\frac{\pi }{l}\sinh \mathrm{\,}\pi & \cosh \mathrm{\,}\pi%
\end{array}%
\right)
\end{equation*}%
satisfies condition (\ref{3b.110}).

Our concluding remark is that the s.a. operators associated with s.a.
differential expressions $\check{H}$ (\ref{3b.10}) with $V=\overline{V},$
conventionally attributed to the quantum-mechanical energy a particle on an
interval $\left[ 0,l\right] $ in a potential field $V$, are specified by the
same boundary conditions if $V$ is integrable on $\left[ 0,l\right] $
because under this conditions, the ends of the interval remain regular. It
is completely clear in the case of a bounded potential, $\left| V\left(
x\right) \right| <M$, because the addition of a bounded s.a. operator
defined everywhere to a s.a. operator with a certain domain yields a s.a.
operator with the same domain.

Theorem \ref{t3b.6} yields also a modified version of Theorem \ref{t3b.4}.
Because the appropriate consideration is completely similar to the previous
one resulting in Theorem \ref{t3b.7}, we directly formulate this
modification.

\begin{theorem}
\label{t3b.8}Any s.a. operator $\hat{f}_{U}$ associated with an even s.a.
deferential expression $\check{f}$ of order $n$ on an interval $\left(
a,b\right) $ with the regular end $a$ and the singular end $b$ in the case
where the initial symmetric operator $\hat{f}^{\left( 0\right) }$ has the
minimum deficiency indices $m_{+}=m_{-}=n/2$, $U\in U\left( n/2\right) $,
can be defined as%
\begin{equation}
\hat{f}_{U}:\;\left\{ 
\begin{array}{l}
D_{f_{U}}=\left\{ \psi _{U}\left( x\right) \in D_{\ast }:\;A_{1/2}^{+}%
\mathcal{E}\psi _{U}\left( a\right) =0\right\} \,, \\ 
\hat{f}_{U}\psi _{U}=f\psi _{U}\,,%
\end{array}%
\right.  \label{3b.120}
\end{equation}%
where $A_{1/2}$ is some rectangular $n\times n/2$ matrix satisfying the
conditions%
\begin{equation}
\mathrm{rank}\,A_{1/2}=n/2  \label{3b.121}
\end{equation}%
and%
\begin{equation}
A_{1/2}^{+}\mathcal{E}A_{1/2}=0\,,  \label{3b.122}
\end{equation}%
the matrix $\mathcal{E}$ and the column $\Psi _{U}\left( a\right) $ are
respectively given by (\ref{3b.81}), and (\ref{3b.84}).

Conversely, any $n\times n/2$ matrix $A$ satisfying (\ref{3b.121}) and (\ref%
{3b.122}) define some s.a. operator associated with the s.a. differential
expression $\check{f}$ by (\ref{3b.120}).

If the end $a$ is singular while the end $b$ is regular, $A_{1/2}$ and $a$
in (\ref{3b.120})--(\ref{3b.122}) are replaced by the respective $B_{1/2}$
and $b$.
\end{theorem}

Similarly to Theorem \ref{t3b.7}, this theorem is accompanied by the remark
on the hidden $U\left( n/2\right) $-nature of boundary conditions (\ref%
{3b.120}):\ the matrices $A_{1/2}$ and $A_{1/2}Z$, where $Z$ is some
nonsingular $n/2\times n/2$ matrix, yield the same s.a. operator.

We illustrate this theorem by the example of the deferential expression $%
\check{H}$ (\ref{3b.10}) with $V=\overline{V}$, on the semiaxis $\left[
0,\infty \right) $ (the quantum-mechanical energy of a particle on the
semiaxis in the potential field $V$). We begin with the differential
expression $\check{H}_{0}$ (\ref{3b.48a}) (a free particle) already
considered as an illustration after Theorem \ref{t3b.4} where it was shown
that the corresponding deficiency indices are $\left( 1,1\right) $ and show
how the known result is obtained without evaluating the deficient subspaces,
which allows extending the results to the case $V\neq 0$. In this case, $n=2$%
, $n/2=1$, the matrix $A_{1/2}$ is a two column with elements $a_{1},a_{2},$
where at least one of the numbers in the pair $a_{1},a_{2}$ is nonzero,
which is required by condition (\ref{3b.121}), while condition (\ref{3b.122}%
) requires $\overline{a_{1}}a_{2}=a_{1}\overline{a_{2}}$. Formula (\ref%
{3b.120}) then yields the already known s.a. boundary conditions%
\begin{equation}
\psi ^{\prime }\left( 0\right) =\lambda \psi \left( 0\right) \,,\;\lambda
\in \mathbb{R}^{1}\,,  \label{3b.123}
\end{equation}%
where $\lambda =\overline{a_{2}}/\overline{a_{1}}=a_{2}/a_{1}$ is an
arbitrary, but fixed, real number, $-\infty \leq \lambda \leq \infty $, $%
-\infty \backsim \infty $; $\lambda =\pm \infty $ correspond to the boundary
condition $\psi \left( 0\right) =0$. These boundary conditions defining s.a.
operators $\hat{H}_{0\lambda }$ (\ref{3b.96}) \ are well known in physics.

It is evident that the same boundary conditions specify the s.a. operators $%
\hat{H}_{\lambda }$ associated with the s.a. differential expressions $%
\check{H}$ (\ref{3b.10}) in the case where the potential is bounded, $\left|
V\left( x\right) \right| <M$, and $\hat{H}_{\lambda }$ is then defined on
the some domain as $\hat{H}_{0\lambda }$.

We now shortly discuss the physically interesting question of under which
conditions on the potential $V\left( x\right) $, the s.a. differential
expression $\check{H}$ (\ref{3b.10}) on $\left[ 0,\infty \right) $ also
falls under Theorem \ref{t3b.8} as the differential expression $\check{H}%
_{0} $ (\ref{3b.48a}), and is therefore specified by the same s.a. boundary
conditions (\ref{3b.123}).

For the left end to remain regular, it is necessary that $V\left( x\right) $
be integrable at the origin, i.e., integrable on any segment $\left[ 0,a%
\right] $, $a<\infty $. We know that if the left end is regular, the
deficiency indices of the associated symmetric operator $\hat{H}^{\left(
0\right) }$ can be $\left( 1,1\right) $ or $\left( 2,2\right) $, and we need
criteria for these be minimum, but not maximum. At this point, we address to
some useful general result on the maximum deficiency indices $\left(
n,n\right) $ for the symmetric operator $\hat{f}^{\left( 0\right) }$
associated with an even s.a. differential expression $\check{f}$ of order $n$
with one regular end and one singular end. It appears that the occurrence of
the maximum deficiency indices is controlled by the dimension of the kernel
of the adjoint $\hat{f}^{\ast }$, $\dim \ker f^{\ast }$, or by the number of
the linearly independent square-integrable solutions of the homogeneous
equation $\check{f}u=0$. Namely, $\hat{f}^{\left( 0\right) }$ has maximum
deficiency indices $m_{+}=m_{-}=n$ iff the homogeneous equation has the
maximum number $n$ of linearly independent square-integrable solutions, in
other words, iff the whole fundamental system $\left\{ u_{i}\right\}
_{1}^{n} $ of solutions of the homogeneous equation lies in $L^{2}\left(
a,b\right) ;$ the same is true for the homogeneous equation $\check{f}%
u=\lambda u$ with any real $\lambda $.

It follows from this general statement that in order to have the deficiency
indices $\left( 1,1\right) $ in our particular case where $n=2$, it is
sufficient to point out the conditions on $V$ under which the homogenous
equation $-u^{\prime \prime }+Vu=0$ has at lest one non-square-integrable
solution. A few of such conditions are known since Weyl \cite{Weyl10}. We
cite the two which seem rather general and also simple from the application
standpoint and formulate them directly in the form relevant to the
deficiency indices:\ the symmetric operator $\hat{H}^{\left( 0\right) }$ has
deficiency indices $\left( 1,1\right) $ if%
\begin{equation}
1)\mathrm{\;}V\left( x\right) \in L^{2}\left( 0,\infty \right) \,,
\label{3b.124}
\end{equation}%
i.e., the\ potential\ $V$\ is\ square-integrable,\ \cite{Putna49}, or%
\begin{equation}
2)\;V\left( x\right) >-Kx^{2}\,,\;K>0\,,  \label{3b.125}
\end{equation}%
for sufficiently large $x$ \cite{HarWi48}. Condition (\ref{3b.125}) is a
particular case of a more general condition \cite{Levin49}. For the proofs,
other conditions and details, see \cite{Naima69}. We here don't dwell on the
proofs because we independently obtain the same results in another context
later, but make several remarks concerning physical applications.

We consider it useful, in particular, for further references, to repeat once
more that under conditions (\ref{3b.124}) or (\ref{3b.125}) on the potential 
$V$ integrable at the origin, all s.a. Hamiltonians associated with the s.a.
differential expression $\check{H}$ (\ref{3b.10}) on the semiaxis $\left[
0,\infty \right) $ form a one-parameter family $\left\{ \hat{H}_{\lambda
}\right\} $, $-\infty \leq \lambda \leq \infty $, $-\infty \backsim \infty $%
, and any $\hat{H}_{\lambda }$ is specified by s.a. boundary conditions (\ref%
{3b.123}):%
\begin{equation}
\hat{H}_{\lambda }:\left\{ 
\begin{array}{l}
D_{\lambda }=\left\{ \left\{ \psi _{\lambda }:\psi _{\lambda },\psi
_{\lambda }^{\prime }\;\text{\textrm{a.c.\thinspace on} }\left[ 0,\infty
\right) \right. ;\psi _{\lambda },-\psi _{\lambda }^{\prime \prime }+V\psi
_{\lambda }\in L^{2}\left( 0,\infty \right) ;\psi _{\lambda }^{\prime
}\left( 0\right) =\lambda \psi _{\lambda }\left( 0\right) \right\} , \\ 
\hat{H}_{\lambda }\psi _{\lambda }=-\psi _{\lambda }^{\prime \prime }+V\psi
_{\lambda }\,.%
\end{array}%
\right.  \label{3b.126}
\end{equation}%
Condition (\ref{3b.124}) covers the conditions of the regularity of the left
end because it automatically implies that $V$ is integrable at the origin.
By the way, this conditions does not at all imply that $V$ vanishes at
infinity, the potential can have growing peaks of any sign with growing $x$.

The majority of potentials encountered in physics, in particular, the
potentials vanishing or growing at infinity, satisfy condition (\ref{3b.125}%
). Criterion (\ref{3b.125}) is optimal in the sense that if $V\backsim
-Kx^{2\left( 1+\varepsilon \right) }$ as $x\rightarrow \infty $, where $%
\varepsilon >0$ can be arbitrarily small, the both linearly independent
solutions $u_{1,2}$ of the homogenous equation $-u^{\prime \prime }+Vu=0,$
and also of the equation $-u^{\prime \prime }+Vu=\lambda u$ with any real $%
\lambda ,$ are square integrable:%
\begin{equation*}
u_{1,2}\left( x\right) \backsim \frac{1}{x^{(1+\varepsilon )/2}}\exp \left[
\pm i\frac{K^{1/2}}{2+\varepsilon }x^{2+\varepsilon }\right]
\,,\;x\rightarrow \infty \,,
\end{equation*}%
therefore, the deficiency indices of the symmetric operator $\hat{H}^{\left(
0\right) }$ are $\left( 2,2\right) $ and the s.a. boundary conditions
include boundary conditions at $\infty .$ This circumstance is crucial in
the sense that its neglecting leads to some ``paradox''. From the naive
standpoint, the situation where the stationary Schr\"{o}dinger equation $%
-\psi ^{\prime \prime }+V\psi =E\psi $ has only square-integrable solutions
for any real energy $E$, apparently implies that all the eigenstates in such
a potential are bound, and what is more, the discrete energy spectrum turns
out to be continuous, which is impossible. This situation is quite similar
to the case of a ``fall to the center'' for a particle of negative energy in
a strongly attractive potential\footnote{%
It can be respectively called a ``fall to infinity'' because a classical
particle escapes to infinity in a finite time.} $V\left( x\right) <-\frac{1}{%
4x^{2}}$ as $x\rightarrow 0$. The resolution of the paradox is in the
obligatory boundary conditions at infinity; without these boundary
conditions, we actually deal with the Hamiltonian $\hat{H}^{\ast }$ that is
non-s.a.. Only taking s.a. boundary conditions at infinity into account, we
get a s.a. Hamiltonian all the eigenstates of which are bound, but the
spectrum is really discrete.

We must also emphasize that the condition of the integrability of the
potential $V$ at the origin providing the regularity of the left end is also
crucial. The case where the potential is singular and nonintegrable at the
origin requires a special consideration.

The last remark concerns the Hamiltonian for a particle moving along the
real axis in the potential field $V$. If $V\left( x\right) $ is a locally
integrable function\footnote{%
That is, if $V\left( x\right) $ is integrable on any segment $\left[ a,b%
\right] $, $-\infty <a<b<\infty $.}, the Hamiltonian is defined as a s.a.
operator associated with the previous differential expression $\check{H}$ (%
\ref{3b.10}), but now on the whole real axis $\mathbb{R}^{1}=\left( -\infty
,+\infty \right) $ and with the both singular ends, $-\infty $ and $+\infty $%
. Let $\hat{H}^{\left( 0\right) }$ be the initial symmetric operator
associated with $\check{H}$. The crucial remark is that according to formula
(\ref{3b.64}), its deficiency indices $m_{+}=m_{-}=m$ are defined by the
deficiency indices $m_{+}^{\left( -\right) }=m_{-}^{\left( -\right)
}=m^{\left( -\right) }$ and $m_{+}^{\left( +\right) }=m_{-}^{\left( +\right)
}=m^{\left( +\right) }$ of the respective symmetric operators $\hat{H}%
_{-}^{\left( 0\right) }$ and $\hat{H}_{+}^{\left( 0\right) }$ associated
with the same differential expression $\check{H}$ restricted to the
respective negative semiaxis $\mathbb{R}_{-}^{1}=(-\infty ,0]$ and positive
semiaxis $\mathbb{R}_{+}^{1}=\left[ 0,+\infty \right) $:%
\begin{equation}
m=m^{\left( -\right) }+m^{\left( +\right) }-2\,.  \label{3b.127}
\end{equation}%
Let the potential $V$ satisfy one of the conditions that are the extensions
of conditions (\ref{3b.124}) and (\ref{3b.125}) to the whole real axis $%
\mathbb{R}^{1}$,%
\begin{equation}
1)\;V\left( x\right) \in L^{2}\left( -\infty ,+\infty \right) \,,
\label{3b.128}
\end{equation}%
i.e., $V$\ is\ square\ integrable\ on$\;\mathbb{R}^{1}\,,$ or%
\begin{equation}
2)\;V\left( x\right) >-Kx^{2}\,,\;K>0\,,  \label{3b.129}
\end{equation}%
for sufficiently large $\left| x\right| .$ Then the symmetric operator $%
\hat{H}_{+}^{\left( 0\right) }$ satisfies conditions (\ref{3b.124}) or (\ref%
{3b.125}), and therefore, its deficiency indices are $\left( 1,1\right) $,
i.e., $m^{\left( +\right) }=1$; the same is evidently true for the symmetric
operator $\hat{H}_{-}^{\left( 0\right) }$, it is sufficient to change the
variable $x\rightarrow -x$, i.e., $m^{\left( -\right) }=1$ also. It follows
by (\ref{3b.127}) that $m=1+1-2=0$, i.e., the deficiency indices of the
symmetric operator $\hat{H}^{\left( 0\right) }$ are $\left( 0,0\right) $.
This means that $\hat{H}^{\left( 0\right) }$ is essentially s.a., and its
unique s.a. extension is $\hat{H}=\hat{H}^{\ast }$.

We note that the same result follows from a consideration of the asymmetry
form $\omega _{\ast }$ for $\hat{H}^{\ast }$. According to (\ref{3b.44}) and
(\ref{3b.45}), it is given by%
\begin{equation}
\omega _{\ast }\left( \chi _{\ast },\psi _{\ast }\right) =\left. \left[ \chi
_{\ast },\psi _{\ast }\right] \right| _{-\infty }^{\infty }\,,\;\forall \chi
_{\ast },\psi _{\ast }\in D_{\ast }\,,  \label{3b.130}
\end{equation}%
where $\left[ \chi _{\ast },\psi _{\ast }\right] =-\overline{\chi _{\ast }}%
\psi _{\ast }^{\prime }+\overline{\chi _{\ast }^{\prime }}\psi _{\ast }$.
The crucial remark is then that the restrictions of the functions $\psi
_{\ast }\in D_{\ast }$ to the respective semiaxis $\mathbb{R}_{-}^{1}$ and $%
\mathbb{R}_{+}^{1}$ evidently belong to the domains of the respective
adjoints $\hat{H}_{-}^{\ast }=\left( \hat{H}_{-}^{\left( 0\right) }\right)
^{+}$ and $\hat{H}_{+}^{\ast }=\left( \hat{H}_{+}^{\left( 0\right) }\right)
^{+}$ and therefore $\left[ \chi _{\ast },\psi _{\ast }\right] $ have the
same boundary values at infinity as the respective boundary values for $%
\hat{H}_{-}^{\ast }$ and $\hat{H}_{+}^{\ast }$. But if the deficiency
indices of $\hat{H}_{-}^{\left( 0\right) }$ and $\hat{H}_{+}^{\left(
0\right) }$ are $\left( 1,1\right) $, the corresponding boundary values are
identically zero. It follows that $\omega _{\ast }$ (\ref{3b.130}) in this
case is identically zero as well, and the adjoint $\hat{H}^{\ast }$ is
symmetric, and therefore is s.a. We return to these arguments later where we
independently prove the vanishing of the boundary values $\left[ \chi _{\ast
},\psi _{\ast }\right] \left( \infty \right) $ for $\hat{H}_{+}^{\left(
0\right) }$.

The final conclusion is that under conditions (\ref{3b.128}) or (\ref{3b.129}%
), there is a unique s.a. operator $\hat{H}$ associated with a s.a.
differential expression $\check{H}$ (\ref{3b.10}) on the real axis $\mathbb{R%
}^{1}$ and defined on the natural domain:%
\begin{equation*}
\hat{H}:\left\{ 
\begin{array}{l}
D_{H}=\left\{ \psi \left( x\right) :\psi ,\,\psi ^{\prime }\;\mathrm{a.c.\,in%
}\;\mathbb{R}^{1};\;\psi ,\,-\psi ^{\prime \prime }+V\psi \in L^{2}\left(
-\infty ,+\infty \right) \right\} \,, \\ 
\hat{H}\psi =-\psi ^{\prime \prime }+V\psi \,.%
\end{array}%
\right.
\end{equation*}

This fact is implicitly adopted in the majority of textbooks on quantum
mechanics for physicists and considered an unquestionable common place. In
particular, it concerns the one-dimensional Hamiltonians with bounded
potentials like a potential barrier, a finite well, a solvable potentials
like $V_{0}\mathrm{ch}^{-2}\left( ax\right) \mathrm{\,}$, the Hamiltonians
with growing potentials, for example, the Hamiltonian for a harmonic
oscillator where $\check{H}=-d^{2}/dx^{2}+x^{2}$, and even the Hamiltonians
with linear potential $V=kx$, which goes to $-\infty $ at one of the ends,
but only linearly, not faster than quadratically.

As to the harmonic oscillator Hamiltonian, it follows from the
Akhiezer--Glazman theorem (Theorem \ref{AGT}) that its standard
representation $\hat{H}=\hat{a}^{+}\hat{a}+1$ implies that $\hat{a}$ is the
closed operator associated with the non-s.a. differential expression $\check{%
a}=d/dx+x$ and defined by%
\begin{equation*}
\hat{a}:\left\{ 
\begin{array}{l}
D_{a}=\left\{ \psi \left( x\right) :\psi \;\text{\textrm{a.c.\thinspace in} }%
\left( -\infty ,+\infty \right) ;\;\psi ,\left( d/dx+x\right) \psi \in
L^{2}\left( -\infty ,+\infty \right) \right\} , \\ 
\hat{a}\psi =\left( d/dx+x\right) \psi \,,%
\end{array}%
\right.
\end{equation*}%
while $\hat{a}^{+}$ is its adjoint, it is the operator associated with the
non-s.a. differential expression $\check{a}^{+}=-d/dx+x$ and defined by%
\begin{equation*}
\hat{a}^{+}:\left\{ 
\begin{array}{l}
D_{a^{+}}=\left\{ \psi _{\ast }\left( x\right) :\psi _{\ast }\;\text{\textrm{%
a.c.\thinspace in} }\left( -\infty ,+\infty \right) ;\;\psi _{\ast },\left(
-d/dx+x\right) \psi _{\ast }\in L^{2}\left( -\infty ,+\infty \right)
\right\} , \\ 
\hat{a}^{+}\psi _{\ast }=\left( -d/dx+x\right) \psi _{\ast }\,.%
\end{array}%
\right.
\end{equation*}%
These subtle points are usually omitted in the physical literature. To be
true, they are irrelevant for finding the eigenfunctions of $\hat{H}$
because the latter are smooth functions exponentially vanishing at infinity.

The other remarks on the physical applicability of conditions (\ref{3b.124})
and (\ref{3b.125}) are naturally and practically literally extended to
conditions (\ref{3b.128}) and (\ref{3b.129}). In particular, if condition (%
\ref{3b.129}) is violated, and, for example, $V\left( x\right) <-Kx^{2\left(
1+\varepsilon \right) }$, $\varepsilon >0$, as $x\rightarrow -\infty $
or/and $x\rightarrow \infty $, we respectively have $m^{\left( -\right) }=2$
or/and $m^{\left( +\right) }=2$ and consequently $m=1$ or $m=2$. In this
case, we have the respective one-parameter $U\left( 1\right) $-family or
four-parameter $U\left( 2\right) $-family of s.a. Hamiltonians $\left\{ 
\hat{H}_{U}\right\} $, $U\in U\left( 1\right) $ or $U\in U\left( 2\right) $,
that are specified by some s.a. boundary conditions at infinity, $x=-\infty $
or/and $x=\infty $. To be true, such potentials are considered apparently
nonphysical at present (unless they emerge in some exotic cosmological
scenarios).

\subsection{Alternative way of specifying self-adjoint differential
operators in terms of explicit self-adjoint boundary conditions}

The description of s.a. extensions of symmetric differential operators in
terms of s.a. boundary conditions due to the above presented conventional
methods is sometimes of an inexplicit character, especially for the case of
singular ends, such that the $U\left( m\right) $ nature of the whole family
of s.a. extensions is not evident.

We now discuss a possible alternative way of specifying s.a. differential
operators associated with a given s.a. differential expression in terms of
explicit, in general asymptotic, s.a. boundary conditions, the $U\left(
m\right) $ nature of this specification is evident. The idea of the method
is a result of two observations. The both equally concerns the asymmetry
forms $\omega _{\ast }$ and $\Delta _{\ast }$. For definiteness, we speak
about the quadratic asymmetry form $\Delta _{\ast }$, although the all to be
said applies to $\omega _{\ast }$:\ we recall that $\Delta _{\ast }$ and $%
\omega _{\ast }$ define each other.

For the first observation, we return to the previous section, but use the
notation adopted in this section for differential operators where the
elements of the Hilbert space $L^{2}\left( a,b\right) $ are denoted by $\psi 
$ with an appropriate subscript, the closure of the initial symmetric
operator $\hat{f}^{\left( 0\right) }$ is denoted by $\hat{f}$, $\overline{%
\hat{f}^{\left( 0\right) }}=\hat{f}$, the deficient subspaces are denoted by 
$D_{+}$ and $D_{-}$ with $z=i\kappa ,$ and etc.

By first von Neumann formula (\ref{3a.4}), any $\psi _{\ast }\in D_{\ast }$
is uniquely represented as%
\begin{equation*}
\psi _{\ast }=\psi +\psi _{+}+\psi _{-}\,,\;\psi \in D_{f}\,,\;\psi _{+}\in
D_{+}\,,\;\psi _{-}\in D_{-}\,.
\end{equation*}%
By von Neumann formula (\ref{3a.18}), the asymmetry form $\Delta _{\ast }$
is nontrivial only on the direct sum $D_{+}+D_{-}$ of the deficient
subspaces and expressed in terms of $D_{+}$ and $D_{-}$ components of $\psi
_{\ast }$ as%
\begin{equation*}
\Delta _{\ast }\left( \psi _{\ast }\right) =2i\kappa \left( \left| \left|
\psi _{+}\right| \right| ^{2}-\left| \left| \psi _{-}\right| \right|
^{2}\right) \,.
\end{equation*}%
Let $\left\{ e_{+,k}\right\} _{1}^{m_{+}}$ and $\left\{ e_{-,k}\right\}
_{1}^{m-}$ be some orthobasises in the respective $D_{+}$ and $D_{-}$ such
that%
\begin{equation*}
\psi _{+}=\sum_{k=1}^{m_{+}}c_{+,k}e_{+,k}\,,\;\psi
_{-}=\sum_{k=1}^{m_{-}}c_{-,k}e_{-,k}\,,
\end{equation*}%
where $c_{\pm ,k}$ are the respective expansion coefficients, then the
asymmetry form $\Delta _{\ast }$ becomes%
\begin{equation}
\Delta _{\ast }\left( \psi _{\ast }\right) =2i\kappa \left(
\sum_{k=1}^{m_{+}}\left| c_{+,k}\right| ^{2}-\sum_{k=1}^{m_{-}}\left|
c_{-,k}\right| ^{2}\right) \,.  \label{3b.133.4}
\end{equation}

The problem of symmetric and s.a. extensions of the initial symmetric
operator $\hat{f}^{\left( 0\right) }$ can be considered in terms of the
expansion coefficients. The deficient subspaces $D_{+}$ and $D_{-}$ reveals
itself as the respective complex linear spaces $\mathbb{C}_{+}^{m_{+}}$ of
the $m_{+}$-columns $\left\{ c_{+,k}\right\} _{1}^{m_{+}}$ and $\mathbb{C}%
_{-}^{m_{-}}$ of the $m_{-}$-columns $\left\{ c_{-,k}\right\} _{1}^{m_{-}}$.
The quadratic form $\frac{1}{i}\Delta _{\ast }$ becomes a Hermitian diagonal
form, canonical up to the factor $2\kappa $, in the complex linear space $%
\mathbb{C}^{m_{+}+m_{-}}$ that is a direct sum of $\mathbb{C}_{+}^{m_{+}}$
and $\mathbb{C}_{-}^{m_{-}}$, $\mathbb{C}^{m_{+}+m_{-}}=\mathbb{C}%
_{+}^{m_{+}}+\mathbb{C}_{-}^{m_{-}}$, giving contributions to $\frac{1}{i}%
\Delta _{\ast }$ of the opposite signs. The deficiency indices $m_{+}$ and $%
m_{-}$ define the signature of this quadratic form $\mathrm{sign\,}\frac{1}{i%
}\Delta _{\ast }=\left( m_{+},m_{-}\right) \,,$ being its inertia indices.
In this terms, we can repeat all the arguments of the previous section
leading to the main theorem with the same conclusions. We repeat them in the
end of the present consideration in new terms.

We now note that we can choose an arbitrary mixed basis $\left\{
e_{k}\right\} _{1}^{m_{+}+m_{-}}$ in the direct sum $D_{+}+D_{-}$ such that%
\begin{equation*}
\psi _{+}+\psi _{-}=\sum_{k=1}^{m_{+}+m_{-}}c_{k}e_{k}
\end{equation*}%
which respectively changes the basis in $\mathbb{C}^{m_{+}+m_{-}},$ and the
form $\Delta _{\ast }$ becomes%
\begin{equation}
\Delta _{\ast }\left( \psi _{\ast }\right) =2i\kappa \sum_{k=1}^{m_{+}+m_{-}}%
\overline{c_{k}}\omega _{kl}c_{l}\,,\;\omega _{kl}=\overline{\omega }_{kl}\,,
\label{3b.133.7}
\end{equation}%
such that $\frac{1}{i}\Delta _{\ast }$ becomes the general Hermitian
quadratic form, of course, with the same signature. We then diagonalize this
form and repeat the above arguments with the known conclusions.

To be true, the second observation includes a suggestion. We know that in
the case of differential operators, the asymmetry form $\Delta _{\ast }$ is
determined by the finite boundary values of the local form $\left[ \psi
_{\ast },\psi _{\ast }\right] $ that is a form in terms of $\psi _{\ast
}\left( x\right) $ and its derivatives, see (\ref{3b.18}) and (\ref{3b.23}), 
\begin{eqnarray*}
&&\Delta _{\ast }\left( \psi _{\ast }\right) =\left[ \psi _{\ast },\psi
_{\ast }\right] \left( b\right) -\left[ \psi _{\ast },\psi _{\ast }\right]
\left( a\right) \,, \\
&&\left[ \psi _{\ast },\psi _{\ast }\right] \left( a\right) =\underset{%
x\rightarrow a}{\lim }\left[ \psi _{\ast },\psi _{\ast }\right] \left(
x\right) \,,\;\left[ \psi _{\ast },\psi _{\ast }\right] \left( b\right) =%
\underset{x\rightarrow b}{\lim }\left[ \psi _{\ast },\psi _{\ast }\right]
\left( x\right)
\end{eqnarray*}%
we repeat (\ref{3b.46}) and (\ref{3b.47}). For brevity, we call $\left[ \psi
_{\ast },\psi _{\ast }\right] \left( a\right) $ and $\left[ \psi _{\ast
},\psi _{\ast }\right] \left( b\right) $ the boundary forms. We certainly
know that the boundary form at a regular end is a finite nonzero form of
order $n$ with respect to finite boundary values of functions and their
derivatives of order up to $n-1$ for a differential expression $\check{f}$
of order $n$. For a singular end, the evaluation of the respective boundary
form is generally nontrivial. The suggestion is that the boundary form is
expressed in terms of finite number coefficients in front of generally
divergent or infinitely oscillating leading asymptotic terms of functions
and their derivatives at the end. Therefore, in the general case, boundary
forms are expressed in terms of boundary values and the the coefficients
describing the asymptotic boundary behavior of functions. For brevity, we
call the whole set of the relevant boundary values and the above-introduced
coefficients the abv-coefficients (asymptotic boundary value coefficients).
Let the $p$-column $\left\{ c_{k}\right\} _{1}^{p}$ denote the
abv-coefficients for $\psi _{\ast }\in D_{\ast }.$These columns form a
complex linear space $\mathbb{C}^{p}$, and $\Delta _{\ast }$ is a finite
quadratic anti-Hermitian form in this space%
\begin{equation}
\Delta _{\ast }\left( \psi _{\ast }\right) =2i\kappa \sum_{k=1}^{p}\overline{%
c}_{k}\omega _{kl}c_{l}\,,\;\omega _{kl}=\overline{\omega }_{kl}.
\label{3b.133.8}
\end{equation}%
It is now sufficient to compare (\ref{3b.133.8}) with (\ref{3b.133.7}) and
repeat the above consideration with the known conclusions on the possibility
of s.a. extensions of $\hat{f}^{\left( 0\right) }$ and their specification
in terms of the abv-coefficients by passing to linear combinations $\left\{
c_{+,k}\right\} _{1}^{m_{+}}$ and$\;\left\{ c_{-,k}\right\} _{1}^{m_{-}}$, $%
p=m_{+}+m_{-}$, diagonalizing form (\ref{3b.133.8}). We call them the
diagonal abv-coefficients. All the just said is quite natural. Of course,
the nonzero contributions to $\Delta _{\ast }$ are due to the deficient
subspaces, but only the abv-coefficients of functions in $D_{+}+D_{-}$ are
relevant, the deficiency indices are evidently identified with the signature
of the form $\frac{1}{i}\Delta _{\ast }$, and the isometries $\hat{U}%
:D_{+}\rightarrow D_{-}$ reveal themselves as isometries of one set of
diagonal boundary values, for example, $\left\{ c_{+,k}\right\} _{1}^{m_{+}}$
to another set $\left\{ c_{-,k}\right\} _{1}^{m_{-}}$. We formulate the
conclusions in terms of abv-coefficients in the end of our consideration.

The alternative method is a result of obviously joining the two
observations. We outline the consecutive steps of the method for a
differential expression $\check{f}$ of order $n$.

The first step is evaluating the behavior of functions $\psi _{\ast }\left(
x\right) \in D_{\ast }$ near the singular ends and either proving that the
respective boundary forms vanish identically by establishing the asymptotic
behavior of functions at the ends or establishing the asymptotic terms that
give nonzero contributions to the respective boundary forms. Unfortunately,
there is no universal recipe for performing the both procedures at present.
We only give some instructive examples below. As we already said above, the
result must be a representation (\ref{3b.133.8}) of $\Delta _{\ast }$ in
terms of abv-coefficients $\left\{ c_{k}\right\} _{1}^{p}$.

The next step consists in diagonalizing the obtained form (\ref{3b.133.8}),
i.e., diagonalizing the Hermitian matrix $\omega $. As a result, $\Delta
_{\ast }$ becomes a diagonal quadratic form (\ref{3b.133.4}) in terms of
diagonal abv-coefficients, $\left\{ c_{+,k}\right\} ^{m_{+}}$ and $\left\{
c_{-,k}\right\} ^{m_{-}}$, $m_{+}+m_{-}=p$. The resulting conclusions are
actually a repetition of the main theorem in the case of finite deficiency
indices. Namely, if the inertia indices $m_{+}$ and $m_{-}$ of form (\ref%
{3b.133.4}) are different, $m_{+}\neq m_{-}$, there is no s.a. operators
associated with a given s.a. differential expression $\check{f}$. If $%
m_{+}=m_{-}=0$, i.e., if $\Delta _{\ast }=0$, the initial symmetric operator 
$\hat{f}^{\left( 0\right) }$ is essentially s.a., and there is a unique s.a.
operator associated with $\check{f}$ that is given by the closure $\hat{f}$
of $\hat{f}^{\left( 0\right) }$ coinciding with the adjoint $\hat{f}^{\ast }:%
\hat{f}=\hat{f}^{+}=\hat{f}^{\ast }$. If $m_{+}=m_{-}=m>0$, there is an $%
m^{2}$-parameter $U\left( m\right) $-family $\left\{ \hat{f}_{U}\right\}
,\;U\in U\left( m\right) ,$ of s.a. operators associated with $\check{f}$.
Any s.a. $\hat{f}_{U}$ is specified by s.a. boundary conditions defined by a
unitary $m\times m$ matrix $U$ relating the diagonal boundary values $%
\left\{ c_{+,k}\right\} _{1}^{m}$ and $\left\{ c_{-,k}\right\} _{1}^{m}$ and
given by%
\begin{equation}
c_{-,k}=U_{kl}c_{+,k}\,,\;k=1,\ldots ,m\,.  \label{e.1}
\end{equation}%
In the case of singular ends, these boundary conditions have a form of
asymptotic boundary conditions prescribing the asymptotic form of functions $%
\psi _{U}\in D_{f_{U}}$ at the singular ends.

A comparative advantage of the method is that it avoids explicitly
evaluating the deficient subspaces and deficiency indices, the deficiency
indices are obtained by passing. Unfortunately, it is not universal because
at present we don't know a universal method for evaluating the asymptotic
behavior of functions in $D_{\ast }$ at singular ends.

We now consider possible applications of the proposed alternative method.

We first show in detail , maybe superfluous, how simply the problem of s.a.
differential expression $\check{p}$ (\ref{3b.9}) on an interval $\left(
a,b\right) $ is solved by the alternative method. We recall that the
illustration of the conventional methods by the example of $\check{p}$
presented at the end of the previous section was rather extensive. In this
case, $\left[ \psi _{\ast },\psi _{\ast }\right] =-i\left| \psi _{\ast
}\right| ^{2},$ see (\ref{3a.43}), therefore, the quadratic asymmetry form $%
\Delta _{\ast }$ is $\Delta _{\ast }\left( \psi _{\ast }\right) =-i\left|
\psi _{\ast }\left( b\right) \right| ^{2}+i\left| \psi _{\ast }\left(
a\right) \right| ^{2}\,,$ and $\psi _{\ast }\in D_{\ast }$ implies $\psi
_{\ast },\psi _{\ast }^{\prime }\in L^{2}\left( a,b\right) $.

Let $\left( a,b\right) =\left( -\infty ,\infty \right) $, the whole real
axis. The finiteness of the boundary form $\left[ \psi _{\ast },\psi _{\ast }%
\right] \left( \infty \right) $ means that $\left| \psi _{\ast }\right|
^{2}\rightarrow C\left( \psi _{\ast }\right) $,\ $x\rightarrow \infty
\,,\;\left| C\left( \psi _{\ast }\right) \right| <\infty ,$ where $C\left(
\psi _{\ast }\right) $ is a finite constant. But this constant must be zero,
because $C\left( \psi _{\ast }\right) \neq 0$ contradicts the square
intergability of $\psi _{\ast }$. It is easy to see that for the validity of
this conclusion, $\psi _{\ast }\rightarrow 0$ as $x\rightarrow \infty $, it
is sufficient that $\psi _{\ast }$ be square integrable at infinity together
with its derivative $\psi _{\ast }^{\prime }$; actually, we repeat the
well-known assertion that if the both $\psi _{\ast }$ and $\psi _{\ast
}^{\prime }$ are square integrable at infinity, then $\psi _{\ast }$
vanishes at infinity. Similarly, we prove that $\psi _{\ast }\rightarrow 0$
as $x\rightarrow -\infty $ and therefore $\left[ \psi _{\ast },\psi _{\ast }%
\right] \left( -\infty \right) =0$ also for any $\psi _{\ast }\in D_{\ast }$%
. We finally have that $\Delta _{\ast }\left( \psi _{\ast }\right) \equiv 0$%
, in particular, $\mathrm{sign\,}\frac{1}{i}\Delta _{\ast }=\left(
0,0\right) $. This means that there is a unique s.a. operator $\hat{p}$
associated with $\check{p}$ on the real axis and given by (\ref{3a.52}),
which is in a complete agreement with the known fact established here in
passing that the deficiency indices of the initial symmetric operator $\hat{p%
}^{\left( 0\right) }$ are $\left( 0,0\right) $ and therefore, $\hat{p}%
^{\left( 0\right) }$ is essentially s.a. and $\hat{p}=\overline{\hat{p}%
^{\left( 0\right) }}=\hat{p}^{\ast }$.

Let $\left( a,b\right) =\left[ 0,\infty \right) $. By the previous
arguments, we have $\left[ \psi _{\ast },\psi _{\ast }\right] \left( \infty
\right) =0$, while $\left[ \psi _{\ast },\psi _{\ast }\right] \left(
0\right) =-i\left| \psi \left( 0\right) \right| ^{2}\neq 0$ in general.
Consequently, the Hermitian quadratic form $\frac{1}{i}\Delta _{\ast }\left(
\psi _{\ast }\right) =\left| \psi _{\ast }\left( 0\right) \right| ^{2}$ is
positive definite and $\mathrm{sign\,}\frac{1}{i}\Delta _{\ast }=\left(
1,0\right) $. This means that there is no s.a. operators associated with $%
\check{p}$ on a semiaxis, which is in complete agreement with the known fact
that the deficiency indices of $\hat{p}^{\left( 0\right) }$ in this case are 
$\left( 1,0\right) $.

Let $\left( a,b\right) =\left[ 0,l\right] $, a finite segment. In this case,
we have $\frac{1}{i}\Delta _{\ast }=\left| \psi _{\ast }\left( 0\right)
\right| ^{2}-\left| \psi _{\ast }\left( l\right) \right| ^{2}\,,$ a
nontrivial Hermitian quadratic form with $\mathrm{sign\,}\frac{1}{i}\Delta
_{\ast }=\left( 1,1\right) $, which confirms the known fact that the
deficiency indices of $\hat{p}^{\left( 0\right) }$ in this case are $\left(
1,1\right) $. The corresponding s.a. boundary conditions are%
\begin{equation*}
\psi \left( l\right) =e^{i\vartheta }\psi \left( 0\right) \,,\;0\leq
\vartheta \leq 2\pi \,,
\end{equation*}%
they define the one-parameter $U\left( 1\right) $-family $\left\{ \hat{p}%
_{\vartheta }\right\} $ of s.a. operators associated with $\check{p}$ on a
segment $\left[ 0,l\right] $, the family given by (\ref{3a.57}).

The case of an even s.a. differential expression with the both regular ends
is completely fall into the framework of the alternative method. Let $\check{%
f}$ be an even s.a. differential expression of order $n$ on a finite
interval $\left( a,b\right) ,$ the both ends being regular. In this case, we
have representation (\ref{3b.114}) for the sesquilinear asymmetry form $%
\omega _{\ast }$\thinspace , while the quadratic asymmetry form $\Delta
_{\ast }$ is represented as%
\begin{equation}
\Delta _{\ast }\left( \psi _{\ast }\right) =\Psi _{\ast }^{+}\left( b\right) 
\mathcal{E}\Psi _{\ast }\left( b\right) -\Psi _{\ast }^{+}\left( a\right) 
\mathcal{E}\Psi _{\ast }\left( a\right) \,,  \label{3b.138}
\end{equation}%
where the matrix $\mathcal{E}$ is given by (\ref{3b.81}) and $\Psi _{\ast
}\left( b\right) ,\,\Psi _{\ast }\left( b\right) $ are the columns whose
components are the respective boundary values of functions $\psi _{\ast }\in
D_{\ast }$ and their (quasi)derivatives of order up to $n-1,$%
\begin{equation}
\Psi _{\ast }\left( a\right) =\left( 
\begin{array}{c}
\psi _{\ast }\left( a\right) \\ 
\psi _{\ast }^{[1]}\left( a\right) \\ 
\vdots \\ 
\psi _{\ast }^{[n-1]}\left( a\right)%
\end{array}%
\right) ,\;\Psi _{\ast }\left( b\right) =\left( 
\begin{array}{c}
\psi _{\ast }\left( b\right) \\ 
\psi _{\ast }^{[1]}\left( b\right) \\ 
\vdots \\ 
\psi _{\ast }^{[n-1]}\left( b\right)%
\end{array}%
\right) ,  \label{e.2}
\end{equation}%
or $\Psi _{\ast k}\left( a\right) =\psi _{\ast }^{[k-1]}\left( a\right)
\,,\;\Psi _{\ast k}\left( b\right) =\psi _{\ast }^{[k-1]}\left( b\right) \,,$
$k=1,...,n\,,$ an analogue of (\ref{3b.84}).

An important preliminary remark concerning dimensional considerations is in
order here. In the mathematical literature, the variable $x$ is considered
dimensionless, such that $\psi _{\ast },\psi _{\ast }^{[1]},...,\psi _{\ast
}^{[n-1]}$ have the same zero dimension as well as the differential
expression $\check{f}$ itself. Therefore, comparing (\ref{3b.138}) with (\ref%
{3b.133.8}), where $p=2n$ and $\kappa =1,${\Large \ }as it is conventionally
adopted in the mathematical literature, we could immediately identify the
set $\left\{ c_{k}\right\} _{1}^{2n}$ with the set $\left\{ \psi _{\ast
}^{[k-1]}\left( a\right) \right\} _{1}^{n}\cup \left\{ \psi _{\ast
}^{[k-1]}\left( b\right) \right\} _{1}^{n}\,,$ the matrix $\omega $ is then
given by%
\begin{equation}
\omega =\frac{1}{2i}\left( 
\begin{array}{cc}
-\mathcal{E} & 0 \\ 
0 & \mathcal{E}%
\end{array}%
\right) \,.  \label{e.3}
\end{equation}

But in physics, the variable $x$ is usually assigned a certain dimension,
the dimension of length, which we write as $\left[ x\right] =\left[ \mathrm{%
length}\right] ,$ while functions $\psi _{\ast }$ have dimension of the
square root of inverse length, $\left[ \psi _{\ast }\right] =\left[ \mathrm{%
length}\right] ^{-1/2}.$ Therefore, $\psi _{\ast }^{[k]}\left( x\right) $
has the dimension $\left[ \psi _{\ast }^{[k]}\right] =\left[ \mathrm{length}%
\right] ^{-k-1/2},$ and if the coefficient function $f_{n}\left( x\right) $
in $\check{f}$ is taken dimensionless, the $\check{f}$ itself is assigned
the dimension $\left[ \check{f}\right] =\left[ \mathrm{length}\right] ^{-n}$%
. It is convenient to have all variables $c_{k}$, $k=1,\ldots ,2n$, in (\ref%
{3b.133.8}) of equal dimension in order the matrix elements of the unitary
matrix $U$ in (\ref{e.1}) be dimensionless. This can be done as follows.

We introduce arbitrary, but fixed, parameter $\tau $ of dimension of length, 
$\left[ \tau \right] =\left[ \mathrm{length}\right] $, and represent $\Delta
_{\ast }\left( \psi _{\ast }\right) $ as\footnote{%
The dimension of $\Delta _{\ast }$ is $[\Delta _{\ast }]=[\mathrm{length}%
]^{-n}\,.$}%
\begin{equation}
\Delta _{\ast }\left( \psi _{\ast }\right) =\tau ^{-n+1}\left[ \Psi _{\tau
}^{+}\left( b\right) \mathcal{E}\Psi _{\tau }\left( b\right) -\Psi _{\tau
}^{+}\left( a\right) \mathcal{E}\Psi _{\tau }\left( a\right) \right] \,,
\label{e.4}
\end{equation}%
where%
\begin{equation*}
\Psi _{\tau }\left( a\right) =\left( 
\begin{array}{c}
\psi _{\ast }\left( a\right) \\ 
\tau \psi _{\ast }^{\left[ 1\right] }\left( a\right) \\ 
\vdots \\ 
\tau ^{n-1}\psi _{\ast }^{\left[ n-1\right] }\left( a\right)%
\end{array}%
\right) \,,\;\Psi _{\tau }\left( b\right) =\left( 
\begin{array}{c}
\psi _{\ast }\left( b\right) \\ 
\tau \psi _{\ast }^{\left[ 1\right] }\left( b\right) \\ 
\vdots \\ 
\tau ^{n-1}\psi _{\ast }^{\left[ n-1\right] }\left( b\right)%
\end{array}%
\right)
\end{equation*}%
or in components%
\begin{equation*}
\Psi _{\tau k}\left( a\right) =\tau ^{k-1}\psi _{\ast }^{\left[ k-1\right]
}\left( a\right) \,,\;\Psi _{\tau k}\left( b\right) =\tau ^{k-1}\psi _{\ast
}^{\left[ k-1\right] }\left( b\right) \,,\;k=1,\ldots ,n\,,
\end{equation*}%
the dimension of $\Psi _{\tau }\left( a\right) $ and $\Psi _{\tau }\left(
b\right) $ is $\left[ \Psi _{\tau }\right] =\left[ \mathrm{length}\right]
^{-1/2}$.

We can now identify the set $\left\{ c_{k}\right\} _{1}^{2n}$ with $\Psi
_{\tau }\left( a\right) \cup \Psi _{\tau }\left( b\right) $ and proceed to
diagonalizing quadratic form (\ref{e.4}) or matrix $\omega $ (\ref{e.3}).
Diagonalizing is evidently reduced to separately diagonalizing the quadratic
form $\Psi _{\tau }^{+}\left( a\right) \mathcal{E}\Psi _{\tau }\left(
a\right) $ and $\Psi _{\tau }^{+}\left( b\right) \mathcal{E}\Psi _{\tau
}\left( b\right) $ or to diagonalizing the matrix $\mathcal{E}$. But this
was already done above, see formulas (\ref{3b.112}), (\ref{e.5}), and (\ref%
{e.6}). The final result is%
\begin{equation}
\Delta _{\ast }\left( \psi _{\ast }\right) =2i\kappa \left[ \Psi _{\tau
\left( +-\right) }^{+}\Psi _{\tau \left( +-\right) }-\Psi _{\tau \left(
-+\right) }^{+}\Psi _{\tau \left( -+\right) }\right] \,,  \label{e.7}
\end{equation}%
where $\kappa =1/4\tau ^{-n+1}$ and $\Psi _{\tau \left( +-\right) }$, $\Psi
_{\tau \left( -+\right) }$ are the $n$-columns%
\begin{equation*}
\Psi _{\tau \left( +-\right) }=\left( 
\begin{array}{c}
\Psi _{\tau +}\left( b\right) \\ 
\Psi _{\tau -}\left( a\right)%
\end{array}%
\right) \,,\;\Psi _{\tau \left( -+\right) }=\left( 
\begin{array}{c}
\Psi _{\tau -}\left( b\right) \\ 
\Psi _{\tau +}\left( a\right)%
\end{array}%
\right) \,,
\end{equation*}%
where $\Psi _{\tau \pm }\left( a\right) $, are the $n/2$-columns 
\begin{eqnarray}
\Psi _{\tau +}\left( a\right) &=&\left( 
\begin{array}{l}
\psi _{\ast }\left( a\right) +i\tau ^{n-1}\psi _{\ast }^{\left[ n-1\right]
}\left( a\right) \\ 
\tau \psi _{\ast }^{\left[ 1\right] }\left( a\right) +i\tau ^{\left[ n-2%
\right] }\psi _{\ast }^{\left[ n-2\right] }\left( a\right) \\ 
\vdots \\ 
\tau ^{n/2-1}\psi _{\ast }^{\left[ n/2-1\right] }\left( a\right) +i\tau
^{n/2}\psi _{\ast }^{\left[ n/2\right] }\left( a\right)%
\end{array}%
\right) \,,  \label{3b.145} \\
\Psi _{\tau -}\left( a\right) &=&\left( 
\begin{array}{l}
\tau ^{n/2-1}\psi _{\ast }^{\left[ n/2-1\right] }\left( a\right) -i\tau
^{n/2}\psi _{\ast }^{\left[ n/2\right] }\left( a\right) \\ 
\vdots \\ 
\tau \psi _{\ast }^{\left[ 1\right] }\left( a\right) -i\tau ^{n-2}\psi
_{\ast }^{\left[ n-2\right] }\left( a\right) \\ 
\psi _{\ast }\left( a\right) -i\tau ^{n-1}\psi _{\ast }^{\left[ n-1\right]
}\left( a\right)%
\end{array}%
\right) \,,  \label{3b.146}
\end{eqnarray}%
or in components 
\begin{eqnarray}
&&\Psi _{\tau +,k}\left( a\right) =\tau ^{k-1}\psi _{\ast }^{\left[ k-1%
\right] }\left( a\right) +i\tau ^{n-k}\psi _{\ast }^{\left[ n-k\right]
}\left( a\right) \,,\;k=1,\ldots ,n/2\,,  \label{3b.147} \\
&&\Psi _{\tau -,k}\left( a\right) =\tau ^{n/2-k}\psi _{\ast }^{\left[ n/2-k%
\right] }\left( a\right) -i\tau ^{n/2+k-1}\psi _{\ast }^{\left[ n/2+k-1%
\right] }\left( a\right) \,,\;k=1,\ldots ,n/2\,.  \label{3b.148}
\end{eqnarray}%
We note that $\Psi _{\tau -,k}\left( a\right) $ are obtained from $\Psi
_{\tau +,k}\left( a\right) $ by the change $i\rightarrow -i,$ and $%
k\rightarrow n/2+1-k.$ The $n/2$-columns $\Psi _{\tau \pm ,k}\left( b\right) 
$ are given by similar formulas with the change $a\rightarrow b$. In other
words, the components of the $n$-columns $\Psi _{\tau \left( +-\right) }$
and $\Psi _{\tau \left( -+\right) }$ are respectively given by%
\begin{equation*}
\Psi _{\tau \left( +-\right) k}=\left\{ 
\begin{array}{l}
\tau ^{k-1}\psi _{\ast }^{\left[ k-1\right] }\left( b\right) +i\tau
^{n-k}\psi _{\ast }^{\left[ n-k\right] }\left( b\right) \,,\;k=1,\ldots
,n/2\,, \\ 
\tau ^{n-k}\psi _{\ast }^{\left[ n-k\right] }\left( a\right) -i\tau
^{k-1}\psi _{\ast }^{\left[ k-1\right] }\left( a\right)
\,,\;\,,\;k=n/2+1,\ldots ,n\,,%
\end{array}%
\right.
\end{equation*}%
and%
\begin{equation*}
\Psi _{\tau \left( -+\right) k}=\left\{ 
\begin{array}{l}
\tau ^{n/2-k}\psi _{\ast }^{\left[ n/2-k\right] }\left( b\right) -i\tau
^{n/2+k-1}\psi _{\ast }^{\left[ n/2+k-1\right] }\left( b\right)
\,,\;k=1,\ldots ,n/2\,, \\ 
\tau ^{k-n/2-1}\psi _{\ast }^{\left[ k-n/2-1\right] }\left( a\right) +i\tau
^{3n/2-k}\psi _{\ast }^{\left[ 3n/2-k\right] }\left( a\right)
\,,\;\,,\;k=n/2+1,\ldots ,n\,.%
\end{array}%
\right.
\end{equation*}%
It follows from (\ref{e.7}) that the s.a. boundary conditions defining a
s.a. operator $\hat{f}_{U}$ associated with $\check{f}$ are given by%
\begin{equation}
\Psi _{\tau \left( -+\right) }=U\Psi _{\tau \left( +-\right) }\,,
\label{e.8}
\end{equation}%
where $U$ is an $n\times n$ unitary matrix, $U\in U\left( n\right) $. When $%
U $ ranges over all $U\left( n\right) $ group, we cover the whole $n^{2}$%
-parameter $U\left( n\right) $-family $\left\{ \hat{f}_{U}\right\} $ of s.a.
operators associated with a given s.a. deferential expression $\check{f}$ or
order $n$ on a finite interval $\left( a,b\right) $ with the both regular
ends.

We conclude this item with some evident remarks.

1)\ We use the same symbol $\hat{f}_{U}$ for the notation of s.a. extensions
as before, although the subscript $U$ has now another meaning. In the
previous context, the subscript $U$ was a symbol of a an isometry $\hat{U}%
:D_{+}\rightarrow D_{-}$, in the present context, it is a symbol of a
unitary mapping (\ref{e.8}) of one set of boundary values to another one.

2)\ We could organize the column $\Psi _{\tau \left( -+\right) }$ in another
way, for example,%
\begin{equation*}
\Psi _{\tau \left( -+\right) }=\left( 
\begin{array}{c}
\Psi _{\tau -}\left( b\right) \\ 
\Psi _{\tau +}\left( a\right)%
\end{array}%
\right) \rightarrow \Xi \Psi _{\tau \left( -+\right) }=\left( 
\begin{array}{c}
\Psi _{\tau +}\left( a\right) \\ 
\Psi _{\tau -}\left( b\right)%
\end{array}%
\right) ,
\end{equation*}
where the unitary matrix $\Xi $ is $\Xi =\left( 
\begin{array}{cc}
0 & I \\ 
I & 0%
\end{array}%
\right) $, here, $I$ is the $n/2\times n/2$ unit matrix, $\Xi ^{2}=I$. Then $%
U$ in (\ref{e.8}) would change to $\Xi U$, which is also a unitary matrix.

3)\ It is evident that we can specify s.a. boundary conditions by $\Psi
_{\tau \left( +-\right) }=U\Psi _{\tau \left( -+\right) }\,.$ It is
sufficient to make the change $U\rightarrow U^{-1}$ in (\ref{e.8}).

4)\ If a matrix $U$ in (\ref{e.8}) is of a specific block-diagonal form%
\begin{equation}
U=\left( 
\begin{array}{cc}
U\left( b\right) & 0 \\ 
0 & U^{-1}\left( a\right)%
\end{array}%
\right) \,,  \label{3b.153}
\end{equation}%
where $U\left( a\right) $ and $U\left( b\right) $ are $n/2\times n/2$
unitary matrices\footnote{%
For convenience, we take the down right block in r.h.s. of (\ref{3b.153}) in
the form $U^{-1}\left( a\right) $ rather than $U\left( a\right) $; see below
(\ref{3b.154}).}, we obtain the so-called splitted s.a. boundary conditions%
\begin{equation}
\Psi _{\tau -}\left( a\right) =U\left( a\right) \Psi _{\tau +}\left(
a\right) \,,\;\Psi _{\tau -}\left( b\right) =U\left( b\right) \Psi _{\tau
+}\left( b\right) \,,  \label{3b.154}
\end{equation}%
For illustration, we consider the familiar second-order differential
expression $\check{H}$ (\ref{3b.10}) on a segment $\left[ 0,l\right] $ with
an integrable potential $V$ which implies that the both ends are regular.
This includes the case of a free particle where $V=0$ and $\check{H}=\check{H%
}_{0}$ (\ref{3b.48a}). The s.a. boundary conditions in this case are given by%
\begin{equation*}
\left( 
\begin{array}{c}
\psi \left( l\right) -i\tau \psi ^{\prime }\left( l\right) \\ 
\psi \left( 0\right) +i\tau \psi ^{\prime }\left( 0\right)%
\end{array}%
\right) =U\left( 
\begin{array}{c}
\psi \left( l\right) +i\tau \psi ^{\prime }\left( l\right) \\ 
\psi \left( 0\right) -i\tau \psi ^{\prime }\left( 0\right)%
\end{array}%
\right) \,,
\end{equation*}%
where $U$ is an $2\times 2$ unitary matrix, to our knowledge, they were
first given in \cite{BonFaV01} with $\tau =l.$

Choosing $U=I$, we obtain s.a. boundary conditions (\ref{3b.89}): $\psi
^{\prime }\left( 0\right) =\psi ^{\prime }\left( l\right) =0\,.$

With $U=-I$, we reproduce s.a. boundary conditions (\ref{3b.88}): $\psi
\left( 0\right) =\psi \left( l\right) =0\,.$

If%
\begin{equation*}
U=\left( 
\begin{array}{cc}
e^{i\theta } & 0 \\ 
0 & e^{-i\vartheta }%
\end{array}%
\right) \,,\;-\pi \leq \theta ,\;\vartheta \leq \pi \,,
\end{equation*}%
we obtain splitted s.a. boundary condition (\ref{3b.117}): $\psi ^{\prime
}\left( 0\right) =\lambda \psi \left( 0\right) \,,\;\psi ^{\prime }\left(
l\right) =\lambda \psi \left( l\right) \,,$ where $\lambda =-\frac{1}{\tau }%
\tan \frac{\vartheta }{2}\,,\;\mu =-\frac{1}{\tau }\tan \frac{\theta }{2}%
\,,\;-\infty \leq \lambda \,,\;\mu \leq \infty \,,\;-\infty \backsim \infty
\,.$

Choosing%
\begin{equation*}
U=\left( 
\begin{array}{cc}
0 & e^{i\vartheta } \\ 
e^{-i\vartheta } & 0%
\end{array}%
\right) \,,
\end{equation*}%
we obtain modified periodic s.a. boundary conditions (\ref{3b.118}): $\psi
\left( l\right) =e^{i\vartheta }\psi \left( 0\right) \,,\;\psi ^{\prime
}\left( l\right) =e^{i\vartheta }\psi ^{\prime }\left( 0\right) \,.$

Finally, taking $\tau =l/\pi $ and%
\begin{equation*}
U=-\frac{1}{\cosh \mathrm{\,}\pi }\left( 
\begin{array}{cc}
i\sinh \mathrm{\,}\pi & 1 \\ 
1 & i\sinh \mathrm{\,}\pi%
\end{array}%
\right) \,,
\end{equation*}%
we reproduce ``exotic'' s.a. boundary conditions (\ref{3b.87}).

Another case where the alternative method is efficient is the case of an
even differential expression $\check{f}$ of order $n$ with one regular end,
let it be $a$, and one singular end, $b$, if the boundary forms vanish
identically at the singular end, in particular, $\left[ \psi _{\ast },\psi
_{\ast }\right] \left( b\right) \equiv 0$. In this case, the quadratic
asymmetry form $\Delta _{\ast }$ is, see (\ref{e.7}) with $\Psi _{\tau \pm
}\left( b\right) =0,$%
\begin{equation}
\Delta _{\ast }\left( \psi _{\ast }\right) =2i\kappa \left[ \Psi _{\tau
-}^{+}\left( a\right) \Psi _{\tau -}\left( a\right) -\Psi _{\tau
+}^{+}\left( a\right) \Psi _{\tau +}\left( a\right) \right] \,,
\label{3b.156}
\end{equation}%
where the $n/2$-columns of boundary values are given by (\ref{3b.145})--(\ref%
{3b.148}). It follows from (\ref{3b.156}) that s.a. boundary conditions
defining a s.a. operator $\hat{f}_{U}$ associated with $\check{f}$ are given
by\footnote{%
Of course, we can interchange $\Psi _{\tau _{-}}\left( a\right) $ and $\Psi
_{\tau +}\left( a\right) $ in (\ref{3b.157}). We can also repeat the remark
after formula (\ref{e.8}) concerning the new meaning of the symbol $\hat{f}%
_{U}$.}%
\begin{equation}
\Psi _{\tau -}\left( a\right) =U\Psi _{\tau +}\left( a\right) \,,
\label{3b.157}
\end{equation}%
where $U$ is an unitary matrix, $U\in U\left( n/2\right) $. When $U$ ranges
over all group $U\left( n/2\right) $, we cover the whole $\left( n/2\right)
^{2}$-parameter $U\left( n/2\right) $-family $\left\{ \hat{f}_{U}\right\} $
of associated s.a. operators in the case under consideration.

We know from the above considerations, see Lemma \ref{l3b.6} that the
sufficient condition for vanishing the boundary forms at the singular end is
that the deficiency indices of the initial associated symmetric operator $%
\hat{f}^{\left( 0\right) }$ be minimum, $\left( n/2,n/2\right) $. But
formula (\ref{3b.156}) explicitly shows that conversely, if the boundary
form $\left[ \psi _{\ast },\psi _{\ast }\right] \left( b\right) $ vanishes
identically, the signature of the Hermitian form $\frac{1}{i}\Delta _{\ast }$
is%
\begin{equation}
\mathrm{sign\,}\frac{1}{i}\Delta _{\ast }=\left( n/2,n/2\right) \,.
\label{3b.158}
\end{equation}%
which means that the deficiencies indices of $\hat{f}^{\left( 0\right) }$
are $\left( n/2,n/2\right) $. In other words, we can state that for an even
s.a. differential expression $\check{f}$ of order $n$ with one regular and
one singular end, the deficiency indices of the associated initial symmetric
operator $\hat{f}^{\left( 0\right) }$ are $\left( n/2,n/2\right) $ iff the
boundary forms at the singular end identically vanish. Therefore, for such
differential expressions, the description of the associated s.a.
differential operators by s.a. boundary conditions (\ref{3b.157}) is in
complete agreement with the previous description given by Theorem \ref{t3b.4}
and Theorem \ref{t3b.8}. We only note that the application of Theorem \ref%
{t3b.4} requires evaluating the deficient subspaces and that the matrix $%
A_{1/2}$ in Theorems \ref{t3b.8} is defined up to the change $%
A_{1/2}\rightarrow A_{1/2}Z$, where $Z$ is a nonsingular matrix, while s.a.
boundary conditions (\ref{3b.157}) avoid evaluating the deficient subspaces
and contain no arbitrariness.

For illustration, we consider the same differential expression $\check{H}$ (%
\ref{3b.10}) on the semiaxis $\left[ 0,\infty \right) $ with a potential $V$
integrable at the origin, such that the left end is regular. We know the two
criteria given by respective (\ref{3b.124}) and (\ref{3b.125}) for the
initial symmetric operator $\hat{H}^{\left( 0\right) }$ to have the
deficiency indices $\left( 1,1\right) $ and, therefore, the boundary form $%
\left[ \psi _{\ast },\psi _{\ast }\right] \left( \infty \right) $ to vanish
identically. In the spirit of the alternative method, we now directly,
without addressing to deficiency indices, show that under either of
conditions (\ref{3b.124}) or (\ref{3b.125}), we have $\left[ \psi _{\ast
},\psi _{\ast }\right] \left( \infty \right) \equiv 0$.

We begin with condition\footnote{%
We have already mentioned that this condition implies the integrability of $%
V\left( x\right) $ at the origin.} (\ref{3b.124}). The corresponding
assertion is based on the observation that under this condition, the
function $x^{-1/2}\psi _{\ast }^{\prime }$ is bounded for $x>a>0$,%
\begin{equation}
\left| x^{-1/2}\psi _{\ast }^{\prime }\right| <C^{\prime }\left( \psi _{\ast
}\right) <\infty \,,\;x>a>0,\;\forall \psi _{\ast }\in D_{\ast }\,.
\label{3b.159}
\end{equation}%
It follows that the function $x^{-1/2}\overline{\psi _{\ast }}\psi _{\ast
}^{\prime }$ is square integrable at infinity as well as $\psi _{\ast }$,
therefore, the function $x^{-1/2}\left[ \psi _{\ast },\psi _{\ast }\right]
=x^{-1/2}\left( \overline{\psi _{\ast }^{\prime }}\psi _{\ast }-\overline{%
\psi _{\ast }}\psi _{\ast }^{\prime }\right) $ is also square integrable at
infinity. On the other hand, the finiteness of the boundary form $\left[
\psi _{\ast },\psi _{\ast }\right] \left( \infty \right) $,%
\begin{equation}
\left[ \psi _{\ast },\psi _{\ast }\right] \rightarrow C\left( \psi _{\ast
}\right) ,\;x\rightarrow \infty \,,\;\left| C\left( \psi _{\ast }\right)
\right| <\infty \,,  \label{3b.160}
\end{equation}%
implies $x^{-1/2}\left[ \psi _{\ast },\psi _{\ast }\right] \rightarrow
x^{-1/2}C\left( \psi _{\ast }\right) \,,\;x\rightarrow \infty \,.$ But the
function in l.h.s. is square integrable at infinity, whereas the function in
r.h.s. is not unless $C\left( \psi _{\ast }\right) =0$, which proves that $%
\left[ \psi _{\ast },\psi _{\ast }\right] \left( \infty \right) \equiv 0$.
It remains to prove (\ref{3b.159}).

For this, we recall that $\psi _{\ast }\in D_{\ast }$ implies $\psi _{\ast
},-\psi _{\ast }^{\prime \prime }+V\psi _{\ast }\in L^{2}\left( 0,\infty
\right) $ which in turn implies that the function $\int_{a}^{x}d\xi \left|
\chi _{\ast }\right| ^{2}$, where $\chi _{\ast }=-\psi _{\ast }^{\prime
\prime }+V\psi _{\ast },$ is bounded, $\int_{a}^{x}d\xi \left| \chi _{\ast
}\right| ^{2}<C_{1}\left( \chi _{\ast }\right) <\infty \,,$ therefore,%
\begin{equation}
\left| \int_{a}^{x}d\xi \chi _{\ast }\right| <C_{1}^{1/2}\left( \chi _{\ast
}\right) \sqrt{x-a}\,,\;x>a\,,  \label{3b.161}
\end{equation}%
by the Cauchy--Bounjakowsky inequality. If $V\in L^{2}\left( 0,\infty
\right) $ as well as $\psi _{\ast }$, the function $V\psi _{\ast }$ is
integrable on $\left[ 0,\infty \right) $, and therefore, the function $%
\int_{a}^{x}d\xi V\psi _{\ast }$ is bounded on $\left[ 0,\infty \right) $,%
\begin{equation}
\left| \int_{a}^{x}d\xi V\psi _{\ast }\right| <C_{2}\left( \chi _{\ast
}\right) <\infty \,.  \label{3b.162}
\end{equation}%
Integrating the equality $-\psi _{\ast }^{\prime \prime }+V\psi _{\ast
}=\chi _{\ast }$, we have%
\begin{equation*}
\psi _{\ast }^{\prime }\left( x\right) =\int_{a}^{x}d\xi V\psi _{\ast
}-\int_{a}^{x}d\xi \chi _{\ast }+\psi _{\ast }^{\prime }\left( a\right) \,,
\end{equation*}%
and then using (\ref{3b.161}) and (\ref{3b.162}), we obtain the inequality%
\begin{equation*}
\left| \psi _{\ast }^{\prime }\right| <C_{2}\left( \psi _{\ast }\right)
+C_{1}^{1/2}\left( \psi _{\ast }\right) \sqrt{x-a}+\left| \psi _{\ast
}^{\prime }\left( a\right) \right| \,,\;x>a\,,\;\forall \psi _{\ast }\in
D_{\ast }\,,
\end{equation*}%
which yields (\ref{3b.159}) and proves the assertion.

The important concluding remark is that as the given proof shows, in order
that the boundary form $\left[ \psi _{\ast },\psi _{\ast }\right] \left(
\infty \right) $ vanish identically, condition (\ref{3b.124}) can be
weakened:\ it is sufficient that the potential $V$ be square integrable at
infinity, i.e., $V\in L^{2}\left( a,\infty \right) $ with some $a>0$.

We now turn to condition (\ref{3b.125}). The corresponding assertion is
based on the observation that under this condition, the function $\psi
_{\ast }^{\prime }/x$ is square integrable at infinity as well as $\psi
_{\ast }$,%
\begin{equation}
\int_{a}^{\infty }d\xi \left| \frac{\psi _{\ast }^{\prime }}{\xi }\right|
^{2}<\widetilde{C}^{\prime }\left( \psi _{\ast }\right) <\infty
\,,\;a>0\,,\;\forall \psi _{\ast }\in D_{\ast }\,.  \label{3b.163}
\end{equation}%
It follows that the function $\overline{\psi _{\ast }}\psi _{\ast }^{\prime
}/x$ is integrable at infinity, and, therefore, the function $x^{-1}\left[
\psi _{\ast },\psi _{\ast }\right] =x^{-1}\left( \overline{\psi _{\ast
}^{\prime }}\psi _{\ast }-\overline{\psi _{\ast }}\psi _{\ast }^{\prime
}\right) $ is also integrable at infinity. On the other hand, the finiteness
of the boundary form $\left[ \psi _{\ast },\psi _{\ast }\right] \left(
\infty \right) $, see (\ref{3b.160}), implies that $x^{-1}\left[ \psi _{\ast
},\psi _{\ast }\right] \rightarrow C\left( \psi _{\ast }\right) x^{-1}\,$as\ 
$x\rightarrow \infty .$ But the function in l.h.s. is integrable at
infinity, whereas the function in r.h.s. is not unless $C\left( \psi _{\ast
}\right) =0$, which proves that $\left[ \psi _{\ast },\psi _{\ast }\right]
\left( \infty \right) =0$. It remains to prove (\ref{3b.163}).

The proof is by contradiction. We first make some preliminary estimates, as
in the proof of the previous assertion, based on the conditions $\psi _{\ast
}\,,-\psi _{\ast }^{\prime \prime }+V\psi _{\ast }=\chi _{\ast }\in
L^{2}\left( 0,a\right) \,.$ These conditions imply that $\int_{a}^{x}d\xi
\left| \psi _{\ast }\right| ^{2}<C_{1}\left( \psi _{\ast }\right) <\infty
\,, $ we already used this estimate before, and that%
\begin{eqnarray}
&&\int_{a}^{x}d\xi \frac{\left| \psi _{\ast }\right| }{\xi ^{3}}%
^{2}<C_{3}\left( \psi _{\ast }\right) <\infty \,,\;\int_{a}^{x}d\xi \frac{%
\left| \psi _{\ast }\right| }{\xi ^{4}}^{2}<C_{4}\left( \psi _{\ast }\right)
<\infty \,,\;a>0\,,  \notag \\
&&\left| \int_{a}^{x}d\xi \left( \overline{\chi _{\ast }}\frac{\psi _{\ast }%
}{\xi ^{2}}\psi _{\ast }+\frac{\overline{\psi _{\ast }}}{\xi ^{2}}\chi
_{\ast }\right) \right| <2C_{1}^{1/2}\left( \chi _{\ast }\right)
C_{4}^{1/2}\left( \psi _{\ast }\right) .  \label{3b.164}
\end{eqnarray}

The condition (\ref{3b.125}) means that there exist some $a>0$ such that $%
V\left( x\right) /x^{2}>-K$, $K>0$, and, therefore,%
\begin{equation}
\int_{a}^{x}d\xi \frac{V}{\xi ^{2}}\left| \psi _{\ast }\right|
^{2}>-K\int_{a}^{x}d\xi \left| \psi _{\ast }\right| ^{2}>-KC_{1}\left( \psi
_{\ast }\right) \,.  \label{3b.165}
\end{equation}%
On the other hand, we have%
\begin{equation*}
\overline{\psi _{\ast }}\chi _{\ast }+\psi _{\ast }\overline{\chi _{\ast }}=-%
\frac{d^{2}}{dx^{2}}\left| \psi _{\ast }\right| ^{2}+2\left| \psi _{\ast
}^{\prime }\right| ^{2}+2V\left| \psi _{\ast }\right| ^{2}\,.
\end{equation*}%
Multiplying this equality by $1/x^{2}$ and integrating with integrating the
term $-x^{-2}d^{2}/dx^{2}\left| \psi _{\ast }\right| ^{2}$ by parts, we
obtain that \ \ 
\begin{eqnarray*}
&&\,\frac{1}{x^{2}}\frac{d}{dx}\left| \psi _{\ast }\right|
^{2}=2\int_{a}^{x}d\xi \left| \frac{\psi _{\ast }^{\prime }}{\xi }\right|
^{2}+2\int_{a}^{x}d\xi \frac{V}{\xi ^{2}}\left| \psi _{\ast }\right|
^{2}-6\int_{a}^{x}d\xi \frac{\left| \psi _{\ast }\right| ^{2}}{\xi ^{4}}%
-\int_{a}^{x}d\xi \left( \overline{\chi _{\ast }}\psi _{\ast }+\overline{%
\psi _{\ast }}\chi _{\ast }\right) \\
&&\,-2\frac{\left| \psi _{\ast }\right| ^{2}}{x^{3}}+C_{5}\left( \psi _{\ast
}\right) \,,\;C_{5}\left( \psi _{\ast }\right) =\left. \left( \frac{1}{x^{2}}%
\frac{d}{dx}\left| \psi _{\ast }\right| ^{2}+\frac{2}{x^{3}}\left| \psi
_{\ast }\right| ^{2}\right) \right| _{x=a}\,.
\end{eqnarray*}%
In view of (\ref{3b.164}) and (\ref{3b.165}), this yields the inequality%
\begin{equation*}
\frac{d}{dx}\left| \psi _{\ast }\right| ^{2}>x^{2}\left( 2\int_{a}^{x}d\xi
\left| \frac{\psi _{\ast }^{\prime }}{\xi }\right| ^{2}-C_{6}\left( \psi
_{\ast }\right) \right) -2\frac{\left| \psi _{\ast }\right| ^{2}}{x^{3}}\,,
\end{equation*}%
where $C_{6}=2KC_{1}\left( \psi _{\ast }\right) +6C_{4}\left( \psi _{\ast
}\right) +2C_{1}^{1/2}\left( \chi _{\ast }\right) C_{4}^{1/2}\left( \psi
_{\ast }\right) -C_{5}\left( \psi _{\ast }\right) $. Let now the integral $%
I\left( x\right) =\int_{a}^{x}d\xi \left| \frac{\psi _{\ast }^{\prime }}{\xi 
}\right| ^{2}$ diverge as $x\rightarrow \infty $. Then for sufficiently
large $x$, $x>b>a$, we have $2I(x)-C_{6}(\psi _{\ast })>C_{7}\left( \psi
_{\ast }\right) >0$, and, therefore, we obtain the inequality%
\begin{equation*}
\frac{d}{dx}\left| \psi _{\ast }\right| ^{2}>x^{2}C_{7}\left( \psi _{\ast
}\right) -2\frac{\left| \psi _{\ast }\right| ^{2}}{x^{3}}\,.
\end{equation*}%
Again, integrating this inequality and taking (\ref{3b.164}) into account,
we find $\left| \psi _{\ast }\right| ^{2}>C_{7}\left( \psi _{\ast }\right)
x^{3}/3-C_{8}\left( \psi _{\ast }\right) \,,$ where $C_{8}\left( \psi _{\ast
}\right) =2C_{3}\left( \psi _{\ast }\right) +C_{7}\left( \psi _{\ast
}\right) b^{3}/3-\left| \psi _{\ast }\right| ^{2}\left( b\right) ,\ $whence
it follows that $\left| \psi _{\ast }\right| ^{2}\rightarrow \infty $ as $%
x\rightarrow \infty $, which contradicts the square integrability of $\psi
_{\ast }$ at infinity. This contradiction proves that the function $\psi
_{\ast }^{\prime }/x$ is square integrable at infinity, i.e., (\ref{3b.163})
holds, and thus proves the assertion. We should not forget that because the
sesquilinear and quadratic forms define each other, the vanishing of the
boundary form $\left[ \psi _{\ast },\psi _{\ast }\right] $ implies the
vanishing of the sesquilinear boundary form $\left[ \chi _{\ast },\psi
_{\ast }\right] $, and vice versa.

The proved criteria for vanishing the boundary forms at infinity allows
formulating the assertion that the s.a. operators associated with s.a.
differential expression $\check{H}$ (\ref{3b.10}) on the semiaxis $\left[
0,\infty \right) $ with a potential $V$ integrable at the origin and
satisfying either the condition that it is also square integrable at
infinity or the condition that $V\left( x\right) >-Kx^{2}$, $K>0$, for
sufficiently large $x$ are specified by s.a. boundary conditions given by%
\begin{equation}
\psi \left( 0\right) -i\tau \psi ^{\prime }\left( 0\right) =e^{i\vartheta } 
\left[ \psi \left( 0\right) +i\tau \psi ^{\prime }\left( 0\right) \right]
\,,\;-\pi \leq \vartheta \leq \pi \,,  \label{3b.165a}
\end{equation}%
which is equivalent to%
\begin{equation*}
\psi ^{\prime }\left( 0\right) =\lambda \psi \left( 0\right) \,,\;\lambda =-%
\frac{1}{\tau }\tan \frac{\vartheta }{2},\;-\infty \leq \lambda \leq \infty
\,,
\end{equation*}%
the both $\lambda =\pm \infty $ yield the same s.a. boundary condition $\psi
\left( 0\right) =0$: the whole family $\left\{ \hat{H}_{\lambda }\right\} $
of s.a. operators associated with $\check{H}$ is not the real axis, but a
circle. We thus reproduce the previous result given by (\ref{3b.126}).

The above criteria are evidently extended to the case of the same
differential expression $\check{H}$ (\ref{3b.10}), but now on the whole real
axis $\mathbb{R}^{1}=\left( -\infty ,\infty \right) $, providing the
vanishing of the boundary forms on the both infinities. This allows
immediately formulating a similar assertion for this case: if a potential $%
V\left( x\right) $ is locally integrable and satisfies the two alternative
conditions that $V$ is either square integrable at minus infinity or $%
V\left( x\right) >-K_{-}x^{2}$, $K_{-}>0$, for sufficiently large negative $%
x $ and $V$ is either square integrable at plus infinity or $V\left(
x\right) >-K_{+}x^{2}$, $K_{+}>0$, for sufficiently large $x$ (generally $%
K_{-}$ and $K_{+}$ may be different), then there is a unique s.a. operator $%
\hat{H}$ associated with $\check{H}$, it is given by the closure of the
initial symmetric operator $\hat{H}^{\left( 0\right) }$ defined on the
natural domain, $\hat{H}=\overline{\hat{H}^{\left( 0\right) }}=\hat{H}^{\ast
}$.

The case of a free particle where $V=0$ and $\check{H}=\check{H}_{0}$
certainly falls under the above conditions, such that $\hat{H}_{0}$ defined
on natural domain (\ref{3b.49}) is really s.a. as we said in advance in
subsec. 3.4. It may be also useful to mention that in this case we can
strengthen the estimates on the asymptotic behavior of functions $\psi
_{\ast }\left( x\right) \in D_{0\ast }$, namely, $\psi _{\ast }\left(
x\right) ,\psi _{\ast }^{\prime }\left( x\right) \rightarrow 0$ as $\left|
x\right| \rightarrow \infty $. For this, it is sufficient to prove that $%
\psi _{\ast }^{\prime }$ is square integrable both at plus and minus
infinity, which means that $\psi _{\ast }^{\prime }\in L^{2}\left( -\infty
,\infty \right) $ as well as $\psi _{\ast }$ and $\psi _{\ast }^{\prime
\prime }$. It then remains to refer to the assertion that we obtained when
considering the case of the differential expression $\check{p}$ (\ref{3b.9}%
):\ if $\psi _{\ast },\psi _{\ast }^{\prime }$ are square integrable at
infinity, plus or minus, this implies that $\psi _{\ast }$ vanishes at
infinity, and to apply this assertion to the respective pairs $\psi _{\ast
},\psi _{\ast }^{\prime }$ and $\psi _{\ast }^{\prime },\psi _{\ast
}^{\prime \prime }$. We only prove that if $\psi _{\ast }\in D_{0\ast }$, $%
\psi _{\ast }^{\prime }$ is square integrable at $+\infty $; the proof for $%
-\infty $ is completely similar. The proof is by contradiction. The
condition $\psi _{\ast }\in D_{0\ast }$ implies that $\psi _{\ast }$ and $%
\psi _{\ast }^{\prime \prime }$ are square integrable at infinity;
therefore, the integral $\int_{a}^{x}d\xi \left( \overline{\psi _{\ast }}%
\psi _{\ast }^{\prime \prime }+\overline{\psi _{\ast }^{\prime \prime }}\psi
_{\ast }\right) $ is convergent as $x\rightarrow \infty $,%
\begin{equation*}
\int_{a}^{x}d\xi \left( \overline{\psi _{\ast }}\psi _{\ast }^{\prime \prime
}+\overline{\psi _{\ast }^{\prime \prime }}\psi _{\ast }\right) \rightarrow
C\left( \psi _{\ast }\right) \,,\;x\rightarrow \infty \,,\;\left| C\left(
\psi _{\ast }\right) \right| <\infty \,.
\end{equation*}%
On the other hand%
\begin{equation*}
\int_{a}^{x}d\xi \left( \overline{\psi _{\ast }}\psi _{\ast }^{\prime \prime
}+\overline{\psi _{\ast }^{\prime \prime }}\psi _{\ast }\right) =\frac{d}{dx}%
\left| \psi _{\ast }\right| ^{2}-2\int_{a}^{x}d\xi \left| \psi _{\ast
}^{\prime }\right| ^{2}-\left. \frac{d}{dx}\left| \psi _{\ast }\right|
^{2}\right| _{x=a}\,.
\end{equation*}%
and if $\int_{a}^{x}d\xi \left| \psi _{\ast }^{\prime }\right|
^{2}\rightarrow \infty $ as $x\rightarrow \infty $, then $\frac{d}{dx}\left|
\psi _{\ast }\right| ^{2}\rightarrow \infty $ as $x\rightarrow \infty $
also, and therefore, $\left| \psi _{\ast }\right| ^{2}\rightarrow \infty $
as $x\rightarrow \infty $, which contradicts the square intergability of $%
\psi _{\ast }$ and proves the required.

We have thus completely paid our debt since subsec. 3.4.

We again note that in the above consideration related to $\check{H}$ on the
whole axis, we escape evaluating the deficient subspaces and deficient
indices, but, in passing, we obtain that the deficiency indices of $\hat{H}%
^{\left( 0\right) }$ are $\left( 0,0\right) $, and therefore, $\hat{H}%
^{\left( 0\right) }$ is essentially s.a..

It remains to demonstrate how the alternative method can work in the case of
singular ends. For illustration, we take the differential expression $\check{%
H}=-d^{2}/dx^{2}-\alpha /x^{2}$ on the positive semiaxis $\left[ 0,\infty
\right) $ with the dimensionless coupling constant $\alpha >1/4$. This
differential expression can be identified with the radial Hamiltonian $%
\check{H}_{l}$ (\ref{3b.50}), (\ref{3b.51}) for a three-dimensional particle
in the field of a strongly attractive central potential $V=-\frac{\alpha }{%
r^{2}}$ with $l=0$ (the $s$-wave) or $V=-\frac{\alpha +l\left( l+1\right) }{%
r^{2}}$ with $l\neq 0$ (the higher waves); such a potential yields a
phenomenon known as the ``fall to a center''. Historically, this was the
first case where the standard textbook approach did not allow constructing
scattering states and even raised the question on the applicability of
quantum mechanics to strongly singular potentials \cite{MotMa33}.

The potential $V=-\frac{\alpha }{x^{2}}$ satisfies the both criteria for
vanishing the boundary form $\left[ \psi _{\ast },\psi _{\ast }\right]
\left( \infty \right) $, and the problem of constructing s.a. operators
associated with $\check{H}$ reduces to the problem of evaluating the
boundary form $\left[ \psi _{\ast },\psi _{\ast }\right] \left( 0\right) $.
It is solved by the following arguments that can be extended to another
cases, and maybe, to the general case, the idea was already stated above, in
the consideration related to formula (\ref{3b.48}). By the definition of the
domain $D_{\ast }$, the functions $\psi _{\ast }$ and $\chi _{\ast }=-\psi
_{\ast }^{\prime \prime }-\alpha /x^{2}\psi _{\ast }$ belong to $L^{2}\left(
0,\infty \right) $. This means that $\psi _{\ast }\in D_{\ast }$ can be
considered as a square-integrable solution of the inhomogenous differential
equation%
\begin{equation}
-\psi _{\ast }^{\prime \prime }-\alpha /x^{2}\psi _{\ast }=\chi _{\ast }
\label{3b.167}
\end{equation}%
with a square-integrable, and therefore, locally integrable, inhomogenous
term $\chi _{\ast }$. Therefore, as any solution of (\ref{3b.167}), the
function $\psi _{\ast }$ can be represented as%
\begin{equation}
\psi _{\ast }=c_{+}u_{+}+c_{-}u_{-}-\frac{1}{2i\mu _{0}\varkappa }\left[
u_{+}\int_{0}^{x}d\xi u_{-}\left( \xi \right) \chi _{\ast }\left( \xi
\right) -u_{-}\int_{0}^{x}d\xi u_{+}\left( \xi \right) \chi _{\ast }\left(
\xi \right) \right]  \label{3b.168}
\end{equation}%
in terms of the two linearly independent solutions $u_{\pm }=\left( \mu
_{0}x\right) ^{1/2\pm i\varkappa }$ of the homogenous equation $-u_{\pm
}^{\prime \prime }-\alpha /x^{2}u_{\pm }=0$, where $\varkappa =\sqrt{\alpha
-1/4}>0$ and $\mu _{0}$ is an arbitrary, but fixed, dimensional parameter of
dimension of inverse length introduced by dimensional considerations, the
factor -$1/2i\mu _{0}\varkappa $ is the inverse Wronskian of the solutions $%
u_{+}$ and $u_{-}$, and $c_{\pm }$ are some constants.

Representation (\ref{3b.168}) allows easily estimating the asymptotic
behavior of $\psi _{\ast }$ as $x\rightarrow 0$. Using the
Cauchy--Bounjakowsky inequality in estimating the integral term in (\ref%
{3b.168}), we obtain that the asymptotic behavior of $\psi _{\ast }^{\prime
} $ and $\psi _{\ast }$ near the origin is given by%
\begin{eqnarray}
\psi _{\ast } &=&c_{+}u_{+}+c_{-}u_{-}+\left( \mu _{0}x\right)
^{3/2}\varepsilon \left( x\right) \,,  \notag \\
\psi _{\ast }^{\prime } &=&\left( \frac{1}{2}+i\varkappa \right) \mu
_{0}c_{+}u_{+}^{-1}+\left( \frac{1}{2}-i\varkappa \right) \mu
_{0}c_{-}u_{-}^{-1}+\left( \mu _{0}x\right) ^{1/2}\widetilde{\varepsilon }%
\left( x\right) \,,  \label{3b.169}
\end{eqnarray}%
where $\varepsilon \left( x\right) $, $\widetilde{\varepsilon }\left(
x\right) \backsim \int_{0}^{x}d\xi \left| \chi _{\ast }\right|
^{2}\rightarrow 0$ as $x\rightarrow 0$. We note that it is the equally
vanishing of the solutions $u_{\pm }$ of the homogenous equation at the
origin that caused difficulties in the choice of an acceptable scattering
state. Formulas (\ref{3b.169}) yield $\left[ \psi _{\ast },\psi _{\ast }%
\right] \left( 0\right) =2i\mu _{0}\varkappa \left( \left| c_{+}\right|
^{2}-\left| c_{-}\right| ^{2}\right) \,,$ whence it immediately follows the
s.a. boundary conditions $c_{-}=c_{+}e^{i\vartheta }\,,\;0\leq \vartheta
\leq 2\pi \,,$ or in the unfolded form,%
\begin{equation}
\psi =c\left( \mu _{0}x\right) ^{1/2}\left[ \left( \mu _{0}x\right)
^{i\varkappa }+e^{i\vartheta }\left( \mu _{0}x\right) ^{-i\varkappa }\right]
+\left( \mu _{0}x\right) ^{3/2}\varepsilon \left( x\right) \,,
\label{3b.172}
\end{equation}%
which have the form of asymptotic boundary conditions.

This asymptotic s.a. boundary conditions can be rewritten as%
\begin{equation}
\psi =c\left( \mu _{0}x\right) ^{1/2}\cos \left[ \varkappa \ln \left( \mu
_{0}x\right) -\vartheta /2\right] +\left( \mu _{0}x\right) ^{3/2}\varepsilon
\left( x\right) \,,  \label{3b.173}
\end{equation}%
then the extension parameter $\vartheta $ can be treated as the phase of the
scattering wave at the origin, or as%
\begin{equation}
\psi =c\left( \mu _{0}x\right) ^{1/2}\left[ \left( \mu x\right) ^{i\varkappa
}+\left( \mu x\right) ^{-i\varkappa }\right] +\left( \mu _{0}x\right)
^{3/2}\varepsilon \left( x\right) \,,  \label{3b.174}
\end{equation}%
where the dimensional parameter $\mu =\mu _{0}e^{-\vartheta /2\varkappa }$, $%
\mu _{0}e^{-\pi /\varkappa }\leq \mu \leq \mu _{0}$, plays the role of the
extension parameter and manifests a ``dimensional transmutation'' and also,
as can be shown, the breaking of a ``naive'' scale symmetry of the system:\ $%
x\rightarrow x/l\Longrightarrow \hat{H}\rightarrow l^{2}\hat{H}$.$\ $We also
note that by passing, we obtain that the deficiency indices of $\hat{H}%
^{\left( 0\right) }$ are $\left( 1,1\right) .$

The conclusion is that in the case under consideration, we have a
one-parameter $U\left( 1\right) $-family $\left\{ \hat{H}_{\vartheta
}\right\} =\left\{ \hat{H}_{\mu }\right\} $ of s.a. Hamiltonians associated
with $\check{H}$, these are parameterized either by the angle $\vartheta $
or the dimensional parameter $\mu $ and are specified by asymptotic boundary
conditions (\ref{3b.172}), or (\ref{3b.173}), or (\ref{3b.174}). The
parameters $\vartheta $ or $\mu $ enter the theory as additional parameters
specifying the corresponding different quantum mechanical systems.

One of the physical consequences of this conclusion for three-dimensional
system is that we should realize that if we describe interaction in terms of
strongly attractive central potentials, a complete description requires
additional specification in terms of new parameters that mathematically
reveal itself as extension parameters.

-----------------------------------------------------------------------------------------------------------------------

\begin{acknowledgement}
Gitman is grateful to the Brazilian foundations FAPESP and CNPq for
permanent support; Voronov thanks FAPESP for support during his stay in
Brazil; Tyutin thanks RFBR 05-02-17217 and LSS-1578.2003.2 for partial
support. Voronov also thanks RFBR (05-02-17451) and LSS-1578.2003.2.
\end{acknowledgement}

\end{document}